\documentclass[hidelinks,11pt]{article}
\usepackage{mymacros}
\usepackage{hyperref}
\usepackage[normalem]{ulem}
\usepackage{enumitem}
\usepackage{etoolbox}

\usepackage{orcidlink}

\usepackage{graphicx}
\usepackage{subcaption}

\usepackage{bbm} 

\DeclareMathAlphabet{\mathpzc}{OT1}{pzc}{m}{it}

\definecolor{darkmagenta}{rgb}{0.55, 0.0, 0.55}

\newcommand{\dkm}{\color{darkmagenta}}

\newcommand{\1}{\mathbbm{1}}

\newcommand{\pt}{\operatorname{pt}}
\newcommand{\one}{{\1}}
\newcommand{\eps}{\varepsilon}
\usepackage[titles]{tocloft}
\renewcommand{\bar}[1]{\overline{#1}}

\newcommand{\rg}{\operatorname{RG}}

\newcommand{\Loc}{\operatorname{Loc}}
\newcommand{\Eplus}{\E_+}

\newcommand{\ellplus}{\ell_{+}}
\newcommand{\s}{s}

\newcommand{\Fcal}{\mathcal{F}}
\newcommand{\uf}{f}

\newcommand{\CleN}{C_{\le N}}

\newcommand{\bbb}{B}
\newcommand{\hh}{\mathpzc{h}}
\newcommand{\lp}{\mathpzc{l}}
\newcommand{\Fi}{\operatorname{Fi}}
\newcommand{\rexp}{\hat{\theta}}
\newcommand{\domRG}{\mathbb{D}}
\newcommand{\Wkappa}{\cW^{\kappa}}
\newcommand{\Gj}{G}
\newcommand{\Gplus}{G_+}
\newcommand{\blocks}{\cB}
\newcommand{\blocksplus}{\cB_+}
\newcommand{\hf}{\mathfrak{h}}
\newcommand{\free}{{\rm F}}
\newcommand{\per}{{\rm P}}
\newcommand{\II}{\mathbb{I}}
\newcommand{\A}{\mathbb{A}}
\newcommand{\ka}{a}
\newcommand{\kaa}{\mathfrak{a}}
\newcommand{\kb}{\mathfrak{b}}

\newcommand{\kp}{\mathfrak{p}}

\newcommand{\qLap}{q}
\newcommand{\scale}{\rho}

\newcommand{\cst}{c^{\rm st}}
\renewcommand{\geq}{\ge}
\renewcommand{\leq}{\le}

\newcommand{\bscale}{\mathfrak{b}}
\newcommand{\pp}{\mathpzc{p}}

\newcommand{\constF}{c^{\rm F}}

\newcommand{\assumPhi}{(\textbf{A}_{\Phi})}
\makeatletter
\newcommand{\customlabel}[2]{%
   \protected@write \@auxout {}{\string \newlabel {#1}{{#2}{\thepage}{#2}{#1}{}} }%
   \hypertarget{#1}{#2}
}
\makeatother

\title{Boundary conditions and universal finite-size scaling \\
for the hierarchical $|\varphi|^4$ model in dimensions 4 and higher}

\author{
  Emmanuel Michta\,\orcidlink{0000-0001-7222-0422}\footnote{Department of Mathematics,
     University of British Columbia,
     Vancouver BC, Canada V6T 1Z2.
      {\tt michta@math.ubc.ca}.}
  \and
  Jiwoon Park\,\orcidlink{0000-0002-1159-2676}\footnote{Department of Pure Mathematics and
   Mathematical Statistics, University of Cambridge, Centre for
   Mathematical Sciences, Wilberforce Road, Cambridge, CB3 0WB, UK.
     {\tt jp711@cantab.ac.uk}.}
  \and
  Gordon Slade\,\orcidlink{0000-0001-9389-9497}\footnote{Department of Mathematics,
     University of British Columbia,
     Vancouver BC, Canada V6T 1Z2.   {\tt slade@math.ubc.ca}.}
}

\begin{document}

\date{}  
\vspace{-5ex}

\maketitle

\begin{abstract}
We analyse and clarify
the finite-size scaling of the weakly-coupled
hierarchical $n$-component $|\varphi|^4$ model for all integers $n \ge 1$
in all dimensions $d\ge 4$,
for both free and periodic boundary conditions.
For $d>4$, we prove that for a volume of size $R^{d}$ with periodic boundary
conditions the infinite-volume critical point is an effective finite-volume
critical point, whereas for free boundary conditions the effective
critical point is shifted smaller by an amount of order $R^{-2}$.
For both boundary conditions, the   average field has the same
non-Gaussian limit within a critical
window of width $R^{-d/2}$ around the effective critical point,
and in that window we compute
the universal scaling profile for the susceptibility.
In contrast, and again for both boundary conditions, the average field
has a massive Gaussian limit when above the effective critical point by an
amount $R^{-2}$.  In particular, at the infinite-volume
critical point the susceptibility scales as $R^{d/2}$ for periodic boundary
conditions and as $R^{2}$ for free boundary conditions.
We identify a mass
generation mechanism for free boundary conditions that is responsible for
this distinction and which we believe has wider validity, in particular to
Euclidean (non-hierarchical) models on $\mathbb{Z}^d$ in dimensions $d \ge 4$.
For $d=4$ we prove a similar picture with logarithmic corrections.
Our analysis is based on the rigorous renormalisation group
method of Bauerschmidt, Brydges and Slade,
which we improve and extend.
\end{abstract}

\setcounter{tocdepth}{1}
\tableofcontents

\section{Introduction and main results}
\label{sec:intro_main_results}
\subsection{Critical behaviour of the \texorpdfstring{$|\varphi|^4$}{phi4} model}

The $\varphi^4$ model on the Euclidean lattice $\Z^d$ is a close relative of the Ising model
and has been studied for
decades as one of the most fundamental
spin models in statistical mechanics and Euclidean quantum field theory \cite{GJ87,FFS92}.
Given $n \in \N$, $g>0$, $\nu \in \R$, a finite set $\Lambda \subset \Z^d$, and a spin field
$\varphi : \Lambda \to \R^n$, the finite-volume $|\varphi|^4$ \emph{Hamiltonian} is
\begin{align}
    H (\varphi) = \frac{1}{2} (\varphi,  (-\Delta ) \varphi)
    +
    \sum_{x\in \Lambda}\Big( \frac{1}{4} g |\varphi_x|^4
    + \frac{1}{2} \nu  |\varphi_x|^2   \Big).
	\label{eq:Hamiltonian}
\end{align}
Our interest here is the theory's critical behaviour,
which occurs for $\nu$ at and near a critical value $\nu_c$ which is negative and corresponds
to a specific double-well potential when $n=1$.
The disordered phase is $\nu > \nu_c$
(this would be $\beta < \beta_c$ for the Ising model), while $\nu < \nu_c$
is the ordered phase.

Recent advances
include a proof that the spontaneous magnetisation vanishes at
the critical point in all dimensions $d \ge 3$ \cite{GPPS23},
and a proof of the critical theory's Gaussian nature in the upper critical dimension $d=4$ \cite{AD21} following a long history for dimensions $d \ge 4$
\cite{Aize82,AG83,Froh82} which in particular
established mean-field critical behaviour for $d>4$.
The critical scaling and logarithmic corrections to mean-field behaviour at the upper critical dimension $d=4$
also have a long history via rigorous renormalisation group (RG) analysis
in the case of weak coupling, as we discuss in detail in Section~\ref{sec:Gaussianlimit}.

Finite-size scaling in dimensions $d \ge 4$ has been widely discussed
in the physics literature, both via scaling arguments and numerical simulations
(e.g., \cite{BEHK22,DGGZ22,LB97,LM16,WY14,ZGFDG18}),
and it is desirable to have rigorous results which specify the behaviour definitively.
Our purpose here is to analyse the critical finite-size scaling of the weakly-coupled $n$-component $|\varphi|^4$ model in dimensions
$d \ge 4$, to indicate the
differing effects of free boundary conditions (FBC) vs periodic boundary conditions
(PBC),
to elucidate the universal scaling profile in the vicinity of the \emph{effective}
finite-volume critical point
(sometimes called a \emph{pseudocritical} point),
and to compute the logarithmic corrections present for $d=4$.

Following a long tradition in rigorous RG analysis
going back to Dyson \cite{Dyso69}, we work with
the $d$-dimensional \emph{hierarchical} model rather than the Euclidean lattice $\Z^d$,
which provides for a simpler analysis yet still exhibits behaviour which we believe
to apply exactly also for $\Z^d$.
Among the copious previous work on hierarchical models, we mention \cite{ACG13,GK82,HHW01,BBS-brief,BI03c,BI03d,Wata04,BM87}.
Hierarchical models have also recently attracted attention in
percolation theory, e.g., \cite{Hutc24,Hutc25-PLMS}.
Our approach is based on the rigorous RG method for the
$4$-dimensional hierarchical $|\varphi|^4$ model in \cite{BBS-brief}.
We improve and extend the method of \cite{BBS-brief}
in order to apply it to the finite-size scaling.
We also extend the method to apply to dimensions $d>4$.
No additional complication arises in treating $n$-component fields for general $n \ge 1$  compared to the treatment of $n=1$.
Our RG method is inspired by Wilson's progressive integration over
scales, but with no uncontrolled approximations.

Finite-size scaling has a long history in physics, which is natural given that
laboratory systems are finite by definition.
Numerical simulations of course also involve only finite systems, and finite-size
effects are important for the interpretation of simulation data.
The early history
of finite-size scaling is summarised
in \cite{Card98}, an introduction is provided in \cite[Section~4.4]{Card96},
and an extensive account is given in \cite[Chapter~32]{Zinn21}.
In brief, the physics picture is as follows:
\begin{itemize}
\item
The exact critical point for an infinite system is replaced by a critical window
whose size scales with the volume of
$\Lambda$ as $|\Lambda|^{-\theta}$
with
\emph{window} or \emph{rounding exponent} $\theta$, and with a logarithmic
correction at the upper critical dimension.
\item
Within the critical window, the scaling of the susceptibility and other
moments of the average field
is governed by universal profiles.
\item
The location of the critical window is affected by FBC vs PBC.
\end{itemize}

\begin{figure}
\centering
\begin{subfigure}{.5\textwidth}
  \centering
	\includegraphics[scale=0.75]{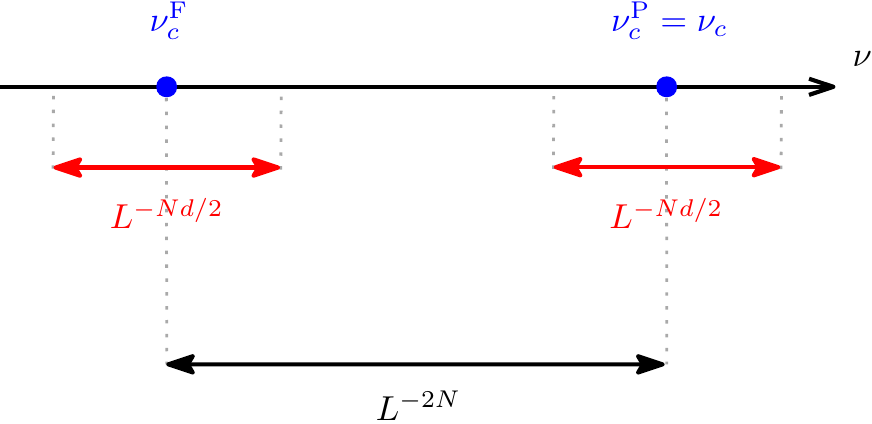}
	\caption{$d>4$.}
\end{subfigure}
~
\begin{subfigure}{.5\textwidth}
  \centering
	\includegraphics[scale=0.75]{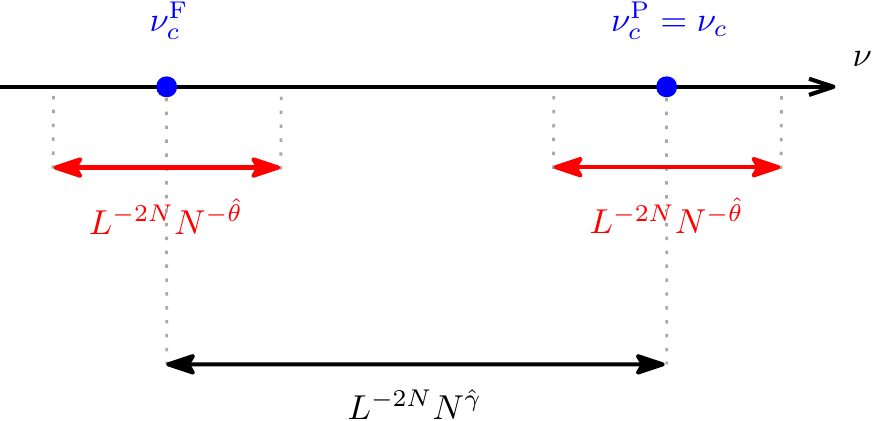}
	\caption{$d=4$.}
\end{subfigure}
\caption{
Effective critical points $\nu_c^\free$, $\nu_c^\per$
and
scaling windows (red) for FBC and PBC for (a) $d>4$ and (b) $d=4$.
The scaling windows for FBC and PBC do not overlap.
At the infinite-volume critical point $\nu_c$,
the   average field has a Gaussian limit with FBC and a non-Gaussian
limit with PBC.}
\label{fig:window}
\end{figure}

Our results give a complete description of the above three points for the
weakly-coupled hierarchical $|\varphi|^4$ model in a finite volume $\Lambda_N$
of size $L^{dN}$,
in dimensions $d \ge 4$, with $L$ fixed
and $N$ large.  A summary is given in Figure~\ref{fig:window}.
We write the average field as $\Phi_N = L^{-dN}\sum_{x\in\Lambda_N}\varphi_x$.

\medskip\noindent
For $d>4$, we prove that for all $n \ge 1$:
\begin{itemize}
\item
For PBC,
there is a critical window of width $L^{-Nd/2}$ containing the infinite-volume
critical point $\nu_c$, within which the rescaled  average field
$L^{Nd/4}\Phi_N$ converges to a non-Gaussian distribution with density proportional to
$e^{-\frac 14 |x|^4-\frac 12 s|x|^2}$.
The variable $s \in (-\infty,\infty)$ parametrises the location
within the critical window.  We explicitly compute the universal profiles of the susceptibility
and other
moments of the average field
within the window.  Within the window,
the susceptibility scales as $L^{Nd/2}$.
\item
For FBC, the effective critical point is not
$\nu_c$ (as it is for PBC) but instead is shifted left to a value $\nu_c^\free$ with $\nu_c-\nu_c^\free \asymp L^{-2N}$.
Around the shifted effective critical point $\nu_c^\free$ there is again a critical window
of width $L^{-Nd/2}$
within which the   average field and the susceptibility
scale \emph{exactly} as they do for PBC, with the same universal profiles.
\item
For both FBC and PBC,  when $\nu$ is above the effective critical point
by an amount $sL^{-2N}$ with $s>0$,  the rescaled average field $L^{N(d-2)/2} \Phi_N$ converges to a Gaussian distribution with density proportional to $e^{-\frac 12 s|x|^2}$.
The susceptibility scales as $L^{2N}$.  For FBC, the infinite-volume
critical point $\nu_c$ lies in this Gaussian range and hence the susceptibility at $\nu_c$
scales as $L^{2N}$, unlike for PBC where the scaling is $L^{Nd/2}$.
\end{itemize}

\medskip\noindent
For $d=4$,
let
\begin{equation}
\label{eq:hatexponents}
    \hat\gamma     = \frac{n+2}{n+8},
    \qquad
    \hat\theta = \frac{4-n}{2(n+8)}  = \frac 12 -\hat\gamma.
\end{equation}
We prove that the $d>4$ picture applies also for $d=4$ but with logarithmic corrections
(polynomial in $N$) for all $n \ge 1$:
\begin{itemize}
\item
For PBC,
there is a critical window of width $L^{-2N}N^{-\hat\theta}$ containing the infinite-volume
critical point $\nu_c$, within which the
rescaled average field $N^{-1/4} L^N \Phi_N$
converges to a non-Gaussian distribution with density proportional to
$e^{-\frac 14 |x|^4-\frac 12 s|x|^2}$.
The universal profile of the susceptibility
and other moments of the average field
within the window are \emph{identical} to those for $d>4$.
Within the window the susceptibility scales as $L^{2N}N^{1/2}$.
\item
For FBC, the effective critical point is shifted
left to a value $\nu_c^\free$ with $\nu_c-\nu_c^\free \asymp L^{-2N}N^{\hat\gamma}$.
Around the shifted effective critical point $\nu_c^\free$ there is  a critical window
of width $L^{-2N}N^{-\hat\theta}$
within which the   average field, the susceptibility and other
moments of the average field
scale \emph{exactly} as they do in the critical window for PBC.
\item
For both FBC and PBC, above their effective critical points
by an amount $sL^{-2N}N^{\hat\gamma}$ with $s>0$,
the rescaled average field $L^N\Phi_N$
converges to a Gaussian distribution with density proportional to
$e^{-\frac 12 s|x|^2}$.
The susceptibility scales as $L^{2N}$.  For FBC, the infinite-volume
critical point $\nu_c$ lies in this Gaussian range and hence the susceptibility at $\nu_c$
scales as $L^{2N}$, unlike for PBC where the scaling is $L^{2N}N^{1/2}$.
\end{itemize}

For dimensions $d>4$ (\emph{above}
the upper critical dimension)  with PBC,
the physics predictions for
the universal profile, the window width, and the size of the susceptibility
in the window are stated in \cite[Section~32.3.1]{Zinn21}.
For both FBC and PBC in dimensions $d>4$, \cite[Table~1]{F-SBKW16}
presents predictions which are consistent with our theorems for the scaling of
the susceptibility, the scaling of the average field, and the shift in the
effective critical point for FBC (see also \cite[p.38]{BEHK22}).
For
$d=4$, the physics predictions for
the logarithmic correction to the susceptibility and the window size are given in \cite[(3.6)]{Kenna04} and \cite[(4.3)]{Kenna04}.

Our results provide a rigorous justification of the physics predictions, for the hierarchial lattice in all dimensions
$d \ge 4$ and for all $n \ge 1$.
Our proof presents a clear mechanism responsible for the shift
in the effective critical point for FBC: the free boundary condition generates an effective
mass in the Hamiltonian which must be compensated by a shift in the value of $\nu$ to attain effective FBC critical behaviour.
In addition, although our results are proved only for the hierarchial model,
they lead to precise conjectures for the behaviour
for Euclidean models on $\Z^d$ for all $d \ge 4$, which we
spell out in Section~\ref{sec:open-problems}, not just for the $n$-component $|\varphi|^4$
model but also for $O(n)$ models (e.g., Ising, $XY$, Heisenberg models)
and self-avoiding walk models ($n=0$).
A summary of further results and conjectures in this direction
is given in \cite{LPS25-universal}.

\medskip\noindent
{\bf Notation.}  We write $f_N \sim g_N$ to denote $\lim_N f_N/g_N =1$,
and use $f_N \asymp g_N$
to denote the existence of $C>0$ such that $C^{-1}g_N \le f_N \le Cg_N$.
We also use the notation $\sim$ and $\asymp$
in limits when $f,g$ depend on a different variable than $N$.

\subsection{The hierarchical Laplacian}
\label{sec:hierdefs}

The hierarchial $|\varphi|^4$ model is defined using the hierarchical Laplacian
in the Hamiltonian \eqref{eq:Hamiltonian}.
We define the hierarchical Laplacian
in this section as the generator for a random walk with FBC or
PBC (see also \cite{Hutc25-PLMS} for discussion of hierarchical BC).
The definitions in this section apply to any dimension $d\ge 1$.
Also, we introduce a parameter $\alpha$ (in \eqref{eq:pxy})
even though we only use the case $\alpha=2$, in order to emphasise its role.

\subsubsection{The hierarchical group}
\label{sec:hierarchical_group_def}

\begin{figure}[h]
\begin{center}
\input{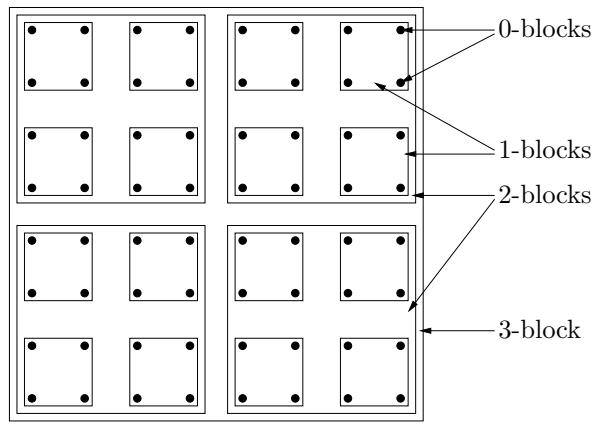}
\end{center}
\caption{\label{fig:reblock}
The block $\Lambda_3$ with its blocks in ${\cal B}_0,{\cal B}_1,{\cal B}_2,{\cal B}_3$ for
the case $d=L=2$.}
\end{figure}

Given integers $L>1$ and $N \ge 1$, let $\Lambda_N$ denote the cube
$\Lambda_N =\{x \in \Z^d: 0 \le x_i < L^N \; \text{for}\; i=1,\ldots,d\}$,
which has \emph{volume} (cardinality) $L^{dN}$.
The thermodynamic limit is the limit $N \to \infty$, for which
$\Lambda_\infty = \{x \in \Z^d: 0 \le x_i \; \text{for}\; i=1,\ldots,d\}$.
Given an integer $0 \le j\le N$, we can partition $\Lambda_N$ into $L^{d(N-j)}$
disjoint \emph{blocks} which are each translates of $\Lambda_j$ and which each contain
$L^{dj}$ vertices.  We denote the set of such
$j$-\emph{blocks} by $\cB_j$.  See Figure~\ref{fig:reblock}.

An equivalent representation of $\Lambda_\infty$ is as follows.
With $\Z_L = \Z/L\Z$ the cyclic group, let
\begin{equation}
    \mathbb{H}_\infty
    =
    \{\tilde x = (\tilde x_1,\tilde x_2,\ldots):
    \text{$\tilde x_i \in \Z_L^d$ with only finitely many nonzero $\tilde x_i$}\}.
\end{equation}
This is an abelian group with coordinatewise addition mod-$L$; we denote
the group addition as $+$.
We define $\mathbb{H}_N$ to be the subgroup of $\mathbb{H}_\infty$
with $\tilde x_i=0$ for $i>N$.
The map $\sigma : \mathbb{H}_\infty \to \Lambda_\infty$ defined
by
\begin{equation}
\label{eq:hier-bijection}
    \sigma(\tilde x) = \sum_{j=1}^\infty L^{j-1} x_j
\end{equation}
is a bijection, where $x_j$ is the representative of $\tilde x_j$ in $\Lambda_1 \subset \Z^d$.
Addition on the right-hand side of \eqref{eq:hier-bijection} is in $\Z^d$.
Note that $\sigma$ restricts to a bijection $\sigma_N:\mathbb{H}_N \to \Lambda_N$.
The bijection induces an addition and a group structure on $\Lambda_N$ via
$x \oplus y =
\sigma_N (\sigma_N^{-1} x + \sigma_N^{-1} y)$ (with $+$ the addition on $\mathbb{H}_N$);
this makes $\sigma_N$ and $\sigma$ into group isomorphisms.

For $x,y\in \Lambda_N$,
we define the \emph{coalescence scale} $j_{xy}$ to be the
smallest $j$ such that $x$ and $y$ lie in the same $j$-block.
In terms of $\mathbb{H}_N$, $j_{xy}$ is the largest coordinate $i$ such
that $(\sigma_N^{-1}x)_i$ differs from $(\sigma_N^{-1}y)_i$, and from this
observation we see that
\begin{equation}
\label{eq:joplus}
    j_{x\oplus z,y\oplus z} = j_{xy} \qquad (x,y,z \in \Lambda_N).
\end{equation}
In particular, with subtraction in $\Lambda_N$ denoted by $\ominus$,
\begin{equation}
\label{eq:jx-y}
    j_{xy} = j_{0,y\ominus x} \qquad (x,y \in \Lambda_N).
\end{equation}

\subsubsection{Hierarchical random walk}

\noindent{\bf Random walk on the infinite hierarchial group.}
Given
$\alpha >0$,
we define a discrete-time
random walk on the group $\Lambda_\infty$ via the transition probabilities $J_{xx}=0$ ($x \in\Lambda_\infty$) and
\begin{equation}
\label{eq:pxy}
    J_{xy}
    = \frac 1z \frac{1}{L^{(d+\alpha)j_{xy}}}
    \qquad
    (x,y\in\Lambda_\infty, \; x \neq y),
\end{equation}
with the constant $z=z(d,L,\alpha)$ chosen so that $\sum_{x \neq 0}J_{0x}=1$.
By \eqref{eq:jx-y}, the transition probabilities are translation invariant
in the sense that $J_{x,y}=J_{0,y\ominus x}$.
Since the number of sites $x \in \Lambda_{\infty}$ with $j_{0x}=k$ is
\begin{equation}
    n_k=L^{dk}-L^{d(k-1)} = L^{dk}(1-L^{-d}),
\end{equation}
the constant $z$ is given by
\begin{align}
\label{eq:Jz}
    z
    & =
    \sum_{x\neq 0}L^{-(d+\alpha)j_{0x}}
    =
    \sum_{k=1}^\infty n_k L^{-(d+\alpha)k} = \frac{1-L^{-d}}{L^\alpha-1}.
\end{align}

\medskip\noindent{\bf Random walk with FBC.}
We define a random walk on $\Lambda_N \times \{\dagger\}$ which is killed when it
lands on the cemetery state $\dagger$.  The killing occurs when a random walk on the infinite
hierarchical lattice $\Lambda_\infty$ exits $\Lambda_N$ for the first time.
This corresponds to FBC.
Explicitly, for $x \in \Lambda_N$,
\begin{align}
    J^{\rm F}_{x\dagger}
    &=
    \sum_{y : j_{xy}>N}J_{xy}
    =
    \frac 1z \sum_{k = N+1}^\infty n_k L^{-(d+\alpha)k}
    =
    L^{-\alpha N},
\end{align}
whereas for $x,y \in \Lambda_N$ we maintain the infinite-volume transition probability
from \eqref{eq:pxy}, namely $J^{\free}_{xx}=0$ and
\begin{align}
    J^\free_{xy}
    &=
    \frac 1z L^{-(d+\alpha)j_{xy}}
    \qquad
    (x,y\in\Lambda_N,\; x \neq y).
\end{align}
This produces a defective
(unnormalised)
 distribution $J^{\rm F}$ on $\Lambda_N$ for which
\begin{equation}
    \sum_{x \in \Lambda_N } J^{\rm F}_{0x} = 1- L^{-\alpha N}.
\end{equation}

\medskip\noindent{\bf Random walk with PBC.}
With $N$ fixed, we define an equivalence relation on $\Lambda_\infty$
by $x \sim y$ if $\tilde x_i = \tilde y_i$ for $i=1,\ldots,N$,
where $\tilde x= \sigma^{-1}x$, $\tilde y = \sigma^{-1}y$.
The quotient space $\Lambda_\infty/ \!\!\sim$ can be identified with
$\Lambda_N$.
The random walk on $\Lambda_\infty$ with transition matrix \eqref{eq:pxy}
projects via the quotient map to a random walk on $\Lambda_N$.
This projection corresponds to PBC.
Explicitly, since for
$k>N$ the block $\Lambda_k$ contains exactly $n_kL^{-dN}$ points with $j_{0x} = k$ that are equivalent to $y\in\Lambda_N$,
for the same normalising constant $z$ as in \eqref{eq:Jz} the quotient walk has transition probabilities
\begin{align}
	J^\per_{xy}
	&
    = \frac{1}{z} L^{-(d+\alpha)j_{xy}}
    + L^{-dN} \sum_{k = N+1}^{\infty} n_k \frac{1}{z} L^{-(d+\alpha)k}
    \nnb
	&= \frac{1}{z} L^{-(d+\alpha)j_{xy}}
    + L^{-(d+\alpha)N}
    \qquad
    (x,y\in\Lambda_N,\; x \neq y)
\end{align}
and $J^\per_{xx} = L^{-(d+\alpha)N}$.

\subsubsection{The hierarchical Laplacian with boundary conditions}
\label{sec:LaplacianBC}

In general, given a transition matrix $J_{xy}$ (defective or not), we define
a Laplacian by $-\tilde{\Delta} = 1-J$ and we let
$-\tilde \Delta$ act on an $n$-component spin field
$\varphi:  \Lambda_N \to \R^n$ component-wise.
Here $1$ denotes the identity matrix.
If $J$ is defective then the Laplacian is massive (i.e., has a spectral gap),
and otherwise it is massless.  For the $|\varphi|^4$ model, the critical behaviour
is not affected by replacing the Laplacian term in the Hamiltonian by a multiple of
the Laplacian.
We use this flexibility in the following definitions
in order to achieve coherence between our Laplacian
with PBC and the hierarchial Laplacian defined in
\cite[Definition 4.1.7]{BBS-brief}.
Since the random walks with FBC and PBC are translation
invariant (with respect to $\oplus$) by definition, the Laplacians they define obey
\begin{equation}
\label{eq:Lap_translation_invar}
    \tilde \Delta_{xy}=\tilde \Delta_{0,y\ominus x} \qquad (x,y\in\Lambda_N).
\end{equation}
In particular,  it suffices to specify the
matrix elements $\tilde \Delta_{0x}$ to define the Laplacians.

For $*={\rm P,F}$, and with $J^*$ the matrix of transition probabilities
with boundary condition $*$, we define
\begin{align}
    -\tilde\Delta^{*} &= 1 - J^{*}.
\end{align}
For $x,y \in \Lambda_N$ we define $Q_{N;xy} =L^{-dN}$.
It follows from the definitions of $J^*$ that
\begin{align}
\label{eq:1-PF}
    1-J^{\rm F}
    &=
    1-J^{\rm P}+ L^{-\alpha N}Q_{N} .
\end{align}
The matrix $-\tilde\Delta^{\rm P}$ obeys
$\sum_{x\in\Lambda_N}(1-J^{\rm P})_{0x}=0$, whereas
$-\tilde\Delta^{\rm F}$ obeys
$\sum_{x\in\Lambda_N}(1-J^{\rm F})_{0x}=L^{-\alpha N}$.

Multiplication of the Laplacian by a scalar amounts to a rescaling of the field.
We do the rescaling in the same way for both boundary conditions, by defining
\begin{align}
\label{eq:Deltastar}
    -\Delta^{*} &= \qLap (-\tilde\Delta^{*}) =  \qLap(1 - J^{*})
\end{align}
with
\begin{equation}
\label{eq:qLapdef}
    \qLap = \frac{1-L^{-d}}{1-L^{-(d+\alpha)}}.
\end{equation}
With this definition, $\Delta^{\per}$
is exactly the Laplacian in \cite[Definition 4.1.7]{BBS-brief}, as we
indicate in Section~\ref{sec:hier}.
By \eqref{eq:1-PF},
\begin{align}
\label{eq:newDeltaF}
    -\Delta^{\rm F}
    & =
    -\Delta^{\rm P}  + \qLap L^{-\alpha N}Q_{N}.
\end{align}

The Green function (inverse Laplacian) on $\Lambda_\infty$ obeys
\begin{equation}
\label{eq:GFFdecay2}
  (-\Delta)^{-1}_{0x}   \asymp |x|^{-(d-\alpha)} \qquad (0<\alpha <d).
\end{equation}
On the right-hand side of \eqref{eq:GFFdecay2}, $|x|$ denotes the Euclidean norm of the
point in $\Z^d$ under the identification of $\Lambda_\infty$ with an
orthant of $\Z^d$ (as in Section~\ref{sec:hierarchical_group_def}).
The proof of \eqref{eq:GFFdecay2}
given in \cite[(4.1.29)]{BBS-brief} for $\alpha=2$ applies also for $0<\alpha<d$.
For $\alpha=2$, the right-hand side of
\eqref{eq:GFFdecay2} is the decay of the Green function for the standard
nearest-neighbour Laplacian on $\Z^d$.  This indicates a close
connection between the hierarchical Laplacian and the Euclidean
Laplacian.

\subsection{The hierarchical \texorpdfstring{$|\varphi|^4$}{phi4} model}
\label{sec:hierarchical_def}

Let $-\Delta^*$ represent either $-\Delta^{\free}$ or $-\Delta^{\per}$
on $\Lambda_N$, so that
either boundary condition can be used in the following definitions.
We \emph{always} take $\alpha=2$ in the definition of the Laplacian.
Given $n \in \N$, $g>0$, $\nu \in \R$, and a spin field
$\varphi : \Lambda_N \to \R^n$, we define the hierarchical \emph{Hamiltonian}
\begin{align}
    H_{\nu,N}^* (\varphi)
    =
    \frac{1}{2} (\varphi,  (-\Delta^* ) \varphi)
    + \sum_{x\in \Lambda_N}\Big(
     \frac{1}{4} g  |\varphi_x|^4
     +
     \frac{1}{2} \nu  |\varphi_x|^2  \Big),
\label{eq:Hamiltonian_definition}
\end{align}
the \emph{partition function}
\begin{equation}
Z_{\nu,N}^* = \int_{(\R^{n})^{\Lambda_N}}  e^{-H_{\nu,N}^* (\varphi)} d\varphi,
\end{equation}
and the associated expectation
\begin{equation}
    \langle F \rangle_{\nu,N}^*
    =
    \frac{1}{Z_{\nu,N}^*}
    \int_{(\R^{n})^{\Lambda_N}}F(\varphi) e^{-H_{\nu,N}^* (\varphi)} d\varphi
    .
    \label{eq:associated_expectation}
\end{equation}
In \eqref{eq:Hamiltonian_definition}, and in what follows, we use the
notation
$(\varphi,\psi) = \sum_{x\in\Lambda_N} \varphi_x \cdot \psi_x$, where
the dot product is for $\R^n$.
Also, $|\varphi_x|=(\varphi_x \cdot \varphi_x)^{1/2}$ denotes the Euclidean norm of $\varphi_x\in \R^n$.
The definitions \eqref{eq:Hamiltonian_definition}--\eqref{eq:associated_expectation}
depend on $g$ but since we regard $g$ as fixed
(and small) in the following, we do not make this dependence explicit.  We will
however be interested in varying $\nu$.
By definition, the distribution of the field $\varphi$ under \eqref{eq:associated_expectation}
is invariant under the map
$(\varphi_x)_{x\in\Lambda_N} \mapsto (R\varphi_x )_{x\in\Lambda_N}$ for any
orthogonal transformation $R \in O(n)$.

Our main objects of study are the distribution of the \emph{average field}
\begin{equation}
\label{eq:totalfield}
    \Phi_{N} = \frac{1}{L^{dN}} \sum_{x \in \Lambda_N}  \varphi_x \in \R^n
\end{equation}
and the expectations
$\langle |\Phi_N|^{2p} \rangle_{\nu,N}$ for $p \in \N$.
An important  special case is $2p=2$, the
finite-volume \emph{susceptibility}:
\begin{align}
\label{eq:chiPhi}
    \chi_N^* (\nu )
    =
    \sum_{x\in \Lambda_N}\langle \varphi_{0}^{(1)}\varphi_{x}^{(1)} \rangle_{\nu,N}^*
    =
    \frac 1n \sum_{x\in \Lambda_N} \langle \varphi_0 \cdot \varphi_x \rangle_{\nu,N}^*
    =
    \frac{|\Lambda_N|}{n} \langle |\Phi_N|^{2} \rangle_{\nu,N}^*,
\end{align}
where $\varphi_x^{(j)}$ denotes the $j^{\rm th}$ component of $\varphi_x \in \R^n$.
The second equality  of \eqref{eq:chiPhi} is a consequence of the $O(n)$ invariance.
The third follows from the fact that, by \eqref{eq:Lap_translation_invar},
under the probability measure given by \eqref{eq:associated_expectation}
the pair $(\varphi_x,  \varphi_y )$ has the same distribution as
$(\varphi_0,   \varphi_{y \ominus x}  )$
for any  $x, y \in \Lambda_N$, so $(\varphi_x,\Phi_N)$ and $(\varphi_y,\Phi_N)$
have the same distribution too.
The susceptibility is more naturally defined using a truncated expectation
in which the square of the infinite-volume
spontaneous magnetisation is subtracted from the expectation
$\langle \varphi_0 \cdot \varphi_x \rangle_{\nu,N}$.
We discuss the distinction between these two definitions in more detail at
the end of Section~\ref{sec:chi}.

\subsection{Main results}
\label{sec:main_results_state}
\label{sec:mr}

Our main results are valid in all dimensions $d \ge 4$ and for all $n \ge 1$.
They are of three types:
\begin{enumerate}
\item
In Theorem~\ref{thm:book_main_theorem_bis},
we identify the infinite-volume
critical point $\nu_c$, and for
$\nu=\nu_c+\eps$ with $\eps >0$ fixed,
we prove a Gaussian limit theorem for the average field,
and compute moments of the average field.
In this limit theorem, the average field is scaled by $L^{Nd/2}$ (Gaussian scaling).
The results apply for both FBC and PBC, and there are logarithmic corrections
for $d=4$.
\item
In Theorem~\ref{thm:mr-nongaussian}, we identify effective finite-volume critical points
$\nu_{c,N}^*$ for FBC and PBC, with $\nu_{c,N}^\per=\nu_c$, and with $\nu_c -\nu_{c,N}^\free \asymp L^{-2N}$ for $d>4$ and $\nu_c-\nu_{c,N}^\free  \asymp N^{\frac{n+2}{n+8}}L^{-2N}$ for $d=4$.
We prove a non-Gaussian limit for the average
field for $\nu=\nu_{c,N}^* + sw_N$ for $s \in (-\infty,\infty)$, where the window
scale $w_N$ is of order $L^{-Nd/2}$ for $d>4$ and of order
$ N^{\frac{4-n}{2(n+8)}} L^{-2N}$
for $d=4$.  The average field is scaled by $L^{Nd/4}$ for $d>4$ and by $N^{-1/4}L^N$ for $d=4$
(non-Gaussian scaling for all $d \ge 4$).
Universal profiles for the moments of the average field
are identified, and there
is a logarithmic correction when $d=4$.  The same non-Gaussian limit applies
for both FBC and PBC, and the FBC and PBC windows do not overlap.
\item
In Theorem~\ref{thm:mr-gaussian},
for FBC and PBC we prove a Gaussian limit for the average
field for $\nu=\nu_{c,N}^* + sv_N$ with $s \in (0,\infty)$, where
$v_N$ is of order $L^{-2N}$ for $d>4$ and of order $N^{\frac{n+2}{n+8}}L^{-2N}$
for $d=4$.
The field is scaled by
$L^{N(d-2)/2}$ for all $d \ge 4$, which is non-Gaussian scaling.
Theorem~\ref{thm:mr-gaussian} probes the vicinity of the critical point on a finer
scale than Theorem~\ref{thm:book_main_theorem_bis}, but on a scale that is less fine
than Theorem~\ref{thm:mr-nongaussian}.
\end{enumerate}
The proofs of the three theorems are given in the bulk of the paper.

\subsubsection{Gaussian limit and critical behaviour}
\label{sec:Gaussianlimit}

The following theorem identifies a \emph{critical point} $\nu_c$, for
large $L$ and for small
$g>0$, at which the infinite-volume hierarchial susceptibility diverges.
Although our proof requires $L$ to be large and $g$ to be small,
we believe the theorem remains true for all $L \ge 2$ and all $g>0$.
The theorem also describes the manner of the susceptibility's divergence.
For $d=4$ the theorem is partly proved in \cite[Theorem~4.2.1]{BBS-brief},
and we adapt that proof here to include dimensions $d>4$.
For the statement of the theorem, for $s>0$
we define a Gaussian measure on $\R^n$, together with its moments, by
\begin{align}
\label{eq:muGaussian}
    d\mu_{n,s} = \frac{e^{-\frac12 s|x|^2} dx}{\int_{\R^n}e^{-\frac12 s|x|^2} dx }, \qquad
    M_{n,k}(s)= \int_{\R^n}|x|^k d\mu_{n,s}
    .
\end{align}

\begin{theorem}
\label{thm:book_main_theorem_bis}
{\rm (Massive Gaussian limit.)}
Let $d\ge 4$,
let $n \in \N$, let $L$ be sufficiently large, and
let $g>0$ be sufficiently small (depending on $L$).
There is a critical value $\nu_c \in \R$ (depending on $d,n,g,L$),
constants $A_d$ (depending on $d,g,n,L$),
and a strictly positive continuously differentiable
function $m_\eps^2$ defined for all $\eps >0$
with
\begin{equation}
\label{eq:renorm_big_mass}
	m^2_\eps \sim
		\begin{cases}
			A_4^{-1}
            \,\eps\,(\log \eps^{-1})^{-\hat \gamma} & (d=4) \\
			A_{d}^{-1}\,\eps & (d>4)
		\end{cases}
\qquad
	\text{as } \eps \downarrow 0,
\end{equation}
with $\hat{\gamma}$ given by \eqref{eq:hatexponents},
such that the following infinite-volume limits exist
for $\eps >0$ and have the values indicated,
independent of the boundary condition $*=\free$ or $* =\per$:
\begin{enumerate}
\item For any $J \in \R^n$,
\begin{equation}
\label{eq:subcrit_gaussian_limit}
	\lim_{N\to \infty}
	\langle e^{J\cdot \Phi_N L^{Nd/2}}\rangle_{\nu_c + \eps,N}^*
	= \int_{\R^n}e^{J\cdot x} d \mu_{n,m^2_\eps}
	.
\end{equation}
\item
For $p \in \N$,
\begin{equation}
\label{eq:subcrit_gaussian_moment}
	\lim_{N \to \infty}
	L^{dNp}\langle  |\Phi_N|^{2p}\rangle_{\nu_c + \eps,N}^*
	= M_{n,2p}(m^2_\eps)
	.
\end{equation}
\end{enumerate}
Also, as $g \downarrow 0$,
the amplitude in \eqref{eq:renorm_big_mass} and the critical point satisfy
\begin{align}
    A_{4} &\sim
    \Big( \frac{(n+8) (1- L^{-d}) g}{\log L^{2}} \Big)^{\hat\gamma},
    \quad A_{d} =1+O(g) \;\;\;(d>4),
    \quad
    \nu_c \sim - (n+2) g
    (-\Delta)^{-1}_{00} ,
    \label{eq:A,B_definition_bis}
\end{align}
with $(-\Delta)^{-1}_{00}$  the diagonal matrix element of the
Green function of the infinite-volume
hierarchical Laplacian.
\end{theorem}

Equation~\eqref{eq:subcrit_gaussian_limit}
states that the total field
$\sum_{x \in \Lambda_N}\varphi_x$,
after a Gaussian rescaling $L^{-Nd/2}$, converges in
distribution under $\langle \,\cdot\, \rangle_{\nu_c +\eps, N}$
to a Gaussian random variable with mean zero
and variance $1/ m^2_\eps$.
The moments in \eqref{eq:subcrit_gaussian_moment} can be explicitly evaluated,
so that for
$p \ge 1$, $\eps >0$,
and $*=\free$ or $* =\per$,
\begin{equation}
\label{eq:subcrit_gaussian_moment_cor}
	\lim_{N \to \infty}
	L^{dNp}\langle  |\Phi_N|^{2p}\rangle_{\nu_c + \eps,N}^*
    = \left(\frac{2}{m^2_\eps}\right)^p
    \frac{\Gamma(\frac{n+2p}{2})}{\Gamma(\frac{n}{2})}
    .
\end{equation}
By the definition of the susceptibility in \eqref{eq:chiPhi},
together with \eqref{eq:renorm_big_mass} and the $p=1$ case
of \eqref{eq:subcrit_gaussian_moment_cor},
the infinite-volume susceptibility $\chi_{\infty}^* = \lim_{N\rightarrow \infty} \chi_N^* (\nu_c   + \epsilon)$ obeys,  independent of the boundary condition,
\begin{align}
\label{eq:chi_N_asymptotic_bis}
    \chi_\infty (\nu_c   + \epsilon)
    = \frac{1}{m_\eps^2}
    \sim
    \begin{cases}
    A_{4}\,
    \epsilon^{-1} \big( \log \epsilon^{-1} \big)^{\hat\gamma} & (d=4)
    \\
    A_{d}\, \epsilon^{-1} & (d>4)
    \end{cases}
    \qquad
	\text{as } \eps \downarrow 0.
\end{align}
For $d=4$ and PBC, this recovers \cite[Theorem 4.2.1]{BBS-brief}.
We emphasise that the equality in \eqref{eq:chi_N_asymptotic_bis} is
not a definition---we compute the infinite-volume limit
of the susceptibility defined by
\eqref{eq:chiPhi} and find that it is equal to a function governed by
the asymptotic formula \eqref{eq:renorm_big_mass}.

Several related results have been proved for the
$|\varphi|^4$ model on the Euclidean lattice $\Z^4$, which we now summarise.
For this, we write $c$ for a $g$- and $n$-dependent constant that varies from line to line.
\begin{itemize}
\item
For $n \ge 1$, the asymptotic formula \eqref{eq:chi_N_asymptotic_bis}
has been proved
with $A_{4}\sim [(n+8)g/(16\pi^2)]^{\hat\gamma}$  as $g \downarrow 0$ \cite{BBS-phi4-log},
as part of a large RG literature for weakly-coupled $|\varphi|^4$ models
on $\Z^4$.
\item
For $1$-component spins, it has been proved
that the critical two-point function decays as $\langle \varphi_0\varphi_x\rangle  \sim c |x|^{-2}$ as $|x| \rightarrow \infty$ \cite{GK85,FMRS87,ST-phi4}.
The susceptibility and correlation length have logarithmic corrections
$\chi \asymp \eps^{-1}|\log \eps|^{1/3}$ and $\xi \asymp \eps^{-1/2}|\log \eps|^{1/6}$ as $\eps \downarrow 0$  \cite{HT87} (these are
also proved for $d>4$ with no logarithmic corrections in \cite{HT87}).
The critical magnetisation vanishes with the magnetic field
{$h$ as $M \sim c h^{1/3}|\log h|^{1/3}$} \cite{Lohm20}.
\item
For general $n \ge 1$, the  $|x|^{-2}$ decay of the
critical two-point function is proved in \cite{ST-phi4}.
As $\eps \downarrow 0$, the susceptibility obeys
$\chi \sim c \eps^{-1}|\log \eps|^{\frac{n+2}{n+8}}$  \cite{BBS-phi4-log}.
The correlation length of order $p$
obeys $\xi_p \sim c_p \eps^{-1/2}|\log \eps|^{\frac{n+2}{2(n+8)}}$ for all $p>0$
\cite{BSTW-clp}.
The specific heat obeys
$c_H \sim c |\log \eps|^{\frac{4-n}{n+8}}$ ($n =1,2,3$),
$c_H \sim c \log  | \log \eps |$ ($n=4$),
and $c_H \sim c$ ($n \ge 5$) \cite{BBS-phi4-log}.

In \cite{BBS-saw4,BBS-saw4-log,  BSTW-clp}, the above asymptotic formulas
for the critical two-point function,
the susceptibility, and the correlation length of order $p$
are extended to $n=0$.  The case $n=0$ corresponds
to the continuous-time weakly self-avoiding walk via an exact supersymmetric representation.
\end{itemize}
\noindent
Still missing from the above
catalogue is a proof that the spontaneous magnetisation
obeys $M_+ \sim c \eps^{1/2}|\log \eps|^{\frac{3}{n+8}}$ as
$\eps=\nu_c-\nu$ (sign change) approaches zero from the ordered phase.
The critical behaviour from
the ordered phase is beyond the scope of current RG methods.

Apart from \cite{HT87}, we are not aware of rigorous RG results for $\varphi^4$
in dimensions $d>4$.  On the other hand, lace expansion methods have been used
to prove $|x|^{-(d-2)}$ decay of the critical two-point function in dimensions
$d>4$ for $1$-component $\varphi^4$ \cite{Saka15}, for $1$- and $2$-component
$|\varphi|^4$ \cite{BHH21}, and for the (spread-out) Ising model \cite{Saka07}.
A different proof of $|x|^{-(d-2)}$ decay for the Ising
and $\varphi^4$ models for $d>4$ is given in \cite{DP25-Ising}.

Scaling relations among the many logarithmic exponents for $\Z^4$ are presented in
\cite{Kenn12}.

\subsubsection{Non-Gaussian limit inside the critical window}

To state our results for the non-Gaussian limit,  we need several definitions.

\smallskip\noindent{\bf The non-Gaussian limiting measure.}
Given $n\in\N$ and $s \in \R$ we define a
non-Gaussian probability measure on $\R^n$ and its $k^{\rm th}$ moments
($k > -n$) by
\begin{equation}
\label{eq:Signks}
    d\sigma_{n,s}
    =
    \frac{e^{-\frac 14 |x|^4 - \frac s2 |x|^2} dx}{\int_{\R^n} e^{-\frac 14 |x|^4 - \frac s2 |x|^2} dx},
\qquad
    \Sigma_{n,k}(s) = \int_{\R^n} |x|^k d\sigma_{n,s}.
\end{equation}

\smallskip\noindent {\bf The quartic coupling constant.}
Our RG analysis involves a critical (massless) quartic running coupling constant $g_j$
whose value at scale $j=N$ plays an important role.
The sequence $g_j$ is nonnegative and depends on $g$, $L$, $n$ and $d$.
For $d=4$, $g_N \sim (\bbb N)^{-1}$, and for $d>4$, $\lim_{N\to\infty} g_N = g_\infty$.
The positive constants $\bbb$ and $g_\infty$ satisfy
\begin{equation}
\label{eq:Bdef}
     \bbb =   (n+8) (1- L^{-d}), \qquad g_\infty =g+O(g^2).
\end{equation}

\smallskip\noindent{\bf The window  scale.}
Recall the logarithmic correction exponents
$\hat\gamma  = \frac{n+2}{n+8}$ and $\hat\theta =  \frac 12 -\hat\gamma$ from
\eqref{eq:hatexponents}, and the amplitudes $A_{d}$ from
Theorem~\ref{thm:book_main_theorem_bis}.
We define the \emph{window scale} (or \emph{rounding} scale)
\begin{equation}
\label{eq:window_choice}
	w_N =
    \begin{cases}
        A_{4} (\log L^2)^{\hat\gamma}\bbb^{-1/2}N^{-\hat\theta} L^{-2N} & (d=4)
        \\
        A_{d}\, g_\infty^{1/2} L^{-Nd/2}
        & (d>4).
    \end{cases}
\end{equation}
For $d=4$, the factors $(\log L^2)^{\hat\gamma}\bbb^{-1/2}N^{-\hat\theta}$
can equivalently be written as the product of $(\log L^{2N})^{\hat\gamma}$, which
is a correction logarithmic in the volume, multiplied by $(\bbb N)^{-1/2}\sim g_N^{1/2}$ which is
the $4$-dimensional counterpart of the $g_\infty^{1/2}$ appearing for $d>4$.
To understand the choice of window scale via a rough computation,
for $d>4$ we start with the ansatz $\chi_N^{\rm P}(\nu_c) \asymp L^{Nd/2}$
and choose $w_N$ to scale in such a manner that $\chi_\infty(\nu_c+w_N) \asymp
\chi_N^{\rm P}(\nu_c)$.  Since $\chi_\infty(\nu_c+w_N) \asymp w_N^{-1}$
by \eqref{eq:chi_N_asymptotic_bis}, we conclude that the window should obey
$w_N \asymp L^{-Nd/2}$ for $d>4$.  For $d=4$, similar reasoning leads
to $w_N \asymp L^{-2N}N^{-\rexp}$ when we assume that
$\chi_N^{\rm P}(\nu_c) \asymp L^{2N} N^{1/2}$
(a prediction of \cite{Kenna04} for $\Z^4$, which we prove and extend in
Corollary~\ref{cor:suscept_crit_window} for the hierarchical model).

\smallskip\noindent{\bf The effective critical points.}
For PBC we define the \emph{effective critical point}
to equal the infinite-volume critical point from Theorem~\ref{thm:book_main_theorem_bis}:
\begin{equation}
    \nu^{\rm P}_{c,N} = \nu_c .
\end{equation}
Although $\nu^{\rm P}_{c,N}$ is independent of the volume parameter $N$, we include
the subscript to emphasise its role as a finite-volume critical point.
For FBC we first define
\begin{align}
\label{eq:vNdef}
    v_N=
    \begin{cases}
        A_{4} (\log L^2)^{\hat{\gamma}}N^{\hat\gamma}L^{-2N}
                    & (d=4)
        \\
        A_{d}L^{-2N} & (d>4).
    \end{cases}
\end{align}
By definition, $v_N$ is much larger than $w_N$, namely,
\begin{align}
\label{eq:vNwN}
    v_N=
    \begin{cases}
        (\bbb N)^{1/2} w_N  & (d=4)
        \\
        g_\infty^{-1/2} L^{N(d-4)/2}w_N  & (d>4).
    \end{cases}
\end{align}
For FBC, we specify a volume-dependent effective
critical point $\nu_{c,N}^{\rm F}$,
which we compute exactly when $d=4,5$ and approximately when $d>5$, by
\begin{align}
\label{eq:nu_c,N_def}
    \nu_{c,N}^{\free}
    &=
    \begin{cases}
    \nu_{c}-\qLap v_N (1+\constF N^{-\hat \gamma}) & (d=4)
    \\
    \nu_{c}-\qLap v_N & (d=5)\\
	\nu_{c}-\qLap v_N(1+ O(L^{-N}) ) &  (d>5) ;
    \end{cases}
\end{align}
here $\qLap=1+O(L^{-d})$ is the constant from \eqref{eq:qLapdef}
and $\constF  = O(g)$
is determined in the proof.
The error term $O(L^{-N}v_N)$ for $d>5$ is
the same order as $w_N$ for $d=6$ and is larger than $w_N$ for $d>6$.
Although we do not precisely identify $\nu_{c,N}^\free$ for $d \ge 6$, we
do prove its existence as an anchor for the critical window.
See Remark~\ref{rk:FBC_crit_point} for further discussion.

\smallskip\noindent{\bf The large-field scale.}
In terms of the constants $\bbb$
and  $g_\infty$ in \eqref{eq:Bdef}, we define
the \emph{large-field scale}
\begin{equation}
\label{eq:hhdef}
    \hh_N =
    \begin{cases}
        (\bbb N)^{1/4}L^{-N} & (d=4)
        \\
        g_\infty^{-1/4}L^{-Nd/4} & (d>4).
    \end{cases}
\end{equation}
The prefactors for $L^{-Nd/4}$ in \eqref{eq:hhdef} are each essentially equal to $g_N^{-1/4}$, which is large.
The definition of $\hh_N$ ensures that if
$|\varphi_x| \asymp \hh_N$, then $\sum_{x\in \Lambda_N} g_N |\varphi_x|^4
\asymp 1$ is positive uniformly in
$N,L,g$, so the exponential decay in the Hamiltonian due to $\exp[-\frac 14 g_N
\sum_{x\in \Lambda_N} |\varphi_x|^4]$ becomes significant at the large-field
scale.

\smallskip
Recall that the average field $\Phi_N$ is defined by \eqref{eq:totalfield}.
The following theorem concerns the finite-size near-critical
scaling of the weakly-coupled hierarchical $|\varphi|^4$ model in dimensions $d \ge 4$.
It is part of the statement of the theorem that there are constants
$g_\infty$ and $c^\free$ as above,
and a sequence $\nu_{c,N}^\free$ obeying \eqref{eq:nu_c,N_def}, such that the conclusions
of the theorem hold.
The $d$-dependence has not
been made explicit in the notation of the theorem.
For part~(ii) of the theorem, we introduce two different
restrictions on the real sequence $s_N$:
\begin{align}
\label{eq:sNbds}
    s_N \le \begin{cases}
        L^{\sqrt{N}} & (d=4)
        \\
        L^{N(d-3)/2} & (d \ge 5) ,
    \end{cases}
\qquad\quad
    |s_N| \le \begin{cases}
        (\log N)^{1/4} & (d=4)
        \\
        N^{1/4} & (d \ge 5).
    \end{cases}
\end{align}
The definitions of the measure $d\sigma_{n,s}$ and its moments $\Sigma_{n,k}(s)$ are given in \eqref{eq:Signks}.
Note that the integral and the moments on the right-hand sides
of \eqref{eq:mr-nongaussian_law} and \eqref{eq:mrPBC-nongaussian_moments} are
\emph{independent} of $g$ and of the dimension $d \ge 4$.

\begin{theorem}
\label{thm:mr-nongaussian}
{\rm (Non-{G}aussian limit in the critical window.)}
Let $d\ge 4$,
let $n \in \N$, let $L$ be sufficiently large, and
let $g>0$ be sufficiently small (depending on $L$).
The following statements hold for the $n$-component hierarchical
model, for boundary conditions $*={\rm F}$ and $*={\rm P}$:
\begin{enumerate}
\item
For any $J \in \R^n$ and any real sequence $s_N$ converging to $s\in (-\infty,\infty)$,
\begin{equation}
\label{eq:mr-nongaussian_law}
    \lim_{N \to \infty}
    \big\langle e^{J\cdot \Phi_{N}/\hh_N } \big\rangle_{g, \nu_{c,N}^* + s_Nw_N, N}^{*}
    =
    \int_{\R^n} e^{J\cdot x }\, d\sigma_{n,s}.
\end{equation}
\item
Let $p \ge 1$,
let $s_N$ be any real sequence such that either: (a) $s_N$ converges,
or (b) $\lim s_N = +\infty$ in such a manner that
the first inequality of \eqref{eq:sNbds} holds,
or (c) $\lim s_N = -\infty$ in such a manner that
the second inequality of \eqref{eq:sNbds} holds.
Then, as $N \to \infty$,
\begin{align}
    \big\langle
    |\Phi_{N} |^{2p}
    \big\rangle_{g, \nu_{c,N}^* +s_Nw_N,N}^{*}
    & =
    \hh_N^{2p}
    \Sigma_{n,2p}(s_N)
    \big( 1+ o(1)  \big).
\label{eq:mrPBC-nongaussian_moments}
\end{align}
\end{enumerate}
\end{theorem}

Theorem~\ref{thm:mr-nongaussian}(i) is a statement of convergence of moment generating functions
and implies that $\Phi_{N}/\hh_N$ converges in distribution to a random variable on $\R^n$
with distribution $d\sigma_{n,s}$.
In other words,
\begin{align}
\label{eq:CW4}
    (\bbb N)^{-1/4} L^N \Phi_{N} & \Rightarrow d\sigma_{n,s} \quad (d=4),
    \\
\label{eq:CWd}
    g_\infty^{1/4} L^{Nd/4} \Phi_{N}& \Rightarrow d\sigma_{n,s} \quad (d>4).
\end{align}
The convergence in \eqref{eq:mr-nongaussian_law}
may be compared with the problem of determining the limiting
distribution of the average spin of the Curie--Weiss model
(the Ising model on the complete graph $\mathbb{K}_V$
on $V$ vertices).
Several authors have considered this problem at the critical point
\cite{SG73,EN78,CJL78,Elli85,DM23},
which corresponds to $n=1$ and the choice $s=0$, and identify
the non-Gaussian distribution $d\sigma_{1,0}$ as the limiting
distribution of the rescaled   average spin $V^{1/4}S_V$.
The scaling window is also treated in
\cite{CS11a,EL10}, and in particular it is shown in \cite[Theorem~3.2]{EL10}
that the limiting distribution of $V^{1/4}S_V$ is $d\sigma_{1,s}$ for $\beta$ of
the form $1+sV^{-1/2}$ (critical window)
for all $s \in \R$.
Related results for the Ising model on Erd\H{os}--R\'enyi random
graphs are obtained in \cite{KLS20}.
The scaling we prove for dimensions $d>4$ in \eqref{eq:CWd}
similarly involves the fourth root of the volume $L^{Nd/4}$, whereas for
$d=4$ in \eqref{eq:CW4} there is an additional
logarithmic (in the volume)
correction in the factor $N^{-1/4}$.

Theorem~\ref{thm:mr-nongaussian}(ii) states that, within a
window of $\nu$ values of order $w_N$ around the effective critical point
$\nu_{c,N}^*$, the expectation of $|\Phi_{N}|^{2p}$ is given by a factor of $\hh_N$ for each field,
multiplied by a $p$-dependent but $g$-independent universal profile $\Sigma_{n,2p}(s)$.
This holds even for the crossover out of the critical window,
when $s_N \to \pm\infty$ in the manner specified in
part~(ii).
We have not attempted to obtain optimal restrictions on how rapidly
$s_N \to \pm\infty$.
The case (ii)(a) with $s_N \rightarrow 0$ indicates that there
can be no ``hidden'' window smaller than the critical window $w_N$ we have identified.

\subsubsection{Gaussian limit above the critical window}

We recall the definitions of $d\mu_{n,s}$ and $M_{n,k}(s)$ from \eqref{eq:muGaussian}
and also define $\lp_N$ as
\begin{align}
\label{eq:lpdef}
    d\mu_{n,s} = \frac{e^{-\frac12 s|x|^2} dx}{\int_{\R^n}e^{-\frac12 s|x|^2} dx }, \qquad
    M_{n,k}(s)= \int_{\R^n}|x|^k d\mu_{n,s},
    \qquad
    \lp_N=  L^{-N(d-2)/2}.
\end{align}
These definitions parallel $d\sigma_{n,s}$, $\Sigma_{n,k}(s)$, $\hh_N$.
The sequences $v_N$  and $w_N$
are related as in \eqref{eq:vNwN}, so $v_N \gg w_N$ and
the $\nu$ values treated in the following theorem are well above
the critical window.
We again introduce growth restrictions on the sequence $s_N$,
this time as
\begin{align}
\label{eq:sNbds2}
    s_N \le \begin{cases}
        L^{\sqrt{N}} & (d=4)
        \\
        L^{N/2} & (d\geq 5).
        \end{cases}
\end{align}
As in Theorem~\ref{thm:mr-nongaussian}, the
limiting distribution
and the moments in the conclusions
of Theorem~\ref{thm:mr-gaussian} are independent of $g$ and of $d \ge 4$.

Although the limiting distribution
in Theorem~\ref{thm:mr-gaussian} is Gaussian, in its statement the scaling $\lp_N^{-1}\Phi_N
= L^{-N(d+2)/2}  \sum_{x\in \Lambda_N} \varphi_x$ is non-Gaussian,  as opposed to
the central limit theorem scaling
$L^{-Nd/2}\sum_{x\in \Lambda_N} \varphi_x$ of Theorem~\ref{thm:book_main_theorem_bis}.

\begin{theorem}
\label{thm:mr-gaussian}
{\rm (Gaussian limit above the critical window.)}
Let $d\ge 4$,
let $n \in \N$, let $L$ be sufficiently large, and
let $g>0$ be sufficiently small (depending on $L$).
The following statements hold for the $n$-component hierarchical
model, for boundary conditions $*={\rm F}$ and $*={\rm P}$:
\begin{enumerate}
\item
For any $J \in \R^n$ and any real sequence $s_N$ converging to $s>0$,
\begin{equation}
\label{eq:gaussian_measure}
    \lim_{N \to \infty}
    \big\langle e^{J\cdot \Phi_{N}/\lp_N } \big\rangle_{g, \nu_{c,N}^* + s_Nv_N, N}^{*}
    =
    \int_{\R^n} e^{J\cdot x }\, d\mu_{n, s}.
\end{equation}
\item
Let $p \ge 1$,
let $s_N$ be any real sequence such that
$s_N\ge \epsilon$ for some fixed $\epsilon >0$ and $s_N$ obeys \eqref{eq:sNbds2}.
Then, as $N\rightarrow \infty$,
\begin{align}
    \big\langle
    |\Phi_{N} |^{2p}
    \big\rangle_{g, \nu_{c,N}^*+s_Nv_N,N}^{*}
    & =
    \lp_N^{2p}
    M_{n,2p}(s_N)
    \big( 1+ o(1)  \big).
\label{eq:SNp-window-free}
\end{align}
\end{enumerate}
\end{theorem}

We have chosen to present Theorem~\ref{thm:mr-gaussian} as such for comparison with the
statement of Theorem~\ref{thm:mr-nongaussian}, but since the limiting
distribution is Gaussian we can make
\eqref{eq:gaussian_measure}--\eqref{eq:SNp-window-free} more explicit:
\begin{align}
	\label{eq:FBC_gaussian_measure_2}
	    \lim_{N \to \infty}
    \big\langle e^{J\cdot \Phi_{N}/\lp_N } \big\rangle_{g, \nu_{c,N}^* + sv_N, N}^{*}
    &= e^{|J|^2/(2s)}, \\
    \label{eq:SNp-window_2}
        \big\langle
    |\Phi_{N} |^{2p}
    \big\rangle_{g, \nu_{c,N}^* +s_Nv_N,N}^{*}
    & =
    \lp_N^{2p}
    \left(\frac{2}{s_N}\right)^p
    \frac{\Gamma(\frac{n+2p}{2})}{\Gamma(\frac{n}{2})}
    \big( 1+ o(1)  \big),
\end{align}
under the hypotheses of Theorem~\ref{thm:mr-gaussian}.

Theorem~\ref{thm:mr-gaussian} covers a range of $\nu$ values that
is outside of the scaling window around the effective critical point
$\nu_{c,N}^*$, where we know from Theorem~\ref{thm:mr-nongaussian}
that the limiting law is the non-Gaussian
measure $d\sigma_{n,s}$ rather than $d\mu_{n,s}$.
It follows from \eqref{eq:nu_c,N_def} that there is
a solution $\tilde s_N$ to the equation $\nu_{c,N}^\free + \tilde s_N v_N = \nu_c$,
which obeys $\tilde s_N \sim \qLap$ for all $d \ge 4$.
With FBC, Theorem~\ref{thm:mr-gaussian}
therefore gives a Gaussian limit at the infinite-volume critical point $\nu_c$,  unlike the
non-Gaussian limit established in Theorem~\ref{thm:mr-nongaussian}
for the critical window around the effective critical point
$\nu_{c,N}^{\rm F}$, or around $\nu_c$ with PBC.
In particular, $\nu_c$ is outside of the critical window centred at $\nu_{c,N}^{\rm F}$.

An exact analogue of \eqref{eq:gaussian_measure} is proved in
\cite[Theorem~1.3(ii)]{BBS-phi4-log} for the Euclidean (non-hierarchical)
model with PBC in dimension $d=4$.

\subsection{Discussion of main results}

In this section we
analyse the susceptibility and its universal profile,
compute universal ratios of
moments of the average field
in the critical
window, and discuss the case $n=0$ of self-avoiding walk.
This discussion is useful for the interpretation of our results,
but none of it plays a role in the proofs of
Theorems~\ref{thm:book_main_theorem_bis}--\ref{thm:mr-gaussian}
which are given in the bulk of the paper.

\subsubsection{The susceptibility}
\label{sec:chi}

The following corollary of Theorem~\ref{thm:mr-nongaussian} shows that the
universal profile $n^{-1}\Sigma_{n,2}(s)$ governs
the susceptibility in the critical window
around the effective critical point, both for FBC and for PBC, and for all dimensions
$d \ge 4$.

\begin{corollary}
\label{cor:suscept_crit_window}
Let $d \ge 4$, let $n \geq 1$,
let $L$ be sufficiently large, and let $g>0$ be sufficiently small
(depending on $L$).
Then for boundary conditions $*={\rm F}$ and $*={\rm P}$,
and for $s_N$ obeying the hypothesis of Theorem~\ref{thm:mr-nongaussian},
as $N \to \infty$
the susceptibility obeys
\begin{equation}
\label{eq:chiwindow}
    \chi_N^*(\nu_{c,N}^* + s_Nw_N) =
    \begin{cases}
        n^{-1} \Sigma_{n,2}(s_N) (\bbb N)^{1/2} L^{2N} [1+o(1)] & (d=4)
        \\
        n^{-1} \Sigma_{n,2}(s_N) g_\infty^{-1/2} L^{Nd/2} [1+o(1)] & (d>4).
    \end{cases}
\end{equation}
\end{corollary}

\begin{proof}
By  \eqref{eq:chiPhi} and by
\eqref{eq:mrPBC-nongaussian_moments} with $p=1$,
\begin{align}
\label{eq:chiNcor}
    \chi_N^*(\nu_{c,N}^* +s_Nw_N)
    &=
    \frac{1}{n}L^{dN}
    \big\langle
    |\Phi_{N} |^{2}
    \big\rangle_{g, \nu_{c,N}^* +s_Nw_N,N}^*
    =
    L^{dN}\hh_N^2
    \frac{1}{n} \Sigma_{n,2}(s_N)
    \big( 1+ o(1) ) \big),
\end{align}
which is the same as \eqref{eq:chiwindow} due to the definition of $\hh_N$ in \eqref{eq:hhdef}.
\end{proof}

It follows from \eqref{eq:chi_N_asymptotic_bis} and the definition of $w_N$ that,
for $s_N >0$ with $s_Nw_N \to 0$
(with the added restriction $s_N\ge\eps >0$ for $d=4$), as $N \to \infty$
the infinite-volume susceptibility obeys
\begin{equation}
\label{eq:chiinfsinf}
    \chi_\infty(\nu_c+s_Nw_N) \sim
    \begin{cases}
    s_N^{-1} (\bbb N)^{1/2} L^{2N}  & (d=4)
    \\
    s_N^{-1}g_\infty^{-1/2} L^{Nd/2} & (d>4).
    \end{cases}
\end{equation}
When $d>4$, the above
right-hand side is simply $1/s_N w_N$, whereas for $d=4$, there is a logarithmic correction.
To compare this with Corollary~\ref{cor:suscept_crit_window}, we
use the elementary fact (see \eqref{eq:f_n_asymptotics}) that
\begin{equation}
\label{eq:Sigasy}
    n^{-1}\Sigma_{n,2}(s)
    \sim
    \begin{cases}
    s^{-1}
    & (s \to +\infty)
    \\
    n^{-1} |s|
    & (s \to -\infty),
    \end{cases}
\end{equation}
so for $s_N \to +\infty$
obeying the first inequality of \eqref{eq:sNbds}, it follows from \eqref{eq:chiwindow} that
\begin{equation}
\label{eq:chiNsinf}
    \chi_N^*(\nu_{c,N}^* + s_Nw_N) \sim
    \begin{cases}
        s_N^{-1} (\bbb N)^{1/2} L^{2N} & (d=4)
        \\
        s_N^{-1} g_\infty^{-1/2} L^{Nd/2} & (d>4).
    \end{cases}
\end{equation}
Comparison of \eqref{eq:chiinfsinf} and
\eqref{eq:chiNsinf} shows that, when $s_N\to +\infty$ subject to
the first inequality of \eqref{eq:sNbds},
the finite-volume and infinite-volume susceptibilities are asymptotically equivalent:
\begin{equation}
    \chi_N^*(\nu_{c,N}^*+s_Nw_N) \sim \chi_\infty(\nu_c+s_Nw_N)
    .
    \label{eq:crossover}
\end{equation}
This demonstrates a crossover out of the critical window.

On the other hand, if $s_N \to -\infty$ obeys the second inequality of \eqref{eq:sNbds}, then
\eqref{eq:chiwindow} and \eqref{eq:Sigasy} give
\begin{equation}
\label{eq:chisneg}
	\chi_N^*(\nu_{c,N}^* + s_Nw_N) \sim
    \frac{1}{\Sigma_{n,2}(0)}|s_N|\chi_N^*(\nu_{c,N}^*).
\end{equation}
This should be contrasted with the behaviour of the susceptibility when
it is more naturally defined
in terms of the truncated two-point function---this truncated susceptibility should have a peak
inside the scaling window and decrease, not grow, as $s_N$ becomes increasingly negative
(see, e.g., \cite[Section~4.4]{Card96}).
For $\nu < \nu_c$, our definition of
$\chi_N$ does not subtract a term $L^{dN}M_+(\nu)^2$ with $M_+$ the spontaneous magnetisation
which is predicted to obey $M_+ \sim |\eps|^{1/2}|\log \eps|^{\frac{3}{n+8}}$ for $d=4$  and
$M_+ \sim |\eps|^{1/2}$
for $d>4$ (in infinite volume).
The linear growth of $\chi_N^*(\nu_c^* + s w_N)$
as $s \to -\infty$, shown in \eqref{eq:chisneg}, is a manifestation
of the growing magnetisation we expect to contribute to $\chi_N$ on the low-temperature
side of the window, namely (for $d=4$)
$L^{4N} [(|s|w_N)^{1/2}N^{\frac{3}{n+8}}]^2 \asymp |s|L^{2N}N^{-\hat\theta + \frac{6}{n+8}}
= |s|L^{2N}N^{1/2}
\asymp |s|\chi_N(\nu_c)$.
There is related discussion in \cite[p.~768]{Zinn21}.
Thus \eqref{eq:chisneg}
provides
an intriguing peek into the
low-temperature phase, whose full critical behaviour remains an outstanding open problem.
For the $4$-dimensional Ising model it has been proved that the spontaneous magnetisation
obeys $c|\eps|^{1/2} \le M_+ \le c'|\eps|^{1/2}|\log |\eps|\,|^{3/2}$---see \cite{AF86}
and \cite[Section~14.4]{FFS92}---but to our knowledge
this has not been proved for the $|\varphi|^4$ model, even for $n=1$.

\subsubsection{Universal profile for the susceptibility}
\label{sec:profile}

Next, we establish basic properties of the universal profile
for the susceptibility in \eqref{eq:chiwindow}, which we denote here by
\begin{equation}
    \uf_n(s) = \frac{1}{n}  \Sigma_{n,2}(s)
    =
    \frac{\int_{\R^n} |x|^2 e^{-\frac 14 |x|^4 - \frac s2 |x|^2} dx}
    {n\int_{\R^n} e^{-\frac 14 |x|^4 - \frac s2 |x|^2} dx}.
\end{equation}
The fact that $\uf_n(s)$ is
strictly decreasing, both as a function of $s$ and as a function of $n$, is proved
in Lemma~\ref{lemma:fn}.
Plots of $\uf_n$ are also given in Appendix~\ref{sec:ufmono}.

For $k \in ( -1,\infty)$ we define the integral
\begin{align}
    I_k (s) = \int_{0}^{\infty} x^{k} e^{-\frac 14 x^4  -   \frac 12 s x^2}dx ,
\label{eq:Ik_definition}
\end{align}
and rewrite the integrals defining $\Sigma_{n,k}(s)$ and $\uf_n(s)$ in polar form to obtain
\begin{align}
    \Sigma_{n,k}(s)=  \frac{I_{k+n-1}(s)}{I_{n-1}(s)}, \qquad
    \uf_n (s) =   \frac{ I_{n+1}  (s)}{n I_{n-1}  (s)} .
\label{eq:uf_n_definition}
\end{align}
Although our main concern is for positive integers $n$, the
formula \eqref{eq:uf_n_definition}
for $\uf_n(s)$ remains meaningful for real values $n \in (0,\infty)$.
Furthermore, integration by parts in the denominator of
$\uf_n(s)$ gives
\begin{align}
n I_{n-1} (s) = \int_0^{\infty} (x^{n+3} + sx^{n+1} ) e^{-\frac{1}{4} x^4 - \frac{1}{2} sx^2} dx = I_{n+3} (s) + sI_{n+1} (s),
\label{eq:nI_n-1_integration_by_parts}
\end{align}
and therefore, for all $n>0$,
\begin{align}
\label{eq:newuf}
    \uf_n (s) =   \frac{ I_{n+1}  (s)}{n I_{n-1}  (s)}
    =
    \frac{I_{n+1}(s)}{I_{n+3} (s) + sI_{n+1} (s)} .
\end{align}
The right-hand side of \eqref{eq:newuf} extends the definition of $\uf_n$ from
$n \in (0,\infty)$ to $n \in (-2,\infty)$.

In particular, for $n=0$ the denominator becomes
\begin{equation}
I_{3} (s) + sI_{1} (s)
=
\int_0^{\infty} (x^{3} + sx  ) e^{-\frac{1}{4} x^4 - \frac{1}{2} sx^2} dx
= -
\int_0^\infty d \big( e^{-\frac{1}{4} x^4 - \frac{1}{2} sx^2} \big) = 1,
\end{equation}
so \eqref{eq:newuf} gives
\begin{equation}
    \uf_0(s) = I_1(s) = \int_{0}^{\infty} x  e^{-\frac 14 x^4  -   \frac 12 s x^2}dx.
\end{equation}
As we discuss in Section~\ref{sec:n0},
the case $n=0$, although not included in our main results,
has significance for the self-avoiding walk.  In particular,
$\uf_0$ arises as the profile for both strictly and weakly self-avoiding
walk on the complete graph.

For $n >-2$, \eqref{eq:newuf} provides the recursion relation
\begin{equation}
\label{eq:ufn-recursion}
    f_n(s) = \frac{1}{(n+2)f_{n+2}(s) + s}.
\end{equation}
It follows from \eqref{eq:ufn-recursion}
that $\lim_{n\to -2}f_n(s) = s^{-1}$ for $s>0$,
and that this limiting function $f_{-2}$
is maximal in the sense that $f_n(s) < s^{-1}$ for all $s>0$ and $n>-2$.
This profile $f_{-2}(s)=s^{-1}$ is consistent with the value $n=-2$ corresponding to a Gaussian model
($n=-2$ corresponds to a two-component fermion field
for which the fourth power vanishes), since the susceptibility of the Gaussian
model is simply $\chi(\nu)=\nu^{-1}$.
Values of $n$ in the interval $[-2,0)$ are not completely understood.
The model with $n=-2$ is identified as
loop-erased random walk in \cite{WF19} (which is also Gaussian
for $d \ge 4$), and
there is speculation about walk interpretations for $n \in (-2,0)$ in \cite[Section~2]{WF19}.

By the change of variables $t = x^4/4$, the integral
$I_k (s)$ can be rewritten in terms of
the \emph{Fax\'en integral} (see \cite[p.~332]{Olve97})
\begin{equation}
\label{eq:Faxen}
    \Fi (\alpha, \beta ; y) = \int_0^{\infty} e^{-t + y t^{\alpha}} t^{\beta - 1} dt
    \qquad
    (0 \le \alpha <1 ,  \;\;\beta>0)
\end{equation}
as
\begin{align}
\label{eq:IFi}
I_k (s) = 2^{\frac{k-3}{2}} \Fi ({\textstyle \frac{1}{2}, \frac{k+1}{4} ; -s }).
\end{align}
Since $I_k (0) = 2^{\frac{k-3}{2}} \Fi(\frac{1}{2}, \frac{k+1}{4} ; 0) = 2^{\frac{k-3}{2}} \Gamma(\frac{k+1}{4})$ for $k>-1$,  $f_n (0)$ is given explicitly by
\begin{equation}
\label{eq:ufn0-2}
	\uf_n (0)
    =
    \frac{\Gamma(\frac{n+2}{4})}{2 \Gamma(\frac{n+4}{4})}
    \qquad (n > -2),
\end{equation}
where we used \eqref{eq:ufn-recursion} for $n \in (-2,0]$.
The known asymptotic behviour of the Fax\'en integral
(see \cite[Ex.~7.3, p.~84]{Olve97}) gives, for $k>-1$,
\begin{align}
\label{eq:IkFi}
    I_k (s)
    \sim
    \begin{cases}
    2^{\frac{k-1}{2}}\Gamma(\frac{k+1}{2})s^{-(k+1)/2} & (s \to \infty)
    \\
    \sqrt{\pi} |s|^{(k-1)/2}e^{s^2/4} & (s \to -\infty),
    \end{cases}
\end{align}
and hence, by \eqref{eq:uf_n_definition},
\begin{equation}
\label{eq:Mnkasy}
    \Sigma_{n,k}(s)
    \sim
    \begin{cases}
    \frac{\Gamma(\frac{n+k}{2})}{\Gamma(\frac{n}{2})}\left( \frac{2}{s}\right)^{k/2}
    & (s \to +\infty)
    \\
    |s|^{k/2}
    & (s \to -\infty).
    \end{cases}
\end{equation}
In particular, by setting $k=2$ and by the recursion relation for the Gamma function,
\begin{align}
	\uf_n(\s) &\sim \frac{1}{s} \qquad (n>-2,\; \s \to +\infty), \hspace{12mm}
	\uf_n(\s) \sim \frac{|s|}{n} \qquad (n>0,\; \s \to -\infty),
\label{eq:f_n_asymptotics}
\end{align}
where again we used \eqref{eq:ufn-recursion} for $n \in (-2,0]$ and $s \to +\infty$.

\subsubsection{Universal ratios}

This section applies for
all dimensions $d \ge 4$.
All statements in this section are valid for either choice $*={\rm P}$ or $*={\rm F}$
for the boundary condition,
but to declutter the notation
we do not place asterisks on the expectations or on $\nu_{c,N}$, which should
always be interpreted as $\langle\,\cdot\,\rangle^*$ and $\nu_{c,N}^*$.

Theorem~\ref{thm:mr-nongaussian}
allows for explicit computation of the universal ratios
\begin{equation}
\label{eq:R2p}
    \cR^{(2p)}_n(s) =
    \lim_{N\to\infty}
    \frac{\big\langle
    |\Phi_{N} |^{2p}
    \big\rangle_{g, \nu_{c,N}+sw_N,N}}
    {\big\langle
    |\Phi_{N} |^{2}
    \big\rangle_{g, \nu_{c,N}+sw_N,N}^{p}}
    =
    \frac{\Sigma_{n,2p}(s)}{[\Sigma_{n,2}(s)]^p}
\end{equation}
for $n \ge 1$, $p \ge 1$, and $s\in\R$.
For $s=0$, by \eqref{eq:uf_n_definition}
and with the value of $I_k(0)$ stated above \eqref{eq:ufn0-2},
this gives
\begin{align}
    \cR^{(2p)}_n(0)
    &
    =
    \frac{ \Gamma(\frac{n+2p}{4})
    [\Gamma(\frac{n}{4})]^{p-1}}
    {[ \Gamma(\frac{n+2}{4})]^{p}},
\label{eq:cR2p0}
\end{align}
which is the same as the value stated in \cite[(32.36)]{Zinn21} for
$\Z^d$ with PBC in dimensions $d>4$.

In particular, by \eqref{eq:cR2p0},
$\cR_n^{(4)}(0)$ (known in statistics as the \emph{kurtosis}) takes the universal value
\begin{equation}
\label{eq:Binder}
    \cR_n^{(4)}(0)
    =
    \frac{ \Gamma(\frac{n+4}{4})
    \Gamma(\frac{n}{4})}
    {[ \Gamma(\frac{n+2}{4})]^{2}}
    =
       \frac 4n
    \left(
    \frac{\Gamma(\frac{n+4}{4})}
    { \Gamma(\frac{n+2}{4})}
    \right)^{2}
    =
    \frac{1}{nf_n(0)^2}.
\end{equation}
For $n=1$ the reciprocal, $Q = 1/\cR_1^{(4)}(0)$
is $Q=4 [\Gamma(\frac{3}{4})/\Gamma(\frac{1}{4}) ]^2 =0.456\, 947\ldots$.
 Related to $\cR_n^{(4)}(0)$ is
the \emph{Binder cumulant} $U_n(s) =1 - \frac 13 \cR_n^{(4)}(s)$
(in statistics $-3U_n(s)$ is the \emph{excess kurtosis}).

A fundamental quantity is the \emph{renormalised coupling constant},
which in finite volume and for general
$n \ge 1$ is defined by
\begin{equation}
    \lambda_{n,N} (s)
    =
    -
    \frac{\langle |\Phi_{N}|^4 \rangle_{g,\nu_{c,N}+sw_N,N}
    -
    \frac{n+2}{2}
    \langle |\Phi_{N}|^2\rangle_{g,\nu_{c,N}+sw_N,N}^2}
    {\langle |\Phi_{N}|^2\rangle_{g,\nu_{c,N}+sw_N,N}^2}
    .
\end{equation}
For a Gaussian distribution, $\lambda_{N}=0$.
In the limit as $N \to \infty$,
\eqref{eq:R2p} gives
\begin{equation}
\label{eq:u4Q}
    \lambda_n(s) = \lim_{N\to\infty} \lambda_{n,N} (s)
    =
    \frac{n+2}{2} - \cR_n^{(4)}(s)
    =
    \frac{n+2}{2} -
    \frac{\int_{\R^n} |x|^4 d\sigma_{n,s}}{(\int_{\R^n} |x|^2 d\sigma_{n,s})^2}
    .
\end{equation}
In particular, for $n=1$,
\begin{equation}
    \lambda_1(0)
    =
    3 - 4\left( \frac{\Gamma(5/4)}{\Gamma(3/4)} \right)^2
    =
    0.81156\ldots
    \qquad
    (d \ge 4).
\end{equation}
Also, the asymptotic formulas \eqref{eq:Mnkasy} imply that
\begin{align}
    \lim_{s\to\infty} \lambda_{n}(s)
    & = \frac{n+2}{n} - \frac{n+2}{n} =0,
    \\
    \lim_{s\to-\infty} \lambda_{n}(s) & = \frac{n+2}{n}-1= \frac{2}{n}.
\end{align}
By the proof of Lemma~\ref{lemma:fn} (put $n=1$ in \eqref{eq:nf_n_derivative}),
the right-hand side of \eqref{eq:u4Q} is strictly monotone decreasing, so the
renormalised coupling constant varies monotonically over the range $[0,2/n]$ as the
window is traversed.
For $n=1$, the Lebowitz and Griffiths inequalities imply that
that $\lambda(s) \in [0,2]$, so as $s$ decreases through the window we see a
value of the renormalised coupling constant that increases from the Gaussian
value of zero to the maximum possible non-Gaussian value $2$.

\subsubsection{Self-avoiding walk and \texorpdfstring{$n=0$}{n=0}}
\label{sec:n0}

The value $n=0$ is not part of our
main results, which are restricted to $n \ge 1$ components.
However, as we discuss in this section, there is compelling evidence of
the universal nature of the profile $\uf_0$ for self-avoiding walk in dimensions $d \ge 4$.
In particular, we confirm that $\uf_0$ is the profile for:
(i)~self-avoiding walk on the
complete graph, and
(ii)~the continuous-time weakly self-avoiding walk (a.k.a.\ discrete Edwards model)
on the complete graph.

\bigskip\noindent{\bf Self-avoiding walk on the complete graph.}
 The number of $n$-step self-avoiding walks on the complete graph $\mathbb{K}_N$
 on $N$ vertices, starting
from a fixed origin, is simply $\prod_{j=1}^n(N-j)$ for $n \le N$.
The susceptibility is the generating function for this sequence, i.e.,
\begin{equation}
    \chi_N^{\mathbb{K}}(z) = \sum_{n=0}^{N-1}  z^{ n} \prod_{j=1}^n(N-j)
    \qquad
    (z \in \R).
\end{equation}
The effective critical point is  $N^{-1}$.
The following proposition, which is proved in Appendix~\ref{sec:SAW},
shows that the profile $f_0(s)$ applies in this setting,
with window and susceptibility scaling as square root of the volume $N$
(this corresponds to $L^{Nd/2}$ in \eqref{eq:chiwindow}).
The minus sign on the left-hand side of \eqref{eq:chiKn}
compensates for the different monotonicity of $z$ compared to $e^{-\nu}$.

\begin{proposition}
\label{prop:SAW-KN}
For self-avoiding walk on the complete graph, as $N \to \infty$,
\begin{align}
\label{eq:chiKn}
     \chi_N^{\mathbb{K}}(N^{-1}(1- s(2N)^{-1/2}))
    &\sim (2N)^{1/2} f_0( s)
    \qquad
    (s \in \R).
\end{align}
\end{proposition}

\bigskip\noindent{\bf Continuous-time weakly self-avoiding walk on the complete graph.}
Let $X$ denote the continuous-time random walk on the complete graph $\mathbb{K}_N$.
The walk takes steps
at the events of a rate-$(1-N^{-1})$ Poisson process, with steps taken to a vertex chosen
uniformly at random from all other vertices (the choice being independent of the
Poisson process).
The \emph{local time} of $X$ at a vertex $v\in\mathbb{K}_N$, up to time $T$, is the random variable
\begin{equation}
    L_{T,v} = \int_0^T \1_{X(t)=v}\, dt
    .
\end{equation}
For $x,y\in\mathbb{K}_N$,
for fixed $g>0$, and for $\nu\in \R$, the \emph{two-point function} is
\begin{equation}
\label{eq:2ptfcndef}
    G_{xy}(\nu)
    =
    \int_0^\infty E_x
    \Big( e^{-g\sum_{v\in \mathbb{K}_N} L_{T,v}^2} \1_{X(T)=y}\Big)
    e^{-\nu T} \, dT
    .
\end{equation}
By symmetry, the \emph{susceptibility} is
\begin{equation}
    \chi_N^{\rm WSAW}(\nu) = \sum_{y\in \mathbb{K}_N} G_{xy}(\nu)
    =
    G_{00}(\nu) + (N-1)G_{01}(\nu)
    ,
\end{equation}
where $0,1$ denote two distinct vertices of $\mathbb{K}_N$.

As we discuss in more detail in Appendix~\ref{sec:SAW}, this model has a
critical point $\nu_c=\nu_c(g)$.
The following proposition shows that the universal profile $f_0(s)$
arises for the susceptibility in this model in its critical window.
Note that the profile is independent of the value of $g>0$, as also seen for
$n \ge 1$ in Corollary~\ref{cor:suscept_crit_window}.
The proof of Proposition~\ref{prop:WSAW}, which
applies results of \cite{BS20}, is given in Appendix~\ref{sec:SAW}.

\begin{proposition}
\label{prop:WSAW}
Let $g>0$.
There are positive constants $\lambda_1,\lambda_2$ (explicit and
$g$-dependent) such that
\begin{align}
    \lim_{N \to \infty} N^{-1/2} \chi^{\rm WSAW}_N (\nu_c+sN^{-1/2}) & =
    \lambda_1 f_0(\lambda_2 s)
    \qquad
    (s \in \R).
\end{align}
\end{proposition}

\subsection{Open problems}
\label{sec:open-problems}

Our results suggest several open problems.

\subsubsection{Intermediate scales and the plateau}

We have studied the average field $\Phi_N$, which is averaged over the entire
finite volume.  It would be of interest to analyse the average field $\Phi_M$
on $\Lambda_M$ for $1 \ll M \ll N$, i.e., for intermediate scales.

\smallskip\noindent{\bf Problem 1.}  Compute the scaling limit of the average field
on intermediate scales.

\smallskip
Intermediate scales have been recently studied at the infinite-volume
critical point for the $1$-component $\varphi^4$ model on $\Z^4$ with FBC  \cite{AD21},
where a generalised Gaussian limit is obtained.  For the hierarchical model
in dimensions $d \ge 4$ and for all $n \ge 1$,
we expect that an extension of our method could be used to analyse the average
field on intermediate scales at the effective critical point $\nu_{c,N}^*$ (for
either FBC or PBC), and to prove that there is a crossover from Gaussian to
non-Gaussian behaviour as $M$ increases.

\smallskip \noindent{\bf Problem 2.}
Prove the ``plateau'' phenomenon for the two-point function for all dimensions
$d \ge 4$ and for all $n \ge 1$.

\smallskip
More precisely, prove that within the critical windows, for both FBC and PBC
and for all $d \ge 4$ and $n \ge 1$,
the finite-volume two-point function obeys (with $s$-dependent constants)
\begin{equation}
\label{eq:plateau}
    \langle \varphi_0\cdot \varphi_x\rangle_{\nu_{c,N}^*+s_Nw_N,N}
    \asymp
    |x|^{-(d-2)} +
    	\begin{cases}
    		N^{1/2}L^{-2N} &(d=4) \\
    			   L^{-Nd/2} &(d>4).
    	\end{cases}
\end{equation}
The constant term $N^{1/2}L^{-2N}$ or $L^{-Nd/2}$ is the ``plateau'' term.
Note that in summation of the right-hand side of \eqref{eq:plateau} over $x \in \Lambda_N$,
the constant term dominates the sum, and also the resulting sum has the same scaling
as the susceptibility in \eqref{eq:chiwindow}.
A related problem is to prove that, for FBC and general $n$, at the infinite-volume
critical point (which is outside the FBC window) the constant plateau term
is absent.  (For $n=1,2$,  the FBC two-point function is dominated by the infinite volume one, by Griffith's inequality, so the absence follows in this case.)
For the hierarchical model with $n \ge 1$, we expect that an extension of our method
could be used to prove \eqref{eq:plateau}.

\smallskip\noindent
{\bf Note added before publication:}  Problem~2 has now been solved
by an extension of our method \cite{PS25}; the universal profile $\uf_n$
also gives the amplitude for the plateau term.

\smallskip
The effect of boundary conditions
on the decay of the two-point function of
statistical mechanical models in finite-volume above the upper critical
dimension has been widely discussed, e.g., in \cite{ZGFDG18,WY14,LM16},
and there has been some debate about the plateau.  Early numerical evidence for
a plateau with PBC can be seen in \cite[Figure~4]{LB97}.
For the Ising model in dimensions $d>4$ at the infinite-volume critical
point, with FBC the absence of the plateau has been proved
in \cite{CJN21}, and with PBC a plateau lower bound  is
proved in \cite{LPS25-Ising} (for PBC a matching upper bound remains unproved).
For (spread-out) percolation in dimensions $d>6$ with PBC, an analogous plateau
is proved to exist
throughout the critical window in \cite{HMS23}, whereas with
FBC
the absence of the plateau at the infinite-volume critical point is proved in \cite{CH20}.
For (spread-out) lattice trees and lattice animals in dimensions $d>8$,
the existence of a plateau is established in \cite{LS25a}.  A general
theory of the plateau is presented in \cite{LPS25-universal}.

\subsubsection{Extension to Euclidean models}

\smallskip\noindent{\bf Problem 3.}
Prove the results of Theorems~\ref{thm:mr-nongaussian} and \ref{thm:mr-gaussian}
for the $|\varphi|^4$ model on $\Z^d$ for dimensions $d \ge 4$ and for $n \ge 1$.
Analyse intermediate scales and
prove the ``plateau'' phenomenon for the two-point function in this Euclidean
setting.
Do the same for the Ising, XY, and Heisenberg models,
and more generally for all $N$-vector models.

\smallskip
We expect that the conclusions of Theorems~\ref{thm:mr-nongaussian} and \ref{thm:mr-gaussian}
hold \emph{verbatim} in the Euclidean setting for all these models, with the only change being
a need to  modify constant prefactors in the large-field scale $\hh_N$, the window
width $w_N$, and the FBC shift $v_N$ of the effective critical point.
However, to extend our methods to the Euclidean setting in order to prove this
would require new ideas, despite the fact that a large part of our analysis
has already been extended \cite{BS-rg-IE,BS-rg-step}.  One
challenge
in the Euclidean setting would be to
improve the large-field regulator used in \cite{BS-rg-IE,BS-rg-step}, which bounds
the non-perturbative RG coordinate by an exponentially \emph{growing}
factor (see  \cite[(1.38)]{BS-rg-IE}) rather than the exponentially \emph{decaying} factor
that we exploit
(see \eqref{eq:TG-norm-def}).  Also,
 the covariance decomposition we use for the hierarchial
model has a constant covariance at the final scale, whereas the decomposition used
in \cite{BS-rg-IE,BS-rg-step} does not have this useful feature.
This issue could likely be overcome using the decomposition in \cite[Section~3]{BPR24}
which does have a constant covariance at scale $N$ and which obeys similar estimates
to the decomposition used in \cite{BS-rg-IE,BS-rg-step}
(see \cite[Corollary~4.1]{BPR24} for $d=2$ and \cite[Proposition~3.4]{BPR24} for the
general case).

For the Ising model at the infinite-volume critical point in dimensions $d >4$,
for PBC
a lower bound on the susceptibility of order $L^{Nd/2}$ is proved in \cite{LPS25-Ising}
(a matching upper bound has not yet been proved), and for
FBC the $L^{2N}$ behaviour of the susceptibility is proved in \cite{CJN21}.

\subsubsection{Self-avoiding walk}

\noindent{\bf Problem 4.}
Prove the statements of both Theorems~\ref{thm:mr-nongaussian} and \ref{thm:mr-gaussian}
with  $n=0$ for the
weakly self-avoiding walk in dimension $d \ge 4$, both on the hierarchical lattice
and on $\Z^d$.
Much more ambitiously, prove this for the strictly self-avoiding walk on $\Z^d$
in dimensions $d \ge 4$.

\smallskip
For $n=0$,
the values of the logarithmic correction exponents in \eqref{eq:hatexponents}
are $\hat\gamma = \frac 14$ and $\rexp = \frac 14$, which indicates
window scaling $L^{-2N}N^{-1/4}$ for $d=4$
and $L^{-Nd/2}$ for $d>4$.
The scaling of $\hh_N$ remains as it is for $n \ge 1$,
namely $L^{-2N}N^{1/4}$ for $d=4$
and $L^{-Nd/4}$ for $d>4$.

It is a consequence of \cite[Theorem~1.4]{MS23} that  for $d>4$ with PBC,
the susceptibility
of weakly self-avoiding walk is at least
of order $L^{Nd/2}$ at the $\Z^d$ critical point.
A matching upper bound has not been proved.
For FBC, numerical results are presented in
\cite{ZGFDG18}.
A related conjecture for the
universality of the profile for the expected length of self-avoiding walks in dimensions $d>4$
is investigated numerically in \cite{DGGZ22}.

\subsection{Structure of proof and guide to the paper}
\label{sec:guide}

The proof of our main results in Theorems~\ref{thm:book_main_theorem_bis}--\ref{thm:mr-gaussian}
is based on a rigorous renormalisation group analysis.  The RG analysis is a multiscale
analysis based on a finite-range decomposition of the resolvent of the hierarchial
Laplacian, which is used to perform expectations via progressive integration over
scales.  The finite-range decomposition is described in Section~\ref{sec:Veff}, along with
the form of the integration at the final scale.  The difference between FBC and PBC
appears only at this final scale, where there is a mass generation effect for FBC
which is responsible for shifting $\nu_{c,N}^\free$ smaller than $\nu_{c,N}^\per$.
The mass is produced due to the fact that the random walk with the FBC Laplacian is transient on $\Lambda_N$,
with a defective distribution as discussed in Section~\ref{sec:LaplacianBC}.
The random walk with PBC is recurrent
and its distribution is not defective.  The mass generation with FBC
compared to PBC is apparent
in Lemma~\ref{lem:chi_N_good_form-v3}.
At the final scale, which is scale $\hat N$ and follows scale $N$,
the remaining field to integrate is a constant $\R^n$-valued field.

In Section~\ref{sec:effpot},  we introduce the scale-$N$ effective potential
$V_N(\varphi) = |\Lambda_N|(\frac 14 g_N |\varphi|^4 + \frac 12 \nu_N |\varphi|^2)$ and its
non-perturbative counterpart $K_N$.
In Theorems~\ref{thm:rg_main_theorem_large_m} and \ref{thm:rg_main_theorem_new}
we summarise properties of $g_N$, $\nu_N$, $K_N$
that are proved in later sections.  These theorems and further results in
Section~\ref{sec:effpot} also reveal properties of
a renormalised (squared) mass which can in some cases be negative---negative mass is what
permits negative values of $s$ in the critical window in Theorem~\ref{thm:mr-nongaussian}.

In Section~\ref{sec:pf}, we prove our main results
Theorems~\ref{thm:book_main_theorem_bis}--\ref{thm:mr-gaussian} assuming the deep
results of Section~\ref{sec:effpot}.  Because this requires only the integration
over the final scale, which is an integral over $\R^n$, the analysis in
Section~\ref{sec:pf} involves
only
calculus.

The principal part of our effort, which occupies approximately half the paper
in Sections~\ref{sec:pf1}--\ref{sec:pf2},
is to establish the properties listed in Section~\ref{sec:effpot}
for the renormalised mass and for $V_N$ and $K_N$.  This entails the consideration of
effective
potentials $V_j$ and non-perturbative coordinates $K_j$ for each scale $j=1,\ldots,N$,
and of the RG map that advances $(V_j,K_j)$ to $(V_{j+1},K_{j+1})$.
Our RG analysis in Sections~\ref{sec:pf1}--\ref{sec:pf2} follows the general approach laid out
by Bauerschmidt, Brydges and Slade in
\cite{BBS-brief}.
Sections~\ref{sec:pf1}--\ref{sec:pf2} depend in important ways on the results of \cite{BBS-brief}, and we frequently refer to and rely on \cite{BBS-brief}.

We develop important improvements to the method of \cite{BBS-brief}.
Specific improvements include:
(i) the
introduction of a large-field regulator which
captures
the decay of $K_j$ for large fields,
(ii) the control of mass derivatives of $V_j$ and $K_j$, and
(iii) the inclusion
of dimensions $d>4$ and not only $d=4$.

The latter point may appear to be minor,
since for $d>4$ the monomial $|\varphi|^4$ is irrelevant (in the RG sense)
and there are no logarithmic corrections to deal with.   However, $|\varphi|^4$ is a so-called \emph{dangerous irrelevant monomial},
and its inclusion in the analysis requires care and innovation.
The concept of a dangerous irrelevant variable was introduced in \cite{Fish83}.
In our context, this concept is reflected by the fact that,
although irrelevant, $g$ gets renormalised even when $d>4$: the constant $g_\infty$
of \eqref{eq:Bdef} accumulates contributions of all orders.
This $g_\infty$ occurs in the window scale
$w_N$ in \eqref{eq:window_choice}
and in the large-field scale $\hh_N$ in \eqref{eq:hhdef}.
It is therefore centrally important in Theorem~\ref{thm:mr-nongaussian}.
Also, the irrelevance of $|\varphi|^4$
for $d>4$ means that its contribution to the Hamiltonian scales
down exponentially in the RG flow.  Since the quartic term in the Hamiltonian
is what gives stability to integrals, and since we are working in
a regime where the quadratic term can have a bad sign
(recall that $\nu_c<0$), care is
required in order to exploit the quartic term despite its irrelevance.
We therefore must track the RG flow of the irrelevant
quartic term carefully when $d>4$.  This makes the extension of the RG method from
$d =4$ to $d>4$ nontrivial.
For Euclidean models it would be worse:
$d> 4$ also involves dangerously irrelevant terms that involve gradients of
$\varphi$.
These gradient terms do not arise in the hierarchical model.

In Section~\ref{sec:pf1}, we set up the necessary spaces and norms used
to formulate the RG map, and we state estimates which control the RG map.
The main bounds on the RG map are stated in
Theorems~\ref{thm:Phi^U_estimate}--\ref{thm:Phi^K_estimate}.

In Section~\ref{sec:pf3}, the main bounds on the RG map are used
to prove three propositions, Propositions~\ref{prop:stable_manifold}--\ref{prop:nu_c,N_mass_derivative},
which themselves are used to prove
Theorems~\ref{thm:book_main_theorem_bis}--\ref{thm:mr-gaussian}.
The proofs of Propositions~\ref{prop:stable_manifold}--\ref{prop:nu_c,N_mass_derivative}
are given in Section~\ref{sec:pf3}, subject to Proposition~\ref{prop:nu-derivs}
which captures properties of derivatives of the RG flow with respect to the initial value
of $\nu$ and with respect to the mass.
It then remains to prove the main RG results
Theorems~\ref{thm:Phi^U_estimate}--\ref{thm:Phi^K_estimate},
as well as Proposition~\ref{prop:nu-derivs}.

Section~\ref{sec:RGderivs} is devoted to the proof of Proposition~\ref{prop:nu-derivs}.
The proof relies entirely on Theorems~\ref{thm:Phi^U_estimate}--\ref{thm:Phi^K_estimate},
and, although lengthy, is essentially computational.

Section~\ref{sec:fluctation_integral_properties} proves the first of the
main RG theorems, Theorem~\ref{thm:Phi^U_estimate}.
This theorem controls the perturbative part of the RG map, by estimating the
corrections to the perturbative flow
as the effective potential $V_j$ is advanced to $V_{j+1}$.
The proof involves the estimation of polynomials measured in the norms defined in Section~\ref{sec:pf1}.  Related estimates that are needed to control the
advance of the non-perturbative
coordinate from $K_j$ to $K_{j+1}$ are also performed
in Section~\ref{sec:fluctation_integral_properties}.

Section~\ref{sec:pf2} concludes the proof of Theorem~\ref{thm:Phi^K_estimate}
and thereby concludes the proof of our main results
Theorems~\ref{thm:book_main_theorem_bis}--\ref{thm:mr-gaussian}.
It is in Section~\ref{sec:pf2} that our analysis becomes the most sophisticated.
This is not surprising, since it requires the control of an essentially infinite-dimensional
problem in a complicated space whose norm tracks derivatives of $K_{j+1}$ with
respect to $V_j$, $K_j$, the mass, and the field $\varphi$.
It is here that we rely most heavily on results from \cite{BBS-brief}, although as
mentioned previously we include new features:  the large-field regulator, mass derivatives,
and dimensions $d>4$.

The paper ends with two appendices.
Appendix~\ref{sec:ufmono} proves a monotonicity theorem for the universal profile
$\uf_n$.  Appendix~\ref{sec:SAW} proves the two
Propositions~\ref{prop:SAW-KN} and \ref{prop:WSAW} for self-avoiding walk
which demonstrate its connection to the profile $\uf_0$.

\section{Integration to the final scale}
\label{sec:Veff}

\subsection{Covariance decomposition and the hierarchical field}
\label{sec:hier}

The progressive integration underlying the multi-scale analysis of the
RG method has its basis in a decomposition of the
resolvent of the hierarchical Laplacian, which we discuss next.
Recall the definitions and notations of the hierarchical
Laplacian from Section \ref{sec:hierdefs}, which we specialise here to $\alpha=2$.
The following discussion applies for all dimensions $d \ge 1$.

Given $x\in \Lambda_N$ and given $j$ between $0$ and $N$,
let $B_j (x)$ denote the unique block in $\cB_j$ that contains $x$.
We define the matrices of symmetric operators $Q_j$ and $P_j$ on $\ell^{2} (\Lambda_N)$ by
\begin{align}
    \label{eq:Qj-def}
    Q_{j;xy}
    &=
    \begin{cases}
    L^{-dj} & B_j(x)=B_j(y) \\
    0 & B_j(x) \neq B_j(y)
    \end{cases}
    \quad
    (j=0,1,\ldots,N),
\\
    \label{eq:Pj-def}
    P_{j}
    &=
    Q_{j-1} - Q_{j}
    \quad
    (j = 1,\dots ,N).
\end{align}
As is shown in \cite[Lemma~4.1.5]{BBS-brief}, the operators $P_{1},\dots ,P_{N}, Q_{N}$ are
orthogonal projections whose ranges are disjoint and provide a direct
sum decomposition of $\ell^2(\Lambda)$.

The PBC Laplacian defined in \eqref{eq:Deltastar} can be seen after some arithmetic
to have matrix elements
\begin{align}
\label{eq:DeltaP}
    -\Delta^{\rm P}_{N;0x}
    &=
    \begin{cases}
    \frac{1-L^{-d}}{1-L^{-(d+2)}}(1-L^{-(d+2)N})  & (x=0)
    \\
    -\frac{L^2-1}{1-L^{-d}} L^{-(d+2)j_{0x}}
    - \frac{1-L^{-d}}{1-L^{-(d+2)}}  L^{-(d+2)N}
    & (x \in \Lambda_N\setminus\{0\}).
    \end{cases}
\end{align}
The above is identical to the representation of the hierarchical Laplacian in \cite[((4.1.8)]{BBS-brief}.
Thus \cite[(4.1.7)]{BBS-brief} provides a representation of the Laplacian
as a sum over scales, namely
\begin{equation}
    -\Delta_{N}^{\rm P} = \sum_{j=1}^N L^{-2(j-1)}P_j.
\end{equation}
By \eqref{eq:newDeltaF}, the FBC Laplacian can therefore be written as
\begin{equation}
    -\Delta_{N}^{\rm F} =
    \sum_{j=1}^N L^{-2(j-1)}P_j+ qL^{-2 N}Q_N.
\end{equation}

Let
\begin{align}
\label{eq:gamma_j_per}
	\gamma_j(\ka)
	&=\frac{L^{2 (j-1)}}{1+\ka L^{2(j-1)}}.
\end{align}
It follows exactly as in \cite[Proposition~4.1.9]{BBS-brief} that
the resolvents of $-\Delta_N^{\per}$ and $-\Delta_N^{\free}$ have the decompositions
\begin{align}
\label{eq:PBCdecomp}
    (-\Delta_N^{\rm P}+\ka)^{-1}
    &= \sum_{j=1}^N \gamma_j(\ka)
    P_j + \ka^{-1} Q_{N}, \\
\label{eq:fbc_resolvent}
    (-\Delta_N^{\rm F}+\ka)^{-1}
    &=
    \sum_{j=1}^N \gamma_j(\ka)
    P_j
    + (\ka+qL^{-2 N})^{-1} Q_{N}.
\end{align}
In order to ensure that $\gamma_j(\ka)>0$
for all $j\in\{1,2,\cdots,N\}$,
we assume that $a>-L^{-2(N-1)}$.
We also assume for now
that $a\neq 0$ and $a \neq -\qLap L^{-2N}$ for the coefficients of $Q_N$ terms,  respectively,  though we
will find a way to relax this in Section~\ref{sec:integration-final-scale}.
We use the variable $\ka$, rather than $m^2$ as in \cite[Proposition~4.1.9]{BBS-brief},
because unlike in \cite{BBS-brief}
we allow both positive and negative values of the ``mass'' parameter $\ka$.
The $qL^{-2N}$ term in the coefficient of $Q_N$ for FBC, which is absent for PBC,
is ultimately the source of the shift in the effective critical point for FBC.

The matrices defined by
\begin{align}
\label{eq:Cjdef}
	C_j(\ka) &= \gamma_j(\ka)P_j, \qquad
	C_{\leq N}(\ka) = \sum_{j=1}^N C_j(\ka),
\end{align}
are symmetric, real, and positive semi-definite. At the last scale we define
\begin{align}
	C^*_{\hat N}(\ka) &=
	Q_N \times
	\begin{cases}
	\ka^{-1}  \quad & (* = \per) \\
	(\ka + qL^{-\alpha N})^{-1} &(* = \free),
	\end{cases}
\end{align}
so that
\begin{equation}
 (-\Delta_N^{*}+\ka)^{-1}
    = \CleN + C^*_{\hat N}.
\end{equation}
The matrix $C^{\per}_{\hat N}$ is positive semi-definite for $\ka > 0$
and negative semi-definite for $\ka<0$, whereas
$C^{\free}_{\hat N}$ is positive semi-definite for $\ka > -qL^{-2 N}$ and negative semi-definite for $\ka < -qL^{-2N}$.
Note that $-L^{-2(N-1)}<-qL^{-2N}$ when $L$ is large (then $\qLap$ is close to $1$),
so $\ka > -qL^{-2 N}$ is
a greater restriction than the requirement $\ka >-L^{-2(N-1)}$ imposed in the
previous paragraph.

By definition, $P_j\one=0$
where $\one$ is the constant field $\one(x) = 1$ for all $x \in \Lambda_N$,
and therefore also $C_j \one =0$ for $j=1,\ldots,N$.
The non-interacting ($g=0$) massive
susceptibility $\chi_{N}^{*,0}(\ka) =(-\Delta^{*}+\ka)^{-1}\one$
is therefore equal to $C^*_{\hat N}\1$, and hence
\begin{equation}
    \chi^{{\rm P},0}_N(\ka) =\frac{1}{\ka},
    \qquad
    \chi^{\free,0}_N(\ka) = \frac{\qLap^{-1} L^{2N}}{1+ \ka \qLap^{-1} L^{2N}}.
\end{equation}
The infinite-volume critical value for $g=0$ is $\ka=0$, and
\begin{equation}
    \chi^{{\rm P},0}_N(0) =\infty,
    \qquad
    \chi^{\free,0}_N(0) = q^{-1}L^{2N},
    \qquad
    \chi^{\free,0}_N(-qL^{-2N}) = \infty.
\end{equation}
Thus we see that the non-interacting PBC critical value agrees with the infinite-volume value,
whereas the FBC critical value is shifted by $-\qLap L^{-2N}$.
This is a harbinger of the shift in \eqref{eq:nu_c,N_def} for the interacting
model, and it is ultimately responsible for that shift in the effective FBC critical
value and its window.

Now an insignificant adaption of \cite[Proposition~4.1.9]{BBS-brief}
from $m^2>0$ to $\ka > -L^{-2 (N-1)}$ shows that, for
independent Gaussian fields
$\zeta_j$  with covariance $C_j$,
the field $\varphi=\zeta_1+\cdots+\zeta_N$ is a \emph{hierarchical field}
in the sense that:
\begin{enumerate}
\item
 on any two blocks $B,B'\in \cB_j$ that are not identical,
$\zeta_j|_B$ and $\zeta_j|_{B'}$ are independent
 for all $j=1,\ldots,N$, and
\item
on any block $b \in \cB_{j-1}$
 the field $\zeta_j$ is constant, i.e.,
$\zeta_{j,x}=\zeta_{j,y}$ for all $x,y\in b$ and for all $j=1,\ldots,N$.
\end{enumerate}
In addition, by \cite[Exercise~4.1.6]{BBS-brief}, the field $\zeta_j$ vanishes
with probability $1$ when summed over a block at scale $j$, namely:
\begin{equation}
\label{eq:zeta-zero-sum}
    \sum_{b \in \cB_{j-1}(B)}\zeta_j(b) = 0 \quad \text {a.s. for all $B \in \cB_{j}$,}
\end{equation}
where $\cB_{j- 1}(B)$ is the set of $(j-1)$-blocks inside $B$.
Degenerate Gaussian fields (with non-invertible covariance) are discussed
in \cite[Section~2.1]{BBS-brief}.
In the above decomposition of the field $\varphi$ we exclude a last field corresponding to $C_{\hat{N}}^*$,
which is dealt with in Section~\ref{sec:integration-final-scale}.

\subsection{Integration at the final scale}
\label{sec:integration-final-scale}

Throughout this section we consider both FBC and PBC
but to lighten the notation we indicate the boundary condition only sometimes for emphasis.
Also, we often do not make the dependence on $\ka$ or $g$ explicit in the notation.
To keep the covariance matrices $C_j(\ka)$
and $C_{\hat N}^*(\ka)$
positive semi-definite, we restrict $\ka$ initially to $\ka >0$
for PBC, and to $\ka+qL^{-2N}>0$ for FBC, though later we will show that these restrictions
can be relaxed.

Our goal is to study expectations (with either FBC or PBC)
\begin{equation}
\label{eq:def_phi_4_meas}
    \langle F \rangle_{\nu,N}
    =
    \frac{1}{Z_{\nu,N}}
    \int_{(\R^{n})^{\Lambda_N}}F(\varphi) e^{-H_{ \nu,N} (\varphi)} d\varphi,
\end{equation}
where the Hamiltonian is
\begin{align}
\label{eq:def_phi_4_hamilt}
H_{ \nu,N} (\varphi) = \frac{1}{2} (\varphi,  (-\Delta_{N} ) \varphi)
+  V_0(\varphi),
\end{align}
with
\begin{equation}
\label{eq:V0def}
    V_0(\varphi)
    =
    \sum_{x\in \Lambda_N}
    \Big(\frac{1}{4} g  |\varphi_x|^4
    +
    \frac{1}{2} \nu |\varphi_x|^2
    \Big) .
\end{equation}
The hierarchical Laplacian in \eqref{eq:def_phi_4_hamilt} acts component-wise
on the field $\varphi$.
Let
\begin{align}
    &Z_0 (\varphi) = \exp (- V_0(\varphi)).
\label{eq:Z_0_definition}
\end{align}
With $\E_C$ denoting expectation with respect to the Gaussian measure with
covariance $C = (-\Delta_N + \ka)^{-1}$, the definitions lead to
\begin{equation}
\label{eq:EF}
    \langle F \rangle_{\nu+ \ka,N}
    =
    \frac{\E_C FZ_0}{\E_C Z_0}.
\end{equation}

The Gaussian expectation $\E_C$ can be computed in two steps as
\begin{equation}
\label{eq:2int}
    \E_C F = \E_{C_{\hat N}} \E_{C_{\le N}} F(\varphi + \hat\varphi),
\end{equation}
where $\E_{C_{\le N}}$ involves integration over $\varphi$, and $\E_{C_{\hat N}}$
involves integration over $\hat\varphi$.
In particular,
\begin{equation}
    \E_C   Z_0(\psi)
    e^{(J\one ,\psi)}
    =
    \E_{C_{\hat N}} \E_{C_{\le N}}  Z_0(\varphi+\hat\varphi) e^{(J\one ,\varphi+\hat\varphi)},
\end{equation}
where on the left-hand side $\psi$ is a dummy integration variable,
and on the right-hand side $\one$ is (again) the constant field with value $1$.
By completing the square in the integral $\E_{C_{\le N}}$
(as in \cite[Exercise~2.1.10]{BBS-brief}),
and since $\CleN \one = 0$, we see that
\begin{align}
    \E_{C_{\le N}}  Z_0(\varphi+\hat\varphi) e^{(J\one ,\varphi+\hat\varphi)}
    &=
    e^{(J\one,\hat\varphi)}e^{\frac12(J \one ,C_{\leq N}J\one)}
    \E_{C_{\le N}}Z_0(\varphi + \hat\varphi + C_{\leq N}J\one)
    \nnb & =
    e^{(J\one,\hat\varphi)}
    \E_{C_{\le N}}Z_0( \varphi + \hat\varphi ).
\end{align}
With the definition
\begin{equation}
\label{eq:Z_N_definition}
    Z_N(\hat\varphi ; \ka) =  Z_N(\hat\varphi) = \E_{C_{\le N}}Z_0(\varphi + \hat\varphi ),
\end{equation}
this leads to the identity
\begin{equation}
\label{eq:ECh}
    \E_C   Z_0(\psi) e^{(J\one ,\psi)} =
    \E_{C_{\hat N}}e^{(J\one,\hat\varphi)} Z_N(\hat\varphi).
\end{equation}

For $\ka >0$ with PBC, and for $\ka >-qL^{-2N}$ with FBC, the
matrix $C^*_{\hat N}(\ka)$ has rank $1$ so the expectation $\E_{C_{\hat N}}$
is supported on constant fields
(the \emph{zero mode})
and therefore reduces to an integral over $\R^n$.
This is reflected by our notation in
Lemma~\ref{lem:chi_N_good_form-v3} (and also later) where for simplicity inside integrals
we evaluate $Z_N$ on $y\in \R^n$ rather than on a constant field $y\in (\R^n)^{\Lambda_N}$.
The lemma
relaxes the restrictions on $\ka$ to the interval
$\ka \in (-L^{-2(N-1)},\infty)$ on which the covariance $C_{\leq N}(\ka)$
is positive semi-definite.
We occasionally write
\begin{equation}
    \Omega_N = L^{dN}
\end{equation}
for the volume of $\Lambda_N$.

\begin{lemma}
\label{lem:chi_N_good_form-v3}
Let $d \ge 1$, $g >0$, $\nu \in \R$,
$\ka \in (-L^{-2(N-1)},\infty)$,
and $J \in \R^n$.
Then
\begin{align}
\label{eq:S_p_good_1-2}
\big\langle e^{J\cdot \Phi_{N} } \big\rangle_{\nu+ \ka, N}^{\per}
&=  \frac{\int_{\R^n} e^{J\cdot y} \; Z_N (y) e^{-\frac{1}{2} \ka \Omega_N |y|^2} dy}{\int_{\R^n} Z_N (y) e^{-\frac{1}{2} \ka \Omega_N |y|^2} dy}
,   \\
\label{eq:S_p_good_1-2_fbc}
 \big\langle e^{J\cdot \Phi_{N} } \big\rangle^{\free}_{ \nu+ \ka, N}
&=  \frac{\int_{\R^n} e^{J\cdot y} \; Z_N (y) e^{-\frac{1}{2} (\ka +  q L^{-2N}  ) \Omega_N |y|^2} dy}{\int_{\R^n} Z_N (y) e^{-\frac{1}{2} (\ka +  q L^{-2N} ) \Omega_N |y|^2 dy}}
,
\end{align}
where $Z_N$ is defined in \eqref{eq:Z_N_definition} using the covariance $C_{\leq N}(\ka)$
and with $\nu$ in \eqref{eq:V0def}.
\end{lemma}

\begin{proof}
We give the proof for PBC; the proof for FBC is completely analogous.
Consider first the case $\ka>0$ (or $\ka +qL^{-2N}>0$  for FBC),
and set $\sigma^2 = \ka^{-1}\Omega_N^{-1}$.
Our starting point is \eqref{eq:EF} and \eqref{eq:ECh}, from which we see that
\begin{equation}
\label{eq:LTratio}
    \big\langle e^{J\cdot \Phi_{N} } \big\rangle_{g, \nu+ \ka, N}
    =
    \frac{\E_{C_{\hat N}}e^{(J\one,\hat\varphi)/\Omega_N}
    Z_N(\hat\varphi)}{\E_{C_{\hat N}} Z_N(\hat\varphi)}.
\end{equation}
Now we use the fact that under $C_{\hat N}$ the field
$\hat\varphi$ has the same distribution as the constant field $y\one$ where $y\in\R^n$
has independent components each with a normal distribution with mean zero and variance
$\sigma^2$.  We write this Gaussian measure on $\R^n$ as $d\mu_{\sigma^2}(y)$.
The exponent in the right-hand side of \eqref{eq:LTratio} simplifies to $J \cdot y$,
so
\begin{equation}
\label{eq:proof_S_p_good_inter_2}
	\E_{C_{\hat N}}e^{(J\one,\hat\varphi)} Z_N(\hat\varphi)
	=
	\int_{\R^n}e^{J\cdot y}  Z_N(y) d\mu_{\sigma^2}( y).
\end{equation}
For $J = 0$ this is simply
\begin{equation}
\E_C   Z_0(\varphi) =  \int_{\R^n}Z_N(y) d\mu_{\sigma^2}( y),
\end{equation}
and the above three equations give the desired result when $\ka>0$.

To extend the result to $\ka \in (-L^{-2(N-1)},0]$,
we use an analytic continuation argument,
as follows.
By its definition in \eqref{eq:def_phi_4_meas}, the
function $f\colon \ka \mapsto \langle F \rangle_{\nu + \ka,N}$ is meromorphic in $\ka\in\C$,
since both numerator and denominator in
\eqref{eq:def_phi_4_meas} are entire functions of $\nu$.
It is sufficient to show that the function of $\ka$ appearing
in the right-hand side of \eqref{eq:S_p_good_1-2}, which we
denote by $g$, is meromorphic on the domain
$\mathcal{H}_N=\{z \in \C, \; {\rm Re}(z) >  -  L^{-2(N-1)}\}$,
since this implies that the difference $f-g$ is
meromorphic on $\mathcal{H}_N$ and equal to zero on $[0, \infty)$, so
identically zero on $\mathcal{H}_N\supset(-L^{-2(N-1)},\infty]$.

To see that $g$ is meromorphic, we first extend the definition of $Z_N(y)$
from a function of real $\ka$
to a function of complex $\ka$ by noting that the measure
\begin{equation}
\label{eq:complex_measure}
	\exp\big(-\frac{1}{2} (\varphi,\CleN(\ka)^{-1}\varphi) \big)\,d\varphi
\end{equation}
is integrable over $\R^n$ provided that $a\in\mathcal{H}_N$, since this
 ensures that ${\rm Re}(\gamma_j(\ka))>0$
for all $j\leq N$.
This measure is
supported on the orthogonal complement of the kernel of $\CleN (\ka)$ as in
\cite[(2.1.3)]{BBS-brief}, which is the same as
${\rm Im}(Q_N)^{\perp}$ by the observation below \eqref{eq:Pj-def}.
It then follows that $Z_N(y;\ka)$ is holomorphic
on $\mathcal{H}_N$ by
application of Morera's (and Fubini's) theorem together with the facts that
$\ka \mapsto \exp(- \frac{1}{2} (\varphi,\CleN(\ka)^{-1} \varphi))$ is
holomorphic on $\mathcal{H}_N$ and that
$\exp(-\frac14 g\sum_x|\varphi_x|^4 - \frac 12\nu \sum_x|\varphi_x|^2)$
is bounded and integrable
with respect to the measure \eqref{eq:complex_measure}.
We now wish to conclude that the two functions
\begin{align}
	g_1(\ka) &= \int_{\R^n}e^{J \cdot y}Z_N(y;\ka)e^{-\frac12 \ka \Omega_Ny^2} dy,
\qquad
	g_2(\ka) = \int_{\R^n}Z_N(y;\ka)e^{-\frac12 \ka \Omega_Ny^2} dy
\end{align}
are holomorphic in $\ka\in\mathcal{H}_N$, since then their ratio $g = g_1/g_2$ is meromorphic on $\mathcal{H}_N$, as desired.
To see this, by another application of Morera's theorem
it is sufficient to prove that
\begin{equation}
	\int_{\R^n}F(y)Z_N(y;\ka)e^{-\frac12 \ka \Omega_N  y^2}dy
\end{equation}
is integrable for any function $F$ growing at most exponentially. This last fact follows from
\begin{align}
 \Big|
 \int_{\R^n \times {\rm Im}(Q_N)^{\perp}}
	F(y)Z_0(\varphi + y\1)&e^{-\frac12(\varphi,\CleN(a)^{-1}\varphi)
	 -
	\frac12\ka \Omega_N y^2}d (y \otimes \varphi) \Big| \nnb
	&\leq
    \int_{\R^n \times {\rm Im}(Q_N)^{\perp}}
    |F(y)|Z_0(\varphi + y\1)e^{\frac12|\ka |\Omega_N y^2}d (y \otimes \varphi),
\end{align}
by noting that $\R^n \times \text{Im}(Q_N)^\perp \cong
\R^{n\Lambda_N}$ (under $(y, \varphi) \mapsto y\one + \varphi$)
and that the integrand is integrable over $\R^{n\Lambda_N}$ since
$g>0$ and
$\sum_{x} |\varphi_x + y |^4 \ge  \sum_{x} ( |\varphi_x |^4 + |y |^4 )$
(the dot product $\varphi\cdot y \1$ vanishes).
This completes the proof.
\end{proof}

We  use multi-index notation.  For $p \ge 1$, let $A_{p}$ denote the
set of $p$-tuples $\alpha = (\alpha_1,\ldots,\alpha_p)$ with each $\alpha_i\in \{1,\ldots,n\}$.
Given $\varphi : \Lambda_N \to \R^n$ and $\alpha \in A_p$, we write
\begin{equation}
    \Phi_{N}^\alpha
    = \frac{1}{\Omega_N^{p}}\sum_{x_1,\ldots,x_p\in \Lambda_N}
    \varphi_{x_1}^{(\alpha_1)} \cdots \varphi_{x_p}^{(\alpha_p)},
\end{equation}
where $\varphi_{x_i}^{(\alpha_i)}$ denotes the $\alpha_i^{\rm th}$ component
of $\varphi_{x_i}\in \R^n$.
Similarly, for $y \in \R^n$ we write $y^\alpha = y^{(\alpha_1 )}\cdots y^{(\alpha_p)}$.

\begin{corollary}
\label{cor:chi_N_good_form-v2}
Let $d\geq 1$, $g >0$, $\nu \in \R$,
$p\geq1$, $\alpha \in A_p$,  and $\ka\in (-L^{-2(N-1)},\infty )$.
Then
\begin{align}
\label{eq:S_p_good_1_pbc}
\big\langle \Phi_N^\alpha \big\rangle_{\nu+ \ka, N}^{\per}
&=  \frac{\int_{\R^n} y^\alpha \, Z_N (y)
e^{-\frac{1}{2} \ka \Omega_N |y|^2} dy}{\int_{\R^n} Z_N (y) e^{-\frac{1}{2} \ka \Omega_N |y|^2} dy}
,   \\
\label{eq:S_p_good_1_fbc}
 \big\langle \Phi_N^\alpha \big\rangle^{\free}_{ \nu+ \ka, N}
&=  \frac{\int_{\R^n} y^\alpha \, Z_N (y)
e^{-\frac{1}{2} (\ka +  q L^{-2N}  ) \Omega_N |y|^2} dy}{\int_{\R^n} Z_N (y) e^{-\frac{1}{2} (\ka +  q L^{-2N} ) \Omega_N |y|^2 dy}}
,
\end{align}
where $Z_N$ is defined in \eqref{eq:Z_N_definition} using the covariance $C_{\leq N}(\ka)$
and with $\nu$ in \eqref{eq:V0def}.
\end{corollary}

\begin{proof}
The identities \eqref{eq:S_p_good_1_pbc}--\eqref{eq:S_p_good_1_fbc}
follow simply by differentiation with respect to $J$ in
Lemma~\ref{lem:chi_N_good_form-v3}.
\end{proof}

\section{The effective potential and the renormalised mass}
\label{sec:effpot}

The proof of our main results is given in Section~\ref{sec:pf},
based on Theorems~\ref{thm:rg_main_theorem_large_m},
\ref{thm:mass_trick_big_mass}, and
\ref{thm:rg_main_theorem_new} which are proved by extending the RG analysis of the 4-dimensional hierarchical $n$-component $|\varphi|^4$ model from \cite{BBS-brief}.
The proofs of Theorems~\ref{thm:rg_main_theorem_large_m}--\ref{thm:rg_main_theorem_new} are
given in Sections~\ref{sec:pf1}--\ref{sec:pf2}.

\subsection{The effective potential at the final scale}

To apply Lemma~\ref{lem:chi_N_good_form-v3} and its corollary,
we need a good understanding of $Z_N$.
Since the choice of BC enters only in the last integral in
Lemma~\ref{lem:chi_N_good_form-v3}, $Z_N$ itself is independent of the boundary condition.
Theorems~\ref{thm:rg_main_theorem_large_m}--~\ref{thm:rg_main_theorem_new},
concern the behaviour of $Z_N$, and are thus free of any mention to BC.
We express $Z_N$ in terms of a scale-$N$ effective potential
$V_N$, together with a non-perturbative coordinate $K_N:\R^n \to \R$.

Theorem~\ref{thm:rg_main_theorem_large_m} produces a critical value $\nu_c(\ka)$,
for each choice of $\ka \ge 0$, such that $Z_N$ can be controlled for all scales $N$
when $Z_N$ is defined with the initial $\nu$-value in $V_0$ tuned to equal $\nu_c(\ka)$.
It  extends  the results from \cite{BBS-brief} by the
inclusion of an additional large-field decay of the form $e^{-\kappa g_N \Omega_N|\varphi|^4}$ in the bounds on $K_N$
in \eqref{eq:K_bound_at_nu_c_bis-4} and \eqref{eq:K_bound_at_nu_c_bis-5}, by the
addition of the case $d>4$ to the statement, and finally by the control of the
mass dependence of the critical point
via its derivative.
Theorem~\ref{thm:mass_trick_big_mass} proves the existence and provides
the asymptotic behaviour of a
renormalised mass $m_\eps^2 \geq 0$ such that $\nu_c(0) + \eps = m^2_\eps +\nu_c(m^2_\eps)$.
Theorem~\ref{thm:rg_main_theorem_new} is analogous
to Theorem~\ref{thm:rg_main_theorem_large_m}
but now determines the behaviour of $Z_N$ also for negative and
small $N$-dependent mass $\ka$. The counterpart of $\nu_c(\ka)$ in
Theorem~\ref{thm:rg_main_theorem_new} is $\nu_{0,N}(\ka)$ which is also well-understood by
estimates on its derivative.
An adequate notion of renormalised mass, i.e., a version
of Theorem~\ref{thm:mass_trick_big_mass} for $\nu_{0,N}(\ka)$ rather than $\nu_c(\ka)$, is developed later, in Section~\ref{sec:renmass}.
In all three theorems, logarithmic corrections occur
for $d=4$ but not in higher dimensions.

Recall that $\Omega_N=L^{dN}$ denotes the volume.
Given $g_N,\nu_N\in\R$, we define
\begin{align}
V_N (\varphi)
=
\Omega_N (  \textstyle{\frac{1}{4}} g_N |\varphi|^4  + \textstyle{\frac12} \nu_N |\varphi|^2 )
\qquad  (\varphi \in \R^n).
\label{eq:V_N_form}
\end{align}
We also define the volume-dependent mass interval
\begin{align}
\label{eq:INdef}
	\II_{N-1} (0) = ( -\textstyle{\frac12} L^{-2(N-1)} ,
    \textstyle{\frac12} L^{-2(N-1)}  ) \subset \R.
\end{align}
To state estimates on the non-perturbative coordinate $K_N$, we define
\begin{equation}
	\label{eq:eNdef}
	e_N =
	\begin{cases}
		N^{-3/4} & (d=4)\\
		g^{3/4} L^{-3(d-4)N/4} & (4 < d \le 12)
		\\
		g^{3/4}L^{-Nd/2} & (d>12).
	\end{cases}
\end{equation}
We believe that the second option actually applies for all $d > 4$,
and that for $d>12$
our estimates on $K_N$ are not sharp (they are, however, sufficient for our needs).

For the statement of part~(i) of the following theorem,  given squared mass $\ka \in \R$, we define the \emph{mass scale} $j_{\ka}\in \Z$ by
\begin{equation}
\label{eq:mass-scale}
    j_{\ka} = \max\{ j \in \Z :L^{2j} \ka  \leq 1\},
\end{equation}
with the degenerate case $j_{\ka} = \infty$ if $\ka \leq 0$.
By definition, $j_{\ka} <0$ when $\ka > 1$.
We do not state estimates on the \emph{vacuum energy} $u_N$
appearing in \eqref{eq:Z_N_form_bis},  as it cancels in
numerator and denominator when calculating expectations as
in \eqref{eq:S_p_good_1_pbc}--\eqref{eq:S_p_good_1_fbc}.
A detailed computation of $u_N$ is given in \cite[Theorem~2.4]{PS25}.

\begin{theorem}[Final scale for nonnegative mass]
\label{thm:rg_main_theorem_large_m}
Let $d \ge 4$ and $n \ge 1$.
Fix $L$ sufficiently large, $g >0$ sufficiently small.
There exists a continuous strictly increasing function  $\nu_{c} (\ka)$ of $\ka \geq 0$
which is continuously differentiable for $\ka >0$ if $d=4$ and for $\ka \geq 0$ if $d>4$,
and there exist
$(u_N, g_N, \nu_N, K_N)$ (all depending on $\ka$), such that $Z_N$ defined by
\eqref{eq:Z_N_definition} with the choice $\ka$ for the squared mass in the covariance $\CleN$
and with the choice $\nu = \nu_{c} (\ka)$ for the $\nu$ in $Z_0$, satisfies
\begin{align}
Z_N (\varphi) = e^{-u_N |\Lambda_N|} \big( e^{-V_N (\varphi) } + K_N (\varphi) \big)
\label{eq:Z_N_form_bis}
\end{align}
for all $\varphi\in\R^n$ with, uniformly in $\ka \geq 0$ and for some
$\kappa=\kappa(n)$:
\begin{enumerate}
\item
For $d=4$, with $e_N$ defined in \eqref{eq:eNdef}, there exists $A_{4}>0$ such that
\begin{align}
	g_N   &  =
    \begin{cases}
	 (BN)^{-1}(1+O(N^{-1}\log N))
    		& (\ka =0)
	    \\
	 B^{-1} (N\wedge j_a)^{-1}(1 + O((N\wedge j_a)^{-1}\log (N\wedge j_a)))
			& (j_{\ka} \in [1,\infty))
		\\
	 g + O(g^2)
	 		&   (j_{\ka} \leq 0) 	  ,
    \end{cases}
    \label{eq:gNB_bis-4}
	\\
	|\nu_N | &
		\le  O(g_NL^{-2N})   ,
		\label{eq:nu_j_bound_bis-4}  \\
	|K_N (\varphi)|
		& \le  O( e_N)
    e^{-\kappa g_N \Omega_N |\varphi|^4},
			\label{eq:K_bound_at_nu_c_bis-4}\\
	\frac{d}{d \ka} \nu_c (\ka)
		&= A_{4}
		(\log a^{-1})^{\hat{\gamma}} (1+o(1))
		\qquad \text{as } \ka \downarrow 0.
\label{eq:nu_c_deriv_4}
\end{align}
\item
For $d>4$,  with $e_N$ defined in \eqref{eq:eNdef},
there exist $A_{d}>1$ and a continuous function $g_\infty$ of $\ka \geq 0$ satisfying $g_\infty(\ka) = g + O(g^2)$, such that
\begin{align}
	\label{eq:gNB_bis-5}
	g_N
	&= g_\infty
    + O(g^2 L^{-(d-4)N}), \\
	|\nu_N| & \le  O(gL^{-(d-2)N})   ,
	\label{eq:nu_j_bound_bis-5}  \\
	|K_N (\varphi)| & \le  O( e_N )
    e^{-\kappa g_N \Omega_N |\varphi|^4 },
	\label{eq:K_bound_at_nu_c_bis-5} \\
	\frac{d}{d \ka} \nu_c (\ka)
	&= (A_{d}-1)(1+o(1))
	\qquad \text{as } \ka \downarrow 0.
\label{eq:nu_c_deriv_5}
\end{align}
Also, for $\ka \in [0,1]$,
the following bound (useful when $\ka =o(L^{-N})$)
relates the massive $g_N$ and massless $g_\infty$:
\begin{equation}
	\label{eq:gNB_bis_zero}
	g_N(\ka)=
	g_\infty(0) +O(g^2L^{-(d-4)N}) + O(g^2 \ka L^{N}).
\end{equation}
\end{enumerate}
\item
Finally,
the critical value $\nu_c (0)$ and constants  $A_{d}$ (for $d\geq 4$) satisfy the asymptotic
formulas \eqref{eq:A,B_definition_bis} as $g \downarrow 0$.
\end{theorem}

\begin{theorem}
\label{thm:mass_trick_big_mass}
In the setting of Theorem~\ref{thm:rg_main_theorem_large_m},
there exists
a strictly increasing function $m^2_\eps$ of $\eps \ge 0$, with $m_0^2=0$,
such that
\begin{align}
	\nu_c(0) + \eps &= \nu_c(m_\eps^2) + m_\eps^2,
\end{align}
with
\begin{equation}
\label{eq:renorm_big_mass_bis}
	m^2_\eps \sim
		\begin{cases}
			A_{4}^{-1}\,\eps\,(\log \eps^{-1})^{-\hat \gamma} & (d=4) \\
			A_{d}^{-1}\,\eps & (d>4)
		\end{cases}
\qquad
	(\text{as } \eps \downarrow 0).
\end{equation}
\end{theorem}

\begin{theorem}[Final scale for mass  $\ka \in \II_{N-1} (0)$]
\label{thm:rg_main_theorem_new}
Let $d \ge 4$ and $n \ge 1$.
Fix $L$ sufficiently large, $g >0$ sufficiently small.
There exists a continuously differentiable strictly increasing function
$\nu_{0,N} (\ka)$ of  $\ka \in \II_{N-1} (0)$ with $\nu_{0,N}(0) = \nu_c (0)$,
and there exist
$(u_N, g_N, \nu_N, K_N)$  (all depending on $\ka$), such that
$Z_N$ defined by
\eqref{eq:Z_N_definition} with the choice $\ka$ for the mass in the covariance $\CleN $
and with the choice $\nu = \nu_{0,N} (\ka)$ for the $\nu$ in $Z_0$, satisfies
\begin{align}
Z_N (\varphi) = e^{-u_N |\Lambda_N|} \big( e^{-V_N (\varphi) } + K_N (\varphi) \big)
\label{eq:Z_N_form_ter}
\end{align}
for all $\varphi\in\R^n$ with, uniformly in $\ka$ and for some
$\kappa=\kappa(n)$:
\begin{enumerate}
\item
For $d=4$, with $e_N$ defined in \eqref{eq:eNdef} and with $\constF = O(g)$,
\begin{align}
	g_N &  = \frac{1}{BN} + O(N^{-2}\log N),
	\label{eq:gNB_ter}
	\\
	|\nu_N| & \le  O(g_NL^{-2N})  ,
	\label{eq:nu_j_bound_ter}  \\
	|K_N (\varphi)| & \le  O( e_N )
    e^{-\kappa g_N \Omega_N |\varphi|^4},
	\label{eq:K_bound_at_nu_c_ter}
\\
	\frac{d}{d\ka } \nu_{0,N} (\ka) &=
	A_{4} (\log L^2)^{\hat{\gamma}} N^{\hat{\gamma}}
    + \constF -1
    +O(N^{\hat \gamma -1}).
	\label{eq:nu_c,N_derivative}
\end{align}
\item
For $d>4$, with $e_N$ defined in \eqref{eq:eNdef},
\begin{align}
	g_N(a) &  = g_\infty(0) + O(g^2 L^{-N}),
	\label{eq:gNB_d_ter}
	\\
	|\nu_N| & \le  O(gL^{-(d-2)N})   ,
	\label{eq:nu_j_bound_d_ter}  \\
	|K_N (\varphi)| & \le  O( e_N )
    e^{-\kappa g_N \Omega_N |\varphi|^4},
	\label{eq:K_bound_at_nu_c_d_ter}
\\
	\frac{d}{d\ka } \nu_{0,N} (\ka) &=
    (A_{d}-1) ( 1 + O(L^{-N}) ) .
	\label{eq:nu_c,N_derivative_ter}
\end{align}
\end{enumerate}
\end{theorem}

We connect the two critical values $\nu_c$ and $\nu_{0,N}$ by defining
\begin{equation}
\label{eq:nu_1_N_def}
	\nu_{1,N}(\ka) =
	\begin{cases}
		\nu_c(\ka) & (\ka  \geq 0)  \\
		\nu_{0,N}(\ka) & ( \ka \in (-\frac12 L^{-2(N-1)} ,  0) ).
	\end{cases}
\end{equation}
By Theorems~\ref{thm:rg_main_theorem_large_m} and \ref{thm:rg_main_theorem_new},
the function $\nu_{1,N}:(-\frac12 L^{-2(N-1)}, \infty) \to \R$ is continuous,
and it is also differentiable except possibly at  $0$.

\subsection{Effective critical points}
\label{sec:cp}

We define the effective critical points for PBC and FBC by
\begin{align}
\label{eq:nu_PF_def}
	\nu^\per_{c,N}
	&=  \nu_c (0),   	\qquad
	\nu^\free_{c,N}
	  = \nu_{0,N}(-qL^{-2N}) - qL^{-2N}.
\end{align}
The choice of $\nu_c(0)$ as the effective critical point for PBC is natural.
For FBC, we first observe that
$\qLap = 1+O(L^{-d})$ by \eqref{eq:qLapdef}, so $- \qLap L^{-2N} \in \II_{N-1}(0)$ and thus $\nu^\free_{c,N}$ is well-defined.

To explain our choice of $\nu^{\free}_{c,N}$,
we recall from \eqref{eq:S_p_good_1-2_fbc} that
\begin{align}
	\big\langle e^{J\cdot \Phi_{N} } \big\rangle^{\free}_{ \nu+ \ka, N}
	&=  \frac{\int_{\R^n} e^{J\cdot y} \; Z_N (y) e^{-\frac{1}{2} (\ka +  q L^{-2N}  ) \Omega_N |y|^2} dy}{\int_{\R^n} Z_N (y) e^{-\frac{1}{2} (\ka +  q L^{-2N} ) \Omega_N |y|^2 dy}}		.
	\label{eq:S_p_good_1-2_fbc-bis}
\end{align}
The massless choice in
\eqref{eq:S_p_good_1-2_fbc-bis} is $\ka = - q L^{-2N}$, since
it causes the exponential factor to become simply $e^0=1$.
With this choice, to apply Theorem~\ref{thm:rg_main_theorem_new} we must tune $\nu$
to the correct value $\nu = \nu_{0,N}(-qL^{-2N})$ in order to represent
$Z_N$ in terms of $e^{-V_N}$ and $K_N$ with good estimates.
In particular $K_N \to 0$.  With this tuning, we obtain
\begin{equation}
    \nu+ \ka = \nu_{0,N}(-qL^{-2N}) -qL^{-2N},
\end{equation}
which is our definition of the FBC effective critical point $\nu^\free_{c,N}$.

The following corollary and remark justify the formulas for
$\nu_{c,N}^{\free}$ in \eqref{eq:nu_c,N_def},
where we recall from \eqref{eq:vNdef} the definition of $v_N$.

\begin{corollary}
\label{cor:FBC_crit_point}
The effective critical point for FBC satisfies
\begin{align}
\label{eq:nu_F_asymp}
    \nu_{c,N}^{\free}
    &=
    \begin{cases}
    \nu_{c}-qv_N(1 +  \constF  N^{-\hat \gamma} +O(N^{-1}))
    & (d=4)
    \\
    \nu_{c}-qv_N(1+O( L^{-N} )) & (d>4)
    .
    \end{cases}
\end{align}
\end{corollary}

\begin{proof}
By definition and by the Fundamental Theorem of Calculus,
\begin{align}
    \nu_{c,N}^\free
    &= \nu_c (0) + [\nu_{0,N}(-qL^{-2N})  - \nu_{0,N}(0)] -qL^{-2N}
    \nnb
    & =
    \nu_c (0) - \int_{-\qLap L^{-2N}}^0 \big( \frac{d}{d\ka}\nu_{0,N}(s) +1 \big) ds
    .
\end{align}
The desired result follows by substituting
\eqref{eq:nu_c,N_derivative} and \eqref{eq:nu_c,N_derivative_ter} in the right-hand side,
together with \eqref{eq:vNdef}, the definition of $v_N$.
\end{proof}

\begin{remark}
\label{rk:FBC_crit_point}
In \eqref{eq:nu_c,N_def}, for $d=4$ we defined
\begin{equation}
	\nu_{c,N}^\free =
\nu_{c}-qv_N(1 + \constF N^{-\hat \gamma})
\end{equation}
without the
$O(N^{-1})$ error present in \eqref{eq:nu_F_asymp}.
In the rest of the paper,
we use the exact definition of $\nu_{c,N}^\free$ in \eqref{eq:nu_PF_def}
as it enables simplifications later on.
This does not affect the validity of our claims in Section~\ref{sec:main_results_state} since the neglected error term is $O(N^{\hat \gamma -1}L^{-2N})= O(N^{-1/2}w_N)$
which is smaller than the window scale $w_N$.
Similar reasoning applies for $d=5$,
for which the relation between $v_N$ and $w_N$ in \eqref{eq:vNwN}
gives $L^{-N}v_N \asymp L^{-N/2}w_N$.  However, for $d>5$,
by \eqref{eq:vNwN} $O(L^{-N}v_N)$ is of
the same order as $w_N$ for $d=6$ and is larger than $w_N$ for $d>6$.
This issue  is due to the
        fact that our error bound is not sharp for  $d>5$.
		We therefore retain
     the error term $O(L^{-N})$ in \eqref{eq:nu_c,N_def} when $d>5$.
\end{remark}

\subsection{The renormalised masses}
\label{sec:renmass}

The effective critical points
$\nu_{c,N}^\per$ and $\nu_{c,N}^\free$ are defined in Section~\ref{sec:cp}.
For FBC,
$\nu_{c,N}^\free = \nu_{0,N}(-\qLap L^{-2N}) - \qLap L^{-2N}$ is defined
in terms of a \emph{renormalised mass} $- \qLap L^{-2N}$, with $\nu_{0,N}$
evaluated at this renormalised mass.  In this section,
for both FBC and PBC we show how to define
the renormalised mass corresponding to a value of $\nu$ which need not be the
effective critical point, and we state asymptotic properties of the renormalised mass
in Proposition~\ref{prop:a_N_s_intermediate_estimate_new}.
The renormalised mass and its asymptotic properties play an important role in Section~\ref{sec:pf} where we prove our
main results.

The need for the renormalised mass goes back to the formula \eqref{eq:EF}.
After making explicit the dependence on the Hamiltonian's quadratic coefficient $\nu'$
and on the mass $\ka$, \eqref{eq:EF} expresses
$\langle F \rangle_{\nu' + \ka,N}$ in terms
of an expectation $\E_{C(\ka)} FZ_0(\nu')$.
We want to use this formula for
$\langle F \rangle_{\nu,N}$ with a given value $\nu$, which must therefore be expressed
as a sum $\nu=\nu'+a$ in a useful way.  The term ``useful'' means that given
$\ka$ we require that $\nu'$ be the critical value $\nu_c(\ka)$ from
Theorem~\ref{thm:rg_main_theorem_large_m}
(or $\nu_{0,N}(\ka)$
from Theorem~\ref{thm:rg_main_theorem_new}), as needed in order to apply
the theorem.
Values of $\nu$ that we need to consider are of the form
$\nu_{c,N}^*+sw_N$ or $\nu_{c,N}^*+sv_N$.
The renormalised mass achieves the useful decomposition.  As
an instance of this,
for PBC with $s>0$  the renormalised mass
$\ka^\per_N(s)$ will achieve the desired decomposition
$\nu_{c}+sw_N = \nu_{c}(\ka^\per_N(s))+ \ka^\per_N(s)$.

Recall the function $\nu_{1,N}:(-\frac12 L^{-2(N-1)}, \infty) \to \R$  from \eqref{eq:nu_1_N_def},
which is continuous,  strictly increasing, and also differentiable on
$(-\frac12 L^{-2(N-1)}, 0) \cup (0,\infty)$ for all $d\geq 4$.
The same applies to the function $\ka \mapsto \nu_{1,N}(\ka)+\ka$, which must
therefore be a
bijection onto its range, which is
the interval $R_N =(\nu_{0,N}(-\frac12 L^{-2(N-1)})-\frac12 L^{-2(N-1)},\infty)$.
For $*\in \{\free,\per\}$, we define intervals in $\R$ by
\begin{equation}
    \cW_N^* = \{s \in \R : \nu_{c,N}^*+sw_N \in R_N\},
    \qquad
    \tilde{\cW}_N^* = \{s \in \R : \nu_{c,N}^*+sv_N \in R_N\},
    \label{eq:WNVN_definition}
\end{equation}
and define the renormalised masses as the inverse maps to these bijections, as in
the next definition.

\begin{definition} \label{def:mass_trick}
The \emph{renormalised masses} are the functions
$\ka_N^*:\cW^*_N \to (-\frac12 L^{-2(N-1)}, \infty)$
and $\tilde\ka_N^*: \tilde{\cW}^*_N \to (-\frac12 L^{-2(N-1)}, \infty)$
given by the unique solutions to
\begin{equation}
\label{eq:mass_trick}
	\nu_{c,N}^*+sw_N = \nu_{1,N}(\ka^*_N(s))+ \ka^*_N(s),
    \qquad
    \nu_{c,N}^*+sv_N = \nu_{1,N}(\tilde\ka^*_N(s))+ \tilde\ka^*_N(s),
\end{equation}
respectively.
\end{definition}

Since they are inverses of increasing bijections, the two renormalised masses
are also increasing bijections.
In particular, since $\nu_{c,N}^\free < \nu_{c}=\nu_{c,N}^\per$ by
\eqref{eq:nu_F_asymp}, we see that
$\ka_N^\free(s) < \ka_N^\per(s)$ and $\tilde\ka_N^\free(s) < \tilde\ka_N^\per(s)$.

In Definition~\ref{def:mass_trick}, there is a point of non-differentiability
when the mass is zero
due to the non-differentiability of $\nu_{1,N}(\ka)$ at $\ka=0$.
This non-differentiability is
due to the somewhat discretionary definition of $\nu_{1,N}$ as the
 concatenation of $\nu_{0,N}$ and $\nu_c$ at $\ka =0$.  We deal with the two regimes separately.

Since $\nu_{c,N}^\per = \nu_c (0)$, it follows from the monotonicity in $s$ that
$\ka_N^{\per}(s) \leq 0$  if and only if $s \leq 0$.  Similarly,
$\tilde\ka_N^{\per}(s) \leq 0$  if and only if $s \leq 0$.
Thus $s=0$ is the point of regime splitting for PBC.
For FBC, we also define threshold values $s_N^{\free}$ and $\tilde{s}_N^{\free}$, at which $\ka_N^{\free}$ and $\tilde{\ka}_N^{\free}$ change signs.
Again, by monotonicity in $\s$,
these are defined by
\begin{align}
	\label{eq:s_N_NG_def}
	\nu_{c,N}^{\free} + s^{\free}_Nw_N &= \nu_c (0) ,
	\qquad
	\nu_{c,N}^{\free} + \tilde s^{\free}_Nv_N  = \nu_c (0)	.
\end{align}
We summarise the above considerations with the following lemma.

\begin{lemma}
\label{lem:ka_N_s_simple_properties_new}
Let $d \geq 4$, and recall $\cW_N^*$ and $\tilde{\cW}_N^*$ from \eqref{eq:WNVN_definition}.
\begin{enumerate}
	\item
    For $s \in \cW_N^{*}$, the mass $\ka^*_N(s)$
	is strictly increasing in $s$ and satisfies
	$\ka_N^{\per}(s) \leq 0$  if and only if $s \leq 0$, and
	$\ka_N^{\free}(s) \leq 0$  if and only if $s \leq s^{\free}_N$.
	\item
    For  $s \in \tilde{\cW}_N^{*}$, the mass $\tilde \ka^*_N(s)$
	is strictly increasing in $s$ and satisfies
	$\tilde\ka_N^{\per}(s) \leq 0$  if and only if $s \leq 0$, and
	$\tilde\ka_N^{\free}(s) \leq 0$  if and only if $s \leq \tilde s^{\free}_N$.
\end{enumerate}
\end{lemma}

Asymptotic formulas can also be obtained for $s_N^{\free}$ and $\tilde{s}_N^{\free}$.
Indeed,  by directly solving the defining relations \eqref{eq:s_N_NG_def}, and by
using the formula for $\nu_{c,N}^\free$ in \eqref{eq:nu_F_asymp} and
the definitions of $w_N$ and $v_N$ in \eqref{eq:window_choice} and \eqref{eq:vNdef}, we find that, as $N\rightarrow \infty$,
\begin{equation}
\label{eq:s_N_free_asymp-h}
\begin{aligned}
	s_N^\free &\sim \qLap L^{(d-2)N} \hh_N^{2}
    =
	\begin{cases}
		\qLap (BN)^{1/2} & (d=4)
		\\
		\qLap g_\infty^{-\frac12} L^{N(d-4)/2} & (d>4),
	\end{cases}
	\\
	\tilde s_N^\free &\sim \qLap L^{(d-2)N}  \lp_N^{2} = \qLap\qquad (d \geq 4).
\end{aligned}
\end{equation}
Here,
as in \eqref{eq:hhdef} and \eqref{eq:lpdef},
\begin{equation}
    \hh_N =
    \begin{cases}
        (\bbb N)^{1/4}L^{-N} & (d=4)
        \\
        g_\infty^{-1/4}L^{-Nd/4} & (d>4),
    \end{cases}
    \qquad
    \lp_N=  L^{-N(d-2)/2}.
\end{equation}

The next lemma is used in the proof of Proposition~\ref{prop:a_N_s_intermediate_estimate_new}.
Its hypothesis that $s \ge -\frac14 L^{2}\hh_N^2L^{N(d-2)}$
can be written more explicitly as $s \ge -\frac14 L^{2}(\bbb N)^{1/2}$ for $d=4$,
and as $s \ge -\frac14 L^{2}g_\infty^{-1/2}L^{N(d-4)/2}$ for $d>4$,
so for all $d \ge 4$ the
bound diverges to $-\infty$ as $N \to \infty$.
Lemma~\ref{lem:S_to_I_low_bd} provides a range of $s$ values for which
the renormalised masses lie in the interval $\II_{N-1}(0)$ for which
Theorem~\ref{thm:rg_main_theorem_new} applies.
In particular, the interval $[-\frac14 L^{2}\hh_N^2L^{Nd/2},\infty)$ is a subset
of $\cW_N^\free \cap \cW_N^\per$, and the interval $[-\frac14 L^{2},\infty)$ is a subset
of $\tilde{\cW}_N^\free \cap \tilde{\cW}_N^\per$.  This shows that the masses $a_N^*(s)$ and $\tilde a_N^*(s)$
are well-defined for $s$ in these intervals, respectively.

\begin{lemma}
	\label{lem:S_to_I_low_bd}
	Let $d \ge 4$ and $* \in \{\free,\per \}$.
	If we have $s \in [-\frac14 L^{2}\hh_N^2L^{N(d-2)},0]$
for PBC, or  if we have $s \in [-\frac14 L^{2}\hh_N^2L^{N(d-2)},s_N^\free]$
for FBC,  then $\ka_N^*(s) \in \II_{N-1}(0)$.
Also, if $s \in [-\frac14 L^2,0]$ for PBC, or if
$s \in [-\frac14 L^2,s_N^\free]$ for FBC, then $\tilde \ka_N^*(s) \in \II_{N-1}(0)$.
\end{lemma}

\begin{proof}
	By Lemma~\ref{lem:ka_N_s_simple_properties_new},
	our assumptions on $s$ imply that the renormalised masses are not positive,
	so (by monotonicity of the mass in $s$)
	we only need to verify the conclusion for the lower limits
$\underline\sigma = -\frac14 L^{2}\hh_N^2L^{N(d-2)}$
(for $\ka_N^*$) and $\underline\sigma = -\frac 14L^2$ (for $\tilde\ka_N^*$).
	Since the renormalised mass for PBC is larger than the mass
	for FBC, it suffices to consider only FBC.
	We give the details only for $\ka_N^\free$, as
	similar but simpler calculations apply for $\tilde \ka_N^\free$.
	
	Let $\sigma_{\inf} = \inf\{s \in \R: \ka_N^{\free}(s)\in \II_{N-1}(0)\}$.
By continuity of $\ka_N^*$
	and by Lemma~\ref{lem:ka_N_s_simple_properties_new},
	$\sigma_{\inf} < s_N^\free$.
Our goal now is to show that $\underline \sigma > \sigma_{\inf}$.

Suppose first that $d=4$. Given $s \in (\sigma_{\inf}, s_N^\free)$, the
Fundamental Theorem of Calculus and \eqref{eq:nu_c,N_derivative} give
	\begin{align}
		\nu_{0,N}(\ka_N^\free(s))
		-
		\nu_{0,N}(0)
        & =
        \ka_N^\free(s)
		A_4 (\log L^2)^{\hat\gamma}
		N^{\hat \gamma}(1+o(1)).
	\end{align}
Since $\nu_{0,N}(\ka_N^\free(s)) +\ka_N^\free(s) = \nu_{c,N}^{\free}+sw_N$
and $\nu_{0,N}(0)=\nu_c$ by definition, this gives
	\begin{align}
    \nu_{c,N}^{\free}+sw_N
		&=  \nu_c+\ka_N^\free(s) A_4 (\log L^2)^{\hat\gamma}N^{\hat \gamma}(1+o(1))
\nnb
    &= \nu_c+ \ka_N^\free(s)(\bbb N)^{1/2}L^{2N}w_N(1+o(1)),
	\end{align}
where we used the definition of $w_N$ in \eqref{eq:window_choice} for the second equality.
By the formula $\nu_{c,N}^{\free}=\nu_c-\qLap v_N(1+o(1))$ in \eqref{eq:nu_F_asymp},
together with
the relation $v_N=(\bbb N)^{1/2}w_N$ from \eqref{eq:vNwN}, we can rewrite the above equation
as
\begin{align}
    s & = (\bbb N)^{1/2}
    \Big[\qLap  (1+o(1)) + \ka_N^\free(s)L^{2N}(1+o(1)) \Big].
\end{align}
Now we choose $s_0$ so that $\ka_N^{\free}(s_0) = -\frac{3}{8} L^{-2(N-1)}$.
This
$\ka_N^{\free}(s_0)$
is in $\II_{N-1}(0)$ by definition, so $s_0 \ge \sigma_{\rm inf}$ and it
suffices to prove that $\underline \sigma \ge s_0$.  But,
since $\qLap = 1+O(L^{-d})$, $s_0$ obeys
\begin{align}
    s_0 & =  -(\bbb N)^{1/2}
    \Big[\frac{3}{8} L^{2}  -\qLap   \Big](1+o(1))
    <   -\frac{1}{4} (\bbb N)^{1/2}  L^{2} = \underline\sigma
\end{align}
if $L$ is large enough.
This completes the proof for $d=4$.

For $d>4$ the proof is similar.  Now the Fundamental Theorem of Calculus and \eqref{eq:nu_c,N_derivative_ter} lead, as above, to
	\begin{align}
    \nu_{c,N}^{\free}+sw_N - \ka_N^\free (s)
		&=  \nu_c+\ka_N^\free(s) (A_d-1)(1+o(1)).
	\end{align}
In turn,  again by \eqref{eq:nu_F_asymp}, this leads to $s w_N \sim \qLap v_N  + \ka_N^\free(s) A_d $ as $N\rightarrow \infty$.
With $s_0$ chosen such that
$\ka_N^\free(s_0)=-\frac 38 L^2 g_\infty^{-1/2}L^{N(d-4)/2}$, we find that
\begin{align}
    s_0 & = -g_\infty^{-1/2}L^{N(d-4)/2} \Big[  \frac 38 L^2 A_d - \qLap \Big],
\end{align}
and the proof is completed for $d>4$ in the same manner as for $d=4$.
\end{proof}

In addition to the assumption in
Lemma~\ref{lem:S_to_I_low_bd}
that $s$ does
not become too negative,
our analysis of the renormalised mass in the next proposition requires that
$\ka_N^*(s)\to 0$ and $\tilde \ka_N^* (s)\to 0$. By \eqref{eq:mass_trick}, these
requirements are satisfied respectively if $sw_N \to 0$ or $sv_N\to 0$.
In fact, for $d=4$, we assume more.
This leads us to define the $s$-intervals
\begin{align}
	\label{eq:Sigma_N_domain_def}
\begin{split}
	\Sigma_N
	&=
	\begin{cases}
	(-\frac14 L^{2}\hh_N^2L^{N(d-2)}
    ,L^{\sqrt{N}})
	\quad & (d=4) \\
	(-\frac14 L^{2}\hh_N^2L^{N(d-2)}
	,o(w_N^{-1}))
	\quad & (d>4),
	\end{cases}
	\\
	\tilde \Sigma_N
	&=
	\begin{cases}
		(-{\textstyle{\frac14}} L^2, L^{\sqrt{N}})
		\quad & (d=4) \\
		(-{\textstyle{\frac14}} L^2, o(v_N^{-1}))
		\quad & (d>4) .
	\end{cases}
\end{split}	
\end{align}
The upper bound $L^{\sqrt{N}}$ for $d=4$ above has not been optimised but is sufficient for our needs.
For the $d>4$ upper bounds, it is sufficient to fix any sequence which is $o(1)$
as $N\to\infty$, and take the upper bounds to be $o(1)w_N^{-1}$ and $o(1)v_N^{-1}$.
In Proposition~\ref{prop:a_N_s_intermediate_estimate_new}, the
appearance of $L^{-\sqrt{N}}$ as the boundary between
\eqref{eq:a_N_s_estimate_per_new} and \eqref{eq:a_N_s_per_big_O_new}
is a convenient but non-canonical choice
and we invite the reader to focus on \eqref{eq:a_N_s_estimate_free_new} and \eqref{eq:a_N_s_estimate_per_new}.

\begin{proposition}
	\label{prop:a_N_s_intermediate_estimate_new}
    Let $d \geq 4$ and $s \in \Sigma_N$.
	Then $\ka_N^*(s)$ obeys
	the following estimates:
	for FBC,
	\begin{align}
		\label{eq:a_N_s_estimate_free_new}
		\ka_N^\free(s)
		&= 	
		s\hh_N^{-2}L^{-dN}(1+o(1)) - qL^{-2N}.
	\end{align}
	For PBC, if $d\ge 4$ and
		$s \not \in (0,L^{-\sqrt{N}})$,
	\begin{align}
		\label{eq:a_N_s_estimate_per_new}
		\ka_N^\per(s)
		&=
		s\hh_N^{-2}L^{-dN}(1+o(1)).
	\end{align}
	For $d=4$ with $s \in (0,L^{-\sqrt{N}})$,
	\eqref{eq:a_N_s_estimate_per_new}
	is supplemented by the inequality
	\begin{align}
		\label{eq:a_N_s_per_big_O_new}
		0\leq \ka^\per_N(s)
		&\leq
		s
		\hh_N^{-2}L^{-dN}
		(1+o(1)) \qquad (d=4).
	\end{align}
	In addition, \eqref{eq:a_N_s_estimate_free_new}--\eqref{eq:a_N_s_per_big_O_new} all hold with the replacement of
	$\ka_N^*(s)$, $\Sigma_N$, $\hh_N$ by
	$\tilde \ka_N^*(s)$, $\tilde \Sigma_N$, $\lp_N$, respectively.
\end{proposition}

\begin{proof}
	Let $y_N$ denote either $w_N$ or $v_N$.
	In the case $y_N = v_N$,
	$\ka_N(s)$ should be replaced by $\tilde \ka_N(s)$ and
	$s_N^\free$ by $\tilde s_N^\free$.
	We separate the analysis into two parts depending on the sign of $\ka_N^*(s)$,
	which by Lemma~\ref{lem:ka_N_s_simple_properties_new} corresponds to studying
    $\ka_N^\per(s)$ above or below $0$, or  $\ka_N^\free(s)$ above or below $s_N^\free$.
	Note that $\ka_N^*(s) \to 0$ since $sy_N \to 0$ by
	the definition of $y_N$ and by our assumption on $s$.
	
	\smallskip\noindent
	\emph{Case $s \leq s_N^\free$ if $*=\free$ and $s\leq 0$ if $*=\per$.}
	In this case $a_N^*(s)\le 0$ by Lemma~\ref{lem:ka_N_s_simple_properties_new} and $a_N^*(s)$ is
	the unique value of $a$ satisfying
	\begin{equation}
		\nu_{0,N}(a) + a =\nu_{c,N}^*+sy_N.
	\end{equation}
	By the Implicit Function Theorem,
	$\ka_N^*(s)$ is differentiable in $s<0$ ($* = \per$) or in $s<s_N^\free$ ($* = \per$),
	and by the chain rule, its derivative satisfies
	\begin{equation}
		\label{eq:achain}
		\frac{d}{d s} \ka^*_N(s) \big(\frac{d}{d\ka} \nu_{0,N}(\ka) + 1\big) = y_N
	\end{equation}
	for this range of $s$-values. By Lemma~\ref{lem:S_to_I_low_bd},
$\ka_N^*(s) \in \II_{N-1}(0)$.
	Thus, by inserting the derivative estimates
	for $\frac{d}{d\ka} \nu_{0,N}(\ka)$ from
	\eqref{eq:nu_c,N_derivative} and \eqref{eq:nu_c,N_derivative_ter} (valid since $\ka_N^*(s) \to 0$)
	into \eqref{eq:achain},
	we see that
	\begin{align}
		\label{eq:a_N_s_deriv_small_s_new}
		\frac{d}{ds}\ka^*_N(s)
		&=
		\begin{cases}
			A^{-1}_{4} y_N (\log L^2)^{-\hat \gamma}N^{-\hat \gamma}(1+o(1))
			&(d=4)\\
			A_{d}^{-1}y_N(1+o(1))
			&(d>4).
		\end{cases}
	\end{align}
We consider first the case $y_N=w_N$.
For all $d \ge 4$, it follows from the definitions of $w_N$ and $\hh_N$ that,
for $y_N=w_N$, \eqref{eq:a_N_s_deriv_small_s_new} simplifies to
	\begin{align}
\label{eq:aderh}
		\frac{d}{ds}\ka^*_N(s)
		&=
			\hh_N^{-2}L^{-dN}(1+o(1)) .
	\end{align}
	For PBC, we integrate  between $0$ and $s<0$
	and deduce from \eqref{eq:a_N_s_deriv_small_s_new} and the Fundamental Theorem of Calculus that
	\begin{align}
		\ka_N^\per(s) -  \ka_N^\per(0)
		&= - \int_{s}^{0}
		\frac{d}{du}\ka_N^\per(u) du
		= 	s \hh_N^{-2} L^{-dN}(1+o(1)).
	\end{align}
	This gives
	the desired result for PBC because $\ka_N^\per(0) = 0$.
	For FBC and $s < s_N^\free$, since $\ka_N^\free(0) = -\qLap L^{-2N}$,  we can again integrate on the interval $[0,s]$ if $s \ge 0$ or on $[s, 0]$ if $s<0$ to obtain
	\begin{align}
		\ka_N^\free(s)  + \qLap L^{-2N}
		&
		= s \hh_N^{-2} L^{-dN}(1+o(1)).
	\end{align}
This proves \eqref{eq:a_N_s_estimate_free_new}--\eqref{eq:a_N_s_estimate_per_new}
for $w_N$ in this case.
	The computations are analogous for
	$\tilde \ka_N(s)$
	and $s\in \tilde \Sigma_N$,
	with $\lp_N$ instead of $\hh_N$.
	In particular, the definitions of $v_N$ and $\lp_N$ imply that, for all $d \ge 4$,
\eqref{eq:aderh} is replaced by
	\begin{align}
\label{eq:aderl}
		\frac{d}{ds}\tilde \ka^*_N(s)
		&=
			\lp_N^{-2}L^{-dN}(1+o(1)) ,
	\end{align}
and also, by \eqref{eq:mass_trick}, $\tilde\ka_N^\per (0) = 0$ and $\tilde \ka_N^\free (0) = -\qLap L^{-2N}$.
	This completes the proof of
 \eqref{eq:a_N_s_estimate_free_new}--\eqref{eq:a_N_s_estimate_per_new} in this case.
	
	\medskip\noindent
	\emph{Case $s>s_N^\free$ if $*=\free$ and $s>0$ if $*=\per$.}
	In this case $a_N^*(s)>0$ by Lemma~\ref{lem:ka_N_s_simple_properties_new}, so $a_N^*(s)$ is
	the unique value of $a$ satisfying
	\begin{equation}
		\nu_{c}(a) + a =\nu_{c,N}^*+sy_N.
	\end{equation}
	As before, we use the Implicit Function Theorem and the derivative estimates for $\frac{d}{d\ka} \nu_c(\ka)$ in
	\eqref{eq:nu_c_deriv_4} and \eqref{eq:nu_c_deriv_5} to obtain
	\begin{equation}
		\label{eq:a_N_s_deriv_big_s}
		\frac{d}{ds}\ka^*_N(s)
		= \begin{cases}
			A_{4}^{-1} y_N(-\log \ka^*_N(s))^{-\hat \gamma}(1+o(1))  & (d=4)\\
			A_{d}^{-1} y_N(1+o(1)) & (d>4).
		\end{cases}
	\end{equation}
	When $d>4$, we have the same formula as in \eqref{eq:a_N_s_deriv_small_s_new},
	and hence obtain the same conclusion.
	For the remaining case of $d=4$, we define
	$\psi\colon [0,1] \mapsto \R$
	by
	\begin{equation}
		\psi(x) = x(-\log x)^{-\hat\gamma}.
	\end{equation}
	For PBC we integrate over $[0,s]$ and obtain
	(see \cite[p.~261]{BBS-brief} for an elementary proof)
	\begin{equation}
		\ka^\per_N(s) =
		\psi(A_{4}^{-1} sy_N)(1+o(1)).
	\end{equation}
	The logarithm in $\psi$ is
	\begin{align}
		\label{eq:psilog}
		(-\log(A_{4}^{-1} s y_N))^{-\hat \gamma}
		&= (-\log y_N)^{-\gamma}\Big(1+\frac{\log s}{\log y_N} \Big)^{-\gamma}(1+o(1)).
	\end{align}
	It is the $\log s$ that will create the distinction between smaller and larger positive $s$ in the statement of the proposition for PBC.
	By hypothesis our hypothesis that $s \in \Sigma_N$,
    we have $s \leq L^{\sqrt{N}}$.  For $s \in [L^{-\sqrt{N}},L^{\sqrt{N}}]$
	\begin{equation}
		\label{eq:only_a_bound_new}
		\Big| \frac{\log s}{\log y_N} \Big|
		\leq O(N^{-1} \times \sqrt{N}) = o(1).
	\end{equation}
	
	\smallskip \noindent
	We
    first consider the case $y_N=w_N$.
	By definition,
	\begin{equation}
		(-\log w_N)^{-\gamma} = (\log L^2)^{-\hat \gamma} N^{\hat \gamma} (1+o(1)),
	\end{equation}
	so with the above restrictions on $s$,
	\begin{equation}
		\psi(A_{4}^{-1} sw_N)
		= A_{4}^{-1} sw_N (\log L^2)^{-\hat \gamma}N^{-\hat \gamma}(1+o(1))
		= s\hh_N^{-2}L^{-4N}(1+o(1)).
	\end{equation}
	This proves \eqref{eq:a_N_s_estimate_per_new} for $d=4$.
	For $s\in (0,L^{-\sqrt{N}})$ the left-hand side of \eqref{eq:only_a_bound_new} can
	become arbitarily large and we obtain instead the inequality
	\begin{equation}
		0 \leq \ka_N^\per(s) \leq s \hh_N^{-2}L^{-4N}(1+o(1)).
	\end{equation}
	This proves \eqref{eq:a_N_s_per_big_O_new} for PBC and $y_N=w_N$, and the same proof
	applies with minor adjustments for $y_N=v_N$.
	
	For FBC with $y_N=w_N$ and $s-s_N^\free >L^{-\sqrt N}$,
    we instead integrate over $[s_N^\free,s]$ and similarly obtain
	\begin{equation}
		\ka^\free_N(s) =
		\psi(A_{4}^{-1} (s-s_N^\free)y_N)(1+o(1)).
	\end{equation}
	The calculations are then identical with $s$ replaced by $s-s_N^\free$, and give
	\begin{align}
		\ka_N^\free(s)
		&= (s-s_N^\free)\hh_N^{-2}L^{-4N}(1+o(1)).
	\end{align}
	By \eqref{eq:s_N_free_asymp-h},
	\begin{align}
	\label{eq:sFasy_new}
		s_N^\free \hh_N^{-2} L^{-dN} &= \qLap L^{-2N} (1+o(1)),
	\end{align}
	and hence when $s >s_N^\free$ the error $o(\qLap L^{-2N})$ can be absorbed by
	$o(s\hh_N^{-2}L^{-4N})$.  This gives
	\begin{align}
		\ka_N^\free(s)
		=
		-\qLap L^{-2N}
		+ s\hh_N^{-2}L^{-4N}(1+o(1)).
	\end{align}
    For $s\in (s_N^\free,s_N^\free+L^{-\sqrt{N}})$ we have instead the bound
	\begin{align}
		0 \leq \ka_N^\free(s) \leq -qL^{-2N}+ s\hh_N^{-2}L^{-4N}(1+o(1)),
	\end{align}
    which we can rewrite as
	\begin{align}
    \label{eq:aFsmalls}
		\qLap L^{-2N}
        \leq
        \ka_N^\free(s) +\qLap L^{-2N}
		\leq
		 s\hh_N^{-2}L^{-4N}(1+o(1)).
	\end{align}
    But by \eqref{eq:sFasy_new} and our assumption that $s$ is close to $s_N^\free$,
    we can rewrite the lower bound as
    \begin{align}
        \qLap L^{-2N} =
        \frac{s_N^\free}{s} s\hh_N^{-2}L^{-4N}(1+o(1))
        =
        s\hh_N^{-2}L^{-4N}(1+o(1)).
    \end{align}
	The lower and upper bounds in \eqref{eq:aFsmalls} therefore match, and
    this completes the proof of \eqref{eq:a_N_s_estimate_free_new} for FBC, and for $y_N = w_N$.
	The proof is analogous for $y_N = v_N$.
	This completes the proof.
\end{proof}

\subsection{The critical mass domain}

In order to accommodate sequences $s_N \to +\infty$ in Theorems~\ref{thm:mr-nongaussian}
and \ref{thm:mr-gaussian}, we need to work on a larger mass interval than
the interval $\II_{N-1}(0)=(-\frac 12 L^{-2(N-1)},\frac 12  L^{-2(N-1)})$
of Theorem~\ref{thm:rg_main_theorem_new}.
The larger interval we use is
\begin{align}
	\label{eq:I_crit_asymp}	
	\II_{\rm crit}
	&= \begin{cases}
		(-\frac12 L^{-2(N-1)},  L^{-2N}L^{o(N)}  )
        &(d=4)\\
		(-\frac12 L^{-2(N-1)},2 L^{-3N/2} ) &(d > 4).
	\end{cases}
\end{align}
The $L^{o(N)}$
for $d=4$ is
present in order to make the interval larger than if the upper bound were just $L^{-2N}$.
For concreteness, it will appear in the proof of Lemma~\ref{lem:S_to_I_inclusion} that
the choice $L^{\sqrt{N}}$ for $L^{o(N)}$
is permitted  and adequate.
By definition,
$\II_{N-1}(0) \subset \II_{\rm crit}$,
with the inclusion strict due to the larger positive values permitted in
$\II_{\rm crit}$ compared to $\II_{N-1}(0)$.  These larger values are restricted
so that the bounds in the next lemma hold, and those bounds will play a role in Section~\ref{sec:integral-scaling}.
Lemma~\ref{lem:g_N_crit_behaviour} shows that for a mass $\ka \in \II_{\rm crit}$,
the coupling constant $g_N(\ka)$ is close to its massless counterpart.
It is for this reason that the interval bears a subscript indicating criticality.

\begin{lemma}
	\label{lem:g_N_crit_behaviour}
	For $\ka \in \II_{\rm crit}$, let $\nu = \nu_{1,N} (\ka)$, and
	let $g_N$ be
	as in Theorem~\ref{thm:rg_main_theorem_large_m} if $\ka > 0$ and
	as in Theorem~\ref{thm:rg_main_theorem_new} if $\ka \leq 0$.
	Then
	\begin{equation}
		\label{eq:g_N_crit_behaviour_1}
		g_N(a) =
		\begin{cases}
			(BN)^{-1}(1+O(N^{-1/2})) & (d=4) \\
			g_{\infty}(0)(1+O(g L^{-N/2} )) & (d>4).
		\end{cases}
	\end{equation}
\end{lemma}

\begin{proof}
Let $\ka \in \II_{\rm crit}$.
Suppose first that $\ka \leq 0$.  For $d=4$, \eqref{eq:g_N_crit_behaviour_1} is
an immediate consequence of \eqref{eq:gNB_ter}.  For $d>4$,
\eqref{eq:g_N_crit_behaviour_1} is a consequence
of \eqref{eq:gNB_d_ter} since the error term in \eqref{eq:gNB_d_ter}
is better than in \eqref{eq:g_N_crit_behaviour_1}.

Suppose next that $\ka > 0$. For $d>4$,
\eqref{eq:g_N_crit_behaviour_1}
follows from \eqref{eq:gNB_bis_zero} applied to a mass $\ka$ that is the upper bound
$2L^{-3N/2}$	of $\II_{\rm crit}$.
Finally, for $d=4$,
when $\ka \in (0,L^{-2N + N^{1/2}})$ the mass scale $j_a$ (defined in
\eqref{eq:mass-scale}) obeys $N \wedge j_a = N + O(N^{1/2})$,  so
\eqref{eq:g_N_crit_behaviour_1} follows from \eqref{eq:gNB_bis-4}.
This completes the proof.
\end{proof}

Recall the intervals $\Sigma_N$ and $\tilde\Sigma_N$ defined in
\eqref{eq:Sigma_N_domain_def}.
	We now define intervals $S_N\subset \Sigma_N$ and
$\tilde S_N \subset \tilde\Sigma_N$
	by
	\begin{align}
		\label{eq:S_N_domain_def}
		S_N &
		= \begin{cases}
    (-\frac14 L^{2}\hh_N^2 L^{N(d-2)}
    ,L^{\sqrt{N}})
			\quad & (d=4) \\
			(-\frac14 L^{2}\hh_N^2 L^{N(d-2)}
			,L^{N(d-3)/2})
			\quad & (d>4),
		\end{cases}
		\\
		\label{eq:tilde_S_N_domain_def}
		\tilde S_N  &=
		\begin{cases}
			(-{\textstyle{\frac14}} L^2, L^{\sqrt{N}})
			\quad & (d=4) \\
			(-{\textstyle{\frac14}} L^2, L^{N/2})
			\quad & (d>4) .
		\end{cases}
	\end{align}
The intervals $S_N$ and $\tilde S_N$ are balanced to serve two purposes:
(i)
they are small enough to be included in $\Sigma_N$ and $\tilde \Sigma_N$ respectively,  so Proposition~\ref{prop:a_N_s_intermediate_estimate_new} can be applied,
and
(ii)
they are large enough to allow $s_N\rightarrow + \infty$,
and in such a way that $\ka_N(s_N)$ and $\tilde \ka_N(s_N)$ remain inside $\II_{\rm crit}$.
This second point requires the $S$ domains to be strict subsets of the larger $\Sigma$ domains.
The upper bounds for $S_N$ and $\tilde S_N$ are somewhat arbitrary and
have not been optimised;  they match the upper bounds in the restrictions
on $s_N$ in \eqref{eq:sNbds} and \eqref{eq:sNbds2} that occur in our main results.
The lower bounds for $S_N$
have $N$ dependence of order  $-N^{1/2}$ ($d=4$) and $-L^{N(d-4)/2}$
($d>4$) so are considerably more generous than \eqref{eq:sNbds};
they will be further restricted
in the proofs of Theorems~\ref{thm:mr-nongaussian} and \ref{thm:mr-gaussian}, which appear in
Section~\ref{sec:mr-pfs}.

\begin{lemma}
	\label{lem:S_to_I_inclusion}
	Let $d \ge 4$.
	For both choices of BC,
	if $s \in S_N$ then $\ka_N^*(s) \in \II_{\rm crit}$,
	and if $s \in \tilde S_N$ then $\tilde \ka_N^*(s) \in \II_{\rm crit}$.
\end{lemma}

\begin{proof}
By Definition~\ref{def:mass_trick}, $\ka_N^\free(s)$ and $\ka_N^{\per}(s)$ map
into $(-\frac 12 L^{-2(N-1)},\infty)$.
We therefore only need to verify that their range
do not exceed the upper limits of the interval $\II_{\rm crit}$.
We write $\II_{\rm crit}$ as $(\underline{\alpha},\bar{\alpha})$, where
according to \eqref{eq:I_crit_asymp} and the choice $L^{o(N)}=L^{\sqrt{N}}$
indicated below \eqref{eq:I_crit_asymp},
\begin{equation}
    \bar{\alpha} = \begin{cases}
    L^{-2N(1-N^{-1/2})} & (d=4)
    \\
    L^{-3N/2} & (d>4).
    \end{cases}
\end{equation}
We write $y_N$ to represent $w_N$ or $v_N$.
Let $\bar\sigma$
denote the upper limit of $S_N$ or $\tilde S_N$ (as context dictates), i.e.,
		\begin{equation}
			\bar \sigma =
				L^{\sqrt{N}} \quad (d=4),  \qquad		
			\bar \sigma =
			\begin{cases}
				L^{N(d-3)/2} &(d>4,\, y_N = w_N) \\
				L^{N/2} &(d>4,\, y_N = v_N) .
            \end{cases}
		\end{equation}  		
Since $\ka_N^\free(s) < \ka_N^{\per}(s)$, it suffices to prove that
$\ka_N^{\per}(\bar\sigma) < \bar\alpha$ and $\tilde\ka_N^{\per}(\bar\sigma) < \bar\alpha$.

For $d>4$, we see from Proposition~\ref{prop:a_N_s_intermediate_estimate_new} that
$\ka_N^\per(\bar \sigma) \sim \bar \sigma g_\infty^{1/2}L^{-dN/2}$
for $y_N=w_N$, and $\tilde\ka_N^\per (\bar \sigma) \sim \bar \sigma  L^{-2N}$
for $y_N=v_N$.  Therefore
\begin{equation}
			\begin{aligned}
				\ka_N^\per(\bar \sigma) &\sim g_\infty^{1/2} \,L^{-3N/2} &(y_N = w_N)\\
				\tilde\ka_N^\per(\bar \sigma) &\sim L^{-3N/2} &(y_N = v_N).
			\end{aligned}
\end{equation}
Since $g_\infty$ is small, this completes the proof for $d>4$.
For $d=4$, we have instead
\begin{equation}
	\begin{aligned}
			\ka_N^\per(\bar \sigma) &\sim L^{\sqrt{N}}(BN)^{-1/2}L^{-2N}  &(y_N = w_N)\\
			\tilde\ka_N^\per(\bar \sigma) &\sim L^{\sqrt{N}}L^{-2N} &(y_N = v_N),
	\end{aligned}
\end{equation}
which is less than $\bar\alpha$ for $d=4$.
This completes the proof.
\end{proof}

\section{Proof of main results}
\label{sec:pf}

The common point of departure for the proofs of Theorems~\ref{thm:book_main_theorem_bis}, \ref{thm:mr-nongaussian} and \ref{thm:mr-gaussian}
is Lemma~\ref{lem:chi_N_good_form-v3}, which states
that for $n \geq 1$, $g >0$, $\nu \in \R$, and $\ka\in (-L^{-2(N-1)},\infty)$,
\begin{align}
\label{eq:intP1}
\big\langle e^{J \cdot \Phi_N} \big\rangle_{\nu+ \ka, N}^{\per}
&=  \frac{\int_{\R^n} e^{J \cdot y} \; Z_N (y)
e^{-\frac{1}{2} \ka \Omega_N |y|^2} dy}{\int_{\R^n} Z_N (y) e^{-\frac{1}{2} \ka \Omega_N |y|^2} dy}
,   \\
\label{eq:intF2}
 \big\langle e^{J \cdot \Phi_N} \big\rangle^{\free}_{ \nu+ \ka, N}
&=  \frac{\int_{\R^n}e^{J \cdot y} \; Z_N (y)
e^{-\frac{1}{2} (\ka +  q L^{-2N}  ) \Omega_N |y|^2} dy}{\int_{\R^n} Z_N (y) e^{-\frac{1}{2} (\ka +  q L^{-2N} ) \Omega_N |y|^2 dy}}
.
\end{align}
Here $Z_N$ is defined in \eqref{eq:Z_N_definition} using the covariance $C_{\leq N}(\ka)$
and with $Z_0$ given by \eqref{eq:V0def}--\eqref{eq:Z_0_definition}, and its form
is given by
\eqref{eq:Z_N_form_bis} and \eqref{eq:Z_N_form_ter}
with $\nu$ and $\ka$ chosen appropriately.  The common factor
$e^{-u_N|\Lambda_N|}$ cancels in numerator and denominator.

Since we are also interested in different test functions in
Theorems~\ref{thm:book_main_theorem_bis}--\ref{thm:mr-gaussian},
and with different scalings,
we consider integrals with the field rescaled by a parameter $\bscale_N$, of the form
\begin{equation}
\label{eq:sec2_explanat}
	\int_{\R^n}f(x/\bscale_N)e^{-V_N(x)-\frac12 \ka \Omega_N|x|^2}dx
	\quad \text{ and }\quad
	\int_{\R^n}f(x/\bscale_N)K_N(x)e^{-\frac12 \ka \Omega_N|x|^2}dx,
\end{equation}
where $\nu$ and $\ka$ are properly chosen depending on the regimes we want to study.
We use three different field scalings $\bscale_N$, corresponding to the three scalings
in Theorems~\ref{thm:book_main_theorem_bis}--\ref{thm:mr-gaussian}.
We will show that the integral
containing $K_N$ is relatively negligible compared to the
integral containing $e^{-V_N}$, and we will control
$V_N$ precisely using Theorems~\ref{thm:rg_main_theorem_large_m} and \ref{thm:rg_main_theorem_new}.
In Section~\ref{sec:integral-scaling},
we establish the asymptotic behaviour of the integrals \eqref{eq:sec2_explanat},
and in Section~\ref{sec:mr-pfs},
we use the results of Section~\ref{sec:integral-scaling} to prove
Theorems~\ref{thm:book_main_theorem_bis}--\ref{thm:mr-gaussian}.

\subsection{Scaling of integrals}
\label{sec:integral-scaling}

In this section, we prove estimates on the integrals
\eqref{eq:sec2_explanat} under different BCs,  different choices of $\nu$ and $\ka$, and different scalings.
To specify the choice of $\nu$ and $\ka$, recall from
\eqref{eq:mass_trick} that (for $*=\per$ or $*=\free$)
\begin{equation}
\label{eq:mass_trick4}
	\nu_{c,N}^*+sw_N = \nu_{1,N}(\ka^*_N(s))+ \ka^*_N(s),
    \qquad
    \nu_{c,N}^*+sv_N = \nu_{1,N}(\tilde\ka^*_N(s))+ \tilde\ka^*_N(s),
\end{equation}
where, as in \eqref{eq:nu_1_N_def},
\begin{equation}
\label{eq:nu_1_N_def4}
	\nu_{1,N}(\ka) =
	\begin{cases}
		\nu_c(\ka) & (\ka  \geq 0)  \\
		\nu_{0,N}(\ka) & (\ka < 0).
	\end{cases}
\end{equation}
For notational convenience we focus on the case $w_N$ (the case $v_N$ can be obtained by
adding tildes).
Then our choice of $\nu$ and $\ka$ is to take
$\nu=\nu_{1,N}(\ka^*_N(s))$ and $\ka=\ka^*_N(s)$.

The different regimes studied in
Theorems~\ref{thm:book_main_theorem_bis}, \ref{thm:mr-nongaussian} and \ref{thm:mr-gaussian}
correspond respectively to the three
choices $\bscale_N = \pp_N$, $\hh_N$, $\lp_N$  in \eqref{eq:sec2_explanat}, where
\begin{equation}
\label{eq:hlpdef}
    \pp_N = L^{- N d/2},
    \qquad
    \hh_N =
    \begin{cases}
        (\bbb N)^{1/4}L^{-N} & (d=4)
        \\
        g_\infty^{-1/4}L^{-Nd/4} & (d>4),
    \end{cases}
\qquad
    \lp_N = L^{-N(d-2)/2}.
\end{equation}
The sequences $\hh_N$ and $\lp_N$ are as defined in \eqref{eq:hhdef} and
\eqref{eq:lpdef}, and $\pp_N$ is introduced here for the first time.
For $s$ in the domain $S_N$ defined in \eqref{eq:S_N_domain_def},
the asymptotic behaviour of $\ka^*_N$ is given in
Proposition~\ref{prop:a_N_s_intermediate_estimate_new}
as (with a proviso for PBC when $s \to 0$ that
we do not discuss here)
\begin{align}
    \ka_N^\per(s) &=s\hh_N^{-2}L^{-dN}(1+o(1)),
    \qquad
    \ka_N^{\free}(s)+\qLap L^{-2N} =s\hh_N^{-2}L^{-dN}(1+o(1)).
\end{align}
After inserting this into \eqref{eq:intP1}--\eqref{eq:intF2}, we see that
the $|y|^2$ terms in the exponent become  identical for FBC and PBC.
The same occurs for $\tilde\ka_N^*(s)$, with $\lp_N$ in place of $\hh_N$.

We focus on the first integral in \eqref{eq:sec2_explanat}.
The change of variable $y = x / \bscale_N$,
and the choices $m^2$, $\ka_N^{\per}$ and $\ka_{N}^{\free}$ for $\ka$,
lead us to consider the three integrals
\begin{align}
&  \int_{\R^n} f(y)
e^{-V_N(\bscale_N y)} e^{-\frac{1}{2} m^2 \bscale_N^2 \Omega_N |y|^2} dy
,
\\&   \int_{\R^n} f(y)
e^{-V_N(\bscale_N y)} e^{-\frac{1}{2} \ka_N^\per \bscale_N^2 \Omega_N |y|^2} dy
,
\\
&  \int_{\R^n} f(y)
e^{-V_N(\bscale_N y)} e^{-\frac{1}{2} (\ka_N^\free +  q L^{-2N}  )\bscale_N^2  \Omega_N |y|^2} dy
.
\end{align}
For $\bscale_N$ equal to $\pp_N$, $\hh_N$ and $\lp_N$,
and since $\pp_N^2\Omega_N=1$,
the above reduce, respectively, to
\begin{align}
&  \int_{\R^n} f(y)
e^{-V_N(\pp_N y)} e^{-\frac{1}{2} m^2   |y|^2} dy
,
\\&   \int_{\R^n} f(y)
e^{-V_N(\hh_N y)} e^{-\frac{1}{2} s(1+o(1)) |y|^2} dy
,
\\&   \int_{\R^n} f(y)
e^{-V_N(\lp_N y)} e^{-\frac{1}{2} s(1+o(1)) |y|^2} dy
.
\end{align}
For all $d \ge 4$, the coupling constants $g_N$ and $\nu_N$ in
$V_N(y)= L^{dN}(\frac 14 g_N |y|^4 + \frac 12 \nu_N |y|^2)$ are prescribed by
Theorems~\ref{thm:rg_main_theorem_large_m} and \ref{thm:rg_main_theorem_new}.
In particular, $|\nu_N| \le O(g_NL^{-(d-2)N})$.
Thus,
\begin{align}
\label{eq:Vp}
			V_N(\pp_N y)
&= \frac 14 g_N L^{-dN} |y|^4	
			 +O(g_NL^{-(d-2)N}) |y|^2,
\\
\label{eq:Vh}
			V_N(\hh_N y) &
			= \frac14 g_N\hh_N^{4}L^{dN} |y|^4
			+ O(g_N^{1/2}L^{-(d-4)N/2}) |y|^2,
\\
\label{eq:Vl}
			V_N(\lp_N y)
& =
			\frac 14 g_N L^{-(d-4)N} |y|^4
			 + O(g_NL^{-(d-4)N}) |y|^2.
\end{align}
For $V_N(\pp_N y)$, the coefficients of $|y|^4$ and $|y|^2$ both vanish as $N \to \infty$,
and it is $e^{-\frac 12 m^2 |y|^2}$ that survives in the limit.
Similarly,
the coefficients of  $V_N(\lp_N y)$ again go to zero, and it is $e^{-\frac 12 s|y|^2}$
that survives in the limit.
On the other hand, for $V_N(\hh_N y)$ the quadratic term does vanish but now
$g_N\hh_N^4 L^{dN} \to 1$ in the limit precisely due to our choice of $\hh_N$,
and the non-Gaussian limit $e^{-\frac 14 |y|^4 - \frac 12 s|y|^2}$ emerges.

We make the above considerations more precise in the following lemma.
For its statement,
 we introduce $V_N^{(i)}(y)$ and $K_N^{(i)}$ $(i =c,1)$,
to distinguish each of two mass regimes we investigate: $\ka=m^2\geq 0$
fixed and $\ka \in \II_{\rm crit}$, as follows.
\begin{itemize}
\item
$V_N^{(c)}$, $K_N^{(c)}$ are equal to the functions
$V_N$ and $K_N$
obtained from \eqref{eq:Z_N_form_bis}, with parameters
$\nu = \nu_{c}(\ka)$ and mass $\ka \geq 0$.
The estimates of Theorem~\ref{thm:rg_main_theorem_large_m}
then apply uniformly in $\ka \geq 0$.
We do not make explicit the dependence of $V^{(c)}_N,K^{(c)}_N$ on $\ka$.
\item
$V_N^{(1)}$, $K_N^{(1)}$ are equal to the functions $V_N$ and $K_N$
with parameters $\nu = \nu_{1,N}(\ka)$
and mass $\ka \in \II_{\rm crit}$ (defined in \eqref{eq:I_crit_asymp}),
obtained from \eqref{eq:Z_N_form_bis} if $\ka \geq 0$,
or from \eqref{eq:Z_N_form_ter} if $\ka \leq 0$.
The estimates  of Theorem~\ref{thm:rg_main_theorem_large_m}
and Theorem~\ref{thm:rg_main_theorem_new}
then apply uniformly in $\ka \in \II_{\rm crit}$.
We do not make explicit the  dependence of $V^{(1)}_N,K^{(1)}_N$ on $\ka$.
In this case, Lemma~\ref{lem:g_N_crit_behaviour}
applies and gives
\begin{equation}
\label{eq:g_N_crit_behaviour}
	g_N =
	\begin{cases}
		(BN)^{-1}(1+O(N^{-1}\log N)) & (d=4) \\
		g_{\infty}(0)(1+ O(gL^{-N/2}) ) & (d>4)
	\end{cases}
\end{equation}
uniformly in $\ka \in \II_{\rm crit}$.
\end{itemize}
Finally, we define function domains for the next lemma:
\begin{align}
	\tilde \Fcal
	&= \{f : \R^n \to \R \mid f \geq 0,\, |f(x)| \leq Me^{u|x|^\alpha} \text{ for some } u \in \R, \, \alpha \in [0, 2)\},\\
	\Fcal
	&= \{f : \R^n \to \R \mid f \geq 0,\, |f(x)| \leq Me^{u|x|^4}
    \text{ for some } u< {\textstyle{\frac{1}{2}} \min(
     \kappa
    , \textstyle{\frac14} )}\},
\end{align}
where $\kappa$ is the parameter occurring in the exponential
decay rate of $K_N$ in Theorems~\ref{thm:rg_main_theorem_large_m}
and \ref{thm:rg_main_theorem_new}.

\begin{lemma}
\label{lem:V_N_integral_estimates_unified}
Let $d\geq 4$ and let $\lambda_N$ be a real sequence.
For $V^{(c)}$ and $K^{(c)}$ we use $\ka \ge 0$ and $\nu=\nu_c(a)$.
For $V^{(1)}$ and $K^{(1)}$ we use $\ka \in \II_{\rm crit}$ and $\nu=\nu_{1,N}(a)$.
 The following asymptotic
 formulas hold as $N \to \infty$, uniformly for these parameters
 $\ka$ and $\nu$,
 and with $c_N$, $\tilde c_N$ defined by
\begin{equation}
	\label{eq:c_N_c_N_tilde_def}
	c_N =
	\begin{cases}
		N^{1/4} &(d=4) \\
		L^{N /4}  &(d>4),
	\end{cases}
\qquad
	\tilde c_N =
	\begin{cases}
		N^{1/4} &(d=4) \\
		L^{N (d-4)/4 } &(d>4).
	\end{cases}
\end{equation}
\begin{enumerate}
\item
For any $f \in \tilde \Fcal$, if $\lambda_N\geq\eps$ for some $\eps>0$, then
\begin{align}
\label{eq:t_N_limit_p_scaling}
	\int_{\R^n}f(y)e^{-V^{(c)}_N(\pp_Ny)}e^{-\frac12 \lambda_N |y|^2}dy
	&=
		(1+o(1)) \int_{\R^n}f(x)e^{- \frac12 \lambda_N |x|^2}dx
        +
        O(e^{-L^N \lambda_N}),
		\\
\label{eq:t_N_limit_l_scaling}
	\int_{\R^n}f(y)e^{-V^{(1)}_N(\lp_Ny)}e^{-\frac12 \lambda_N |y|^2}dy
	&=
		(1+o(1)) \int_{\R^n}f(x)e^{- \frac12 \lambda_N |x|^2}dx
        +
        O(e^{-\tilde c_N \lambda_N}).
\end{align}
For any $f \in  \Fcal$ and any sequence $\lambda_N$ whose negative part
$\lambda_N^-=|\lambda_N \wedge 0|$ obeys $\lambda_N^- = o(c_N)$,
\begin{align}
\label{eq:t_N_limit_h_scaling}
	\int_{\R^n}f(y)e^{-V^{(1)}_N(\hh_Ny)}e^{-\frac12 \lambda_N |y|^2}dy
	&=
		(1+o(1)) \int_{\R^n}f(x)e^{- \frac14 |x|^4- \frac12 \lambda_N |x|^2}dx
        +
        O(e^{-c_N(c_N + \lambda_N)}).
\end{align}

\item
Recall the definition of $e_N$ in \eqref{eq:eNdef}.
For any $f \in \tilde \Fcal$, if $\lambda_N\geq\eps$ for some $\eps>0$, then
\begin{align}
\label{eq:K_t_N_limit_p_scaling}
	\Big|\int_{\R^n}f(y)K_N^{(c)}(\pp_Ny)e^{-\frac12 \lambda_N  |y|^2}dy\Big|
	&\leq  O(e_N)\int_{\R^n}f(x)e^{-\frac12 \lambda_N |x|^2}dx,
		\\
\label{eq:K_t_N_limit_l_scaling}
	\Big|\int_{\R^n}f(y)K_N^{(1)}(\lp_Ny)e^{-\frac12 \lambda_N |y|^2}dy\Big|
	&\leq  O(e_N)\int_{\R^n}f(x)e^{-\frac12 \lambda_N |x|^2}dx.
\end{align}
For any $f \in  \Fcal$ and for any sequence $\lambda_N$,
\begin{align}
\label{eq:K_t_N_limit_h_scaling}
	\Big|\int_{\R^n}f(y)K_N^{(1)}(\hh_Ny)e^{-\frac12 \lambda_N |y|^2}dy\Big|
	&\leq  O(e_N)\int_{\R^n}f(x)e^{- \frac12 \kappa |x|^4-\frac12 \lambda_N |x|^2}dx
    .
\end{align}
\end{enumerate}
\end{lemma}

\begin{proof}
(i)
We first prove \eqref{eq:t_N_limit_p_scaling}.
By \eqref{eq:Vp},
the exponent (up to sign) becomes
\begin{equation}
    \frac12 \lambda_N |y|^2 + \eta_N(y)
    \quad \text{with} \quad
    \eta_N(y) = \frac 14 g_N L^{-dN} |y|^4	+ O(g_NL^{-(d-2)N})  |y|^2,
\end{equation}
and we rewrite the integral in \eqref{eq:t_N_limit_p_scaling} as
\begin{align}
    \int_{\R^n}f(y)e^{-\frac12 \lambda_N |y|^2} dy
    + I_N(f) \quad \text{with}\quad
	|I_N(f)| \le \int_{\R^n}f(y)e^{-\frac12 \lambda_N |y|^2}\big|e^{-\eta_N(y)}-1\big|dy.
\end{align}

We now make three observations. By our
hypothesis that $f \in \tilde \cF$,
there is a $C_1>0$ depending on $f$ and $C_2,C_3>0$, such that
\begin{align}
\label{eq:int_big_x_remainder}
	\int_{|y|^2\geq c}f(y)e^{-\frac12 \lambda_N |y|^2} dy
	&\leq C_1e^{-\frac 14 c \lambda_N}  , \\
\label{eq:overall_sup}
	\sup_{y \in \R^n}\big|e^{-\eta_N(y)}-1\big|
	&\leq C_2, \\
\label{eq:small_sup}
	\sup_{|y|^2 \leq c}
	\big|e^{-\eta_N(y)}-1\big|
	&\leq C_3 c^2 g_N L^{-(d-2)N}
,
\end{align}
where $c>0$ is arbitrary for \eqref{eq:int_big_x_remainder}, and
 $1 \le c \le 4 L^N$ for \eqref{eq:small_sup}
(the upper bound is convenient but not optimal).
The first point is clear.
The second point
follows from the fact that the minimal value of $\eta_N$ is (by elementary calculus) at worst
of order $-(g_NL^{-(d-2)N})^2/(g_NL^{-dN})= -g_NL^{-(d-4)N}$ (it occurs at $|y|^2 \asymp L^{2N}$ but we do not use this fact).
The third point follows from the bound $|e^w-1| \leq |w|e^{|w|}$.

We now estimate $I_N(f)$ by splitting the integral
at $|y|^2= 4L^N$.
This gives
\begin{align}
    |I_N(f)|
    & \le
    16C_3 g_N L^{-(d-4)N}
    \int_{|y|^2 \le 4L^N}f(y)e^{-\frac12 \lambda_N |y|^2} dy
    \nnb & \qquad +
    C_2 g_NL^{-(d-4)N}\int_{|y|^2 \ge 4L^N}f(y)e^{-\frac12 \lambda_N |y|^2} dy
    \nnb
    & \le
    16C_3 g_N L^{-(d-4)N}
    \int_{\R^n}f(y)e^{-\frac12 \lambda_N |y|^2}dy
    +C_1C_2e^{- L^N \lambda_N}.
\end{align}
This proves the desired estimate \eqref{eq:t_N_limit_p_scaling}.

The proof of \eqref{eq:t_N_limit_l_scaling} follows similarly
using \eqref{eq:Vl}, now with
\begin{equation}
    \eta_N(y) = g_N L^{-(d-4)N} \Big(\frac 14  |y|^4    + O(1)  |y|^2 \Big).
\end{equation}
The inequalities \eqref{eq:int_big_x_remainder}--\eqref{eq:overall_sup}
remain valid, and \eqref{eq:small_sup} is replaced (with the same proof) by
\begin{equation}
	\sup_{|y|^2 \leq c}|e^{-\eta(y)}-1| \le O(1) c^2g_NL^{-(d-4)N}
\end{equation}
with $1 \leq c \leq g_N^{-1/2} L^{N(d-4)/2}$.
With a similar definition for $I_N(f)$,
by splitting the integral at
$4\tilde c_N$ (defined in \eqref{eq:c_N_c_N_tilde_def}),
and by using the asymptotic behaviour of
$g_N$ in \eqref{eq:g_N_crit_behaviour}, we see that
\begin{equation}
	|I_N(f)| \le O(g_N^{1/2})L^{-N(d-4)/2}\int_{\R^n}f(y) e^{-\frac12 \lambda_N |y|^2}dy
	+ O(1) e^{-\lambda_N g_N^{-1/4}L^{N(d-4)/4}},
\end{equation}
from which \eqref{eq:t_N_limit_l_scaling} follows.

For \eqref{eq:t_N_limit_h_scaling}, the proof varies a little
since the $|y|^4$ term survives in the limit.
Also, we now permit $\lambda_N \to -\infty$.
It follows from \eqref{eq:Vh},
the definition of $\hh_N$, and
\eqref{eq:g_N_crit_behaviour}, that
in the exponent we now have
\begin{equation}
    \frac14 |y|^4 + \frac 12 \lambda_N |y|^2 + \eta_N(y)
    \quad \text{with} \quad
    \eta_N(y) = \zeta_N |y|^4+\xi_N  |y|^2,
\end{equation}
with
\begin{equation}
	\zeta_N =
	\begin{cases}
		O(N^{-1/2})  &(d=4)\\
		O(gL^{- N/2} ) &(d>4),
	\end{cases}
\qquad
    \xi_N =  O(g_N^{1/2} L^{-N(d-4)/2}).
\end{equation}
By taking the worst decay of $\zeta_N$ and $\xi_N$, we deduce that
\begin{align}
	\eta_N(y)
	&= \rho_N(|y|^4+|y|^2) \quad\text{ with}  \quad
	\rho_N
	=
	\begin{cases}
		O(N^{-1/2}) &(d=4) \\
		O(L^{-N/2}) &(d>4).
	\end{cases}
\end{align}
We estimate
\begin{equation}
    \int_{\R^n}f(y)e^{-\frac14|y|^4- \frac12 \lambda_N |y|^2}
	\big| e^{-\eta_N(y)}-1\big|dy
\end{equation}
by splitting it at
$|y|^2 = 4c_N$ (defined in \eqref{eq:c_N_c_N_tilde_def},
the $4$ simplifies the final result).
Note that $| e^{-\eta_N(y)}-1|$ is
no longer bounded, but by splitting the integral
we obtain
\begin{equation}
	|I_N(f)| \leq o(1)\int_{\R^n}f(y)e^{-\frac14 |y|^4-\frac12 \lambda_N |y|^2}dy + I_{N,c_N}(f),
\end{equation}
where $I_{N,c_N}(f)$ is an upper bound on the integral $I_N(f)$
restricted to $|y|^2 \geq 4c_N$.
To estimate $I_{N,c_N}(f)$ we
bound $|e^{-\eta_N(y)}-1|$ by
$1+e^{|\rho_N| (|y|^4+|y|^2)}$
to obtain
\begin{align}
	I_{N,c_N}
	&\leq O(1) \int_{|y|^2 \geq 4c_N} f(y) e^{-\frac18 |y|^4}e^{-(\frac18 |y|^4+\frac12 \lambda_N|y|^2)}
	(1+e^{|\rho_N| (|y|^4+|y|^2)})dy\\
	&\leq O(1)  e^{-4c_N( \frac12 c_N+\frac12 \lambda_N -|\rho_N|(1+c_N))} \int_{|y|^2 \geq c_N} f(y) e^{-\frac18 |y|^4}dy \\
	&\leq O(1)  e^{-c_N(c_N + \lambda_N)},
\end{align}
where in the second line we employed our assumption on
the growth of $\lambda_N$, and in the last line we used our hypothesis that $f \in \Fcal$.
This completes the proof of \eqref{eq:t_N_limit_h_scaling}.

\medskip\noindent
(ii)
We now prove the bounds on the $K_N$ integrals in
\eqref{eq:K_t_N_limit_p_scaling}--\eqref{eq:K_t_N_limit_h_scaling}.
By \eqref{eq:K_bound_at_nu_c_bis-4} and  \eqref{eq:K_bound_at_nu_c_bis-5}
(we do not need the large-field
decay of $K_N$ here),
\begin{align}
	\Big|\int_{\R^n}f(y)K_N^{(1)}(y\pp_N)e^{-\frac12 \lambda_N |y|^2}dy\Big|
	&\leq
    O(e_N)\int_{\R^n}f(y)e^{-\frac12 \lambda_N |y|^2} dy,
\end{align}
which is \eqref{eq:K_t_N_limit_p_scaling}.  The inequality \eqref{eq:K_t_N_limit_l_scaling}
is obtained in the same way using \eqref{eq:K_bound_at_nu_c_ter} and
\eqref{eq:K_bound_at_nu_c_d_ter}, again without making use of the large-field decay
in those bounds on $K_N$.

Finally, \eqref{eq:K_t_N_limit_h_scaling} also follows
from \eqref{eq:K_bound_at_nu_c_ter} and
\eqref{eq:K_bound_at_nu_c_d_ter}, but now we do use the large-field decay,
namely
\begin{equation}
    |K_N(y\hh_N)| \le O(e_N) e^{-\kappa g_N L^{dN} \hh_N^4 |y|^4}.
\end{equation}
According to the behaviour of $g_N$ in \eqref{eq:g_N_crit_behaviour}
and the definition of $\hh_N$ in \eqref{eq:hhdef}, the exponent in the above right-hand
side is $-\kappa[1+o(1)]|y|^4$.  This completes the proof.
\end{proof}

\subsection{Proof of Theorems~\ref{thm:book_main_theorem_bis}, \ref{thm:mr-nongaussian} and \ref{thm:mr-gaussian}}
\label{sec:mr-pfs}

In this section we prove our three main results,
Theorems~\ref{thm:book_main_theorem_bis},
 \ref{thm:mr-nongaussian} and  \ref{thm:mr-gaussian},
subject to the results stated in Section~\ref{sec:effpot}.
The proofs are given in parallel. We start with a
joint proof of part~(i) of the three theorems, concerning
the limit of the Laplace transform of the rescaled   average field; the proof
follows directly from the results of the previous
section. We then turn to the proof of part~(ii) of the three theorems, concerning
the asymptotic behaviour of moments of the average field;
this
requires additional effort to accommodate infinite limits for the sequence $s_N$.
Recall that the infinite-volume critical point is $\nu_c=\nu_c(0)$, and
that the effective critical points for PBC and FBC are
\begin{align}
	\nu^\per_{c,N} &= \nu_c(0)  \qquad
	\nu^\free_{c,N}	=\nu_{0,N} (\ka_N^{\free} (0) ) + \ka_N^{\free} (0) .
\end{align}

\begin{proof}[Proof of Theorems~\ref{thm:book_main_theorem_bis}(i), \ref{thm:mr-nongaussian}(i) and \ref{thm:mr-gaussian}(i)]
We prove \eqref{eq:subcrit_gaussian_limit},
 \eqref{eq:mr-nongaussian_law} and \eqref{eq:gaussian_measure}.
Our starting point is \eqref{eq:S_p_good_1-2}--\eqref{eq:S_p_good_1-2_fbc}.

We first prove \eqref{eq:subcrit_gaussian_limit}
for FBC.  Given $\eps \geq 0$, we apply Theorem~\ref{thm:mass_trick_big_mass}
to obtain $m^2 = m_\eps^2 $, satisfying the asymptotic formula \eqref{eq:renorm_big_mass},
such that
\begin{equation}
	\nu_c(0) + \eps =  \nu_c(m^2)+m^2 .
\end{equation}
By \eqref{eq:S_p_good_1-2_fbc}, with $\nu = \nu_c(m^2)$ and $\ka = m^2\geq 0$,
and since $\pp_N^2\Omega_N=1$,
\begin{align}
 \big\langle e^{J\cdot \Phi_{N}/\pp_N } \big\rangle^{\free}_{ \nu_c(m^2)+ m^2, N}
&=  \frac{\int_{\R^n} e^{J\cdot y}
(e^{-V_N(\pp_Ny)}+K_N(\pp_Ny)) e^{-\frac{1}{2} (m^2 +  q L^{-2N}  )  |y|^2} dy}
{\int_{\R^n} (e^{-V_N(\pp_Ny)}+K_N(\pp_Ny)) e^{-\frac{1}{2} (m^2 +  q L^{-2N} )  |y|^2 dy}}
.
\end{align}
For the $e^{-V_N}$ terms we apply \eqref{eq:t_N_limit_p_scaling}
and for the $K_N$ terms we apply \eqref{eq:K_t_N_limit_p_scaling},
with $f(y) = e^{J \cdot y}$ for the numerator
and $f =1$ for the denominator, and
with $\lambda_N^\free = m^2+qL^{-2N}$.
This gives
\begin{equation}
	\langle e^{ J \cdot \Phi_N L^{dN/2}} \rangle^{\free}_{\nu_c(0)+\eps,N}
	= (1+o(1))\int_{\R^n}e^{J \cdot x} d\mu_{\lambda_N^\free}(x).
\end{equation}
Since $\lambda_N^\free \to m_\eps^2$ as $N\to \infty$, this proves \eqref{eq:subcrit_gaussian_limit}.
For PBC, the proof instead uses $\lambda_N^\per = m^2$ but otherwise is identical.

Next, we prove \eqref{eq:mr-nongaussian_law}.  Let $s_N$ be a real sequence with $s_N \to s \in \R$.
Then $s_N$ is eventually in the interval $S_N$ defined in \eqref{eq:S_N_domain_def}.  As discussed in the paragraph above
Lemma~\ref{lem:S_to_I_low_bd}, this implies that $S_N \subset \cW_N^*$.
According to Definition~\ref{def:mass_trick}, we can therefore
define $\ka_N^*(s_N)$ as in \eqref{eq:mass_trick}, so that
\begin{equation}
	\nu_{c,N}^*+s_Nw_N = \nu_{1,N}(\ka_N^*(s_N))+ \ka_N^*(s_N).
\end{equation}
Then we use Lemma~\ref{lem:chi_N_good_form-v3}
with $\nu = \nu_{1,N}(\ka^*_N(s_N))$
and $\ka = \ka^*_N(s_{N})$.
We define a sequence $\lambda^*_N$ by
\begin{align}
\label{eq:lambda-s-N}
	\lambda^*_N\hh_N^{-2} &= \Omega_N\ka^*_N(s_{N})+
    \begin{cases}
    0  & (*=\per)
    \\
    \Omega_N\qLap L^{-2N} & (*=\free).
    \end{cases}
\end{align}
Since $s_N$ converges to $s$, it follows from Proposition~\ref{prop:a_N_s_intermediate_estimate_new} that
\begin{align}
\begin{array}{ll}
	\lambda^*_N = s(1+o(1))  & (*=\free, \text{ or } * = \per \text{ and } s \neq 0),  \\
	|\lambda^*_N| \leq |s_N|(1+o(1))  & (* = \per,\, s=0 ).
\end{array}	
\label{eq:lambda-s}
\end{align}
Note that for FBC there is a cancellation of the $\qLap L^{-2N}$ term due to \eqref{eq:a_N_s_estimate_free_new}.
The different cases in \eqref{eq:lambda-s} are only needed for $d=4$
but are also true for $d>4$ by
Proposition~\ref{prop:a_N_s_intermediate_estimate_new} again.
In any case, we see from \eqref{eq:lambda-s} that $\lambda_N^*$ and $s_N$
both converge to $s$, and therefore
the above dichotomy becomes inconsequential.
As in the proof of \eqref{eq:subcrit_gaussian_limit}, we now apply \eqref{eq:t_N_limit_h_scaling}
and
\eqref{eq:K_t_N_limit_h_scaling}
to obtain
\begin{equation}
	\langle e^{ J \cdot \Phi_N \hh_N^{-1}} \rangle^{*}_{\nu^*_{c,N}+s_Nw_N,N}
	= (1+o(1))\int_{\R^n}e^{J \cdot x} d \sigma_{\lambda^*_N}(x),
\end{equation}
and \eqref{eq:mr-nongaussian_law} then follows since $\lambda^*_N \to s$.

The proof of \eqref{eq:gaussian_measure} uses instead the tilde version of  Proposition~\ref{prop:a_N_s_intermediate_estimate_new},
\eqref{eq:t_N_limit_l_scaling} and
\eqref{eq:K_t_N_limit_l_scaling}.  Since it
involves nothing new, we omit the details.
\end{proof}

\begin{proof}[Proof of Theorems~\ref{thm:book_main_theorem_bis}(ii), \ref{thm:mr-nongaussian}(ii) and \ref{thm:mr-gaussian}(ii)]
The asymptotic formulas \eqref{eq:A,B_definition_bis} are part of the
statement of Theorem~\ref{thm:rg_main_theorem_large_m}, so
we are left to prove \eqref{eq:subcrit_gaussian_moment}, \eqref{eq:mrPBC-nongaussian_moments}, \eqref{eq:SNp-window-free}.

We first prove
\eqref{eq:subcrit_gaussian_moment}, which is equivalent to the statement that
for $p \geq 1$ we have
\begin{equation}
\label{eq:subcrit_gaussian_moment-pf}
	\langle  |\Phi_N|^{2p}\rangle_{\nu_c(0) + \eps,N}^*
	= (1+o(1))\pp_N^{2p} \int_{\R^n} |x|^{2p} d\mu_{m^2_\eps}
	.
\end{equation}
We apply the identities \eqref{eq:S_p_good_1_pbc}--\eqref{eq:S_p_good_1_fbc}
(using linearity to obtain their counterparts with $|\Phi_N|^{2p}$ and $|y|^{2p}$),
with $\nu = \nu_c(m^2)$ and $\ka = m^2$,
and define $\lambda^\per_N = m^2$
and $\lambda_N^\free = m^2+qL^{-2N}$.
Exactly as in the proof of \eqref{eq:subcrit_gaussian_limit}, we apply
\eqref{eq:t_N_limit_p_scaling} and
\eqref{eq:K_t_N_limit_p_scaling}, now with
with $f(y) = |y|^{2p}$ in the numerator.
This leads to \eqref{eq:subcrit_gaussian_moment-pf}.

Next we prove \eqref{eq:mrPBC-nongaussian_moments}.
Recall that $S_N$ is defined in \eqref{eq:S_N_domain_def} by
\begin{align}
	S_N & = \begin{cases}
	(-\frac14 L^{2}(BN)^{1/2},L^{\sqrt{N}})
	\quad & (d=4) \\
	(-\frac14L^{2}g_\infty^{-1/2}L^{N(d-4)/2}, L^{N(d-3)/2})
		\quad & (d \geq 5).
\end{cases}
\end{align}
We are assuming that either:
(a) $s_N$ converges, or
(b) $s_N \to +\infty$ in such a manner that $s_N \leq L^{(d-3)N/2}$ for $d\geq 5$ or $s_N \leq L^{\sqrt{N}}$ for $d =4$,
or
(c) $s_N \to -\infty$ in such a manner that $|s_N|\leq (\log N)^{1/4}$ for $d=4$
or $|s_N| \leq N^{1/4}$ for $d>4$.
In all these cases, $s_N$ remains inside $S_N$ for large enough $N$.
Therefore, by
Lemma~\ref{lem:S_to_I_inclusion},
the solution (recall \eqref{eq:mass_trick}) to
\begin{equation}
	\nu_{c,N}^*+s_Nw_N = \nu_{1,N}(\ka_N^*(s_N))+ \ka_N^*(s_N).
\end{equation}
obeys $\ka_N^*(s_N)\in \II_{\rm crit}$.

We use \eqref{eq:S_p_good_1_pbc}--\eqref{eq:S_p_good_1_fbc}
with $\nu = \nu_{1,N}(\ka^*_N(s_N))$ and  $\ka = \ka^*_N(s_N)$, and
apply \eqref{eq:t_N_limit_h_scaling} and
\eqref{eq:K_t_N_limit_h_scaling} with $f(y) = |y|^{2p}$ in the numerator and
$f(y) = 1$ in the denominator, and
with $\lambda^*_N$  defined as in \eqref{eq:lambda-s-N}.
By assumption, $s_N$ converges to $s \in \R$
or diverges to $\pm \infty$ while staying in the domain $ S_N$.
Thus we can apply Proposition~\ref{prop:a_N_s_intermediate_estimate_new}
as in \eqref{eq:lambda-s} to see that
\begin{align}
\begin{array}{ll}
	\lambda^*_N = s_N(1+o(1))
		& (*=\free, \text{ or } * = \per \text{ and }
    s_N \not\to 0
    ) \\
    \lambda^*_N = o(1)
    	& (* = \per,\, s_N \to 0 ).
\end{array}
\label{eq:lambda-s-2}
\end{align}
The dichotomy in \eqref{eq:lambda-s-2} is only needed for $d=4$, since for $d>4$ we always
have $\lambda_N^* = s_N(1+o(1))$ when $s_N \in S_N$.
Our assumption that
$|s_N|$ (and thus $|\lambda_N^*|$) diverges at worst like $o(c_N)$
when $s_N \to -\infty$ ensures that we can
apply Lemma~\ref{lem:V_N_integral_estimates_unified}.
Overall, this gives (as in \eqref{eq:uf_n_definition} for the integrals $I_k(s)$)
\begin{align}
    \big\langle  |\Phi_{N}|^{2p} \big\rangle_{g, \nu_{c,N}^*+ s_Nw_N, N}^*
    &=  \hh_N^{2p}
    \frac
    {(1+o(1))\int_{\R^n} |y|^{2p} e^{-\frac14 |x|^4-\frac12 \lambda_N^*|x|^2}dx +\eta_{N,2p}}
    {(1+o(1))\int_{\R^n} e^{-\frac14 |x|^4-\frac12 \lambda_N^*|x|^2}dx +\eta_{N,0}}
    \nnb
    & =
    \hh_N^{2p}
    \frac
    {(1+o(1))I_{2p+n-1}( \lambda_N^*) +\eta_{N,2p}}
    {(1+o(1))I_{n-1}( \lambda_N^*) +\eta_{N,0}},
\end{align}
where $\eta_{N,2p}$ comprises error terms
from both \eqref{eq:t_N_limit_h_scaling} and
 \eqref{eq:K_t_N_limit_h_scaling}.
 Explicitly,
\begin{equation}
\label{eq:eta_N_2p_def}
	\eta_{N,2p}
	=
 	O(e^{-c_N(c_N+\lambda_N^*)})
    +O(e_N)
    I_{2p+n-1}(\lambda_N^*/\sqrt{2\kappa})
    ,
\end{equation}
with $c_N$ given in \eqref{eq:c_N_c_N_tilde_def}
and, as defined in \eqref{eq:eNdef},
\begin{equation}
	e_N =
	\begin{cases}
		N^{-3/4} & (d=4)\\
		g^{3/4} L^{-3(d-4)N/4} & (4 < d \le 12)
		\\
		g^{3/4}L^{-Nd/2} & (d>12).
	\end{cases}
\end{equation}
For the case $s_N \to s \in \R$, so that also $\lambda_N^* \to s$,
with \eqref{eq:uf_n_definition}
we immediately conclude that
\begin{equation}
	\big\langle  |\Phi_{N}|^{2p} \big\rangle_{g, \nu_{c,N}^*+ s_Nw_N, N}^*
    = \hh_N^{2p}(1+o(1))\Sigma_{n,2p}(s),
\end{equation}
which proves \eqref{eq:mrPBC-nongaussian_moments} in this case.
It remains to consider infinite limits.
The asymptotic behaviour of the integral on the right-hand side of \eqref{eq:eta_N_2p_def}
is given in \eqref{eq:IkFi}.
Using this, we verify that the first term
in \eqref{eq:eta_N_2p_def} is always much smaller
than the second one and hence
\begin{equation}
	\eta_{N,2p}
	=
 	O(e_N)\times
 	\begin{cases}
	(\lambda_N^*)^{-(2p+n)/2}  & (\lambda_N^* \rightarrow +\infty) \\
	e^{\frac{1}{8 \kappa}
     (\lambda_N^*)^2 } | \lambda_N^*|^{\frac{2p+n -2}{2}}
    & (\lambda_N^* \rightarrow -\infty).
 	\end{cases}
\end{equation}
To obtain \eqref{eq:mrPBC-nongaussian_moments}
it is enough to show that $\eta_{N,2p}$ is always much
smaller than the integral $I_{2p+n-1}(\lambda_N^*)$.
This in turn follows by our assumptions on
the growth of $s_N$ and from \eqref{eq:IkFi} again. Indeed when
$s_N \to + \infty$ this is immediate by
the extra $O(e_N)$ decay.
If $s_N \to - \infty$ then we use our assumption that
$|s_N| \leq (\log N)^{1/4}$ if $d=4$, or $|s_N| \leq N^{1/4}$
if $d>4$, to see that for some unimportant constants $c_\kappa,c_L >0$,
\begin{align}
	\eta_{N,2p}
	= O(e_N)I_{2p+n-2}(\lambda_N^*)e^{c_\kappa (\lambda_N^*)^2}
	= O(1)I_{2p+n-2}(\lambda_N^*)\times
	\begin{cases}
		e^{-\frac34 (1-o(1))\log N} & (d=4)\\
		e^{- c_L(1-o(1)) N} & (d>4).
	\end{cases}
\end{align}
Finally, from \eqref{eq:Mnkasy}, \eqref{eq:lambda-s-2}, and the above considerations,
in the case of either of the two infinite limits we
obtain
\begin{equation}
	\big\langle  |\Phi_{N}|^{2p} \big\rangle_{g, \nu_{c,N}^*+ s_Nw_N, N}^*
    =  \hh_N^{2p}(1+o(1))\Sigma_{n,2p}(\lambda_N^*)
    = \hh_N^{2p}(1+o(1))\Sigma_{n,2p}(s_N),
\end{equation}
which completes the proof of \eqref{eq:mrPBC-nongaussian_moments}.

It remains to prove \eqref{eq:SNp-window-free}. The
structure of the proof is essentially the same as for \eqref{eq:mrPBC-nongaussian_moments}, but
it is simpler since we do not consider $s_N \to -\infty$.
As before, we check that $s_N \in \tilde S_N$ first. Recall that
\begin{equation}
		\tilde S_N  =
	\begin{cases}
		(-{\textstyle{\frac14}} L^2, L^{\sqrt{N}})
		\quad & (d=4) \\
		(-{\textstyle{\frac14}} L^2, L^{N/2}) &(d \geq 5).
	\end{cases}
\end{equation}
Our hypothesis on $s_N$ in \eqref{eq:sNbds2} ensures that
$s_N \in \tilde S_N$
and we assume that $s_N \geq \eps  >0$. We then let
 $\tilde \ka^* = \tilde \ka_N^*(s_N)$ as in
\eqref{eq:mass_trick}, so that
\begin{equation}
	\nu_{c,N}^*+s_Nv_N = \nu_{0,N}(\tilde \ka_N^*(s_N))+ \tilde \ka_N^*(s_N).
\end{equation}
Again, we use \eqref{eq:S_p_good_1_pbc}--\eqref{eq:S_p_good_1_fbc}
with $\nu = \nu_{0,N}(\tilde \ka^*_N(s_N))$ and $\ka = \tilde \ka^*_N(s_N)$,  and
we apply \eqref{eq:t_N_limit_l_scaling} and
\eqref{eq:K_t_N_limit_l_scaling} with $f(y) = |y|^{2p}$ and $f (y)=1$, and
with
\begin{equation}
\label{eq:lambda_N_*_def_bis}
	\lambda^*_N = s_N(1+o(1))
\end{equation}
which follows from Proposition~\ref{prop:a_N_s_intermediate_estimate_new}
and the fact that $s_N \in \tilde S_N$.
We obtain this time the similar but neater equation
\begin{align}
    \big\langle  |\Phi_{N}|^{2p} \big\rangle_{g, \nu_{c,N}^*+ s_N v_N, N}^*
    &=  \lp_N^{2p}
    \frac
    {(1+o(1))\int_{\R^n} |y|^{2p} e^{-\frac12 \lambda_N^*|x|^2}dx +O(e^{-\tilde c_N\lambda^*_N})}
    {(1+o(1))\int_{\R^n} e^{-\frac12 \lambda_N^*|x|^2}dx +O(e^{-\tilde c_N\lambda^*_N})}.
\end{align}
As before, $O(e^{-\tilde c_N\lambda^*_N})$ is always much smaller than the integrals appearing in the numerator and denominator above (whether $\lambda_N^*$ is bounded or $\lambda_N^*\to +\infty$),
so we conclude that
\begin{equation}
	\big\langle  |\Phi_{N}|^{2p} \big\rangle_{g, \nu_{c,N}^*+ s_N v_N, N}^* = \lp_N^{2p}(1+o(1))M_{n,2p}(\lambda_N^*) = \lp_N^{2p}(1+o(1))M_{n,2p}(s_N).
\end{equation}
This completes the proof of \eqref{eq:SNp-window-free}.
\end{proof}

\section{The RG map}
\label{sec:pf1}

To complete the proof of our main results,
it remains to prove Theorems~\ref{thm:rg_main_theorem_large_m}--\ref{thm:rg_main_theorem_new}.
In this section, we formulate our main RG theorems, Theorems~\ref{thm:Phi^U_estimate}--\ref{thm:Phi^K_estimate}.
These are used to prove
Theorems~\ref{thm:rg_main_theorem_large_m}--\ref{thm:rg_main_theorem_new} in Section~\ref{sec:pf3}.

The following are the main improvements in our RG analysis
compared to \cite{BBS-brief}:
\begin{enumerate}
\item
We include dimensions $d>4$ as well as the upper critical dimension $d=4$.
Since $|\varphi|^4$ is a
so-called \emph{dangerous irrelevant monomial} in dimensions $d>4$,
the inclusion of higher dimensions in the analysis requires care and innovation.
\item
The non-perturbative coordinate
$K_N (\varphi)$ has the additional large-field decay
$\exp[-\kappa g_N L^{dN}|\varphi|^4]$,
which is essential for the control of the integration over the last scale in
Lemma~\ref{lem:V_N_integral_estimates_unified}.
This major improvement is accomplished via the introduction of a new large-field regulator.
\item
We permit \emph{negative} mass in Theorem~\ref{thm:rg_main_theorem_new},
which allows for the analysis of the critical window for values of $\nu$
both above and below the effective
critical point, and in particular below the infinite-volume critical point $\nu_c$.
This is feasible because
an initial condition involving a small $N$-dependent deviation
from the infinite-volume critical point will still allow the RG map
to be applied until scale $N$.
The proof requires precise control of
derivatives of the RG flow with respect to the initial
condition $\nu_0$ and the mass $\ka$.
This control of mass derivatives is another new feature of our work.
\end{enumerate}

The notation and definitions used throughout Sections~\ref{sec:pf1}--\ref{sec:pf2}
largely follow those of \cite{BBS-brief}.
In particular, we assume familiarity with \cite[Chapter~4]{BBS-brief} where the
hierarchical $|\varphi|^4$ model is defined and discussed, and on the definition of
the RG map and the perturbative flow discussed in \cite[Chapter~5]{BBS-brief}.
Our proof requires
improvements and extensions to
the arguments
in \cite[Chapters~8--10]{BBS-brief},
and those three chapters play an important role
in Sections~\ref{sec:fluctation_integral_properties} and \ref{sec:pf2}.

\subsection{The renormalisation group map}

We now examine all scales $j$ and not only the final scale $N$.
Thus we write the scale-$(j+1)$ partition function $Z_{j+1}$
in terms of the scale-$j$ partition function $Z_j$ as
\begin{equation}
\label{eq:Z_j_definition}
	Z_{j+1}(\varphi) = \E_{C_{j+1}}Z_{j}(\varphi +\zeta )	,  \qquad  j \in \{0, \cdots, N-1\} ,
\end{equation}
where the Gaussian integral on the right-hand side integrates over $\zeta$ with $\varphi$
held fixed, and the covariance $C_{j+1}$ arises from the decomposition \eqref{eq:Cjdef}.
For notational simplicity, we write \eqref{eq:Z_j_definition} more compactly as
\begin{equation}
\label{eq:Z_j_definition+}
	Z_{+}(\varphi) = \E_{C_{+}}Z(\varphi +\zeta )  =
    \Eplus Z(\varphi +\zeta ).
\end{equation}
This is consistent with the usage in \cite{BBS-brief}, namely that
when two generic scales $j$ and $j+1$
are under discussion,
we drop the subscript $j$ and replace $j+1$ simply by $+$.
As another example, we write $\cB$ for the set of $j$-blocks
(defined at the beginning of Section~\ref{sec:hier}) and $\cB_+$ for the
set of $(j+1)$-blocks.   Such notational reduction will be used throughout the rest
of the paper.

We introduce several definitions needed to define the RG map.
Further details can be found in \cite{BBS-brief}.

\begin{definition} \cite[Definition~5.1.5]{BBS-brief}
A \emph{polymer activity} at scale $j$ is a smooth function
$F : \cB \times (\R^n)^{\cB} \rightarrow \R$ such that $F(b, \varphi)$ only depends on $\varphi(b)$ (but not otherwise on $b\in \cB$)
and $F(b, \varphi) = F(b,  M \varphi)$ for any $M \in O(n)$.
The space of polymer activities is denoted $\cF$.
\end{definition}

A polymer activity $F$ can simply be identified with the function $\tilde{F} : \R^{n} \rightarrow \R$
such that $\tilde{F}(\varphi (b)) = F (b, \varphi)$, so
we will often interpret $F$ as $\tilde F$, with $\varphi\in \R^n$ rather than
in $(\R^n)^{\cB}$,
when the meaning is clear from the context.
Given a constant field $\varphi$ on a block $b \in \cB$,
and given real parameters $g,\nu,u$,
important examples of polymer activities are polynomials in the field of the form
\begin{align}
& V (b, \varphi)
= L^{jd} \Big( \frac{1}{4} g |\varphi|^4 +\frac{1}{2} \nu |\varphi|^2 \Big),
\label{eq:V_j_form}
\\
\label{eq:U_j_form}
& U (b, \varphi)
= L^{jd} u + V (b, \varphi)
.
\end{align}
We denote the vector spaces of such polynomials $V$ and $U$ as $\cV$ and $\cU$ respectively.
By definition, $\cV \simeq \R^2$ and $\cU \simeq \R^3$, and $\cV$ can be identified
with the subspace of $\cU$ which has $u$ equal to $0$.
The spaces $\cV$ and $\cU$ describe the freedom to choose the parameters
in \eqref{eq:V_j_form}--\eqref{eq:U_j_form}.

The following defines the \emph{localisation} operator.
It extracts from a functional of the
field its \emph{relevant} and \emph{marginal} parts; see \cite[Section~5.2.2]{BBS-brief}
for further discussion.
Although $|\varphi|^4$ is irrelevant
in dimensions $d>4$, we nevertheless extract it with $\Loc$.
This allows a unified treatment of dimensions $d>4$ and $d=4$ via second-order
perturbation theory.  It is also important because
the scaling we use for Theorem~\ref{thm:mr-nongaussian}
makes $|\varphi|^4$ appear to be marginal even in dimensions $d>4$.
More generally,
as discussed in Section~\ref{sec:guide}, we need the $|\varphi|^4$
term to stabilise integrals, and by extracting it we can track it for
this purpose.
This is connected with the fact that
$|\varphi|^4$ is a dangerous irrelevant monomial in dimensions $d>4$.

\begin{definition}\cite[Definition~5.2.2]{BBS-brief}
For a $C^4$-function $F : \R^n  \rightarrow \R$,
the \emph{localisation} of $F$ is its fourth order Taylor expansion at $0$:
\begin{align}
\Loc F (\varphi) = \sum_{p=0}^4 \frac{1}{p !} F^{(p)} (0 ; \varphi, \ldots, \varphi),
\end{align}
where $\varphi \in \R^n$ and
$F^{(p)}(\varphi; \dot\varphi_1,\ldots,\dot\varphi_p)
= \partial_{t_1}|_{t_1=0}\cdots\partial_{t_p}|_{t_p=0} F(\varphi + \sum_{i=1}^p t_i\dot\varphi_i)$.
\end{definition}

\begin{definition}
\label{def:thetaF}
{\cite[Definition~2.1.5]{BBS-brief}}
Given $\zeta :\cB \to \R^n$ and $F : (\R^n)^{\cB}
\to \R$, we define
$\theta F : \R^n \to \R$ by $(\theta F)(\varphi) = F(\varphi + \zeta)$, where we interpret
$\varphi$ in the sum on the right-hand side as a constant field in $(\R^n)^{\cB_+}$.
We do not indicate $\zeta$ explicitly in the notation $\theta F$, but it will always
occur inside a convolution integral of the form
\begin{equation}
    (\E_{+}\theta F)(\varphi) = \E_+ F(\varphi + \zeta),
\end{equation}
where $\varphi\in \R^n$ is fixed and the integration on the right-hand side
is over $\zeta$ in $(\R^n)^{\cB}$.
\end{definition}

\begin{definition}\cite[Definition~5.2.5]{BBS-brief}
\label{def:Phipt}
For  $B \in \cB_+$, we define $\Phi_{\pt}:\cU\to \cU_+$ by
\begin{align}
	\Phi_{\pt} (U ; B)
	= \Eplus \theta  U(B) - \frac{1}{2} \Loc \cov_+ [\theta U(B) ,  \theta U(B)]
	\label{eq:Phi_pt_definition}
	,
\end{align}
where $\cov_+[X,Y]=\E_+(XY)-(\E_+X)(\E_+Y) $.
The subscript ``pt'' refers to ``perturbation theory.''
\end{definition}

Explicitly, for $V \in \cV$
with \emph{coupling constants} $g,\nu$, it is shown in
\cite[Proposition~5.3.5]{BBS-brief} that $\Phi_{\pt}(V)$ has coupling constants given
in terms of explicit $j$-dependent coefficients $\beta_j,\eta_j,\xi_j$
by
\begin{align}
	& g_{\pt}  = g - \beta_j g^2
	\label{eq:g_pt_formula}  \\
	& \nu_{\pt} = (1- \hat\gamma \beta_j g) \nu + \eta_j g_j - \xi_j g^2.
	\label{eq:nu_pt_formula}
\end{align}
Here $\hat\gamma = \frac{n+2}{n+8}$, and, as in
\cite[(5.3.7)]{BBS-brief} (we drop the primes from \cite[(5.3.7)]{BBS-brief}),
\begin{align}
\label{eq:betadef}
    \beta_j &=
	(n+8) (1-L^{-d}) (1+\ka L^{2j})^{-2} L^{-(d-4)j},
    \\
\label{eq:etadef}
    \eta_j & =
    (n+2) (1-L^{-d}) (1+\ka L^{2j})^{-1} L^{-(d-2)j} ,  \\
\label{eq:xidef}
    \xi_j & = \xi_0^0 (1+\ka L^{2j})^{-3} L^{-(2d-6) j},
\end{align}
with an (unimportant) $n$- and $L$-dependent constant $\xi_0^0$.
There is also the coupling constant $u_{\pt}$ (vacuum energy)
but since it cancels in ratios
we require only limited information about
it, which we defer to Section~\ref{sec:prelim2} where it begins to play a role.

In the following definition, $V$ represents a polynomial in $\cV$,
and $U_+=(u_+,V_+)$ represents a polynomial
in $\cU$.  The non-perturbative coordinate $K$ is an element of the space
$\cF$ defined with blocks $b\in\cB$ on scale $j$, while $K_+$ lies in the space
$\cF_+$ defined with blocks $B\in \cB_+$ on the next scale $j+1$.

\begin{definition}\cite[Definition~5.2.8]{BBS-brief}
\label{def:RGmap}
For $\ka > -L^{-2j}$, the \emph{renormalisation group map} (RG map)
\begin{equation}
\label{eq:Phi+def}
    \Phi_+ :(V,K) \mapsto (U_+,K_+) = (u_+,V_+,K_+)
\end{equation}
is defined by
\begin{align}
\label{eq:U+def}
    U_+ & = \Phi_{\pt}(V-  \Loc \,(e^V K)),
    \\
\label{eq:K+B}
	K_{+}(B) &= e^{u_{+}|B|} \E_{+}\theta
    \Big(\prod_{b \in \cB(B)} (e^{-V(b)} + K(b)) \Big) - e^{-V_{+}(B)}  ,
\end{align}
where $B \in \cB_+$ and $\cB(B)= \{b\in \cB: b \subset B\}$.
The domain of $\Phi_+$ consists
of those $(V,K)\in \cV \times \cF$ such that the expectation in \eqref{eq:K+B} is convergent.
We write the components of $\Phi_+$ as
\begin{equation}
\label{eq:Phi-components}
    \Phi_+
    =
    (\Phi_+^U,\Phi_+^K)
    =
    (\Phi_+^u,\Phi_+^V,\Phi_+^K)
    .
\end{equation}
\end{definition}

The notation $\Phi_{+} = (\Phi_{+}^U, \Phi_{+}^K)$ is used to distinguish the
RG flow of the coupling constants
$(u_{+}, g_{+}, \nu_{+})$ and the non-perturbative coordinate $K_{+}$.
The map $\Phi_{+}^U$, which gives the flow of coupling constants,
is the extension
of the explicit map $\Phi_{\pt} (V)$ by a remainder term given by
\begin{align}
	R^U_{+} (V, K) = (r_{g,j}, r_{\nu, j}, r_{u, j}) = \Phi_{+}^{U} (V, K) - \Phi_{\pt} (V) .
	\label{eq:remaining_RG_coords_definition}
\end{align}
With the above definition of the RG map,
it follows as in \cite[Section~5.2.1]{BBS-brief} from basic algebra that
$Z_+ = \E_+ \theta Z$ (recall \eqref{eq:Z_j_definition}) is given by
\begin{align}
\label{eq:ZZ+}
    Z_+ = e^{- (u + u_+ ) |\Lambda|}\prod_{B \in \cB_+}(e^{-V_+(B)} + K_+(B))
    = e^{-u|\Lambda|}\E_{+}\theta \prod_{b \in \cB} (e^{-V(b)} + K(b)).
\end{align}
(This slightly differs from \cite[(5.2.6)]{BBS-brief} because in \eqref{eq:ZZ+} $u_+$ is the
increment in the constant term produced by the RG map whereas in \cite[(5.2.6)]{BBS-brief}
it represents the accumulation of increments.  Our usage agrees with
\cite[Definition~5.2.8]{BBS-brief} and \eqref{eq:Phi+def}.)

\subsection{Regulator, norms and domains}
\label{sec:regulator_norms_domain}

In this section, we
formulate several basic elements of our analysis.

Fix $\tilde{\ka} \geq 0$.
As in \cite[Section~8.3]{BBS-brief}, we define a sequence $\tilde g_j$ by
\begin{align}
\label{eq:tilde_g}
  \tilde g_{j+1} &= \tilde g_j - \beta_j(\tilde{\ka}) \tilde g_j^2
  \quad
  \text{with initial condition $\tilde g_{0}=g$.}
\end{align}
Thus the sequence $\tilde g_j$ is defined via the recursion for $g_{\pt}$ of
\eqref{eq:g_pt_formula}, with the mass in $\beta_j$ set equal to $\tilde \ka$.
For $g>0$ sufficiently small,
it follows from \eqref{eq:betadef}
that this sequence obeys the condition $\tilde{g}_{j+1} \in [\frac{1}{2} \tilde{g}_j,  \tilde{g}_j]$.

We define the \emph{large-field} and \emph{fluctuation-field scales}
\begin{align}
h_j  = k_0 \tilde{g}_{j}^{-1/4} L^{ - dj/4 },
\qquad
\label{eq:h_j_definition}
\ell_j = \ell_0 L^{- (d-2)j/2} ,
\end{align}
with $\ell_0=L^{1+d/2}$, and with $k_0$ a constant which is chosen
sufficiently small depending on $n$ (see Remark~\ref{remark:nuca} and Lemma~\ref{lemma:stability_estimate-mass_derivative}).
According to the definition of our Gaussian scaling factor $\lp_N$ in
\eqref{eq:lpdef}, $\ell_N = \ell_0 \lp_N$.  Also, when the mass obeys $\ka \in \II_{\rm crit}$,
it follows from Lemma~\ref{lem:g_N_crit_behaviour} and the definition of our
non-Gaussian scaling factor $\hh_N$ in
\eqref{eq:hhdef} that $h_N$ is essentially a multiple of $\hh_N$.
We require that $g$ be
sufficient small depending on $n$ and $L$ so that $\ell_j < h_j$ for all $j$.

We need to control various derivatives of $\Phi_+$.
For a function $f :(X,\|\cdot\|_X) \rightarrow (Y,\|\cdot\|_Y)$,
its Fréchet derivative at $x \in X$ (when it exists) is the
unique linear operator $Df(x) \colon X \to Y$ satisfying
\begin{equation}
	\lim_{\|h\|_X \to 0}\frac{\|f(x+h)-f(x)-Df(x;h)\|_{Y}}{\|h\|_X} = 0.
\end{equation}
Given a normed space $(\cZ,  |\cdot|_{\cZ})$
and an infinitely
Fr\'echet-differentiable function $F : \cZ \rightarrow \R$,
we define
\begin{align}
	\norm{F}_{T_{z}} = \sum_{p=0}^{\infty} \frac{1}{p!} \norm{F^{(p)}(z)}_{\cZ}
    \quad
    \text{with}
    \quad
	\norm{F^{(p)} (z)}_{\cZ} = \sup \{ F^{(p)}(z; \dot{z}_1, \cdots, \dot{z}_p) : |\dot{z}_i|_{\cZ} \leq 1 \} ,
	\label{eq:T_z_seminorm_definition}
\end{align}
where $F^{(p)}(z; \dot{z}_1, \cdots, \dot{z}_p)$ is the $p^{\rm th}$ Fréchet derivative at
$z$ in directions $\dot{z}_1, \cdots, \dot{z}_p$.
We use properties of this norm which are developed in \cite[Chapter~7]{BBS-brief}.

Although a more elaborate choice of the space $\cZ$ will be used in Section~\ref{sec:pf2},
throughout Section~\ref{sec:pf1}
we take
$\cZ=\R^n$
with
$| \varphi |_{\cZ} = \norm{\varphi}_2 / \mathfrak{h}$
where $\mathfrak{h}$ is either $\ell$ or $h$.
In this case we write the $T_z$ norm as $\norm{\cdot}_{T_\varphi (\mathfrak{h})}$.
For example, for integers $p \ge 0$ the
$T_0(\mathfrak{h})$-norm of a monomial on a block $b\in \cB_j$ is
$\norm {|\varphi|^{2p}}_{T_0(\mathfrak{h})} = \mathfrak{h}^{2p}$.
Note that here in $T_\varphi(\mathfrak{h})$ the subscript is
$\varphi=0$ so derivatives are evaluated at $0$.
In particular, if we keep track only of powers of $L^j$
(permitting constants to depend on $\ell_0$ and $\tilde g_j$)
we see that
\begin{equation}
\label{eq:monomials}
    L^{dj}\norm {|\varphi|^{2p}}_{T_0(\ell)}
    \asymp
    \begin{cases}
    L^{dj} & (2p=0)
    \\
    L^{2j} & (2p=2)
    \\
    L^{-(d-4)j} & (2p=4)
    \\
    L^{-(2d-6)j} & (2p=6),
    \end{cases}
    \qquad
    L^{dj}\norm {|\varphi|^{2p}}_{T_0(h)}
    \asymp
    \begin{cases}
    L^{dj} & (2p=0)
    \\
    L^{dj/2} & (2p=2)
    \\
    L^{0} & (2p=4)
    \\
    L^{-dj/2} & (2p=6).
    \end{cases}
\end{equation}
Thus, monomials with power $2p=0,2$ are relevant,
the monomial with $2p=4$ is marginal for $d=4$ and irrelevant for $d>4$, and the monomial
with power $2p=6$ is irrelevant in dimensions $d \ge 4$.
The $L^0$ for the case $2p=4$ of the $T_0(h)$-norm is a sign that
this monomial will play an important role, in particular for ensuring
stability of integrals, and we need to track it carefully even when $d>4$;
it is a so-called \emph{dangerous irrelevant variable} \cite{Fish83} for $d>4$.
The monomial $|\varphi|^6$ plays a role in the determination of our upper bounds
on the non-perturbative coordinate $K$, which we believe are not sharp for $d\ge 6$
due
to the fact that our estimates do not take into account the decay in coefficients
of $|\varphi|^6$
arising in perturbation theory.  Although not sharp for $d \ge 6$, our estimates on $K$ are
sufficient to prove our main results.

We define the \emph{regulator} at scale $j$ by
\begin{align}
\label{eq:G_j_definition}
	G_j (b,  \varphi) =
    e^{- \kappa |\varphi/h_j|^4}
	\qquad
	(b\in \cB_j , \;\; \varphi \in \R^n),
\end{align}
where the dependence on $b$ merely indicates that $\varphi$ is constant on $b$.
The constant
$\kappa >0$ is chosen to satisfy smallness conditions (depending only on $n$),
see Lemma~\ref{lemma:hat_K_bound}
and the proofs of Propositions~\ref{prop:S0} and \ref{prop:crucial}.
Some of the conditions on $\kappa$ are technical ones,
but the crucial restriction is that
$\kappa$ is smaller than the constant
$\cst$ in Lemma~\ref{lemma:hat_K_bound}.
The latter restriction ensures that the exponent of
$G_j$ remains smaller than the natural decay rate of $\norm{e^{-V_j(b)}}_{T_\varphi (h_j)}$,
namely $L^{dj}\norm{g_j|\varphi|^4}_{T_\varphi (h_j)} \asymp
|\varphi / h_j|^4$.
The regulator is used to define the norm
\begin{align}
	\norm{K}_{T_j^G (h_j) }
	& = \sup_{\varphi \in \R^n} \frac{\norm{K (b)}_{T_{\varphi} (h_j)}}{G_j (b, \varphi)}  .
\label{eq:TG-norm-def}
\end{align}
Often regulators control the divergence of $K_j$ as $|\varphi| \rightarrow \infty$ (e.g.,  \cite{BS-rg-IE,Bryd09}), but it is the opposite here---the
regulator ensures a decay rate on $K_j$ for large fields.

We define
\begin{align}
	\scale_j &= L^{ -(d-4) j}   .
\end{align}
Our bounds on $K_j$ involve $\scale_j$ raised to the
$d$-dependent exponents
\begin{align}
\label{eq:kbd}
	\kaa (d) =&
	\begin{cases}
	3 & (4 \le d < 6) \\
	\frac{2d- 6}{d-4} (1- \epsilon(d))  & (d \ge 6),
	\end{cases}
	\qquad \quad 	
    \kp (d) =
	\begin{cases}
	\frac{3}{4} & (4\le d < 12) \\
	\frac{d}{2(d-4)}  (1- \epsilon(d))  & (d \ge 12),
	\end{cases}
\end{align}
and $\kb (d) = \kaa(d) - \kp (d) >0$,
where $\epsilon(d)> 0$ is such that $\kaa(d) > 2$,  $\kp(d) > \frac{1}{2}$.
These two conditions are satisfied if and only if $\epsilon (d) \in (0, \frac{1}{d-3})$.

Using these parameters, we define
\begin{align}
\label{eq:Wkappa-norm-def}
	\norm{K}_{\Wkappa_j}
    &= \norm{K}_{T_{0} (\ell_j)} + \tilde{g}_j^{9/4}\scale_j^{\kb}  \norm{K}_{T^{G}_j (h_j)}.
\end{align}
We write $\Wkappa_j$ for the space of $K \in \cF_j$ such that $\norm{K}_{\Wkappa_j} < \infty$.
The weight $\tilde{g}_j^{9/4}\scale_j^{\kb}$ serves to equalise our bounds on the
two terms on the right-hand side, since we will eventually establish estimates of
the form
\begin{align}
    \norm{K}_{T_{0} (\ell_j)} \le O(g_j^3 \scale_j^{\kaa}),
    \qquad
    \norm{K}_{T^{G}_j (h_j)} \le  O(g_j^{3/4} \scale_j^{\kp}).
\end{align}
We invite the reader to focus on the cases $d = 4,5$,  for which
$\kaa=3$, $\kb=9/4$, $\kp=3/4$, and the above estimates
become
\begin{align}
    \norm{K}_{T_{0} (\ell_j)} \le O(g_j^3 \scale_j^{3}),
    \qquad
    \norm{K}_{T^{G}_j (h_j)} \le  O(g_j^{3/4} \scale_j^{3/4})
    \qquad (d=4,5).
\end{align}

For $d=4$, the $\Wkappa_j$-norm of \eqref{eq:Wkappa-norm-def} is designed to replace the norm
\begin{equation}
    \norm{K}_{\cW_j} = \norm{K(b)}_{T_0 (\ell_j)}
    + \tilde{g}_{j}^{9/4}
    \sup_{\varphi \in \R^n} \norm{K(b)}_{T_{\varphi}(h_j)}
\end{equation}
of \cite[(8.2.9)]{BBS-brief}, both
by taking into account dimensions $d>4$, and by
imposing a specific exponential decay on $K$ for large fields.
By definition,
\begin{align}
\label{eq:normstronger}
	\norm{K}_{T_{j}^G (h_j)} \geq
    \sup_{\varphi \in \R^n} \norm{K(b)}_{T_{\varphi}(h_j)}
	\qquad \text{and} \qquad
	\norm{K}_{\Wkappa_{ j}} \geq \norm{K}_{\cW_{ j}} .
\end{align}
	The $\cW^\kappa$-norm dominates the $\cW$-norm (which has no regulator),
	so upper bounds on the former imply upper bounds on the latter.

Given $\tilde{\ka} \in \R$, recall the mass scale $j_{\tilde{\ka}}$ given by \eqref{eq:mass-scale}, i.e.,
the greatest integer $j$ such that $L^{2j} \tilde{\ka} \leq 1$, with the degenerate case $j_{\tilde{\ka}} = \infty$ if $\tilde{\ka} \leq 0$.
We define
\begin{equation}
\label{eq:varthetadef}
    \tilde{\vartheta}_j = \begin{cases}
    	2^{-(j-j_{\tilde{\ka}})_+} & ( \tilde{\ka} < 1) \\
	    \tilde{\ka}^{-1} 2^{-j} & ( \tilde{\ka} \geq 1),
     \end{cases}
\end{equation}
which we use to keep track of decay that arises when $j$ is above the mass scale.
As in \cite[(6.1.19)]{BBS-brief}, we define a domain $\II_j$ for
the mass $\ka$ by
\begin{align}
\label{eq:IIdef}
	\II_j (\tilde{\ka}) =
	\begin{array}{ll}
	\begin{cases}
	(- \frac{1}{2} L^{-2j},   \frac{1}{2}  L^{-2 j} )  &  (\tilde{\ka} = 0) \\
	(\frac{1}{2} \tilde{\ka}, 2\tilde{\ka} )  & (\tilde{\ka} > 0).
	\end{cases}
	\end{array}
\end{align}
Note that the interval $\II_j$ actually has no $j$-dependence unless $\tilde\ka =0$.
We emphasise that $\II_j(0)$ permits negative $\ka$;
as mentioned previously this is a major novelty of our analysis which is absent in \cite{BBS-brief} and related RG papers.
Our choice of the domain $\II_{N-1}(0)$ ensures that
the potentially ill-behaved factors $(1+\ka L^{-2j})^{-k}$
appearing in \eqref{eq:betadef}--\eqref{eq:xidef} at scale
$j<N$
are bounded by absolute constants for $\ka \in \II_{N-1}(0)$.
If we define $\vartheta_j$ using $\ka \in \II_j (\tilde{\ka})$ instead of
$\tilde{\ka}$, then by definition
$\vartheta_j \in [\frac{1}{2} \tilde{\vartheta}_j, 2 \tilde{\vartheta}_j]$
so $\vartheta_j$ and $\tilde{\vartheta}_j$ can be used interchangeably in inequalities up
to a usually unimportant factor $2$ or $1/2$.

At scale $j$, derivatives with respect to $\ka$ create factors $L^{2j}$.
This can be inferred, e.g., from the $\gamma_j$ terms in the covariance matrix
in \eqref{eq:gamma_j_per}.
It
is therefore natural to place these scaling factors inside a norm on $\II_j$,
and we equip
$\II_j(\tilde\ka)$ with the norm where $x\in \R$
has norm $|x| L^d L^{2j}$.
The decision to include also the factor $L^d$
has been made for a reason which
arises in the proof of
Lemma~\ref{lemma:mass_derivative_of_expectation},
where mass
derivatives are investigated in detail.
With this definition, the norm of the Fr\'echet derivative of
$f:\R\to\R$, with the above norm on the domain of $f$ (but not the range), is
\begin{equation}
\label{eq:massnorm}
    \|Df(a)\| = \sup_{|x| L^dL^{2j} \le 1} \Big| \frac{d}{dt}\Big|_{t=0} f(a + tx)\Big|
    = \frac{1}{L^dL^{2j}} \Big| \frac{df(a)}{da} \Big|,
\end{equation}
and the prefactor on the right-hand side cancels the growth typical of mass derivatives.

As in \cite[(8.2.1)]{BBS-brief},
with the sequence $\tilde g_j$ given by \eqref{eq:tilde_g}, with
$k_0$ as in \eqref{eq:h_j_definition}, and given
$\tilde{\ka} \ge 0$, we define a domain for $\cV$ by
\begin{align}
\label{eq:domain_D_j}
	& \cD_j = \big\{ (g, \nu) \in \R^2 :
	2k_0 \tilde{g}_j < g < {\textstyle{\frac{1}{2k_0}}}  \tilde{g}_j,
	\;\;  |\nu| < {\textstyle{\frac{1}{2k_0}}}  \tilde{g}_j  \scale_j L^{-2j}  \big\}.
\end{align}
Elements $V\in \cD_j$ are equipped with the norm $\|V(b)\|_{T_0(\ell)}$ for the space $\cV_j$, and we denote the normed space by $\cV(\ell)$. The space $\cU(\ell) \simeq \R^3$ is defined similarly with the norm $\|U(b)\|_{T_0(\ell)} = \|V(b)\|_{T_0(\ell)} + |u||b|$
which at scale $j$ obeys (by \eqref{eq:monomials})
\begin{equation}
\label{eq:Ub}
    \|U(b)\|_{T_0(\ell)} \asymp  |g| \scale_j + |\nu| L^{2j}  + |u|L^{dj}.
\end{equation}
Given $C_{\rg}>0$ (we make a specific choice in Theorem~\ref{thm:Phi^K_estimate}),
we define
a domain in $\cF_j$ by
\begin{align}
\label{eq:Kcaldef}
    \cK_j & =
    \{K \in \cF_j  :
    \|K\|_{\Wkappa} < C_{\rg} \tilde{\vartheta}_j^3 \tilde{g}_j^{3} \scale_j^{\kaa} \}.
\end{align}
Finally, we restrict the domain of $\Phi_+$ of Definition~\ref{def:RGmap},
as in \cite[(8.2.12)]{BBS-brief},
to be the set of $(V,K,\ka)$ in
\begin{equation}
  \label{eq:Phi_+-domain}
  \domRG_j = \cD_j \times  \cK_j \times \II_j.
\end{equation}
By definition, $\domRG$ depends on the choice of $\tilde\ka$ and on the initial
condition $g$ for the recursion \eqref{eq:tilde_g}.
For $d=4$,  $\cD_j \times \cK_j$ is smaller than the domain
that is defined in \cite[(8.2.12)]{BBS-brief}
using the $\cW$ norm instead of the $\Wkappa$ norm, due to \eqref{eq:normstronger},
and also our $\domRG_j$ in \eqref{eq:Phi_+-domain} has an extra coordinate for $\ka$.
It should be noted that the dependence of the RG map $\Phi_{+}$ on the coordinate $\ka$
arises solely from the covariance $C_{+}$ of the Gaussian expectation $\E_+$.

\subsection{Main theorems for the RG map}

To iterate and control the RG map, we need certain estimates involving Fréchet derivatives with respect to $V$ and $K$.
We consider maps $F\colon \cV(\ell)\times \Wkappa \times \II \to \cX$ taking values in a normed space $\cX$, with $\cX$ either $\cU_+(\ell_+)$ or $\Wkappa_+$.
For $(V,K,\ka) \in \cV(\ell) \times\Wkappa \times \II$, the Fréchet derivative with respect to $V$
or $K$ or $\ka$ is denoted $D_V$ or $D_K$ or $D_{\ka}$, respectively, so that $D_V^{p}D_K^{q}
D_{\ka}^{r} F(V,K,a)$ is a multilinear map $\cV(\ell)^{p}
\times \cF^{q} \times \II^r \to \cX$. We write
\begin{align}
\label{eq:V_K_frechet_deriv}
	\|D_V ^{p} D_K^{q} D_{\ka}^{r} F(V,K,a)\|_{\D \to \cX}
	&= \sup_{\dot V^p,\dot K^q,\dot a^r}\|D_V ^{p} D_K^{q} D_{\ka}^{r} F(V,K,a;\dot V^p,\dot K^q,\dot a^r)\|_{\cX}
\end{align}
for the norm of this multilinear map
and the supremum is taken over
all $\dot V^p$, $\dot K^q$, $\dot a^r$ in $\cV(\ell)^{p}
$, $\cF^{q}$ and $\II^r$ of unit norm in their respective spaces. The notation $\D$ for the operator norm on the
left hand-side of \eqref{eq:V_K_frechet_deriv} keeps track of the domain
where $(V,K,a)$ (\emph{not} the directions $\dot V$, $\dot K$, $\dot\ka$) lives.
In bounds that are uniform in $(V,K,a)$, we omit the arguments
appearing in \eqref{eq:V_K_frechet_deriv}.

In Theorems~\ref{thm:Phi^U_estimate}--\ref{thm:Phi^K_estimate}, we use the following hypothesis.
\begin{quote}
\begin{itemize}
\item[\customlabel{quote:assumPhi}{$\assumPhi$}]
Let $\tilde{\ka} \ge 0$, let $L$ be sufficiently large, and let $\kappa$
be sufficiently small
(the latter depending only on $n$).
At scale $j$, let $\tilde{g} = \tilde{g}_j$ be sufficiently small depending on $L$,
and let $(V, K) \in \domRG = \domRG_j$.
\end{itemize}
\end{quote}

The following two theorems lie at the heart of the proof of
Theorems~\ref{thm:rg_main_theorem_large_m},
\ref{thm:mass_trick_big_mass}, and \ref{thm:rg_main_theorem_new}.
Recall from \eqref{eq:kbd} that
\begin{align}
\label{eq:kbd-bis}
	\kaa (d) =&
	\begin{cases}
	3 & (4 \le d < 6) \\
	\frac{2d- 6}{d-4} (1-\epsilon (d))  & (d \ge 6),
	\end{cases}
	\qquad \quad 	
    \kb (d) =
	\begin{cases}
    \kaa(d) - \frac 34 & (4\le d <12) \\
	\frac{3}{2} (1-\epsilon (d))   & (d \ge 12) ,
	\end{cases}
\end{align}
with $\epsilon(d) \in ( 0, \frac{1}{d-3} )$.
In the two theorems, the limitation $p_{\ka}$ on the number of $\ka$-derivatives
is arbitrary but fixed and finite.
For their application to prove Theorems~\ref{thm:book_main_theorem_bis}--\ref{thm:mr-gaussian},
we need $p_{\ka}$ to be at least $2$.

\begin{theorem} \label{thm:Phi^U_estimate}
Let $d \ge 4$.
At scale $j \in \{ 0, \ldots, N-1 \}$, assume \ref{quote:assumPhi}
and let $r \in \{0, \ldots, p_{\ka} \}$.
There exist positive constants
$\tilde M_{p,q}$ (for
integers $p,q\ge 0$),
which depend on $C_{\rg}$, such that the map
$R^U_{+} : \domRG \rightarrow \cU_{+}$ satisfies
\begin{align}
	\norm{ D^p_V D^q_K D_{\ka}^r  R^U_{+} }_{\domRG  \rightarrow \cU_{+}(\ell_+)}
	\leq
	\begin{array}{ll}
	\begin{cases}
	M_{p, 0}  \tilde{\vartheta}_{+}^3  \tilde{g}_+^3 \scale_+ ^{\kaa} & ( p\geq 0,\, q=0) \\
	M_{p,q}  & ( p\geq 0,\, q \geq 1) \\
	0 & ( p\geq 0,\, q\geq 3).
	\end{cases}
	\end{array}
	\label{eq:R^U_derivative_estimate}
\end{align}
\end{theorem}

By \eqref{eq:Ub},
the bound \eqref{eq:R^U_derivative_estimate} on $R_+^U$ implies bounds on
its coupling constants $(r_{g}, r_{\nu})$:
\begin{align}
	\norm{ D^\alpha r_{g} }_{\domRG \rightarrow \R} \leq
	O(\scale_{+}^{-1})
	\norm{D^\alpha R_{+}^U}_{\domRG \rightarrow \cU_+ (\ell_+)},
	\qquad
	\norm{ D^\alpha r_{\nu} }_{\domRG \rightarrow \R}
    \leq O(L^{-2 (j+1)}) \norm{D^\alpha R_{+}^U}_{\domRG \rightarrow \cU_+ (\ell_+)},
	\label{eq:r^U_domination_by_R^U}
\end{align}
for any mixed derivative $D^\alpha$ with respect to $(V, K, \ka)$.

\begin{theorem} \label{thm:Phi^K_estimate}
Let $d \ge 4$.
At scale $j \in \{0,\cdots, N - 1\}$, assume \ref{quote:assumPhi} and let
$r \in \{0, \ldots, p_{\ka} \}$.
There exists a positive constant $C_{\rg}$ such that
$\Phi_{+}^K$ is well-defined as a map $\domRG \rightarrow \Wkappa$ (i.e., the integral \eqref{eq:K+B} converges), and there exist positive $M_{p,q}$ (for
integers $p,q \ge 0$) and $L$-independent $M'_{0,1}$ such that
\begin{align}
	\norm{ D^p_V D^q_K D_{\ka}^r \Phi_{+}^K}_{\domRG  \rightarrow \Wkappa_{+}} \leq
	\begin{array}{ll}
	\begin{cases}
	C_{\rg} \tilde{\vartheta}_{+}^3  \tilde{g}_{+}^3 \scale_+^{\kaa} & (p=0,\, q=0) \\
	M_{p, 0} \tilde{\vartheta}_{+}^{3} \tilde{g}_{+}^{3-p} \scale_+^{\kaa-p} & (p > 0,\, q=0) \\
	M'_{0,1} L^{-\max\{ 2,  (d-4) \kaa \} }    & (p = 0,\, q=1) \\
	M_{p,q} \tilde{g}_{+}^{-p- \frac{9}{4}(q-1)}
    \scale_+^{-p-
                \kb
    (q-1)} & (p\geq 0,\, q\geq 1).
	\end{cases}
	\end{array}
	\label{eq:Phi^K_derivatives_estimates-mass_derivative}
\end{align}
\end{theorem}

The case $(p,q) = (0,1)$ provides a bound on the
$K$-derivative of $\Phi_+^K$ which we refer to as the \emph{crucial contraction}.
It is this latter estimate that requires most of
our work. The case $(p,q)=(0,0)$
(which turns out to follow from the crucial contraction)
implies that under \ref{quote:assumPhi}, the non-perturbative coordinate at the next scale,
$K_+ = \Phi_+^{K}$, satisfies $K_+ \in \cK_+$.
This allows for the RG map to be iterated over scales to
create an $\rg$ flow which we formalise and study in detail
in the next section.

\section{The RG flow: proof of Theorems~\ref{thm:rg_main_theorem_large_m},
\ref{thm:mass_trick_big_mass}, \ref{thm:rg_main_theorem_new}}
\label{sec:pf3}

In this section, we prove Theorems~\ref{thm:rg_main_theorem_large_m},
\ref{thm:mass_trick_big_mass}, \ref{thm:rg_main_theorem_new}, subject to
Theorems~\ref{thm:Phi^U_estimate} and \ref{thm:Phi^K_estimate}.  An ingredient
in the proof is Proposition~\ref{prop:nu-derivs}, and its proof is deferred to
Section~\ref{sec:RGderivs}.
Thus, after Section~\ref{sec:pf3}, what remains
is to prove Theorems~\ref{thm:Phi^U_estimate}--\ref{thm:Phi^K_estimate}
and Proposition~\ref{prop:nu-derivs}.

\subsection{Reduction of proofs of Theorems~\ref{thm:rg_main_theorem_large_m},
\ref{thm:mass_trick_big_mass}, \ref{thm:rg_main_theorem_new}}

In this section, we state
Propositions~\ref{prop:stable_manifold}--\ref{prop:nu_c,N_mass_derivative}
and Lemma~\ref{lemma:g_j_asymptotic},
and use them to prove Theorems~\ref{thm:rg_main_theorem_large_m}--\ref{thm:rg_main_theorem_new}.
The following definition formalises the notion of an RG flow.

\begin{definition}
\label{def:RGflow}
Let $k \in \N$.
Given $\tilde\ka\ge 0$, initial coupling constants $g_0>0$ and $\nu_0\in \R$ defining $V_0$,
and $K_0=0$, we say that
$(V_j,K_j)_{j\leq k}$ is an \emph{RG flow to scale} $k$ \emph{with initial condition} $(g_0,\nu_0)$
\emph{and mass} $\ka$,
if it obeys
\begin{equation}
    (V_{j+1},K_{j+1}) = \big(\Phi_{j+1}^V(V_j,K_j,\ka) , \Phi_{j+1}^K(V_j,K_j,\ka) \big)
    \qquad (0 \le j <k)
\end{equation}
with
\begin{equation}
\label{eq:RGflow_in_D}
    (V_j,K_j,\ka) \in \domRG(\tilde\ka)  \qquad (0 \le j <k) .
\end{equation}
For fixed $g_0$, we write $\mathcal{X}_k$ for the set of $(\nu_0, \ka)$ consisting of
such initial $\nu_0$ and mass $\ka$.
When an RG flow exists to scale $k$ for all $k \in \N$ then we refer to it
as a \emph{global RG flow},
and write
$\mathcal{X}_{\infty}=\cap_{k=1}^\infty \mathcal{X}_k$.
\end{definition}

An important observation is that, by \eqref{eq:RGflow_in_D},
the bounds of Theorems~\ref{thm:Phi^U_estimate}--\ref{thm:Phi^K_estimate} hold
for all scales $j \le k$ for an RG flow to scale $k$.
Also, an RG flow to scale $k+1$ with a given initial condition and mass simply
extends the RG flow to scale $k$ with the same initial condition and mass.
In other words, lengthening an RG flow does not alter its initial portion.
This follows from the definition of the RG map, in which $(V_j,K_j)$ is
determined by scales prior to scale $j$ and is not affected by future scales.

Concerning the relative roles of $\ka$ vs $\tilde\ka$, given any $\ka >0$ we can
always choose (many) $\tilde \ka>0$ such that $\ka \in \II(\tilde\ka) = (\frac 12 \tilde \ka,
2\tilde\ka)$.  On the other hand, if $\tilde\ka =0$ then a requirement that
$\ka \in \II_j(0) = (-\frac 12 L^{-2j},\frac 12 L^{-2j})$ is restrictive, and if
the requirement is to hold for all scales $j$ then we must have $\ka=0$.

Each RG map $\Phi_{j+1}$ depends on $\ka$ through the covariance $C_{j+1}$.
Although these maps do not depend on $\nu_0$,  an RG flow to scale $k$ with initial condition $(g,\nu_0)$ does depend on both $\ka$ and $\nu_0$, as does any function of $(V_j,K_j)$ in an RG flow  with initial
condition $(g,\nu_0)$.
Partial derivatives of a function $f (\nu_0, \ka)$ with respect to
$\nu_0$ and $\ka$ are denoted as
\begin{equation}
    f'=\frac{\partial f}{\partial \nu_0}, \qquad \dot{f}=\frac{\partial f}{\partial \ka} .
\end{equation}
We use this notation heavily.
Moreover, in order to construct an RG flow it is necessary also to choose the initial
value $\nu_0$ to depend on $\ka$.

We equip the subsets $\mathcal{X}_k$ and $\mathcal{X}_\infty$
of $\R^2$ with the subset topology.  Thus,
when we say a function $f: \mathcal{X}_k \rightarrow \R$ is continuous, it means that $f$ is continuous with respect to this topology.
Given a mass $\ka\ge 0$, we will determine a critical value
$\nu_c(\ka)$ from which we can construct a global RG flow, and in this case
the set $\mathcal{X}_\infty$ consists of elements of the form $(\nu_c(\ka),\ka)$,
and continuity reduces to continuity in $\ka$.

\subsubsection{The main propositions}

\begin{proposition}[Existence of critical $\nu_0$]
\label{prop:stable_manifold}
Let $d \ge 4$.  Let $g>0$ and $\kappa >0$ be sufficiently small and let $\ka \geq 0$.
There exists a continuous function $\nu_c: [0,\infty) \to \R$ (depending also on $g$)
such that
$(V_j,K_j)$ is a global RG flow with initial condition $(g,\nu_c(\ka))$ and mass $\ka$,
with $\domRG_j$ defined by any $\tilde\ka$ such that
$\ka \in \II_j(\tilde\ka)$ for all $j \ge 0$.
The critical value $\nu_c(\ka)$ satisfies
$|\nu_c (\ka)| \leq O( \vartheta_{0} g)$, and $\nu_c (0) \sim  - (n+2) g (-\Delta_H)^{-1}_{00}$
as $g\downarrow 0$.
\end{proposition}

The occurrence of $\vartheta_0$ in the last sentence of Proposition~\ref{prop:stable_manifold} has significance only for large mass,
since by definition $\vartheta_0=1$ for $\ka <1$ and $\vartheta_0=a^{-1}$ for $\ka \ge 1$.
For $d>4$, it is shown in the
proof of Theorem~\ref{thm:rg_main_theorem_large_m}(ii)
in Section~\ref{sec:pf-reduction}  that
\begin{equation}
\label{eq:Ad5}
    A_{d} = 1 + \dot\nu_c (0) \qquad (d>4).
\end{equation}
The constant $A_{4}$ in Theorem~\ref{thm:book_main_theorem_bis} appears
in the next proposition.

\begin{proposition}[Derivative of the critical value]
\label{prop:nu_c_derivative_in_mass}
For $d >4$, the critical value $\nu_c (\cdot)$ is a
strictly increasing continuously differentiable function of
$\ka \geq 0$, with $\dot\nu_c (\ka) = O(g)$.
For $d=4$, $\nu_c (\cdot)$ is
strictly increasing and is
continuously differentiable for $\ka >0$,
and there is a constant $A_{4} > 0$ such that
\begin{align}
\label{eq:dotnuc0}
	\dot \nu_c (\ka)
    \sim A_{4}
    (\log  a^{-1} )^{\hat{\gamma}} \qquad \text{as $\ka \downarrow 0$,}
\end{align}
with $A_{4} \sim ( \frac{B g}{\log L^2} )^{\hat{\gamma}}$ as $g \downarrow 0$.
\end{proposition}

The next proposition permits small negative mass $\ka$.

\begin{proposition}
\label{prop:nu_c,N_mass_derivative}
Let $d \ge 4$.  Let $g>0$ and $\kappa >0$ be sufficiently small.
For $N \in \N$, there exists a continuously differentiable function
$\nu_{0,N} : \II_{N-1} (0) \rightarrow \R$,
with $\nu_{0,N} (0) = \nu_c (0)$, such that
$(V_j,K_j)$ is an RG flow to scale $N$ with initial condition $(g,\nu_{0,N}(\ka))$
and mass $\ka\in\II_{N-1} (0)$.
There exists a constant $c^\free=O(g)$ such that the derivative of $\nu_{0,N}$ satisfies
\begin{align}
	\dot \nu_{0,N} (\ka) =
	\begin{cases}
	 A_{4} (\log L^2)^{\hat{\gamma}}
    \big(  N^{\hat{\gamma}} - 1 + c^{\free}  \big) \big( 1 + O(N^{-1}))   & (d=4) \\
	\dot \nu_c (0) + O(g_N \scale_N^{\kaa-2})	 & (d > 4).
	\end{cases}
\end{align}
\end{proposition}

Under the hypotheses of
Proposition~\ref{prop:stable_manifold},
the estimates
of Theorems~\ref{thm:Phi^U_estimate} and \ref{thm:Phi^K_estimate} all hold for all scales $j$.
In particular,
by \eqref{eq:r^U_domination_by_R^U}
the correction $r_{g,j}$ to the perturbative flow of $g_j$ is
$O(\tilde\vartheta_j^3 g_j^3 \scale_j^{\kaa-1})$.
From this we can determine the asymptotic behaviour of $g_j$, as in
the following lemma.  Recall that the mass scale $j_{\ka}$ is defined in \eqref{eq:mass-scale}.
The constant $g_\infty$ which appears in the definition of the window scale $w_N$
in \eqref{eq:window_choice} and
in the formula for $\hh_N$ in \eqref{eq:hhdef} is equal to $g_\infty=g_\infty(0)$.
We postpone the proof of Lemma~\ref{lemma:g_j_asymptotic} to Section~\ref{sec:prop5pfs}
where we prepare the derivative estimates  \eqref{eq:g_j_derivative_in_nu_0_high_d};
the rest of Lemma~\ref{lemma:g_j_asymptotic}
follows easily from the recursion relation for $g_j$.
Derivatives of $g_j$ are also needed for $d=4$, but not yet, and their
analysis is postponed to Section~\ref{sec:RGderivs}.

\begin{lemma}
\label{lemma:g_j_asymptotic}
Let $d \ge 4$.
Let $\tilde{\ka} \ge 0$.
Let $g>0$ and $\kappa >0$ be sufficiently small.
  The following statements also hold when $g_j$ is
replaced by $\tilde g_j$.
\begin{enumerate}
\item
Let $k \in \N$ and suppose that $(V_j,K_j)$
is an RG flow to scale $k$ with initial condition $(g,\nu_0)$ (for some $\nu_0$)
and mass $\ka$. Then, for each $j \le k$,
\begin{align}
\label{eq:gjasy}
	g_j  & \;
    \begin{cases}
    = (Bj)^{-1} + O(j^{-2} \log j) & (d=4, \, \tilde \ka =0)
    \\
    \asymp (j\wedge j_{\ka})^{-1} + O((j\wedge j_{\ka})^{-2} \log (j\wedge j_{\ka})) &  (d=4, \,   j_{\tilde{\ka}} \geq 1 )
    \\
    = g + O(g^2) &  (d=4, \,  j_{\tilde{\ka}} \leq 0  \text{ or }  d>4 )    ,
    \end{cases}
\end{align}
and, for $d>4$, $g_j$ is a differentiable function of
$(\nu_0,\ka) \in \mathcal{X}_k$
with
\begin{align}
	& | g'_j | , \;\;
	| \dot g_j | \leq
    O(L^{2j}) \tilde{\vartheta}_{j}^2  \tilde{g}^2_j   \scale_j^{\kaa -2}
.
	\label{eq:g_j_derivative_in_nu_0_high_d}
\end{align}

\item
For $d>4$, or for $d=4$ with $j_{\tilde{\ka}} < \infty$,
suppose that $(V_j, K_j)$ is a global RG flow with initial condition $(g,\nu_0)$ (for some $\nu_0$)
and mass $\ka$.
Then there exists a continuous function $g_{\infty} (\nu_0,  \ka) = g + O(g^2)$ of $(\nu_0,\ka) \in \mathcal{X}_{\infty}$ such that
\begin{align}
	g_j  =
	\begin{cases}
	g_{\infty}  + O( \vartheta_j^2 g^2) & (d=4, \; j \geq j_{\tilde{\ka}} )
    \\
   	g_{\infty}  + O( g^2 \scale_j )
   & (d > 4) 	
   .
    \end{cases}	
	\label{eq:gjasy-2}
\end{align}
\end{enumerate}
\end{lemma}

\subsubsection{Proofs of Theorems~\ref{thm:rg_main_theorem_large_m},
\ref{thm:mass_trick_big_mass}, \ref{thm:rg_main_theorem_new}}
\label{sec:pf-reduction}

\begin{proof}[Proof of Theorem~\ref{thm:rg_main_theorem_large_m}]
For $\ka \geq 0$, we take $\nu_0 = \nu_c (\ka)$ to be the critical value from Proposition~\ref{prop:stable_manifold}.
The asymptotic formulas \eqref{eq:A,B_definition_bis} for $\nu_c(0)$ and $A_d$
as $g \downarrow 0$, which are claimed in Theorem~\ref{thm:rg_main_theorem_large_m},
are part of the statements of Proposition~\ref{prop:stable_manifold}
and Proposition~\ref{prop:nu_c_derivative_in_mass}.  We proceed to prove the estimates
for $d=4$ in (i) and for $d>4$ in (ii).

\smallskip \noindent (i)
For $d=4$, the three bounds on $g_N$ in \eqref{eq:gNB_bis-4}
are stated in \eqref{eq:gjasy}, and the bounds on $\nu_N$ and $K_N$
in \eqref{eq:nu_j_bound_bis-4}--\eqref{eq:K_bound_at_nu_c_bis-4} follow
from the fact that $(V_N,K_N) \in \domRG_N$ by Proposition~\ref{prop:stable_manifold}.
In fact, the bound on $\nu_N$ is immediate from the definition of $\domRG_N$, while
for $K_N$, for all $d \ge 4$ we have
\begin{equation}
\label{eq:KinD}
    |K_N(\varphi)|
    \le C_{{\rm RG}}\tilde\vartheta_N^3 (g_N \scale_N)^{3/4}
    e^{-\kappa \tilde{g}_N L^{dN} |\varphi|^4}.
\end{equation}
For $d=4$, $\scale_N =1$.  Also, $\tilde\vartheta_N^3 g_N^{3/4} \le O(N^{-3/4})$ due to
the bounds on $g_N$ in \eqref{eq:gNB_bis-4} and
the fact that $\tilde\vartheta_N$ decays exponentially starting from the mass scale.
Since $e_N=N^{-3/4}$ by definition, the bound \eqref{eq:K_bound_at_nu_c_bis-4}
on $K_N$ follows.

The fact that $\nu_c$ is a differentiable function of the mass is part of Proposition~\ref{prop:nu_c_derivative_in_mass}, and the
bound \eqref{eq:nu_c_deriv_4} is given by \eqref{eq:dotnuc0}.
This completes the proof of (i).

\smallskip \noindent (ii)
For $d \geq 5$,
the existence of $g_\infty(\ka)$ and the bound \eqref{eq:gNB_bis-5} on $g_N$
follow from \eqref{eq:gjasy-2}.  The bound
\eqref{eq:K_bound_at_nu_c_bis-4} on $\nu_N$
follows
from the fact that $(V_N,K_N) \in \domRG_N$ by Proposition~\ref{prop:stable_manifold},
and similarly the bound \eqref{eq:K_bound_at_nu_c_bis-5} on $K_N$ follows from
\eqref{eq:KinD} (in fact \eqref{eq:KinD} gives $g^{3/4}L^{-3(d-4)N/4}$ for \emph{all} $d>4$
which is better than $e_N$ for $d>12$).

The fact that $\nu_c$ is a strictly increasing differentiable function of the mass is a part of Proposition~\ref{prop:nu_c_derivative_in_mass}, whose statement includes
the continuity of $\dot\nu_c$ on $[0,\infty)$ with $\dot\nu_c(a)=O(g)$.
Thus \eqref{eq:nu_c_deriv_5} holds with $A_d=1+\dot\nu_c(0)$ as in \eqref{eq:Ad5}.

To verify \eqref{eq:gNB_bis_zero},
we first observe from \eqref{eq:gNB_bis-5} that it suffices to prove that,
when $d\geq 5$,
$|g_N(\ka)-g_N(0)| \le O(g^2 \ka L^{N})$.
To do so,
we temporarily
make the dependence of $g_j$ on $\nu$ explicit,
and use \eqref{eq:g_j_derivative_in_nu_0_high_d}, the Mean Value Theorem,
and Proposition~\ref{prop:nu_c,N_mass_derivative} to see that
\begin{align}
	\label{eq:g_N_bound_proof_1}
	| g_N (\nu_c (\ka), \ka) - g_N (\nu_c (0),  0) |
	& \leq \sup_{\bar{\ka} \in [0, \ka]} |g'_N (\nu_c (\bar{\ka}) , \bar{\ka}) | \,  |\nu_c (\ka) - \nu_c (0)|
    + \sup_{\bar{\ka} \in [0, \ka]} | \dot{g}_N (\nu_c (\bar{\ka}) , \bar{\ka}) | \,  \ka  \nnb
	& \leq O\big( g^2 L^{2N} \scale_N^{\kaa - 2}   \ka   \big)	.
\end{align}
It therefore suffices to prove that $L^{2N} \scale_N^{\kaa - 2} \le L^{N}$, i.e.,
that $2 - (d-4)(\kaa - 2) \leq 1$ with $\kaa$ given by \eqref{eq:kbd} and $\eps(d) \in (0, \frac{1}{d-3})$ small enough when $d\ge 5$.
For $d=5$, since $\kaa=3$ we obtain $2 - (d-4)(\kaa - 2) = 1$,
which is what we want.
For $d\ge 6$, we get $2 - (d-4)(\kaa - 2)= (2d-6)\eps(d)$ and this is sufficient
because we can make the choice $\eps(d) = \frac{1}{2d-6}$.
\end{proof}

\begin{proof}[Proof of Theorem~\ref{thm:mass_trick_big_mass}]
Consider the function $f(\ka) = \nu_c (\ka) + \ka - \nu_c (0)$.
Our goal is to compute the inverse function of $f$,
as $m_{\epsilon}^2 = f^{-1} (\epsilon)$.
By Proposition~\ref{prop:nu_c_derivative_in_mass}, $f(\ka)$ is strictly increasing
with $\dot f(\ka) \geq 1 - O(g)$, so
 $f : [0,\infty) \rightarrow [0, \infty)$ is a bijection and its inverse does exist.

For $d=4$, it follows from Proposition~\ref{prop:nu_c_derivative_in_mass} that
\begin{align}
	f(\ka) \sim A_{4}\,\ka (\log \ka^{-1})^{\hat{\gamma}}
    \qquad \text{as $\ka\downarrow 0$}.
\end{align}
From this, we conclude that
\begin{align}
	 f^{-1} (\epsilon)
    \sim A_{4}^{-1} \epsilon  (\log \epsilon^{-1})^{-\hat{\gamma}}
    \qquad \text{as $\eps\downarrow 0$},
\end{align}
which is the $d=4$ case of \eqref{eq:renorm_big_mass_bis}.

For $d >4$, Proposition~\ref{prop:nu_c_derivative_in_mass} says that $f$ is {\dkm continuously} differentiable with $f' (\ka) = 1 + \dot\nu_c(\ka)$ near 0.
By the Inverse Function Theorem, $m_{\epsilon}^{2} = f^{-1} (\epsilon)$ grows
approximately linearly
for small $\eps$
with slope
$A_{d}^{-1} = (1+ \dot\nu_c(0))^{-1}$,
and this proves the $d>4$ case of \eqref{eq:renorm_big_mass_bis}.
\end{proof}

\begin{proof}[Proof of Theorem~\ref{thm:rg_main_theorem_new}]
Except for \eqref{eq:gNB_d_ter}, the estimates of Theorem~\ref{thm:rg_main_theorem_new}
follow from Proposition~\ref{prop:nu_c,N_mass_derivative} and the
estimates inherent to an RG flow due to the definition of the domain,
together with
Lemma~\ref{lemma:g_j_asymptotic} for $g_N$.
The strategy to prove \eqref{eq:gNB_d_ter} is as in \eqref{eq:g_N_bound_proof_1},
again with $\eps(d) = \frac{1}{2d-6}$.
Indeed, for $\ka \in \II_{N-1}(0)$, and
since (again by Proposition~\ref{prop:nu_c,N_mass_derivative})
\eqref{eq:g_j_derivative_in_nu_0_high_d} holds for this range of $\ka$,
\begin{equation}
	 g_N (\nu_c (\ka), \ka) = g_N (\nu_c (0),  0) +O(g^2L^{N}|a|).
\end{equation}
Finally, since $|\ka| \leq O(L^{-2N})$, the error term is at most
$O(g^2L^{-N})$, which proves \eqref{eq:gNB_d_ter} and completes the proof.
\end{proof}

\subsection{Proofs of Propositions~\ref{prop:stable_manifold},
\ref{prop:nu_c_derivative_in_mass}, \ref{prop:nu_c,N_mass_derivative}
and Lemma~\ref{lemma:g_j_asymptotic}}
\label{sec:prop5pfs}

\subsubsection{Derivatives of \texorpdfstring{$\nu_j$}{nuj}
and proof of Lemma~\ref{lemma:g_j_asymptotic}}

We start with Proposition~\ref{prop:nu-derivs}, which concerns
bounds on derivatives with respect to the mass $\ka$ and the initial
value $\nu_0$.  Its proof is lengthy but is essentially only calculus,
and the proof is deferred to Section~\ref{sec:RGderivs}.
As mentioned previously,
derivatives of $g_j$ are also needed for $d=4$, but we postpone those
to Section~\ref{sec:RGderivs},
alongside derivatives of $K_j$ for $d \ge 4$,
as part of the proof of Proposition~\ref{prop:nu-derivs}.
In Proposition~\ref{prop:nu-derivs}, note
that all statements for $d=4$ agree with those for $d>4$ if we set
$\hat\gamma=0$ and $\scale_j=1$.
Recall that we denote the derivatives of a function $f(\nu_0, \ka)$ as
\begin{equation}
    f'=\frac{\partial f}{\partial \nu_0}, \qquad \dot{f}=\frac{\partial f}{\partial \ka}.
\end{equation}

\begin{proposition}
\label{prop:nu-derivs}
Let $d \ge 4$ and let $g_0$ be sufficiently small.
Let
$N \in \N$
and
$(\nu_{0}, \ka) \in \R \times \II_{N-1} (\tilde{\ka})$.
Suppose that $(V_j,K_j)$ is an RG flow to scale $N$ with initial condition
$(g_0,\nu_{0})$ and mass $\ka$.
Then, for $j \le N$, $(g_j, \nu_j, K_j)$ are twice differentiable functions of $(\nu_{0}, \ka)
\in \mathcal{X}_N$ and satisfy the following bounds.
\begin{enumerate}
\item
For $d=4$, there are $g_0$-dependent constants $c_{0,N}(\nu_0, \ka)= 1 + O(\tilde{\vartheta}_0 g_0)$
and $c_{1,N} (\nu_0, \ka) = 1 + O(\tilde{\vartheta}_0 g_0)$
such that
\begin{align}
	\nu'_j
    =
    \Big( \frac{g_j}{g_0} \Big)^{\hat{\gamma}} \big( c_{0,N}  + O( \tilde{\vartheta}_j \tilde{g}_j) \big) ,
	\qquad
    \dot \nu_j =
    -  1 +  \Big( \frac{g_j}{g_0} \Big)^{\hat{\gamma}}c_{1,N} + O(\tilde{\vartheta}_j \tilde{g}_j) ,
	\label{eq:nu_j_derivative_in_nu_0}
\end{align}
\begin{align}
	| \nu''_{j} | \leq O \Big(L^{2j} \tilde{\vartheta}_{j}^{3} \tilde{g}_j \Big( \frac{\tilde g_j}{g_0} \Big)^{2 \hat{\gamma}} \Big), \quad\;\;
	| \dot{\nu}'_{j} | \leq O \Big(L^{2j} \tilde{\vartheta}_{j}^{3} \tilde{g}_j \Big( \frac{\tilde g_j}{g} \Big)^{\hat{\gamma}} \Big)
	,  \quad\;\;  | \ddot{\nu}_{j} | \leq O(L^{2j} \tilde{\vartheta}_{j}^{3} \tilde{g}_j)
	\label{eq:nu_j_second_derivative_in_nu_0}
\end{align}
Also, there is a constant $C>0$ such that
$\dot{\nu}_j \leq - C \tilde{\vartheta}_0^2 \tilde{g}_j$.
For a global RG flow, the limits $c_{i,\infty}=\lim_{N\to\infty}c_{i,N}$ exist
for $i=0,1$.
Each $c_{i,N}$ is continuous in  $(\nu_0, \ka)\in \mathcal{X}_N$ and
each $c_{i,\infty}$ is continuous in  $(\nu_0, \ka)\in \mathcal{X}_\infty$.

\item
For $d>4$, there are $g_0$-dependent constants
constants $c_{2,N} (\nu_0, \ka)= 1 + O(\tilde{\vartheta}_0 g_0)$ and
positive
$c_{3,N} (\nu, \ka)= O(\tilde{\vartheta}_0 g_0)$
such that
\begin{align}
	& \nu'_j =  c_{2,N} + O(  \tilde{\vartheta}_j \tilde{g}_j \scale_j) ,
\qquad \dot \nu_j =  - c_{3,N} + O (
\tilde\vartheta_j
\tilde{g}_j \scale_j ) ,
	\label{eq:nu_j_derivative_in_nu_0_high_d}
\end{align}
\begin{align}
	& | \nu''_j | , \; | \dot{\nu}'_j |, \; | \ddot{\nu}_j | \leq O(L^{2j}  \tilde{\vartheta}_j^3 \tilde{g}_j \scale_j^{\kaa-2} ) .
	\label{eq:nu_j_second_derivative_in_nu_0_high_d}
\end{align}
Also, there are constants $C_1,C_2>0$ such that
$\dot{\nu}_j \leq - C_1 \tilde{\vartheta}_0^2 \tilde{g}_j$ for sufficiently large $j$ and
\begin{align}
	& | g'_j | \leq C_2  L^{2j}  \tilde{\vartheta}_{j}^2 \tilde{g}^2_j   \scale_j^{\kaa -2},  \qquad
| \dot g_j | \leq C_2 L^{2j} \tilde{\vartheta}_{j}^2  \tilde{g}^2_j   \scale_j^{\kaa -2}
    .
	\label{eq:g_j_derivative_in_nu_0_high_d-bis}
\end{align}
For a global RG flow, the limits $c_{i,\infty}=\lim_{N\to\infty}c_{i,N}$ exist
for $i=2,3$.
Each $c_{i,N}$ is continuous in  $(\nu_0, \ka)\in \mathcal{X}_N$ and
each $c_{i,\infty}$ is continuous in  $(\nu_0, \ka)\in \mathcal{X}_\infty$.
\end{enumerate}
\end{proposition}

\begin{proof}[Proof of Lemma~\ref{lemma:g_j_asymptotic}]
Let $\tilde \ka \ge 0$ and fix $\ka \in \II_{k-1} (\tilde{\ka})$.
By \eqref{eq:remaining_RG_coords_definition}, \eqref{eq:g_pt_formula}, and since
the bound of Theorem~\ref{thm:Phi^U_estimate} applies, for $j<k$ we have
\begin{equation}
    g_{j+1} = g_j - \beta_j g_j^2 + O(\tilde\vartheta_j^3 g_j^3 \scale_j^{\kaa}),
\end{equation}
with $\beta_j(\ka)= (n+8)L^{-(d-4)j}(1+\ka L^{2j})^{-2}(1-L^{-d})$ by \eqref{eq:betadef}.

Suppose first that $d=4$.
In this case, the asymptotic behaviour of solutions to this recursion are given
by \cite[Proposition~6.1.3]{BBS-brief}.  This implies
\eqref{eq:gjasy}
for $j_{\tilde\ka} \ge 1$ (i.e., $\tilde{\ka} \in (0,1)$), and also for
$\tilde{\ka}= \ka=0$
once we observe that the constant $B$ of \eqref{eq:Bdef} is identical to $\beta_0^0$
(see \cite[Lemma~5.3.4]{BBS-brief}).
The extension to the case $\ka \in \II_{k-1} (0)$ follows as in the proof of
\cite[(6.1.20)]{BBS-brief} and we omit the details, other than to mention that
our restriction on $\ka$ guarantees that the factor $(1+\ka L^{2j})^{-2}$ remains
uniformly bounded.
For $\tilde{\ka} >1$,  the third line of \eqref{eq:gjasy} and \eqref{eq:gjasy-2} follow by summing
$g_{j+1}-g_j=O(\beta_j g_j^2)= O(g^2 L^{-4j})$.
The fact that the function $(\nu_0, \ka) \mapsto g_{\infty} (\nu_0, \ka)$
is continuous on
$\cX_{\infty}$ follows from the Dominated Convergence Theorem.

For $d>4$, for both \eqref{eq:gjasy} and \eqref{eq:gjasy-2}, we use
$g_{j+1}-g_j=O(g_j^2 \scale_j) =O(g_j^2L^{-(d-4)j})$ instead, for which the exponentially
decaying error term simplifies matters.
Continuity of $(\nu_0, \ka) \mapsto g_{\infty} (\nu_0, \ka)$ follows from the
Dominated Convergence Theorem and the fact that each $g_j$ is continuous in $(\nu_0, \ka) \in \cX_{j}$.

Finally,  \eqref{eq:g_j_derivative_in_nu_0_high_d} is the same as \eqref{eq:g_j_derivative_in_nu_0_high_d-bis}.
\end{proof}

\subsubsection{The Bleher--Sinai argument}
\label{sec:critical_domain}

The construction of the critical point is based on Lemma~\ref{lemma:critical_domain}.
Lemma~\ref{lemma:critical_domain} is analogous to
the Bleher--Sinai argument of \cite[Proposition~8.3.1]{BBS-brief}, which itself originated in \cite{BS73}.
Unlike \cite[Proposition~8.3.1]{BBS-brief},
Lemma~\ref{lemma:critical_domain} applies for all dimensions $d \ge 4$.
The lemma produces
a sequence of decreasing
intervals of $\nu$-values $I_j(\ka;\tilde\ka)$ for which the RG flow with mass
$\ka$ exists up to scale $j$ when started with initial condition
$\nu \in I_j(\ka;\tilde\ka)$. These intervals shrink to a point which uniquely
identifies the critical point $\nu_c(\ka)$.
The lemma  also gives the rate
at which $|I_j(\ka;\tilde\ka)|\to 0$ when $j \to \infty$.
This turns out to be essential in the proof of Lemma~\ref{lemma:noil},
from which we prove our estimates on the derivative $\dot\nu_c(\ka)$ in  Proposition~\ref{prop:nu_c_derivative_in_mass}.

For the statement of the lemma, given $\tilde\ka \ge 0$, we define the intervals
\begin{align}
	H_j & =  \big( {\textstyle{\frac{1}{2}}} \tilde{g}_j (\tilde{\ka}),  2 \tilde{g}_j (\tilde{\ka}) \big),
 \qquad
	J_j  = ( -  6(n+2) \tilde{\vartheta}_j \tilde{g}_j (\tilde{\ka} ) \scale_j L^{-2j}   ,
    6 (n+2) \tilde{\vartheta}_j \tilde{g}_j (\tilde{\ka} ) \scale_j L^{-2j} ).
    \label{eq:HjJj_definition}
\end{align}
By definition of the domain $\cD_j$ in \eqref{eq:domain_D_j},
if we take any $k_0 \leq \frac{1}{24 (n+2)}$, then
$H_j \times J_j \times \II_j(\tilde\ka) \subset \cD_j$.
Recall from \eqref{eq:Kcaldef} that $\cK_j$ consists of those $K \in \cF_j$
for which $\|K\|_{\Wkappa} < C_{\rg} \tilde{\vartheta}_j^3 \tilde{g}_j^{3} \scale_j^{\kaa}$.
We use \eqref{eq:IIdef} to define $\II_{j}(\tilde \ka)$ for $j \ge -1$.

\begin{lemma}
\label{lemma:critical_domain}
Let $L$ be sufficiently large, $g_0$ be sufficiently small,   and $\tilde{\ka} \geq 0$.
For $0 \le j\leq N$ and $\ka \in \II_{j-1}$, there are
open intervals $I_j (\ka ;\tilde{\ka})\subset \R$,
nested in the sense that $I_{j+1} \subset I_j$, which satisfy the following
for an RG flow to scale $j$ with initial condition $(g_0,\nu_0)$, mass $\ka$,
and $\nu_0 \in I_j$:
\begin{itemize}
\item For all $i\leq j$, we have $(g_{i}, \nu_{i}, K_i)
    \in H_{i} \times J_{i}  \times  \cK_i$.
\item For each fixed $\ka \in \II_{j-1} (\tilde{\ka})$, the
    set $\{\nu_j: \nu_0\in I_j(\ka;\tilde\ka)\}$ of $\nu_j$ values
    with initial condition $\nu_0 \in I_j(\ka;\tilde\ka)$ is exactly $J_j$.
\item For each fixed $\ka \in \II_{j-1} (\tilde{\ka})$,  the
interval $I_j (\ka   ; \tilde{\ka} )$ is connected, has length bounded by
\begin{align}
	| I_j (\ka ;\tilde{\ka} )| \leq
	\begin{cases}
		O\big( \big( \frac{\tilde g_j}{g_0} \big)^{-\hat{\gamma}}|J_j| \big) & (d=4) \\
		O(|J_j|) & (d>4)  ,
	\end{cases}
	\label{eq:I_j_size}
\end{align}
and  $\bar{I}_{j+1} (\ka  ; \tilde{\ka} ) \subset I_{j} (\ka  ; \tilde{\ka})$
(here $\bar I$ is the closure of $I$).
\end{itemize}
\end{lemma}

\begin{proof}
Fix $\tilde{\ka} \geq 0$.
We write $\II_{j-1} = \II_{j-1} (\tilde\ka)$,
$\tilde{g}_j  = \tilde{g}_j(\tilde\ka)$, and set $\mu_j = L^{2j} \nu_j$.
Let $\ka \in \II_{k-1}$.
Fix $k\leq j$.
We make the induction hypothesis that there exists $I_k(\ka;\tilde\ka)$ such that the statement of the lemma holds when $j$ is replaced by $k$.
For $k=0$, the statement does hold with $I_0=J_0$.

To advance the induction,  consider $\nu_0\in I_k (\ka)$.
By Lemma~\ref{lemma:g_j_asymptotic},  $g_{k+1} \in H_{k+1}$.
By Theorem~\ref{thm:Phi^K_estimate}, $K_{k+1}\in \cK_{k+1}$.  By
the perturbative recursion for $\nu_k$ in \eqref{eq:nu_pt_formula} and by
Theorems~\ref{thm:Phi^U_estimate},
\begin{align}
	\mu_{k+1} & = L^2 (\mu_k + e_k)
\label{eq:critical_domain_mu_expression},
\end{align}
with
\begin{align}
	e_k = - \hat{\gamma} \beta_k \mu_k g_k + L^{2k}\eta_k g_k - L^{2k}\xi_k g_k^2 + O( \tilde{\vartheta}_k^3 \tilde{g}_k^3 \scale_k^\kaa).
	\label{eq:e_k_estimate-v2}
\end{align}
Since
$|\hat{\gamma} \beta_k \mu_k g_k | \leq O(\tilde{\vartheta}_k^3 \tilde{g}_k^2 \scale_k^2)$,
$L^{2k} |\eta_k | \leq 2 (n+2) \tilde{\vartheta}_k \scale_k$,
and $L^{2k}\xi_k \le O ( \tilde{\vartheta}_j^3 \scale_k^2)$,  $e_k$ satisfies
\begin{align}
	|e_k | \leq |\eta_k| g_k + O ( \tilde{\vartheta}_k^3 \tilde{g}_k^2 \scale_k^\kaa)
    \leq  \frac{5}{2} (n+2) \tilde{\vartheta}_k g_k \scale_k  \leq  5 (n+2) \tilde{\vartheta}_k \tilde{g}_k \scale_k,
\end{align}
whenever $g_k \leq 2\tilde{g}_k \leq 2 g_0$ is sufficiently small.
By the assumption that the range of $\mu_k$ is exactly $L^{2k} J_k$, and by \eqref{eq:critical_domain_mu_expression} and \eqref{eq:e_k_estimate-v2},
\begin{align}
\begin{split}
	& \sup_{\nu \in I_k (\ka)} \mu_{k+1} \geq \sup_{\nu \in I_k (\ka)} L^2 (\mu_k - 5 (n+2) \tilde{\vartheta}_k  \tilde{g}_k \scale_k) \geq (n+2) L^2 \tilde{\vartheta}_k \tilde{g}_k \scale_k  ,	
	\\
	& \inf_{\nu \in I_k (\ka)} \mu_{k+1} \leq \inf_{\nu \in I_k (\ka)} L^2 (\mu_k + 5 (n+2) \tilde{\vartheta}_k \tilde{g}_k \scale_k) \leq - (n+2) L^2 \tilde{\vartheta}_k \tilde{g}_k \scale_k 	.
\end{split}
\label{eq:BSargument_mu_bounds}
\end{align}
For large $L$, the bounds in \eqref{eq:BSargument_mu_bounds} exceed the
limit $6(n+2)\tilde\vartheta_{k+1}\tilde g_{k+1} \scale_{k+1}$ of the interval
$L^{2(k+1)} J_{k+1}$, since the ratio of $\tilde\vartheta_{k+1}\tilde g_{k+1}$
to $\tilde\vartheta_{k}\tilde g_k$ is bounded and since $\scale_{k+1}\le \scale_k$.
Therefore, by
the continuity of the function $\psi:\nu_{0} \mapsto \mu_{k+1}$, the set $\psi(I_k)$
contains the interval $L^{2(k+1)} J_{k+1}$.
Thus we can define the preimage $\psi^{-1}(L^{2(k+1)} J_{k+1})$ (inside $I_k (\ka  ; \tilde{\ka} )$) to be $I_{k+1} (\ka  ; \tilde{\ka} )$.

This $I_{k+1}$ is an open set because it is the preimage of an open set under the continuous function $\psi$.
By Proposition~\ref{prop:nu-derivs}, $\mu_{k+1}$ is a monotone function of $\nu \in I_{k}$, so since $I_k$ is a connected interval,  $I_{k+1}$ is also
a connected interval.
Since $\psi (I_{k+1})= L^{2(k+1)} J_{k+1}$, it follows from
the estimates on $\mu_{k+1}'=L^{2(k+1)}\nu_{k+1}'$ from
\eqref{eq:nu_j_derivative_in_nu_0} and \eqref{eq:nu_j_derivative_in_nu_0_high_d}
that
\begin{equation}
    |I_{k+1}| \le \frac{1}{\inf_{\nu_0\in I_k}\psi'(\nu_0)} L^{2(k+1)}|J_{k+1}|,
\end{equation}
from which we conclude that \eqref{eq:I_j_size} holds.
 Finally, since $\mu_{k+1}$ is a continuous function of $\nu_0$, the image of
$\bar{I}_{k+1} (\ka  ; \tilde{\ka}  )$ is $L^{2(k+1)} \bar{J}_{k+1}$,  which is contained
in the image of $I_{k} (\ka  ; \tilde{\ka} )$,
so we have $\bar{I}_{k+1} (\ka  ; \tilde{\ka} ) \subset I_{k} (\ka  ; \tilde{\ka} )$.
\end{proof}

\subsubsection{Proofs of Propositions~\ref{prop:stable_manifold},
\ref{prop:nu_c_derivative_in_mass}, \ref{prop:nu_c,N_mass_derivative}}

\begin{proof}[Proof of Proposition~\ref{prop:stable_manifold}]
Let $\ka \ge 0$ and set $\tilde \ka = \ka$ in Lemma~\ref{lemma:critical_domain}.
Then there exists a nested sequence of intervals $I_{j} (\ka  ; \ka )$ satisfying the  properties of Lemma~\ref{lemma:critical_domain},
and we can define the infinite-volume critical point as the unique point in their intersection:
\begin{align}
\label{eq:nuca}
	\{ \nu_c (\ka) \} = \cap_{j\geq 0} I_j (\ka  ; \ka ) .
\end{align}
The intersection is non-empty because $\bar I_{j+1} \subset I_j$ for each $j$, and it contains a unique point since $|I_j| \rightarrow 0$ as $j\rightarrow \infty$.
This holds for every $j \ge 0$, so since $H_j \times J_j \times \II_j(\tilde\ka) \subset \cD_j$ there is a global RG flow
with initial condition $(g,\nu_c(\ka))$ and mass $\ka$.
The properties of the flow $(g_j, \nu_j, K_j)$ with initial condition $\nu_0 = \nu_c (\ka)$ then follow from Lemma~\ref{lemma:critical_domain}.
The proof that $\nu_c (\ka)$ is continuous in $\ka$
is as in \cite[Proposition~8.3.3]{BBS-brief}.

It remains to prove that $|\nu_c (\ka)| \leq O( \vartheta_{0} g)$, and $\nu_c (0) \sim  - (n+2) g (-\Delta_H)^{-1}_{00}$
as $g\downarrow 0$.
The upper bound follows from that $\nu_c (\ka) \in I_0 (\ka) = J_0 (\ka)$.
For $d=4$, the
asymptotic formula  for $\nu_c(0)$
follows from \cite[(4.2.7)]{BBS-brief}.
The proof is similar for $d>4$, as follows.
By solving the recursion relation for $\nu_j$ backwards
as in \cite[(6.1.31)]{BBS-brief}, we obtain
\begin{align}
	\nu_0 = \Pi_{0,j}^{-1} \nu_{j+1} + \sum_{k=0}^j (-\eta_k g_k + \xi_k g_k^2 - r_{\nu, k}) \Pi_{0, k-1}^{-1}
\end{align}
with
\begin{equation}
\label{eq:Pidef-1}
	 \Pi_{i,j}  = \prod_{k=i}^j (1- \hat{\gamma} \beta_k g_k) .
\end{equation}
With the choice $(\nu_0 , \ka) = (\nu_c (\ka), \ka)$,
the infinite product converges since $\beta_j$ decays exponentially,
so $\Pi_{0,j}^{-1} \nu_{j+1} \rightarrow 0$ as $j\rightarrow \infty$.
By \eqref{eq:etadef} and \cite[(5.3.5)]{BBS-brief},
$(n+2)^{-1}\eta_k$ is the $00$ component of
$C_{k+1}$, so
\begin{align}
	\sum_{k=0}^{\infty} \eta_k = (n+2) (-\Delta )^{-1}_{00}.
\end{align}
It follows from the asymptotic behaviour of $g_k$ in Lemma~\ref{lemma:g_j_asymptotic} that
 $\Pi_{0, k-1} = 1+ O(g_0)$, and, with the exponential decay of $\xi_k$ and $r_{\nu, k}$, that
\begin{align}
	\nu_c(0)
    = \sum_{k=0}^{\infty} (-\eta_k g_k + \xi_k g_k^2 - r_{\nu, k}) \Pi_{0, k-1}^{-1}
    = -(n+2) (-\Delta )^{-1}_{00}  g_0 (1+ O(g_0)).
\end{align}
This completes the proof.
\end{proof}

\begin{remark} \label{remark:nuca}
In the above proof,
we made the specific choice $\ka =\tilde\ka$.
Although each $I_j (\ka  ; \tilde{\ka} )$ depends on $\tilde{\ka}$,  as long as   $k_0 \leq \frac{1}{24 (n+2)}$
and $\ka \in (\frac{1}{2} \tilde{\ka}, 2 \tilde{\ka})$, the intersection $\cap_{j\geq 0} I_j (\ka  ; \tilde{\ka} )$ does not depend on $\tilde{\ka}$ by the uniqueness argument of \cite[Proposition~8.3.1]{BBS-brief},
i.e.,
\begin{align}
	\{ \nu_c (\ka) \} = \cap_{j\geq 0} I_j (\ka  ; \tilde{\ka} )
\end{align}
whenever $\ka \in \cap_{j \geq 0} \II_j (\tilde{\ka})$
(the set $\cap_{j \geq 0} \II_j (\tilde{\ka})$
is $(\frac{1}{2} \tilde{\ka}, 2 \tilde{\ka})$ if $\tilde\ka >0$
and is $\{0\}$ if $\tilde\ka=0$).
\end{remark}

The value of $\nu_N$ depends on the initial value $\nu_0=\nu$ and on the mass $\ka$
(also on $g$ but we suppress this dependence since we fix $g$).  We write $\nu_N|_{(\nu_0,\ka)}$
when we need to display the dependence on $(\nu_0,\ka)$.
The following lemma defines the function $\nu_{0,N}(\ka  ; \tilde{\ka})$.
The interval $I_N$ in Lemma~\ref{lemma:nu0N}
is the one determined by Lemma~\ref{lemma:critical_domain}.
We also need the subset of $\R^2$ defined by
$\A_{N}(\tilde\ka) = \{ (\nu, \ka) : \ka \in \II_{N-1}(\tilde\ka), \; \nu\in I_{N} (\ka  ; \tilde{\ka} ) \}$.
By construction of each $I_N (\ka  ; \tilde{\ka} )$ in Lemma~\ref{lemma:critical_domain},
$\A_{N}(\tilde\ka)$ is a preimage of an open set under the continuous map $(\nu_0, \ka) \mapsto \mu_{N}$,  thus is open.

\begin{lemma}
\label{lemma:nu0N}
For $\tilde{\ka} \ge 0$ and $\ka \in \II_{N-1} (\tilde{\ka})$, there is a unique $\nu_{0,N} (\ka  ; \tilde{\ka} ) \in I_N (\ka  ; \tilde{\ka} )$ such that
\begin{align}
\label{eq:410}
	\nu_N |_{(\nu_{0,N} (\ka ;\tilde\ka) , \ka)} = \nu_N |_{(\nu_{c} (\tilde\ka),  \tilde\ka)}	.
\end{align}
This $\nu_{0,N}$ satisfies
\begin{align}
	\nu_c (\ka) = \lim_{N\rightarrow \infty} \nu_{0,N} (\ka, \tilde{\ka}).
	\label{eq:nuclimit-bis}
\end{align}	
The function
$\nu_{0,N} (\cdot ; \tilde{\ka}) : \II_{N-1} (\tilde\ka) \rightarrow \R$ is twice continuously differentiable, obeys
$\nu_{0,N} (\tilde{\ka};\tilde\ka) = \nu_c (\tilde{\ka})$, and,
for all $\ka \in \II_{N-1} (\tilde{\ka})$ its derivative satisfies
\begin{align}
    \dot \nu_{0,N} (\ka ; \tilde{\ka})
    = - \frac{\dot\nu_N}{\nu_N'} \Big|_{(\nu_{0,N} (\ka;\tilde\ka) , \ka)}
    .
\label{eq:nu_0,N_mass_derivative}
\end{align}
\end{lemma}

\begin{proof}
We first verify the existence and uniqueness of $\nu_{0,N}(\ka  ; \tilde{\ka})$ satisfying \eqref{eq:410}.
Let $\tilde{\ka} \geq 0$
 and
$\ka \in \II_{N-1} (\tilde{\ka})$.
By Lemma~\ref{lemma:critical_domain} together with the identification of $\nu_c(\tilde\ka)$ in \eqref{eq:nuca},
we have $(\nu_c (\tilde\ka), \tilde\ka) \in \A_N(\tilde\ka)$,
and hence in particular $\nu_N |_{(\nu_{c} (\tilde\ka),  \tilde\ka)} \in J_N (\tilde{\ka})$.
We have seen in the proof of Lemma~\ref{lemma:critical_domain}
that the map $\nu \mapsto \nu_N |_{(\nu, \ka)}$
from $I_N (\ka ;\tilde\ka)$ to $J_N (\tilde{\ka})$ is surjective.
It is also strictly increasing since $\nu_N'>0$ by Proposition~\ref{prop:nu-derivs},
so the existence and uniqueness of
$\nu_{0,N}(\ka)$ obeying \eqref{eq:410} follows.
Then by Lemma~\ref{lemma:critical_domain},
$(g_j, \nu_j, K_j,  \ka) \in \domRG_j$ for each $j\leq N$ when $\nu = \nu_{0,N} (\ka)$.

The limit \eqref{eq:nuclimit-bis} follows from Remark~\ref{remark:nuca} and
the fact that $\nu_{0,N} (\ka ; \tilde{\ka}) \in I_{N} (\ka ; \tilde{\ka})$ for each $N$.

It remains to establish the claims about differentiability of $\nu_{0,N}(\ka ;\tilde\ka)$.
Given $d \ge 4$, we fix $\tilde{\ka}  \ge 0$ and consider
$\nu_{0,N} (\ka_0 ;\tilde\ka) \in I_N (\ka_0 ; \tilde{\ka})$ for $\ka_0 \in
\II_{N-1} (\tilde{\ka})$.
By Proposition~\ref{prop:nu-derivs}, the map $(\nu_0,  \ka) \mapsto \nu_N$ on $\A_N$ is
twice continuously differentiable on a neighbourhood of $(\nu_{0,N} (\ka_0 ;\tilde\ka),  \ka_0)$, with non-vanishing derivative $\nu_N'$.
By the Implicit Function Theorem,
there exists an open neighbourhood $U \subset \II_{N-1} (\tilde{\ka})$ of $\ka_0$ and $f \in C^2 (U ; \R)$ such that
\begin{align}
	\nu_N |_{(f (\ka), \ka )}
    = \nu_N  |_{(\nu_{0,N} (\ka_0 ;\tilde\ka) ,  \ka_0 )}
    .
\label{eq:nu_c,N_characteristic_eqn}
\end{align}
We choose $U$ smaller if necessary to ensure that $( f (\ka) , \ka ) \in \A_N (\tilde{\ka})$ for
each $\ka \in U$, and then it follows from the uniqueness of $\nu_{0,N} (\cdot  ; \tilde{\ka} )\in I_{N} (\ka  ; \tilde{\ka})$ satisfying \eqref{eq:410} that
\begin{align}
	f (\ka) = \nu_{0,N} (\ka  ; \tilde{\ka}) \qquad (\ka \in U).
\end{align}
The formula \eqref{eq:nu_0,N_mass_derivative}
for the derivative also follows from the Implicit Function Theorem.
\end{proof}

In terms of the functions $c_{i,\infty}$ of Proposition~\ref{prop:nu-derivs}, we define
\begin{align}
	C_{4}^{(1)} (\ka) & =    \frac{1}{ c_{0, \infty}} \Big|_{(\nu_{c} (\ka),  \ka)},   \qquad
	C_{4}^{(2)} (\ka) =  - \frac{c_{1,\infty} }{ c_{0, \infty}} \Big|_{(\nu_{c} (\ka),  \ka)}	,   \qquad
	C_{d} (\ka) = \frac{c_{3,\infty}}{c_{2,\infty}}  \Big|_{(\nu_{c} (\ka),  \ka)} \;\; (d > 4) ,
	\label{eq:C4ds}
\end{align}
which are continuous in $\ka \geq 0$ by Propositions~\ref{prop:nu-derivs} and \ref{prop:stable_manifold}.
(Proposition~\ref{prop:nu-derivs} asserts continuity in $(\nu_0, \ka)$ with the subset topology on $\cX_{\infty}$, so continuity in $\ka$ also uses
the continuity of $\ka \mapsto \nu_c (\ka)$).
The constants in \eqref{eq:C4ds}
are all defined in terms of the infinite-volume critical point $\nu_c(a)$,
and the next lemma shows that the leading behaviour of the
mass derivative of $\nu_{0,N}$ is given in terms of these quantities.
The statement is slightly stronger for $\ka \in \II_{N-1} (0)$,  since
the massless critical point $\nu_c(0)$ appears instead of $\nu_c (\ka)$.

\begin{lemma} \label{lemma:noil}
The mass derivative of $\nu_{0,N}(\cdot;\tilde\ka)$ has the following asymptotic behaviour:
\begin{enumerate}
\item
Let $\tilde{\ka} > 0$ and $\ka \in \II_{N-1} (\tilde{\ka})
=(\frac{1}{2} \tilde{\ka}, 2 \tilde{\ka})$.
Then
\begin{align}
	\dot \nu_{0,N} (\ka  ;\tilde\ka)
    =
	\begin{cases}
		\big(  \big( \frac{g_N}{g} \big)^{-\hat{\gamma}} \big|_{(\nu_{c} (\ka),  \ka)}  C_{4}^{(1)} (\ka)+ C_{4}^{(2)} (\ka) \big) ( 1 + O( \tilde{\vartheta}_N \tilde{g}_N) )	
            & (d=4) \\
		C_{d} (\ka) (1 + O(\tilde{\vartheta}_N \tilde{g}_N \scale_N^{\ka - 1} ))	 & (d > 4),
	\end{cases}
\end{align}
and also $\liminf_{N\rightarrow \infty} \dot{\nu}_{0,N} (\ka) > 0$.

\item
Let $\tilde{\ka} =0$ and $\ka \in \II_{N-1} (0)
= (-\frac 12 L^{-2(N-1)},\frac 12 L^{-2(N-1)})$.
Then
\begin{align}
	\dot \nu_{0,N} (\ka;0)
    =
	\begin{cases}
		\big(  \big( \frac{g_N}{g} \big)^{-\hat{\gamma}} \big|_{(\nu_{c} (0),  0)}  C_{4}^{(1)} (0)+ C_{4}^{(2)} (0) \big) ( 1 + O(\tilde{g}_N) )	
            & (d=4) \\
		C_{d} (0) (1 + O(\tilde{g}_N \scale_N^{\kaa - 2} ))	 & (d > 4) .
	\end{cases}
\end{align}
\end{enumerate}
\end{lemma}

\begin{proof}
Since $\dot{\nu}_{0,N} (\ka) = - \frac{\dot{\nu}_N}{\nu'_N} |_{(\nu_{0,N} (\ka), \ka)}$ by \eqref{eq:nu_0,N_mass_derivative}, our strategy
is to approximate $\dot{\nu}_N$ and $\nu'_N$ by quantities that do not have strong $N$-dependence.
In the proof, we often omit $\tilde{\ka}$ for brevity.

\medskip\noindent(i)
Let $\tilde\ka \ge 0$ and $\ka \in \II_{N-1}(\tilde\ka)$.  We prove the statement
of part~(i) including $\tilde\ka=0$ as we will use this in the proof of
the stronger part~(ii).  To lighten the notation, we leave implicit the dependence
of $\nu_{0,N}$ and $I_N$ on $\tilde\ka$.

We have $\nu_{0,N} (\ka) \in I_N (\ka )$ by construction,
and $\nu_{c} (\ka) \in I_N (\ka)$ by Remark~\ref{remark:nuca}.
By the bound on $|I_N|$ in \eqref{eq:I_j_size},  by
the second-derivative bounds of \eqref{eq:nu_j_second_derivative_in_nu_0} and \eqref{eq:nu_j_second_derivative_in_nu_0_high_d}, and by the Mean Value Theorem,
\begin{align}
\begin{split}
	\big| \nu'_N |_{(\nu_{c} (\ka), \ka)} - \nu'_N |_{(\nu_{0,N} (\ka), \ka)} \big|
	& \leq
	\begin{cases}	
	O(\tilde{\vartheta}_N^4 \tilde{g}_N^2 \big( \frac{\tilde g_N}{g} \big)^{\hat{\gamma}} ) & (d=4)  \\
	O(\tilde{\vartheta}_N^4 \tilde{g}_N^2 \scale_N^{\kaa-1}) & (d>4),
	\end{cases}
	\\
	\big| \dot{\nu}_N |_{(\nu_{c} (\ka), \ka)} - \dot{\nu}_N |_{(\nu_{0,N} (\ka), \ka)} \big|
	& \leq
	\begin{cases}
	O(\tilde{\vartheta}_N^4 \tilde{g}_N^2 ) & (d=4)  \\
	O(\tilde{\vartheta}_N^4 \tilde{g}_N^2 \scale_N^{\kaa-1}) & (d>4) .
	\end{cases}
\end{split}	
	\label{eq:nu_dot_replacement}
\end{align}
Therefore, we can approximate $\dot{\nu}_N$ and $\nu'_N$ by their counterparts
at the critical point $\nu_c(\ka)$.
But by Proposition~\ref{prop:stable_manifold},
when the initial value is tuned to $\nu_c (\ka)$ we can take the limit
as $N \to \infty$ in the formulas for $\nu'_j$ and $\dot{\nu}_j$
in \eqref{eq:nu_j_derivative_in_nu_0} and \eqref{eq:nu_j_derivative_in_nu_0_high_d} to obtain
\begin{align}
	&\nu'_N |_{(\nu_c (\ka),\ka)} = \begin{cases}
	\big( \frac{g_N}{g} \big)^{\hat{\gamma}} (c_{0, \infty} + O(\tilde{\vartheta}_N \tilde{g}_N)) \big|_{(\nu_c (\ka),\ka)} & (d=4) \\
	c_{2,\infty} |_{(\nu_c (\ka),\ka)} + O(\tilde{\vartheta}_N \tilde{g}_N \scale_N)  & (d>4)   ,
	\end{cases}
	\\
	& \dot{\nu}_{N} |_{(\nu_c (\ka),\ka)} = \begin{cases}
	-1 + c_{1,\infty} \big( \frac{g_N}{g} \big)^{\hat{\gamma}} |_{(\nu_c (\ka),\ka)}  + O(\tilde{\vartheta}_N \tilde{g}_N) & (d=4) \\
	-c_{3,\infty} |_{(\nu_c (\ka),\ka)}  + O(\tilde{\vartheta}_N \tilde{g}_N \scale_N)   & (d>4).
	\end{cases}
\end{align}
After division, we find the desired formula for $\dot{\nu}_{0,N} (\ka)$.

For the case $\tilde\ka >0$,
the uniform positivity of $\dot{\nu}_{0,N} (\ka)$ then follows from the facts that $\dot{\nu}_j (\ka) \leq - C \vartheta_0^2 \tilde{g}_j$ and $c_{0,\infty},c_{2,\infty} = 1+ O(\vartheta_0 g)$
by Proposition~\ref{prop:nu-derivs}.
Indeed,
\begin{align}
	\dot{\nu}_{0,N} (\ka) = - \frac{\dot{\nu}_N}{\nu'_N} \Big|_{(\nu_{0,N} (\ka), \ka)}
	\geq \begin{cases}
		C \vartheta_0^2 \tilde{g}_N^{1-\hat{\gamma}} g^{\hat{\gamma}} (1+ O(\vartheta_0 g)) & (d=4) \\
		C \vartheta_0^2 \tilde{g}_N (1+ O(\vartheta_0 g)) & (d>4)
	\end{cases}
\end{align}
and Lemma~\ref{lemma:g_j_asymptotic} gives a strictly positive lower bound on
$\tilde{g}_N$ as $N \rightarrow \infty$ when $\tilde\ka >0$.

\medskip\noindent(ii)
Let $\tilde{\ka} = 0$.
The case $\ka = 0$ has already been established in the proof of part~(i)
and we use it as a reference point.
For $\ka \in \II_{N-1} (0) \backslash \{ 0 \}$,  we compute the second derivative
of $\dot\nu_{0,N}(a)$ using \eqref{eq:nu_0,N_mass_derivative} and the chain rule, and obtain
\begin{align}
	\ddot{\nu}_{0,N} =  \Big( -\frac{\dot{\nu}'_N}{\nu'_N} + \frac{\dot{\nu}_N \nu''_N}{(\nu'_N)^2} \Big)  \dot{\nu}_{0,N} + \Big( - \frac{\ddot{\nu}_N}{\nu'_N} + \frac{\dot{\nu}_N \dot{\nu}'_N}{(\nu'_N)^2} \Big) \Big|_{(\nu_{0,N} (\ka), \ka)} .
\end{align}
Let $M = \sup_{\ka \in \II_{N-1} (0)} |\dot{\nu}_{0,N} (\ka) - \dot{\nu}_{0,N} (0)|$.
By the Mean Value Theorem,
\begin{align}
	M \leq \frac{1}{2} L^{-2(N-1)}  \big( A (M + | \dot{\nu}_{0,N} (0) |) + B \big)	, \label{eq:nu0N-inv-limit-1}
\end{align}
where
\begin{align}
	& A = \sup_{\ka \in \II_{N-1} (0)}  \Big| -\frac{\dot{\nu}'_N}{\nu'_N} + \frac{\dot{\nu}_N \nu''_N}{(\nu'_N)^2} \Big|, \qquad
	B = \sup_{\ka \in \II_{N-1} (0)}  \Big| - \frac{\ddot{\nu}_N}{\nu'_N} + \frac{\dot{\nu}_N \dot{\nu}'_N}{(\nu'_N)^2} \Big| .
\end{align}
Using \eqref{eq:nu_j_second_derivative_in_nu_0} and \eqref{eq:nu_j_second_derivative_in_nu_0_high_d}, we see that
\begin{align}
	|A| \leq \begin{cases}
		O(L^{2N} \tilde{g}_N) & (d=4) \\
		O(L^{2N} \tilde{g}_N \scale_N^{\kaa-2} ) & (d > 4)  ,
	\end{cases} \qquad
	|B| \leq \begin{cases}
		O( (\frac{g_N}{g})^{-\hat{\gamma}}  L^{2N} \tilde{g}_N) & (d=4) \\
		O(L^{2N} \tilde{g}_N \scale_N^{\kaa-2} ) & (d > 4)  .
	\end{cases}
\end{align}
Thus \eqref{eq:nu0N-inv-limit-1} reduces to
\begin{align}
	M \leq O(L^{-2N}) \big( A  | \dot{\nu}_{0,N} (0) | + B \big) \leq \begin{cases}
		O( (\frac{g_N}{g})^{-\hat{\gamma}}  \tilde{g}_N) & (d=4) \\
		O( \tilde{g}_N \scale_N^{\kaa-2} ) & (d > 4)  .
	\end{cases}
\end{align}
This proves that $\dot\nu_{0,N}(\ka)$ is within a small enough error of $\dot\nu_{0,N}(0)$
to give the desired asymptotic formula for part~(ii), once we recall that
we have already established it for $\dot\nu_{0,N}(0)$ in part~(i).
\end{proof}

\begin{proof}[Proof of Proposition~\ref{prop:nu_c_derivative_in_mass}]
We first consider $\tilde{\ka} >0$ and
$\ka \in (\frac{1}{2} \tilde{\ka}, 2\tilde{\ka}) = \cap_{j\geq 0} \II_j (\tilde{\ka})$.
By \eqref{eq:nuclimit-bis},
$\nu_c (\ka)$ is the limit of $\nu_{0,N} (\ka ; \tilde{\ka})$.  Let
$\ka_1, \ka_2 \in (\frac{1}{2} \tilde{\ka}, 2\tilde{\ka})$.
By Lemma~\ref{lemma:noil}(i) and  the
Dominated Convergence Theorem,
\begin{align}
	\nu_c (\ka_1 ) - \nu_c (\ka_2)
    &= \lim_{N\to\infty} \int_{\ka_2}^{\ka_1} \dot\nu_{0,N} (x ; \tilde{\ka})  dx
    =
	\begin{cases}
		\int_{\ka_2}^{\ka_1} [ (\frac{g_{\infty}}{g} )^{-\hat{\gamma}} \big|_{(\nu_c (x),x)} C^{(1)}_{4} (x) + C^{(2)}_{4} (x) ] dx	 & (d=4) \\
		\int_{\ka_2}^{\ka_1} C_d (x) d x	 & (d > 4) .
	\end{cases}
	\label{eq:nu_c_differentiability}
\end{align}
Since this holds for any $\ka_1, \ka_2 \in (\frac{1}{2} \tilde{\ka}, 2\tilde{\ka})$ and $\tilde{\ka} >0$, it in fact holds for any $\ka_1, \ka_2 >0$.
This establishes the continuous differentiability at strictly positive $\ka$.
The positivity of $\dot{\nu}_c (\ka)$ follows from the statement of uniform positivity for $\dot{\nu}_{0,N}$ in Lemma~\ref{lemma:noil}(i).

For $d>4$, since $\nu_c (\ka)$ and $C_{d} (\ka)$ are continuous also at $a=0^+$, taking the limit $\ka_2 \downarrow 0$ gives
\begin{align}
	\nu_c (\ka_1) - \nu_c (0) = \int^{\ka_1}_0 C_d (x) dx.
\end{align}
This extends the continuous differentiability to $\ka =0^+$.
Since $c_{3,\infty} = O(g)$ and $c_{2,\infty} = 1 +O(g)$,
we also have $C_d (x)= \frac{c_{3,\infty}}{c_{2,\infty}} = O(g)$
and hence $\dot\nu_c(\ka)=O(g)$.

For $d=4$, $\nu_c$ is not differentiable at 0.
But since $g_{\infty} (\ka) \sim (Bj_\ka)^{-1}$ as $\ka \downarrow 0$, we see that
\begin{align}
	\dot\nu_c (\ka) \sim
    C_{4}^{(1)} (\ka) g^{\hat{\gamma}}  (B j_{\ka})^{\hat\gamma}
    \sim
    C_{4}^{(1)} (0) (B g)^{\hat{\gamma}} (\log_{L^2} a^{-1} )^{\hat{\gamma}}
    \sim
   	A_{4} (\log a^{-1} )^{\hat{\gamma}}
    \qquad \text{as $\ka \downarrow 0$},
\end{align}
with
\begin{equation}
\label{eq:A4def}
    A_{4} = \frac{(B g)^{\hat{\gamma} } C_{4}^{(1)} (0)}{(\log L^2)^{\hat\gamma}}.
\end{equation}
The asymptotic
formula $A_{4}\sim (Bg/\log L^2 )^{\hat\gamma}$ as $g \downarrow 0$
follows from the fact that $C_{4}^{(1)} (0) = 1+O(g)$.
\end{proof}

\begin{proof}[Proof of Proposition~\ref{prop:nu_c,N_mass_derivative}]
Take $\tilde{\ka} = 0$ and let $\ka \in \II_{N-1} (0)$.
The formula for $\dot\nu_{0,N}(a ; 0)$ in
Lemma~\ref{lemma:noil}(ii) and the asymptotic formula for $g_N$
in Lemma~\ref{lemma:g_j_asymptotic} imply that
\begin{align}
	\dot \nu_{0,N} (\ka) = \dot \nu_{0,N} (\ka ; 0) =
	\begin{cases}
	\big( A_{4}  (\log L^2)^{\hat{\gamma}} N^{\hat{\gamma}} - 1 + c^{\free} \big)  ( 1 + O(N^{-1}) )  & (d=4) \\
	\dot \nu_c (0) + O(g \scale_N^{\kaa-2})	 & (d > 4),
	\end{cases}
\end{align}
where we set $c^{\free} = 1 +C_4^{(2)} (0) = O(g)$.  This completes the proof.
\end{proof}

\section{Derivatives of the RG flow: proof of Proposition~\ref{prop:nu-derivs}}
\label{sec:RGderivs}

In this section, we prove Proposition~\ref{prop:nu-derivs}, which
entails computation of derivatives
of all the RG coordinates.  The proof requires extension of the ideas
in \cite[Section~8.4]{BBS-brief} to include mass derivatives, derivatives
of second order, and dimensions $d>4$.
We divide the proof into two parts: $d=4$ in Section~\ref{sec:nu-deriv-4} and  $d>4$
in Section~\ref{sec:nu-derivs-proof2}.
Section~\ref{sec:RGderivs} is independent of the rest of the paper, apart from proving
Proposition~\ref{prop:nu-derivs},
and it could be omitted in a first
reading.
We always assume Theorems~\ref{thm:Phi^U_estimate} and \ref{thm:Phi^K_estimate} in this section.

\subsection{Leading behaviour}

The recursion
\begin{align}
	\nu_{j+1} & = (1- \hat{\gamma} \beta_j g_j) \nu_j + \eta_j g_j - \xi_j g_j^2 + r_{\nu, j}
\label{eq:rg_coordinates_recursion}
\end{align}
follows from the perturbative recursion \eqref{eq:nu_pt_formula} together with
its nonperturbative remainder $r_{\nu,j}$.
This recursion has derivatives
\begin{alignat}{2}
\label{eq:nuprime}
	\nu'_{j+1} & = (1- \hat{\gamma} \beta_j g_j) \nu'_j
    - \hat{\gamma} \beta_j g'_j \nu_j + \eta_j g'_j -2\xi_j g_j g'_j + r'_{\nu,j}  ,
    && \nu_0'=1,
    \\
    \dot\nu_{j+1} & =
    (1- \hat{\gamma} \beta_j g_j) \dot\nu_j
    -\hat\gamma \beta_j g_j + \eta_j \dot g_j
    - \hat\gamma (\dot\beta_j g_j + \beta_j \dot g_j)\nu_j
    -\dot \xi_j g_j^2 - 2\xi_j g_j \dot g_j + \dot r_{\nu,j}, \quad
    && \dot\nu_0=0,
\label{eq:dotnu}
\end{alignat}
where in \eqref{eq:dotnu} we have used \eqref{eq:betadef}--\eqref{eq:etadef} to replace $\dot\eta_jg_j$ by its equivalent $- \hat{\gamma} \beta_j  g_j$.

A guiding principle throughout Section~\ref{sec:RGderivs} is that
the recursions for $\nu_j'$ and $\dot \nu_j$  have their
leading behaviours governed by the solutions $q_j'$ and $\dot q_j$ to the recursions
\begin{alignat}{2}
\label{eq:qprimerec}
    q_{j+1}' & = (1- \hat{\gamma} \beta_j g_j) q_j', && q_0'=1,
    \\
\label{eq:qdotrec}
    \dot q_{j+1} & =   (1- \hat{\gamma} \beta_j g_j) \dot q_j - \hat{\gamma} \beta_j  g_j,
    \quad
    && \dot q_0=0.
\end{alignat}
The solution to \eqref{eq:qprimerec} is
\begin{align}
\label{eq:qjprime}
    q_{j}' & = \Pi_{0,j-1},
\end{align}
where, as in \eqref{eq:Pidef-1},
\begin{align}
	 \Pi_{k,j} & = \prod_{i=k}^j (1- \hat{\gamma} \beta_i g_i) .
\end{align}
Also, it can be verified that the
recursion \eqref{eq:qdotrec} for $\dot q_j$ is solved by
\begin{align}
	\dot q_{j}  = - 1  + \Pi_{0, j-1} .
\label{eq:qjdot}
\end{align}
Our estimates for $\nu'_j$ and $\dot{\nu}_j$ are stated in terms of
the sequences
\begin{align}
	\Sigma_j = \frac{ \nu'_j - \Pi_{0,j -1} }{\Pi_{0,j-1}} , \qquad F_j = \frac{\dot{\nu}_j +1 - \Pi_{0, j-1}}{\Pi_{0,j-1}} ,
	\label{eq:SigmaFdef}
\end{align}
which respectively compare $\nu_j'$ and $\dot \nu_j$ with the sequences
$q_j'$ and $\dot q_j$ from \eqref{eq:qjprime} and \eqref{eq:qjdot}.  The denominators in \eqref{eq:SigmaFdef}
are included for convenience.
By definition, $\Sigma_1=\Sigma_0=F_0=0$ and
\begin{equation}
\label{eq:F1-F0}
    F_1-F_0=F_1 =
    \Pi_{0,0}^{-1} \big(- \hat\gamma \dot\beta_0 g_0 \nu_0
    -\dot \xi_0 g_0^2   \big).
\end{equation}

\subsection{Proof of Proposition~\ref{prop:nu-derivs} for \texorpdfstring{$d=4$}{d=4}}
\label{sec:nu-deriv-4}

For $d=4$, the important product $\Pi_{k,j}$ is well understood,
according to the following
lemma.

\begin{lemma}
\label{lemma:Pi_asymptotic}
Let $d=4$.
Suppose that $(V_j,K_j)_{j\leq k}$ is an RG flow to scale $k$ with initial condition $(g_0,\nu_0)$
and mass $\ka \ge 0$, with $g_0>0$ sufficiently small.
Then for $t \in (0, 2]$ and $i \le j < k$,
\begin{align}
	\prod_{m=i}^{j} (1- t \beta_m {g}_m) =  \Big( \frac{g_{j+1}}{{g}_i} \Big)^{t} \bar{c}_{i,j} [t] 	,
	\label{eq:Pi_asymptotic}
\end{align}
where $\bar{c}_{i,j} [t] = 1 + O_t (\vartheta_i g_i)$.
For a global RG flow
the limit $\bar{c}_{i } [t] =\lim_{j\to\infty}\bar{c}_{i,j} [t]$
exists, is continuous in $(\nu_0,\ka) \in \cX_{\infty}$,
and obeys $\bar{c}_{i } [t] = 1 + O_t (\vartheta_i g_i)$.
\end{lemma}

\begin{proof}
The arithmetic used in the proof of \cite[Lemma~6.1.6]{BBS-brief} shows that
the formula \eqref{eq:Pi_asymptotic} holds with $\bar{c}_{i,j} [t]$  given by
\begin{align}
	\bar{c}_{i,j} [t] = \prod_{m=i}^j (1+ v_m),  \qquad v_m = - 1 + \Big( 1-  \frac{r_{g,m}}{g_{m+1}}  \Big)^{t} \frac{1- {t} \beta_m g_m}{(1- \beta_m g_m)^{t}}.
	\label{eq:bar_c_ij_definition}
\end{align}
A small calculation, using Theorem~\ref{thm:Phi^U_estimate} to bound $r_{g,m}$,
shows that $|v_m| \leq O(\tilde{\vartheta}_m \tilde{g}_m^2)$.
Since this last upper bound sums to $O(\tilde{\vartheta}_m \tilde{g}_m)$ as in
\cite[(6.1.17)]{BBS-brief}, we conclude that
$\bar{c}_{i,j } [t] = 1 + O_t (\vartheta_i g_i)$.
For a global RG flow the infinite product converges and also obeys
$\bar{c}_{i} [t] = 1 + O_t (\vartheta_i g_i)$.

To see the continuity of $\bar{c}_{i} [t]$, we
first observe that Theorems~\ref{thm:Phi^U_estimate} and \ref{thm:Phi^K_estimate} imply that
each RG map $\Phi_{j+1} : (V_j, K_j, \ka) \mapsto (V_{j+1}, K_{j+1})$ is continuous,
so $(V_j, K_j)$ is continuous in $(\nu_0, \ka)\in \mathcal{X}_k$.
Thus each $\bar{c}_{i,j} [t]$ given by \eqref{eq:bar_c_ij_definition} is continuous in
$(\nu_0, \ka)\in \mathcal{X}_k$, and the continuity of $\bar{c}_i$ in
$(\nu_0, \ka)\in \mathcal{X}_\infty$ follows by the
Dominated Convergence Theorem.
\end{proof}

Proposition~\ref{prop:RG_coordinates_mass_derivative}
is the main ingredient in the proof of Proposition~\ref{prop:nu-derivs} for $d=4$.
It includes derivatives of all the RG coordinates because we must bound them all together.

Recall the definition of an RG flow from Definition~\ref{def:RGflow}, and
the definition $\hat\gamma = \frac{n+2}{n+8}$ from \eqref{eq:hatexponents}.

\begin{proposition}[Case $d=4$]
\label{prop:RG_coordinates_mass_derivative}
Suppose that $L$ is sufficiently large and $g_0$ is sufficiently small.
Suppose that $(V_j,K_j)_{j\leq k}$ is an RG flow to scale $k$ with initial condition $(g_0,\nu_0)$ and mass $\ka \ge 0$.
There exist positive constants $C_0$, $C_1$, $C_2$, $C$  (independent of $k$) such that
for all $j \leq k$,  $(g_j, \nu_j, K_j)$ are continuously differentiable functions of $(\nu_0, \ka)\in \mathcal{X}_k$ and satisfy the following bounds.
\begin{enumerate}
\item The $\nu_0$-derivatives satisfy
\begin{align}
	& | \Sigma_j - \Sigma_{j-1} | \leq C_0 \tilde{\vartheta}_j^3 \tilde{g}^2_j , \quad\;\;
 	|g'_j| \leq C_1 L^{2j} \tilde{\vartheta}_j^3 \tilde{g}_j^{2} \Pi_{0,j-1}   ,  \quad\;\;
	\norm{K'_j}_{\Wkappa_j} \leq C_2 L^{2j} \tilde{\vartheta}_j^3 \tilde{g}_j^{2} \Pi_{0,j-1}	.	
	 \label{eq:nujdnu0}
\end{align}

\item The $\ka$-derivatives satisfy
\begin{align}
	& | F_{j} - F_{j-1} |  \leq C_0 \tilde{\vartheta}_j^3 \tilde{g}_j^2 \Pi_{0,j-1}^{-1} ,
	\quad\;\;  | \dot{g}_j| \leq C_1 L^{2j} \tilde{\vartheta}_j^3 \tilde{g}^{2}_j ,
	\quad\;\;  \big\| \dot K_j \big\|_{\Wkappa_j} \leq C_2 L^{2j} \tilde{\vartheta}_j^3 \tilde{g}^{2}_j,
	\label{eq:nujdka}
\end{align}
and $\dot{\nu}_j \leq - C  \tilde{\vartheta}_0^2 \tilde{g}_j$.

\item The second derivatives satisfy
\begin{align}
	| \nu''_{j} | \leq O \Big(L^{2j} \tilde{\vartheta}^3_{j} \tilde{g}_j \Big( \frac{\tilde g_j}{g_0} \Big)^{2\hat{\gamma}} \Big) , \qquad
	| \dot{\nu}'_{j} | \leq O\Big( L^{2j} \tilde{\vartheta}^3_{j} \tilde{g}_j \Big( \frac{\tilde g_j}{g_0} \Big)^{\hat{\gamma}} \Big)
	, \qquad | \ddot{\nu}_{j} | \leq O(L^{2j} \tilde{\vartheta}^3_{j} \tilde{g}_j ) .
	\label{eq:nujdd}
\end{align}
\end{enumerate}
\end{proposition}

\begin{remark}
\label{remark:MCCC}
It is important that $C_0, C_1$ and $C_2$ are independent of $j$ and $k$.
The constants will eventually be chosen to satisfy
\begin{align}
\label{eq:MCCC}
	M= \max_{0 \le p,q \leq 2} M_{p, q} \ll C_2 \ll C_1 \ll C_0
\end{align}
where $M_{p,q}$ are the $L$-dependent constants in Theorems~\ref{thm:Phi^U_estimate} and \ref{thm:Phi^K_estimate} (including $M_{0,0}=C_{\rg}$) and \eqref{eq:MCCC} defines $M$.

\medskip
Since it will be important to keep track of the constants
$C_i$ in the following, we often write $O(C_1)$, $O(M_{0,1}C_2)$, etc.
In particular, with this notation, $O(1)$ is independent of $L,C_1,C_2,C_3$,
and of course of the scale.  When a constant does depend on $L$, but not
on the others, we instead write $O_L(1)$.
In addition, we will take $(g_0)^{-1}$ to be large compared to these constants.
The same holds for the constants $C'_0, C'_1,
C'_2$ of Lemma~\ref{lemma:nu_double_derivative_IH}, and
for the corresponding constants in Section~\ref{sec:nu-derivs-proof2} with the additional assumption that $C'_0, C'_1,
C'_2$ are themselves much larger than $C_0,C_1,C_2$.
\end{remark}

\begin{proof}[Proof of Proposition~\ref{prop:nu-derivs}(i)]
We assume that $(V_j,K_j)$ is an RG flow to scale $N$ with initial
condition $(g_0,\nu_0)$ and mass $\ka$.

The negative upper bound on $\dot{\nu}_j$ is a part of the statement of Proposition~\ref{prop:RG_coordinates_mass_derivative}(ii), the required
bounds on the second derivatives are restated in \eqref{eq:nujdd}, and
\eqref{eq:g_j_derivative_in_nu_0_high_d-bis} is restated in \eqref{eq:nujdnu0} and \eqref{eq:nujdka}.
It therefore suffices to prove the bounds
\begin{align}
	\nu'_j
    =
    \Big( \frac{g_j}{g_0} \Big)^{\hat{\gamma}} \big( c_{0,N}  + O( \tilde{\vartheta}_j \tilde{g}_j) \big) ,
	\qquad
    \dot \nu_j +1 =
    \Big( \frac{g_j}{g_0} \Big)^{\hat{\gamma}}c_{1,N} + O(\tilde{\vartheta}_j \tilde{g}_j)
	\label{eq:nu_j_derivative_in_nu_0-pf}
\end{align}
on the first derivatives,
as well as facts that the constants $c_{0,N}$ and $c_{1,N}$
are continuous in  $(\nu_0, \ka)\in \mathcal{X}_N$ and that their limits
$c_{i,\infty}$ are continuous in  $(\nu_0, \ka)\in \mathcal{X}_\infty$.

By \eqref{eq:nujdnu0}, which holds by assumption for all $j \le N$,
\begin{align}
	\Sigma_j = \Sigma_N + O \Big( \sum_{k = j+1}^N \tilde{\vartheta}_k^3 \tilde{g}_k^2 \Big) =  \Sigma_N + O ( \tilde{\vartheta}_j^3 \tilde{g}_j ),
\end{align}
where the sum has been bounded as in \cite[Exercise~6.1.4]{BBS-brief}.
Since $\nu'_j = \Pi_{0,j-1} (1 + \Sigma_j)$ by definition,
it follows from the formula for $\Pi_{0,j-1}$ in Lemma~\ref{lemma:Pi_asymptotic} that
\begin{align}
	\nu'_j
    =
    \Big( \frac{g_j}{g} \Big)^{\hat{\gamma}}
    \big( 1 + \Sigma_N  + O ( \tilde{\vartheta}_j^3 \tilde{g}_j ) \big)
    \big(  \bar{c}_{0, N} [\hat{\gamma}] + O ( \vartheta_j g_j ) \big).
\end{align}
This gives the formula \eqref{eq:nu_j_derivative_in_nu_0-pf} for $\nu'_j$
with $c_{0,N} = (1+ \Sigma_N) \bar{c}_{0, N} [\hat{\gamma}] = 1+ O(\tilde{\vartheta}_0 g)$.
The continuity of $\bar{c}_{0, N}$ and $\bar{c}_{0,\infty}$ is provided by
Lemma~\ref{lemma:Pi_asymptotic}.
By definition, $\Sigma_0=0$.
The limit $\Sigma_{\infty} =
\lim_{N\rightarrow \infty} \sum_{i=0}^{N-1} (\Sigma_{i+1} - \Sigma_i)$
is continuous in $(\nu, \ka)$ by the Dominated Convergence Theorem, and therefore
$c_{0, \infty}$ is continuous in $(\nu_0, \ka) \in \cX_{\infty}$ as claimed.

For $\dot{\nu}_j$, we start with $\dot{\nu}_j+1 = \Pi_{0,j-1}(1+F_j)$ and proceed
similarly.
With Lemma~\ref{lemma:Pi_asymptotic}, we obtain
\begin{align}
    F_j & = F_N
    +  O \Big(g_0^{\hat\gamma}  \sum_{k = j+1}^N \tilde{\vartheta}_k^3 \tilde{g}_k^{2-\hat\gamma} \Big)
    =  F_N + O (g_0^{\hat\gamma} \tilde{\vartheta}_j^3 \tilde{g}_j^{1-\hat\gamma} ),
\end{align}
and hence
\begin{equation}
    \dot\nu_j + 1
    =
    \Pi_{0,j-1}(1+F_N) + O (\tilde{\vartheta}_j^3 \tilde{g}_j ).
\end{equation}
This gives \eqref{eq:nu_j_derivative_in_nu_0-pf} for $\dot\nu_j$
with $c_{1,N} = (1+ F_N) \bar{c}_{0,N} [\hat{\gamma}] = 1 + O(\tilde{\vartheta}_0 g)$.
Again the limit $c_{1,\infty}$ exists and is continuous in
$(\nu_0,\ka)\in\mathcal{X}_\infty$.
\end{proof}

We now turn to the proof of
Proposition~\ref{prop:RG_coordinates_mass_derivative}, which is by induction on $j$.
Part~(i) is essentially as in \cite[Theorem~8.4.1]{BBS-brief}, while parts~(ii) and (iii) require new effort.  We begin by applying Theorems~\ref{thm:Phi^U_estimate} and \ref{thm:Phi^K_estimate} to see that when the bounds of  Proposition~\ref{prop:RG_coordinates_mass_derivative} hold, we can inductively
obtain bounds on the derivatives of $K_j$ and $R_j^U$ up to and including scale $k+1$.
Note that we write $\dot{R}^U_{j+1} = \frac{d}{d\ka} ( R^U_{j+1} (V_j, K_j) )$, so $\dot{R}^U_{j+1}$ is not the same as $D_{\ka} R^U_{j+1}$ which includes the scaling
factor present in \eqref{eq:massnorm}.
Similarly, $(R^U_{j+1})'
 = \frac{d}{d\nu_0} ( R^U_{j+1} (V_j, K_j) )$.

\begin{lemma} \label{lemma:derivative_bounds_first_step}
Let $d=4$.
Suppose that $(V_j,K_j)_{j\leq k}$ is an RG flow to scale $k$ with initial condition $(g_0,\nu_0)$
and mass $\ka \ge 0$.  Suppose further that the bounds of
Proposition~\ref{prop:RG_coordinates_mass_derivative}(i,ii) hold for all $i \le j$.
Then
\begin{align}
	\norm{K'_{j+1}}_{\Wkappa_{j +1}} &\leq C_2 L^{2(j+1)} \tilde{\vartheta}_{j+1}^3 \tilde{g}_{j+1}^{2} \Big( \frac{\tilde{g}_{j+1}}{g_0} \Big)^{\hat{\gamma}} ,\\
	\norm{(R_{j+1}^U)'}_{\cU_{j+1} (\ell_{j+1})}
& \leq  O(M C_2)
L^{2(j+1)} \tilde{\vartheta}_{j+1}^3 \tilde{g}_{j+1}^2 \Big( \frac{\tilde{g}_{j+1}}{g_0} \Big)^{\hat{\gamma}},
\\
	\norm{\dot{K}_{j+1}}_{\Wkappa_{j +1}} &\leq C_2 L^{2(j+1)} \tilde{\vartheta}_{j+1}^3  \tilde{g}_{j+1}^{2} ,\\
	\norm{\dot{R}_{j+1}^U}_{\cU_{j+1} (\ell_{j+1})} & \leq
    O(M C_2)
    L^{2(j+1)} \tilde{\vartheta}_{j+1}^3 \tilde{g}_{j+1}^2.
\end{align}
\end{lemma}

\begin{proof}
We fix $j \le k$ and make the indicated assumption for $i \le j$.
Proposition~\ref{prop:nu-derivs}(i) then applies for such $i$, in addition to the assumed
bounds of Proposition~\ref{prop:RG_coordinates_mass_derivative}(i,ii) for $i \le j$.
We apply induction and assume that the bounds of Lemma~\ref{lemma:derivative_bounds_first_step}
hold when $j+1$ is replaced by $i$ for all $i \le j$.   They hold trivially for $i=0$
since $K_0$ and $R_0$ are both zero.
By \eqref{eq:nu_j_derivative_in_nu_0-pf}
and Lemma~\ref{lemma:Pi_asymptotic}, and with the $L$-dependence
due to $\ell_0=L^{1+d/2}$ in $\ell_j$ of \eqref{eq:h_j_definition},
\begin{align}
	\label{eq:V_prime_T_0_norm_pf}
	\norm{ V_{j}' }_{\cV_{j}} (\ell_{j})
	& \le O_L(1)\big(
	|g'_j| + |\nu_j'|L^{2j} \big)
	\nnb
	&\leq
    O_L(1) L^{2j}
    (\frac{\tilde{g}_{j}}{\tilde{g}_0} )^{\hat{\gamma}}
	\big(O(C_1\tilde g_j^2)+\frac12c_{0,N}+O(\tilde \vartheta_j \tilde g_j)\big)
	\leq O_{L}(1)L^{2j}
	(\frac{\tilde{g}_{j}}{\tilde{g}_0})^{\hat{\gamma}},
\end{align}
for $g_0$ small enough.  Also, by \eqref{eq:nujdnu0} and Lemma~\ref{lemma:Pi_asymptotic},
$\norm{K'_j}_{\Wkappa_j} \leq 2 C_2 L^{2j} \tilde{\vartheta}_j^3 \tilde{g}_j^{2}  (\frac{\tilde{g}_{j}}{\tilde{g}_0})^{\hat{\gamma}}$ for small $g_0$.
We can use these bounds with the chain rule, namely
\begin{equation}
	\frac{d}{d\nu_0} \Phi_{j+1} (V_j, K_j)
	= D_K \Phi_{j+1} (V_j, K_j;K'_j)
	+ D_V \Phi_{j+1} (V_j, K_j;V'_j),
\end{equation}
to bound both $K'_{j+1}$ and $R'_{j+1}$.  Indeed,
by Theorem~\ref{thm:Phi^K_estimate},
and for $L$ large enough to overcome the $L$-independent constant $M_{0,1}'$,
\begin{align}
	\label{eq:pf_K_prime_deriv}
	\norm{K'_{j+1}}_{\Wkappa_{j +1}}
&\leq
	M'_{0,1} L^{-2} 2  C_2 L^{2j} \tilde{\vartheta}_j^3 \tilde{g}_{j}^{2} \big( \frac{\tilde{g}_{j}}{\tilde{g}_0} \big)^{\hat{\gamma}}
+ M_{1,0} \tilde{\vartheta}_{j+1}^3 \tilde{g}_{j+1}^{2}
O_{L}(1)L^{2j}\big( \frac{\tilde{g}_{j}}{\tilde g_0} \big)^{\hat{\gamma}}
\nnb
	& \leq C_2 L^{2(j+1)} \tilde{\vartheta}_{j+1}^3 \tilde{g}_{j+1}^{2}
\big( \frac{\tilde{g}_{j+1}}{g_0} \big)^{\hat{\gamma}},
\end{align}
where we used \eqref{eq:MCCC} in the last step.
Similarly, by Theorem~\ref{thm:Phi^U_estimate},  for $C_2$ large enough we have
\begin{align}
	\norm{ (R^U_{j+1} )' }_{\cU_{j+1} (\ell_{j+1})}
    & \leq M_{0,1}   C_2 L^{2j} \tilde{\vartheta}_j^3 \tilde{g}_j^2 \big( \frac{\tilde{g}_{j}}{\tilde{g}_0} \big)^{\hat{\gamma}}
    + M_{1,0} \tilde{\vartheta}_j^3 \tilde{g}_j^3
    O_L(1)L^{2j}\big( \frac{\tilde{g}_{j}}{\tilde{g}_0} \big)^{\hat{\gamma}}
    \nnb
	& \leq
    2MC_2  L^{2(j+1)} \tilde{\vartheta}_{j+1}^3 \tilde{g}_{j+1}^2 \Big( \frac{\tilde{g}_{j+1}}{g_0} \Big)^{\hat{\gamma}}.
\end{align}

For the mass derivative, by the chain rule, $\frac{d}{d\ka} \Phi_{j+1} (V_j, K_j)  =
\dot
\Phi_{j+1} + D_K \Phi_{j+1} (\dot{K}_j) + D_V \Phi_{j+1} (\dot{V}_j)$.
The first term is controlled by Theorem~\ref{thm:Phi^K_estimate} since
our norm contains mass derivatives.
Also \eqref{eq:nujdka} and \eqref{eq:nu_j_derivative_in_nu_0-pf} imply that, as in \eqref{eq:V_prime_T_0_norm_pf}, $\norm{ \dot{V}_{j} }_{\cV (\ell_{j})} \leq  O_{L}(1)L^{2j}$.
Therefore,
\begin{align}
	\norm{\dot{K}_{j+1}}_{\Wkappa_{i +1}}
&\leq L^{2(j+1)}  C_{\rg} \tilde{\vartheta}_{j+1}^3 \tilde{g}_{j+1}^3
+ M'_{0,1} L^{-2} C_2 L^{2j}\tilde{\vartheta}_j^3 \tilde{g}_{j}^{2} + M_{1,0} \tilde{\vartheta}_{j+1}^3 \tilde{g}_{j+1}^{2}O_{L}(1) L^{2j}  \nnb
	& \leq C_2 L^{2(j+1)} \tilde{\vartheta}_{j+1}^3 \tilde{g}_{j+1}^{2}.
\end{align}
The bound on $\dot{R}_{j+1}^U$ is obtained similarly,
again using Theorem~\ref{thm:Phi^U_estimate}.
\end{proof}

We can now prove Proposition~\ref{prop:RG_coordinates_mass_derivative}(i).

\begin{proof}[Proof of Proposition~\ref{prop:RG_coordinates_mass_derivative}(i)]
Fix $j \le k$ and assume the bounds of
Proposition~\ref{prop:RG_coordinates_mass_derivative}(i)
for all scales $i \le j$.
This assumption is satisfied for $j=0$ since $\Sigma_1-\Sigma_0=0-0=0$,
and also $g_0'$ and $K_0'$ are zero.
Lemma~\ref{lemma:derivative_bounds_first_step} gives
the desired bound on $K'_{j+1}$ at scale $j+1$.
For the bound on $g_{j+1}'$, we first use
Lemma~\ref{lemma:derivative_bounds_first_step}
and \eqref{eq:r^U_domination_by_R^U} to obtain
\begin{align}
	|r'_{g,j+1}|
	&\leq O(1)\|(R^U_{j+1})'\|_{\cV(\ell_{j+1})}
	\leq O(MC_2)L^{2(j+1)} \tilde{\vartheta}_{j+1}^3 \tilde{g}_{j+1}^{2}(\frac{\tilde{g}_{j+1}}{g_0})^{\hat{\gamma}}.
\end{align}
Therefore, by Lemma~\ref{lemma:Pi_asymptotic} and the induction hypothesis,
\begin{align}
	|g'_{j+1}|
	&\leq (1- 2\beta_j g_{j}) |g'_j| + |r'_{g,j+1}|
	\leq
	O(C_1 L^{-2}+MC_2)L^{2(j+1)} \tilde{\vartheta}_{j+1}^3 \tilde{g}_{j+1}^{2} \Pi_{0,j},
\end{align}
where we used $1- 2\beta_j g_{j} \le 1$.
With $L$ large, and assuming $C_1 \gg C_2$, this gives the desired bound on $g'_{j+1}$.

Finally,
for the bound on $\Sigma$,  we first use \eqref{eq:SigmaFdef} and
\eqref{eq:nuprime} to obtain
\begin{align}
    \Sigma_{j+1}-\Sigma_j
    &= \Pi_{0,j}^{-1}\big( \nu_{j+1}' - (1- \hat{\gamma} \beta_j g_j) \nu'_j \big)
    \nnb
      &= \Pi_{0,j}^{-1}\big( - \hat{\gamma} \beta_j g'_j \nu_j + \eta_j g'_j - 2 \xi_j g_j g'_j + r'_{\nu, j} \big) .
	\label{eq:Sigma_recursion_relation}
\end{align}
Then we use the bound on $r'_{\nu, j}$ from Lemma~\ref{lemma:derivative_bounds_first_step},
on $\beta_j,\eta_j,\xi_j$ from \eqref{eq:betadef}--\eqref{eq:xidef},
on $g'_j$ from the induction hypothesis, and on $\Pi_{0,j}$ from Lemma~\ref{lemma:Pi_asymptotic}, and we obtain the desired bound
\begin{align}
	| \Sigma_{j+1} - \Sigma_j | &
        \leq C_0 \tilde{\vartheta}^3_{j+1} \tilde{g}_{j+1}^2,
\end{align}
assuming $C_0 \gg C_1$.
\end{proof}

\begin{proof}[Proof of Proposition~\ref{prop:RG_coordinates_mass_derivative}(ii)]
As induction hypotheses,
we assume that the bounds of \eqref{eq:nujdka} hold for scales $i \le j$.
To start the induction, $\dot g_0$ and $\dot K_0$ are both equal to zero,
and, by \eqref{eq:F1-F0} and the fact that $|\nu_0| \leq O(g_0)$,
$F_1-F_0$ is bounded above by a multiple of $\tilde\vartheta_1^3\tilde g_1^2\Pi_{0,0}^{-1}$ as required.

We first prove that
\begin{align}
	|F_{j+1} - F_j | \leq C_0 \tilde{\vartheta}_j^3 \tilde{g}_j^2 \Pi_{0,j}^{-1} .
	\label{eq:F_j_estimate}
\end{align}
By definition, $F_j=\Pi_{0,j-1}^{-1}(\dot \nu_j - \dot q_j)$, where
$\dot q_j$ satisfies the recursion \eqref{eq:qjdot}.  From this, together
with $\dot\eta_j=-\hat\gamma \beta_j$ and some arithmetic, we arrive at
\begin{align}
	F_{j+1} - F_j =  E_j\Pi_{0,j}^{-1}
\label{eq:F_j_recursion}
\end{align}
with
\begin{equation}
	E_k =  -  \hat{\gamma} \frac{d}{d\ka} (\beta_k g_k) \nu_k + \eta_k \dot{g}_k + \frac{d}{d\ka} (- \xi_k g_k^2 + r_{\nu, k}) .
\end{equation}
By \eqref{eq:nujdka} and Lemma~\ref{lemma:derivative_bounds_first_step}, we can bound each constituent of $E_j$ to obtain
\begin{align}
	|E_j  |  \leq O(M(C_1 + C_2)) \tilde{\vartheta}_j^3 \tilde{g}_j^2 .  \label{eq:E_j_bound}
\end{align}
This proves \eqref{eq:F_j_estimate}.

The desired bound on $\dot{K}_{j+1}$ follows from
Lemma~\ref{lemma:derivative_bounds_first_step}.
For the bound on $\dot g_{j+1}$, we apply
Lemma~\ref{lemma:Pi_asymptotic} and \eqref{eq:r^U_domination_by_R^U} to see that
\begin{align}
	| \dot{g}_{j+1} |
	& \leq (1- 2\beta_j g_j ) |\dot{g}_j | + |\dot{\beta}_j g_j^2 | + | \dot{r}_{g,j}| \nnb
	&\leq O(L^{-2}C_1+1+MC_2) L^{2(j+1)}  \tilde{\vartheta}_{j+1}^3\tilde{g}_{j+1}^2 \nnb
	& \leq C_1 L^{2(j+1)}  \tilde{\vartheta}_{j+1}^3 \tilde{g}_{j+1}^2 ,
\end{align}
for sufficiently large $L$ and $C_1$.

It remains to prove that
$\dot{\nu}_{j+1} \leq -C  \tilde{\vartheta}^2_0 \tilde{g}_{j+1}$.  By definition,
\begin{equation}
	\tilde{\vartheta}_j = \begin{cases}
		2^{-(j-j_{\tilde{\ka}})_+} & ( \tilde{\ka} < 1) \\
		\tilde{\ka}^{-1} 2^{-j} & ( \tilde{\ka} \geq 1),
	\end{cases}
\end{equation}
so, in particular, $\tilde\vartheta_0= 1 \vee \tilde\ka^{-1}$.
The proof is in two steps:
we use induction for $j\leq j_{\ka} +1$
and then argue directly for $j >  j_{\ka} +1$.

For the induction, given $j \ge 1$, we assume that
$\dot{\nu}_{j} \leq -C  \tilde{\vartheta}^2_0 \tilde{g}_{j}$,
with $C = \frac{1}{8} \hat{\gamma} \bbb$  where $\bbb=(n+8)(1-L^{-d})$
is the constant defined in \eqref{eq:Bdef}.
Of course $\dot\nu_0=0$, so we start the
induction by showing that $\dot{\nu}_{1} \leq -C  \tilde{\vartheta}^2_0 \tilde{g}_{1}$.
Recall that $\dot\eta_j= -\hat\gamma \beta_j$.
By the recursion \eqref{eq:nuprime} for $\nu'$
(for the first equality),
and by the induction hypothesis and \eqref{eq:E_j_bound}
(for the second equality),
\begin{align}
	\dot{\nu}_{j+1} &= (1-\hat{\gamma} \beta_j g_j) \dot{\nu}_j + \dot{\eta}_j g_j +  E_j
    \nnb
    & = (1-\hat{\gamma} \beta_j g_j) \dot{\nu}_j -\hat\gamma \beta_j g_j + O(\tilde{\vartheta}_j^3 \tilde{g}_j^2).
\end{align}
In particular, for $j=0$ the first term on the right-hand side is zero,
the second is $-\hat\gamma B(1+\ka)^{-2}g_0
\le -\frac 14 \hat\gamma B\tilde\vartheta_0^2g_0$, and the third term is relatively small,
so this proves that $\dot{\nu}_{1} \leq -\frac 18 \hat\gamma B  \tilde{\vartheta}^2_0 g_0$
which is sufficient since $g_0  > g_1$.
To advance the induction, we first consider
$j \le j_a \wedge ( N-1 ) $ (so $a\le 1$ and $\tilde\vartheta_0=1$).
In this case, $\beta_j \ge \beta_{j_{\ka}} \ge \frac 14 B$, and the term $E_j$ is relatively
small since it is of order $g_j^2$.  We use the induction hypothesis to
see that $\dot\nu_j \le 0$.  Together, these facts show that
$\dot\nu_{j+1} \le - \frac 18 \hat\gamma B g_j
\le - \frac 18 \hat\gamma B g_{j+1}$, which advances
the induction to $j+1 \le ( j_{\ka}+1 ) \wedge N$
with $C = \frac{1}{8} \hat{\gamma} B$.

Suppose that $N > j_{\ka} + 1$.
We apply a different argument for $j > j_{\ka}$ since $\beta_j$ decays as $j$ increases
beyond the mass scale.
We again consider the recursion \eqref{eq:qdotrec} for $\dot{q}$,
but now only for $j \ge i \ge j_{\ka} + 1$
with fixed $i$ and with initial condition $\dot\nu_i$.
With the initial $i$ indicated by an additional
subscript, the solution to this recursion is
\begin{align}
    \dot q_{i,j} = \dot\nu_i \Pi_{i,j-1} - 1 + \Pi_{i,j-1}.
\end{align}
Then we consider
\begin{align}
	F_{i,j} = \frac{\dot{\nu}_j -   \dot{q}_{i,j}}{\Pi_{i,j-1}}
    \qquad (j \ge i).
\end{align}
In particular, $\dot q_{0,j}=\dot q_j$ and $F_{j} = F_{0,j}$.
The proof of \eqref{eq:F_j_estimate} similarly gives
\begin{align}
	|F_{i,j} - F_{i,j-1}| \leq C_0 \tilde{\vartheta}_j^3 \tilde{g}_j^2 \Pi_{i,j-1}^{-1}. \label{eq:T_i,j_restatement}
\end{align}
By definition of $F_{i,j}$,
\begin{align}
	\dot{\nu}_j = \dot{\nu}_i \Pi_{i,j-1}  -1 + \Pi_{i, j-1} (1 + F_{i,j} )
    \qquad (j \ge i).
\label{eq:nu_dot_negativitivity_int1}
\end{align}
Now we choose $i =1 + ( j_{\ka} \vee 0 )$ and consider $j>i$.
Since $\vartheta_j^3 $ decays exponentially in $j-j_{\ka}$,
there are positive $c,c'$ such that
$1 - c \tilde\vartheta_i^2 \tilde{g}_i \le \Pi_{i,j-1} \leq 1 - c' \tilde\vartheta_i^2 \tilde{g}_i$.  Also, \eqref{eq:T_i,j_restatement} gives
\begin{align}
	|F_{i,j}|  \leq \sum_{k=i}^{j-1} |F_{i,k} - F_{i,k+1}| \leq O(C_0)
    \tilde{\vartheta}_{i}^3
    \tilde{g}_{i}^2,
\end{align}
so
for sufficiently small $\tilde{g}_i$,
\begin{align}
	-1 + \Pi_{i, j-1} (1 + F_{i,j} ) \leq 0.
\label{eq:nu_dot_negativitivity_int2}
\end{align}
If we insert this bound in \eqref{eq:nu_dot_negativitivity_int2},  use $\Pi_{i,j-1} \geq 1/2$ for small $\tilde g_i$ and take $i = j_{\ka} +1$, then
we deduce from the upper bound on $\dot\nu_i$ that
\begin{align}
	\dot{\nu}_j
    \le \dot\nu_i \Pi_{i,j-1}
	\leq - \frac{1}{8} \hat\gamma B  \tilde{\vartheta}^2_0 \tilde{g}_i	\times \frac 12
	\le - \frac{1}{16} \hat{\gamma} B \tilde{\vartheta}^2_0 \tilde{g}_j ,
\end{align}
and the proof is complete.
\end{proof}

For the proof of Proposition~\ref{prop:RG_coordinates_mass_derivative}(iii), we first prove
a counterpart to Lemma~\ref{lemma:derivative_bounds_first_step} for these
second derivatives.
(The bound on $\nu_j''$ in \eqref{eq:nu_double_derivative_IH} is included in
the bounds of Proposition~\ref{prop:RG_coordinates_mass_derivative}, but we
include it explicitly in \eqref{eq:nu_double_derivative_IH} for easy reference.)
We provide details of the proof of Proposition~\ref{prop:RG_coordinates_mass_derivative}(iii)
only for the second derivatives
with respect to $\nu_0$.
The bounds on $\dot{\nu}'_j$ and $\ddot{\nu}_j$ follow in the same way,
once we observe that the bounds on the first derivative of $(\dot\nu_j, \dot g_j, \dot K_j)$
(as in part~(ii)) simply have one less factor of $( \tilde g_j/g_0)^{ \hat{\gamma}}$ compared to $(\nu_j',  g_j',  K_j')$.\

\begin{lemma}
\label{lemma:nu_double_derivative_IH}
Let $d=4$.
Suppose that $(V_j,K_j)_{j\leq k}$ is an RG flow to scale $k$ with initial condition $(g_0,\nu_0)$
and mass $\ka \ge 0$.  Suppose that the bounds of
Proposition~\ref{prop:RG_coordinates_mass_derivative} hold for all $i \le j$, and that,
in addition,
\begin{align}
	|\nu''_j | \leq C'_0 L^{2j} \tilde{\vartheta}^3_{j} \tilde{g}_j \Big( \frac{\tilde g_j}{ g_0} \Big)^{2 \hat{\gamma}} ,   \quad \;\;
	|g''_j | \leq C'_1 L^{4j} \tilde{\vartheta}^2_{j} \tilde{g}_j \Big( \frac{\tilde g_j}{ g_0} \Big)^{2 \hat{\gamma}}  ,
	 \quad \;\;
	\norm{K''_j}_{\Wkappa_j} \leq C'_2 L^{4j} \tilde{\vartheta}_{j}^3  \tilde{g}_j \Big( \frac{\tilde g_j}{ g_0} \Big)^{2 \hat{\gamma}} .
	\label{eq:nu_double_derivative_IH}
\end{align}
Then for sufficiently large $C'_2$,
\begin{align}
	\norm{ K''_{j+1} }_{\Wkappa_{j+1}} & \leq C'_2 L^{4(j+1)} \tilde{\vartheta}_{j+1}^3  \tilde{g}_{j+1} \Big( \frac{\tilde g_j}{  g_0} \Big)^{2 \hat{\gamma}} \\
	\norm{ ( R_{j+1}^U )'' }_{\cU_{j+1} (\ell_{j+1})} & \leq  O (M C'_2 ) L^{4j} \tilde{\vartheta}_{j+1}^3 \tilde{g}_{j+1} \Big( \frac{\tilde g_j}{  g_0} \Big)^{2 \hat{\gamma}} .
\end{align}
\end{lemma}

\begin{proof}
We use the fact that the second derivative of a function $f (V,K)$ is
\begin{align}
	f'' (V, K) &= D_V f (V, K ; V'' ) + D_K f (V, K ; K'') \nnb
	& \qquad + D_V^2 f (V, K ; V', V') + 2 D_V D_K f (V, K ; V' , K') + D_K^2 f (V, K ; K', K') .
	\label{eq:chain_rule_second_derivative}
\end{align}
We are free to apply Proposition~\ref{prop:RG_coordinates_mass_derivative}(i,ii),
which have been proved already.

By \eqref{eq:nu_double_derivative_IH} with $C'_0\gg C_1'$, we have
\begin{align}
	\norm{V''_j}_{\cV_j (\ell_j)} \leq O(C'_0) L^{4j} \tilde{\vartheta}_j \tilde{g}_j  \Big( \frac{\tilde g_j}{  g_0} \Big)^{2 \hat{\gamma}}.
\end{align}
With our convention that we omit $j$ and write $+$ for $j+1$, and writing $\cW^\kappa_+$ instead of $\D \to \cW^{\kappa}_+$ for the operator norms,
\begin{align}
	\norm{ K''_{+} }_{\Wkappa_{+}}
	&\leq \norm{ D_V \Phi_{+}^K }_{\Wkappa_{+}} \norm{V''}_{\cV}
    + \norm{ D_K \Phi_{+}^K}_{\Wkappa_{+}}  \norm{ K'' }_{\Wkappa}
    \nnb
    & \qquad + \norm{ D_V^2 \Phi_{+}^K}_{\Wkappa_{+}} \norm{ V' }_{\cV}^2
	 + 2 \norm{ D_V D_K \Phi_{+}^K }_{\Wkappa_{+}} \norm{V' }_{\cV} \norm{K'}_{\Wkappa}
    + \norm{ D_K^2 \Phi_{+}^K (V, K )}_{\Wkappa_{+}} \norm{ K'}^2_{\Wkappa} \nnb
	& \leq O (L^{4j} \tilde{\vartheta}_{+}^3 )
	\big(
	O_L(1)M C'_0 \tilde{g}_{+}^{3} + C'_2L^{-2} \tilde{g} +O_L(1) M C_2\tilde{g}_{+}
    + O_L(1) M \tilde{g}_{+}  +M  C_2^2\tilde{g}_{+}^{-\frac{9}{4} + 4}
	\big)
	\Big( \frac{\tilde g_j}{ g_0} \Big)^{2 \hat{\gamma}}  \nnb	
	& \leq C'_2 L^{4(j+1)} \tilde{\vartheta}_{+}^3  \tilde{g}_{+} \Big( \frac{\tilde g_{+}}{ g_0} \Big)^{2 \hat{\gamma}}
,
\end{align}
for sufficiently large $C_2'$ and sufficiently small $g$ (to absorb $C_0'$).
Here we used the induction hypotheses to bound the second derivatives,
the previously noted bound
$\norm{ V_{j}' }_{\cV (\ell_{j})}
\leq O_L(1)L^{2j} ( \frac{\tilde{g}_{j}}{\tilde{g}_0})^{\hat{\gamma}}$
and \eqref{lemma:derivative_bounds_first_step} to bound the first derivatives,
and $M$ is the $L$-dependent constant
which bounds the constants from the main RG Theorems~\ref{thm:Phi^U_estimate}--\ref{thm:Phi^K_estimate}.
Similarly,
\begin{align}
	\norm{ ( R_{+}^U )'' }_{\cU_{+}} &\leq \norm{ D_V R_{+}^U }_{\cU_{+}} \norm{V''}_{\cV} + \norm{ D_K R_{+}^U (V, K) }_{\cU_{+}}  \norm{ K''}_{\Wkappa} + \norm{ D_V^2 R_{+}^U (V, K ) }_{\cU_{+}} \norm{ V' }_{\cV}^2 \nnb
	& \qquad + 2 \norm{ D_V D_K R_{+}^U }_{\cU_{+}} \norm{V' }_{\cV} \norm{K'}_{\Wkappa} + \norm{ D_K^2 R_{+}^U (V, K )}_{\cU_{+}} \norm{ K'}^2_{\Wkappa} \nnb
	& \leq O (M)  L^{4j} \tilde{\vartheta}_{+}^3  \big( C'_0 \tilde{g}_{+}^{4} + C'_2 \tilde{g}_{+} + \tilde{g}_{+}^{3}  \tilde{g}_{+}^{2 \hat{\gamma}} + \tilde{g}_{+}^{2}  + \tilde{g}_{+}^{4}  \big) \Big( \frac{\tilde g_j}{ g_0} \Big)^{2 \hat{\gamma}} \nnb
	&\leq O (M C'_2 ) L^{4j} \tilde{\vartheta}_{+}^3 \tilde{g}_{+} \Big( \frac{\tilde g}{ g_0} \Big)^{2 \hat{\gamma}},
\end{align}
and the proof is complete.
\end{proof}

\begin{proof}[Proof of Proposition~\ref{prop:RG_coordinates_mass_derivative}(iii)]
Suppose that $(V_j,K_j)_{j\leq k}$ is an RG flow to scale $k$ with initial condition $(g_0,\nu_0)$
and mass $\ka \ge 0$.  Given $j \leq k$, we assume as induction hypothesis that
the bounds \eqref{eq:nu_double_derivative_IH} hold
(they hold trivially for $j=0$ as all are zero).
Then the conclusion of
Lemma~\ref{lemma:nu_double_derivative_IH} also holds, and by the recursion for $g_j''$
and  \eqref{eq:r^U_domination_by_R^U},
\begin{align}
	| g_{+}'' | & \leq (1 - \beta  g) |g''| + 2 \beta |g'|^2 + O(1)\norm{  ( R_{+}^U )'' }_{\cU_{+}}
	\nnb
	&
	 \leq C'_1 L^{4 (j+1)} \tilde{\vartheta}^2_{+} \tilde{g}_{+} \Big( \frac{\tilde g_{+}}{g_0} \Big)^{2 \hat{\gamma}},
\end{align}
for sufficiently
large $C_1'$.
For the bound on $\nu_{j+1}''$, we begin with
\begin{align}
	|\nu''_{+}| & \leq  (1- \hat{\gamma} \beta g) |\nu''| +  \eta |g'|
	+ \hat{\gamma} \beta |g' \nu' | +  \hat{\gamma} \beta |g'' \nu | + 2 \xi |g g''| + 2 \xi |g'|^2 + |r''_{\nu}| .
\end{align}
We already have bounds for all terms on the right-hand side.
The constant $C_0'L^{2j}$ enters in our bound on the first term on the right-hand side,
via \eqref{eq:nu_double_derivative_IH}, and this is smaller by $L^{-2}$
than our target $C_0'L^{2(j+1)}$.
This is the only occurrence of $C_0'$ in our bounds on the right-hand side.
All the other terms are either higher order in $\tilde g$
(e.g., $g'\nu'$) or are of the order desired for our upper bound but contain
a constant that can be overcome by a large choice of $C_0'$ (e.g., $r_{\nu}''$).
After bookkeeping, we arrive as desired at
\begin{align}
    |\nu''_{+}|
	& \leq
    C'_0 L^{2(j+1)} \tilde{\vartheta}^3_{+} \tilde{g}_{+} \Big(
    \frac{g_+}{g_0} \Big)^{2 \hat{\gamma}} ,
\end{align}
for sufficiently
large $C_0'$.
This completes the proof.
\end{proof}

\subsection{Proof of Proposition~\ref{prop:nu-derivs} for \texorpdfstring{$d>4$}{d>4}}
\label{sec:nu-derivs-proof2}

The proof for $d>4$ has the same structure as the proof
in Section~\ref{sec:nu-deriv-4} for $d=4$.
For $d>4$, the important product $\Pi_{k,j}$ is much simpler since now
$\beta_j$ contains the exponentially decaying factor $\scale_j=L^{-(d-4)j}$, and this
simplicity leads to the absence of logarithmic corrections.
On the other hand, we must keep track of powers of $\scale_j$ instead.

\begin{lemma}
\label{lemma:Pi_asymptotic-5}
Let $d>4$.
Suppose that $(V_j,K_j)_{j\leq k}$ is an RG flow to scale $k$ with initial condition $(g_0,\nu_0)$
and mass $\ka \ge 0$, with $g_0>0$ sufficiently small.
Then for $t \in (0, 2]$ and $i \le j < k$,
\begin{align}
	\Pi_{i,j}[t] = \prod_{m=i}^{j} (1- t \beta_m {g}_m)
    =  1 + O_t ( \vartheta_i^2 g_i\scale_i )  .
	\label{eq:Pi_asymptotic-5}
\end{align}
For a global RG flow,
the limit $\Pi_{i } [t] =\lim_{j\to\infty}\Pi_{i,j} [t]$
exists, is continuous in $(\nu_0,\ka)\in\mathcal{X}_\infty$,
and obeys $\Pi_{i } [t] = 1 + O_t (\vartheta_i g_i)$.
\end{lemma}

\begin{proof}
This follows from the facts that for $m \geq i$ we have
$g_m \le O(g_i)$ and $\beta_m \le O( \vartheta_m^2 \scale_m)$,
so $\beta_m {g}_m$ is summable and the sum is dominated by the first term.
The continuity statement follows as in the proof of Lemma~\ref{lemma:Pi_asymptotic}.
\end{proof}

Proposition~\ref{prop:RG_coordinates_derivatives-large_d}
is the main ingredient in the proof of Proposition~\ref{prop:nu-derivs} for $d>4$.
Recall the definitions of $\Sigma_j$ and $F_j$ from
\eqref{eq:SigmaFdef}, and
that
$\kaa$ is
given by \eqref{eq:kbd}.  In particular,  $\kaa=3$ for $d=5$ and $\kaa >2$ for all $d>4$.

\begin{proposition}[Case $d>4$] \label{prop:RG_coordinates_derivatives-large_d}
Suppose that $L$ is sufficiently large and $g_0$ is sufficiently small.
Suppose that $(V_j,K_j)_{j\leq k}$ is an RG flow to scale $k$ with initial condition $(g_0,\nu_0)$
and mass $\ka \ge 0$.
Then for all $j \leq k$,  $(g_j, \nu_j, K_j)$ are continuously differentiable functions of $(\nu_0, \ka)\in\mathcal{X}_k$ and satisfy the following bounds.
\begin{enumerate}
\item The $\nu_0$-derivatives satisfy
\begin{align}
\begin{split}
	& | \Sigma_j - \Sigma_{j - 1} | \leq C_0 \tilde{\vartheta}_j \tilde{g}_j^2 \scale_j^{\kaa-1} ,  \quad\;\;
	| g'_j | \leq C_1  L^{2j}  \tilde{\vartheta}_{j}^2 \tilde{g}^2_j   \scale_j^{\kaa -2} ,   \quad\;\;
	\norm{K'_j}_{\Wkappa_j} \leq C_2 L^{2j}  \tilde{\vartheta}_j^3 \tilde{g}_j^2  \scale_j^{\kaa-1}	.
	\label{eq:nujdnu0-high_d}
\end{split}
\end{align}

\item The $\ka$-derivatives satisfy
\begin{equation}
\begin{split}
	& | F_j - F_{j-1} | \leq C_0 \tilde{\vartheta}^2_j \tilde{g}^2_j \scale^{\kaa -1}_j  ,
	\quad\;\;
	| \dot g_j | \leq C_1 L^{2j} \tilde{\vartheta}_{j}^2  \tilde{g}^2_j   \scale_j^{\kaa -2} ,
	\quad\;\;
	\big\|  \dot K_j \big\|_{\Wkappa_j} \leq C_2 L^{2j} \tilde{\vartheta}_j^3 \tilde{g}_j^2  \scale_j^{\kaa-1}	.
	\label{eq:nujdka-high_d}
\end{split}
\end{equation}

\item The second derivatives satisfy
\begin{align}
	| \nu''_j | , \; | \dot{\nu}'_j |, \; | \ddot{\nu}_j | \leq O(1) L^{2j}  \tilde{\vartheta}_j^3 \tilde{g}_j \scale_j^{\kaa-2}	.
	\label{eq:nujdd-high_d}
\end{align}
\end{enumerate}
\end{proposition}

\begin{proof}[Proof of Proposition~\ref{prop:nu-derivs}(ii)]
The bounds on the second derivatives are repeated in \eqref{eq:nujdd-high_d}, and
the bounds on $g_j'$ and $\dot g_j$ are repeated in
\eqref{eq:nujdnu0-high_d} and \eqref{eq:nujdka-high_d}.  Here we only prove
that
\begin{align}
	& \nu'_j =  c_{2,N} + O(  \tilde{\vartheta}_j \tilde{g}_j \scale_j) ,
\qquad \dot \nu_j =  - c_{3,N} + O ( \tilde{\vartheta}^2_j  \tilde{g}_j \scale_j )
	\label{eq:nuderpf}
\end{align}
with $c_{2,N} (\nu_0, \ka)= 1 + O(\tilde{\vartheta}_0 g_0)$ and
positive
$c_{3,N} (\nu, \ka)= O(\tilde{\vartheta}_0 g_0)$,
obtain the negative upper bound on $\dot \nu_j$ for large $j$,
identify the limiting constants $c_{2,\infty},
c_{3,\infty}$, and show that
each $c_{i,N}$ is continuous in  $(\nu_0, \ka)\in \mathcal{X}_N$ and
each $c_{i,\infty}$ is continuous in  $(\nu_0, \ka)\in \mathcal{X}_\infty$.

It follows from \eqref{eq:nujdnu0-high_d} that
\begin{align}
	\Sigma_j = \Sigma_N + O \Big( \sum_{k = j+1}^N \tilde{\vartheta}_k \tilde{g}_k^2 \scale_{k}^{\kaa-1} \Big) =  \Sigma_N + O ( \tilde{\vartheta}_j \tilde{g}_j^2 \scale_{j}^{\kaa-1} ).
\end{align}
Since $\nu'_j = \Pi_{0,j-1} (1 + \Sigma_j)$ by definition,
it follows from Lemma~\ref{lemma:Pi_asymptotic-5} that
\begin{align}
	\nu'_j
		& = \Pi_{0,N-1}  \big(  1 + O ( \vartheta^2_j \tilde g_j \scale_j ) \big)
\big( 1 + \Sigma_N + O ( \tilde{\vartheta}_j \tilde{g}_j^2 \scale_j^{\kaa-1} ) \big)
.
\end{align}
This proves the first item of \eqref{eq:nuderpf} with
$c_{2,N}= \Pi_{0,N-1}(1 + \Sigma_N)= 1 + O(\tilde{\vartheta}_0 g_0)$,
a continuous function on $\mathcal{X}_N$.
For a global RG flow, the limit $\Sigma_{\infty} = \lim_{N\rightarrow \infty} \sum_{i=0}^{N-1} (\Sigma_{i+1} - \Sigma_i)$ is continuous in $(\nu_0, \ka)$ by the Dominated Convergence
Theorem, and $\Pi_{0,\infty}$ is continuous by Lemma~\ref{lemma:Pi_asymptotic-5}.
Therefore,
$c_{2, \infty} = \lim_{N\rightarrow \infty} c_{2, N} = \Pi_{0,\infty}(1 + \Sigma_{\infty} )$ is
a continuous function of $(\nu_0, \ka) \in \cX_{\infty}$.

The estimate on $\dot{\nu}_j$ follows similarly,
since, $\dot \nu_j = -1 + (1+F_j)\Pi_{0,j-1}$
with
$F_j = F_N + O ( \tilde{\vartheta}^2_j \tilde{g}_j^2 \scale_{j}^{\kaa-1} )$
and
\begin{align}
	\dot{\nu}_j = -1 + \big( 1 + F_N + O ( \tilde{\vartheta}^2_j \tilde{g}_j^2 \scale_j^{\kaa-1} ) \big) \Pi_{0,N-1} \big(  1 + O ( \vartheta^2_j \tilde g_j \scale_j ) \big)	.
\end{align}
This proves the second item of \eqref{eq:nuderpf} with
$-c_{3,N}  = -1 +  (1+ F_N) \Pi_{0,N-1} =  O(\tilde{\vartheta}_0 g)$, a continuous
function on $\mathcal{X}_N$.
Also, since $F_N = O(\tilde{\vartheta}^2_0 g^2)$ and $\Pi_{0,N-1} \leq 1 - \hat{\gamma} \beta_0 g_0$, we see that $c_{3,N} \geq C' \tilde{\vartheta}^2_0 g$ and hence
is strictly positive.
Thus $\dot{\nu}_j \leq - C \tilde{\vartheta}^2_0 g$  for sufficiently large $j$.
For the infinite RG flow, the continuity of $c_{3,\infty} = \lim_{N\rightarrow \infty} c_{3, N} = 1 -  (1+ F_\infty) \Pi_{0,\infty}$ in $(\nu_0, \ka) \in \cX_{\infty}$ follows because $F_{\infty}$ is continuous by the Dominated Convergence
Theorem and $\lim_{N\rightarrow \infty} \Pi_{0,N-1}$ is continuous by Lemma~\ref{lemma:Pi_asymptotic-5}.
\end{proof}

\begin{lemma} \label{lemma:derivative_bounds_first_step_high_dimension}
Let $d>4$, and let $M$ be the constant of \eqref{eq:MCCC}.
Suppose that $(V_j,K_j)_{j\leq k}$ is an RG flow to scale $k$ with initial condition $(g_0,\nu_0)$
and mass $\ka \ge 0$.  Suppose further that the bounds of
Proposition~\ref{prop:RG_coordinates_derivatives-large_d}(i,ii) hold for all $i \le j$.
Then,
for either $\partial_x = \frac{d}{d\nu_0}$ or $\frac{d}{d\ka}$,
\begin{align}
	\norm{\partial_x \Phi^K_{j+1}}_{\Wkappa_{j+1}} & \leq C_2  L^{2(j+1)} \tilde{\vartheta}_{j+1}^3 \tilde{g}_{j+1}^2 \scale_{j+1}^{\kaa-1}	,  \\
\norm{\partial_x R^U_{j+1}}_{\cU_{j+1} (\ell_{j+1}) } & \leq O(MC_2) L^{2j} \tilde{\vartheta}_{j+1}^3 \tilde{g}_{j+1}^2  \scale_{j}^{\kaa-1}.
\end{align}
\end{lemma}

\begin{proof}
For $\tilde{g}_j$ sufficiently small,  \eqref{eq:nujdnu0-high_d},
\eqref{eq:nujdka-high_d} and \eqref{eq:nuderpf} imply that
$\norm{\partial_x V_j}_{\cV_j  (\ell_j)} \leq O(L^{2j})$
(the cited bounds are implied by our hypothesis).
We give the proof only for $\frac{d}{d\ka}$,  since
$\frac{d}{d\nu_0}$ is treated similarly.
By the chain rule and  Theorems~\ref{thm:Phi^U_estimate}--\ref{thm:Phi^K_estimate},
for sufficiently large $C_2$ and $L$ we have
\begin{align}
	\Big\| \frac{d}{d\ka} \Phi^K_{j+1} \Big\|_{\Wkappa_{j+1}}
	& \leq
    \norm{ \dot
    \Phi^{K}_{j+1}}_{\Wkappa_{j+1}}
    + \norm{D_V \Phi_{j+1}^K}_{
    \Wkappa_{j+1}
    }
    \norm{\dot{V}_j}_{\cV_j (\ell_j) }
    +  \norm{D_K \Phi_{j+1}^K}_{\Wkappa_{j+1}} \norm{\dot{K}_j}_{\Wkappa_j}
    \nnb
	& \leq
    \tilde{\vartheta}_{j+1}^3 \tilde{g}_{j+1}^2 \Big( M L^{2(j+1)} \tilde{g}_{j+1} \scale_{j+1}^{\kaa}  + M\scale_{j+1}^{\kaa - 1}  O_L (L^{2j})   + O(L^{- (d-4) \kaa })
    C_2  L^{2j}  \scale^{\kaa - 1}_j  \Big)
    \nnb
	& \leq
    C_2 L^{2(j+1)} \tilde{\vartheta}_{j+1}^3 \tilde{g}_{j+1}^2 \scale_{j+1}^{\kaa-1} ,
\end{align}
where in the second line we chose the $L^{-(d-4)\kaa}$ alternative for the bound on $D_K\Phi_{j+1}^K$,
in the third line we used $L^{-(d-4)\kaa}$ to (more than) advance $\scale_j^{\kaa-1}$
to $\scale_{j+1}^{\kaa-1}$, and we used $O(L^{2j}) \le \frac 12 L^{2 (j+1 )}$ for large enough $L$.
Similarly,
\begin{align}
	\Big\| \frac{d}{d\ka} R^U_{j+1} \Big\|_{ \cU_{j+1} (\ell_{j+1}) }
    & \leq \norm{
    \dot
    R^{U}_{j+1}}_{ \cU_{j+1} (\ell_{j+1})}
    + \norm{D_K R_{j+1}^U}_{\cU_{j+1} (\ell_{j+1}) } \norm{\dot{K}_j}_{\Wkappa_j} + \norm{D_V R_{j+1}^U}_{\cU_{j+1} (\ell_{j+1}) } \norm{\dot{V}_j}_{\cV_j  (\ell_j) }
     \nnb
	&\leq
    M
    \tilde{\vartheta}_{j+1}^3 \tilde{g}_{j+1}^2  \big( L^{2(j+1)}  \tilde{g}_{j+1} \scale_{j+1}^{\kaa}
    + C_2 L^{2j}  \scale^{\kaa-1}_j
    + \tilde{g}_{j+1}\scale_{j+1}^{\kaa}  O_L (L^{2j})
    \big)
    \nnb
	&\leq
    M
    L^{2j} \tilde{\vartheta}_{j+1}^3 \tilde{g}_{j+1}^2 ( L^2\tilde{g}_{j+1} \scale_{j+1}^{\kaa}
    + 2 C_2 \scale^{\kaa-1}_j ).
\end{align}
This completes the proof.
\end{proof}

Recall the definition of $\kaa$ from \eqref{eq:kbd}.   Our choice of $\kaa$ implies that
\begin{equation}
\label{eq:advance}
    L^{2j} \scale_j^{\kaa-2}  = L^{-\eps(2d-6)} L^{2(j+1)} \scale_{j+1}^{\kaa-2}.
\end{equation}
This inequality will be important for the advancement of bounds from one scale
to the next, with its helpful factor $L^{-\eps(2d-6)}$ which is small for large $L$.

\begin{proof}[Proof of Proposition~\ref{prop:RG_coordinates_derivatives-large_d}(i)]
We wish to prove that
\begin{align}
\begin{split}
	& | \Sigma_j - \Sigma_{j - 1} | \leq C_0 \tilde{\vartheta}_j \tilde{g}_j^2 \scale_j^{\kaa-1} ,  \quad\;\;
	| g'_j | \leq C_1  L^{2j}  \tilde{\vartheta}_{j}^2 \tilde{g}^2_j   \scale_j^{\kaa -2} ,   \quad\;\;
	\norm{K'_j}_{\Wkappa_j} \leq C_2 L^{2j}  \tilde{\vartheta}_j^3 \tilde{g}_j^2  \scale_j^{\kaa-1}	.
	\label{eq:nujdnu0-pf}
\end{split}
\end{align}
We assume these bounds and proceed via induction.
This assumption is satisfied for $j=0$ since $\Sigma_1-\Sigma_0= 0$,
and also $g_0'$ and $K_0'$ are zero.
The induction for $K'_j$ is
already advanced by Lemma~\ref{lemma:derivative_bounds_first_step_high_dimension}.

For $g'$, from the recursion $g_{j+1}'=(1-2\beta_jg_j)g_j' + r_{g,j}'$,
the fact that $0 \le 1-2\beta_jg_j \le 1$, the induction
hypothesis, and \eqref{eq:advance}, we obtain
\begin{align}
    |g_{j+1}'|
    &\leq
    C_1  L^{2j}  \tilde{\vartheta}_{j}^2 \tilde{g}^2_j   \scale_j^{\kaa -2}
    +
    |r_{g,j}'|
    \nnb & \le
    \frac 12 C_1  L^{2(j+1)}  \tilde{\vartheta}_{j+1}^2 \tilde{g}^2_{j+1} \scale_{j+1}^{\kaa -2}
    +
    |r_{g,j}'|.
\end{align}
For the remainder term, we apply \eqref{eq:r^U_domination_by_R^U}
and Lemma~\ref{lemma:derivative_bounds_first_step_high_dimension}, and obtain
\begin{align}
	|r_{g,j}'|
	&\leq O(\scale_{j+1}^{-1})\norm{ (R^U_{j+1} )' }_{\cU_{j+1} (\ell_{j+1}) }
	\leq O\Big( MC_2 \frac{\scale_{j}}{\scale_{j+1}} \Big)   L^{2j} \tilde{\vartheta}_{j+1}^3 \tilde{g}_{j+1}^2  \scale_{j}^{\kaa-2}.
\end{align}
For $C_1$ large compared to $MC_2$, this permits the induction to be advanced.

Finally,  we recall \eqref{eq:Sigma_recursion_relation},
and use Lemma~\ref{lemma:Pi_asymptotic-5} to bound $\Pi_{0,j}$ to see that
\begin{align}
	|\Sigma_{j+1} - \Sigma_j|
     & \le
     O(1) \Big[ \big( \beta_j |\nu_j|  + \eta_j  + 2 \xi_j g_j \big) |g'_j| + |r'_{\nu, j}| \Big] .
\end{align}
By \eqref{eq:betadef}--\eqref{eq:xidef} and our bound on $\nu_j$, the factor multiplying
$|g'_j|$ is of order $\tilde{\vartheta}_j^2\scale_j g_j L^{-2j}\scale_j  +
\tilde{\vartheta}_j L^{-2j}\scale_j + \tilde{\vartheta}_j^3 L^{-2j}\scale_j^2 g_j$.
The  main term here is due to $\eta_j$.
Also, it follows from  Lemma~\ref{lemma:derivative_bounds_first_step_high_dimension} that
$|r_{\nu,j}'| \le L^{-2(j+1)} O(MC_2)L^{2j} \tilde{\vartheta}_{j+1}^3 \tilde{g}_{j+1}^2  \scale_{j}^{\kaa-1}$.
With the induction hypothesis for $g'_j$, this gives
\begin{align}
	| \Sigma_{j+1} - \Sigma_j |
	\leq C_0 \tilde{\vartheta}_{j+1} \tilde{g}_{j+1}^2 \scale_{j+1}^{\kaa - 1}
\end{align}
once we choose $C_0$ sufficiently large  compared to $C_1 + C_2$.
\end{proof}

\begin{proof}[Proof of Proposition~\ref{prop:RG_coordinates_derivatives-large_d}(ii)]
We wish to prove that
\begin{equation}
\begin{split}
	& | F_j - F_{j-1} | \leq C_0 \tilde{\vartheta}^2_j \tilde{g}^2_j \scale^{\kaa -1}_j  ,
	\quad\;\;
	| \dot g_j | \leq C_1 L^{2j} \tilde{\vartheta}_{j}^2  \tilde{g}^2_j   \scale_j^{\kaa -2} ,
	\quad\;\;
	\big\|  \dot K_j \big\|_{\Wkappa_j} \leq C_2 L^{2j} \tilde{\vartheta}_j^3 \tilde{g}_j^2  \scale_j^{\kaa-1}	.
	\label{eq:nujdka-pf}
\end{split}
\end{equation}
We assume these bounds and proceed via induction.
To start the induction, $\dot g_0$ and $\dot K_0$ are both equal to zero,
and, by \eqref{eq:F1-F0},
$F_1-F_0$ is bounded above by an $L$-dependent multiple of $\tilde\vartheta_1^3\tilde g_1^2
\scale_1^{\kaa-1}$ as required.
The induction for $\dot K_j$ is already advanced by Lemma~\ref{lemma:derivative_bounds_first_step_high_dimension}

Since $\dot{g}_{j+1} = (1-2\beta_j g_j) \dot{g_j} + \dot{\beta}_j g_j^2 + \dot{r}_{g,j}$
and $\dot\beta_j=O(1)L^{2j}\vartheta_j^3\scale_j$,
\begin{align}
	|\dot{g}_{j+1} |
	& \leq
    |\dot{g}_j| + O(1) L^{2j} \tilde{\vartheta}_j^3 \tilde{g}_j^2 \scale_j
    +
    \scale_{j+1}^{-1}
    O(M C_2) L^{2j} \tilde{\vartheta}_{j+1}^3 \tilde{g}_{j+1}^2 \scale_j^{\kaa - 1}
    \nnb
	& \leq C_1 L^{2(j+1)} \tilde{g}_{j+1}^2  \tilde{\vartheta}_{j+1}^2 \scale_{j+1}^{\kaa -2},
\label{eq:g_derivative_inductive_bound}
\end{align}
where we argue as in the previous proof to obtain a constant $\frac 12 C_1$
(due to change of scale) from the
inductive assumption on $\dot{g}_j$,
and we choose $C_1$ large compared to $MC_2$ to absorb the other terms.

Finally, by \eqref{eq:F_j_recursion},
\begin{align}
	|F_{j+1} - F_j | \leq
     O(1)
     \Big[
     \Big|\frac{d}{d\ka} (\beta_j g_j) \nu_j\Big|
     + |\eta_j \dot{g}_j |
     + \Big|\frac{d}{d\ka} (-\xi_j g_j^2)\Big|
     + |\dot{r}_{\nu, j}) |
     \Big]	.
\end{align}
The main term is $\eta_j \dot{g}_j$, for which $\eta_j=O(L^{-2j}\scale_j)$.
Also, by our hypothesis on $\dot{g}_j$, and since $\kaa>2$,
\begin{align}
	\big| \frac{d}{d\ka} (\beta_j g_j) \nu_j \big|
	& = | \dot{\beta}_j g_j \nu_j + \beta_j \dot{g}_j \nu_j |
	\leq O(C_1) \tilde{\vartheta}_j^2 \tilde{g}_j^2 \scale_j^2 \\
	 \big| \frac{d}{d\ka} (\xi_j g_j^2 ) \big|
    & = | \dot{\xi}_j g_j^2  + 2\xi_j g_j \dot{g}_j |
	\leq O(C_1) \tilde{\vartheta}_j^2 \tilde{g}_j^2 \scale_j^2,
\end{align}
while by \eqref{eq:r^U_domination_by_R^U} and Lemma~\ref{lemma:derivative_bounds_first_step_high_dimension},
\begin{align}
	| \dot{r}_{\nu, j} | \leq O(L^{-2(j+1)} ) \norm{\dot{R}_{j+1}^{U}}_{ \cU_{j+1} (\ell_{j+1})}
\leq O(MC_2) \tilde{\vartheta}_{j+1}^3 \tilde{g}_{j+1}^2 \scale_{j+1}^{\kaa - 1}.
\end{align}
Altogether, since $\kaa \le 3$, this proves that
\begin{align}
	|F_{j+1} - F_j |  \leq C_0 \tilde{\vartheta}_{j+1}^2 \tilde{g}_{j+1}^2 \scale_{j+1}^{\kaa-1}
,
\end{align}
once we choose $C_0$ sufficiently large  compared to $C_1 + C_2$.
\end{proof}

\begin{lemma} \label{lemma:double_derivatives_IH_high_d}
Let $d>4$.
Suppose that $(V_j,K_j)_{j\leq k}$ is an RG flow to scale $k$ with initial condition $(g_0,\nu_0)$
and mass $\ka$.  Suppose that the bounds of
Proposition~\ref{prop:RG_coordinates_derivatives-large_d} hold for all $i \le j$, and that,
in addition,
\begin{align}
	| \partial_{xy} \nu_j | \leq C'_0 L^{2j} \tilde{\vartheta}_j^3 \tilde{g}_j \scale_j^{\kaa-2},
	\quad  \; | \partial_{xy} g_j| \leq  C'_1 L^{4j} \tilde{\vartheta}_j^2 \tilde{g}_j \scale_j^{\kaa-3} ,
	\quad \;  \norm{ \partial_{xy} K_j}_{\Wkappa_j} \leq C'_2 L^{4j} \tilde{\vartheta}_{j}^3 \tilde{g}_j \scale_j^{\kaa-2},
	\label{eq:double_derivatives_IH_high_d}
\end{align}
where $\partial_{xy}$ denotes any second derivative with respect to $\nu_0$ and/or $\ka$.
Then, for sufficiently large $C'_2$,
\begin{align}
	\norm{\partial_{xy} K_{j+1} }_{\Wkappa_{j+1}} & \leq C'_2 L^{4j} \tilde{\vartheta}_{j+1}^3 \tilde{g}_{j+1} \scale_{j+1}^{\kaa - 2} , \\
	\norm{\partial_{xy} R^U_{j+1}}_{\cU_{j+1} (\ell_{j+1})} & \leq O(M C'_2) L^{4j} \tilde{\vartheta}_{j+1}^3 \tilde{g}_{j+1} \scale_{j}^{\kaa -2}
.
\end{align}
\end{lemma}

\begin{proof}
We only consider $\partial_{xy} = \partial_{\nu_0}^2$
as the other cases are similar.
By our hypothesis \eqref{eq:double_derivatives_IH_high_d}, and since $C'_1\ll C'_0$ (recall the paragraph after Remark~\ref{remark:MCCC}),
\begin{equation}
	\|V''_j \|_{\cV_j } = O(C_0')L^{4j}\tilde\vartheta_j^2\tilde g_j\rho_j^{\kaa-2}.
\end{equation}
Also, $\norm{\partial_x V_j}_{\cV_j } \leq O_L (L^{2j})$.
We apply the chain rule as in \eqref{eq:chain_rule_second_derivative}, with the induction hypotheses, to obtain
\begin{multline}
\label{eq:Kpp}
	\norm{\partial_{\nu_0}^2 \Phi^K_{+}}_{\Wkappa_{+}}
	\leq
	 O(L^{4j}\tilde{\vartheta}_{j+1}^3)
	 \Big(  M \tilde{g}_{j+1}^2 \scale_{j+1}^{\kaa-1}C'_0 \tilde{g}_j \scale^{\kaa-2}_j
	 +  M_{0,1}' L^{- (d-4)\kaa} C'_2\tilde{g}_j\scale^{\kaa -2}_j
	\\	
	+
	M \tilde{g}_{j+1} \scale_{j+1}^{\kaa-2} C_0^2
    +
    M\tilde g_{j+1}^{-1} \scale_{j+1}^{-1} C_2\tilde g^2_j \scale^{\kaa-1}_j
    +
    M\tilde{g}_{j+1}^{-9/4} \scale_{j+1}^{-\kb} (C_2 \tilde{g}^2_j \scale^{\kaa-1}_j )^2
    \Big)
    .
\end{multline}
The second term on the right-hand side contains $C_2'$, but this is multiplied by
an inverse power of $L$ from which we see that the contribution of this term
is at most half as big as what is needed to advance the induction.  For the last term,
we use
\begin{equation}
    -\kb +2(\kaa-1) -(\kaa-2)
    =
	-(\kaa -\kp)+2(\kaa-1) -(\kaa-2)
	= \kp>0
\end{equation}
to replace the product $\scale_{j+1}^{-\kb}  \scale_j^{2\kaa-2}$ by
$\scale_{j+1}^{\kaa-2}$.  The other
terms are handled by taking $C_2'$ sufficiently large.  It follows that, as required,
\begin{align}
	\norm{\partial_{\nu_0}^2 \Phi^K_{j+1}}_{\Wkappa_{j+1}}
	& \leq
    C'_2 L^{4j} \tilde{\vartheta}_{j+1}^3 \tilde{g}_{j+1} \scale_{j+1}^{\kaa - 2}.
\end{align}

Similarly,
\begin{align}
	\norm{\partial_{\nu_0}^2 R^U_{j+1}}_{\cU_{j+1} (\ell_{j+1}) }
    &
    \leq O(M) L^{4j} \tilde{\vartheta}_{j+1}^3
    \Big(
	  \tilde{g}_{j+1}^3 \scale_{j+1}^\kaa
    C'_0 \tilde{g}_j   \scale^{\kaa - 2}_j
	+ C'_2 \tilde{g}_j \scale^{\kaa -2}_j
	\nnb
	& \qquad \qquad + \tilde{g}_{j+1}^3 \scale_{j+1}^{\kaa}C_0^2
    + C_2\tilde{g}_j^2 \scale^{\kaa - 1}_j
	+ (C_2\tilde{g}^2_j  \scale^{\kaa-1}_j )^{2}
    \Big)
    \nnb
    & \leq O(M C'_2) L^{4j} \tilde{\vartheta}_{j+1}^3 \tilde{g}_{j+1} \scale_{j+1}^{\kaa -2}
    ,
\end{align}
and the proof is complete.
\end{proof}

\begin{proof}[Proof of Proposition~\ref{prop:RG_coordinates_derivatives-large_d}(iii)]
Our goal is to prove the bounds \eqref{eq:double_derivatives_IH_high_d},
which assert that
\begin{align}
	| \partial_{xy} \nu_j |
    \leq O(1) L^{2j}  \tilde{\vartheta}_j^3 \tilde{g}_j \scale_j^{\kaa-2}	.
	\label{eq:nujdd-high_d-pf}
\end{align}
To do so, we require also bounds on second derivatives of $g_j$, $K_j$, $R_j$.
We proceed by induction, with the induction hypothesis that
the bounds \eqref{eq:double_derivatives_IH_high_d} hold.
The induction holds at scale zero since all the second derivatives are zero.
By Lemma~\ref{lemma:double_derivatives_IH_high_d}, the induction hypothesis
imply the desired bound on $\partial_{xy}K_{j+1}$, as well as a bound on
$\partial_{xy} R^U_{j+1}$, so we are only
left to bound the derivatives of $g_{j+1}$ and $\nu_{j+1}$.

Differentiation of the recursion relation for $g_j$ gives
\begin{align}
	\partial_{xy}  g_{j+1}
    &= (1- 2\beta_j g_j) \partial_{xy} g_j - g_j^2 \partial_{xy} \beta_j - 2 g_j \partial_x g_j \partial_y \beta_j  - 2 g_j \partial_y g_j \partial_x \beta_j
    \nnb & \qquad
    - 2\beta_j \partial_x g_j \partial_y g_j + \partial_{xy} r_{g,j}.
\end{align}
The coefficient $\beta_j$ is independent of $\nu_0$, and we take the worst
case of $\ka$-derivatives of $\beta_j$ in the following.
The terms are bounded by
\begin{align}
    |\partial_{xy} g_j| &
    \leq C'_1 L^{4j} \tilde{\vartheta}_j^2 \tilde{g}_j \scale_j^{\kaa - 3},
    \\
	| g_j^2 \partial_{xy} \beta_j |
    & \leq \tilde{g}_j^2 O(1) L^{4j} \tilde{\vartheta}_j^4 \scale_j ,
    \\
	| g_j \partial_x g_j \partial_y \beta_j |
    & \leq \tilde g_j (C_1L^{2j}\tilde\vartheta_j^2 \tilde g_j^2\scale_j^{\kaa-2})
    (O(L^{2j})\tilde\vartheta_j^3\scale_j)
    =
    O(C_1) L^{4 j} \tilde{\vartheta}_{j}^{5} \tilde{g}_{j}^3 \scale_{j}^{\ka - 1},
    \\
	| \beta_j \partial_x g_j \partial_y g_j |
    & \leq
    (O(1)\tilde\vartheta_j^2 \scale_j)
    (C_1L^{2j}\tilde\vartheta_j^2 \tilde g_j^2\scale_j^{\kaa-2})^2
    =
     O(C_1^2) L^{4j} \tilde{\vartheta}_j^6 \tilde{g}_j^4 \scale_{j}^{2\ka -3},
     \\
	|\partial_{xy} r_{g,j} |
    & \leq \scale_{j+1}^{-1} \norm{\partial_{xy} R_{j+1}^U}_{\cU_{j+1} (\ell_{j+1})} \leq O(MC'_2) L^{4j} \tilde{\vartheta}_{j}^3 \tilde{g}_{j} \scale_{j}^{\ka - 2} \scale_{j+1}^{-1}
    ,
	\label{eq:g_doubleD_highd_final_term}
\end{align}
where we used Proposition~\ref{prop:RG_coordinates_derivatives-large_d}(i,ii) to
bound $\partial_x g_j$ and
Lemma~\ref{lemma:double_derivatives_IH_high_d} to bound $\partial_{xy} r_{g,j}$.
Thus we have the desired bound on $\partial_{xy}  g_{j+1}$ for $C'_1$
sufficiently large compared to $C'_2$.

Each term of
\begin{align}
	\partial_{xy} \nu_{j+1} = \partial_{xy} \big( (1- \hat{\gamma} \beta_j g_j) \nu_j + \eta_j g_j - \xi_j g_j^2 + r_{\nu, j} \big)
\end{align}
is bounded similarly, with the important difference that
\begin{align}
	| \partial_{xy} r_{\nu, j} |  \leq O( L^{-2(j+1)} ) \norm{\partial_{xy} R_{j+1}^U}_{\cU_{j+1} (\ell_{j+1})} \leq O(MC'_2) L^{-2} L^{2j} \tilde{\vartheta}_{j}^3 \tilde{g}_{j} \scale_{j}^{\ka - 2}
\end{align}
instead of \eqref{eq:g_doubleD_highd_final_term};
the first inequality is by \eqref{eq:r^U_domination_by_R^U} and the second by Lemma~\ref{lemma:double_derivatives_IH_high_d}.
This proves the desired bound on $\partial_{xy} \nu_{j+1}$ for $C'_0$
sufficiently large compared to $C'_1$ and $C'_2$.
\end{proof}

\section{Gaussian integrals, the regulator, and proof of Theorem~\ref{thm:Phi^U_estimate}}
\label{sec:fluctation_integral_properties}

In this section, we prove Theorem~\ref{thm:Phi^U_estimate} and prepare for the
proof of Theorem~\ref{thm:Phi^K_estimate}.
In Section~\ref{sec:extended-norm}, we define an extended norm which we
use to mechanise estimates on derivatives with respect to
the effective potential $V$,  the non-perturbative coordinate
$K$, and the mass $\ka$.
In Section~\ref{sec:mass_derivative_exp}, we use the extended norm to obtain
estimates on mass derivatives of Gaussian expectations.
In Section~\ref{sec:regulator}, we
obtain properties of the regulator $G(b,\varphi)$ of
\eqref{eq:G_j_definition}, which are needed in order
to obtain the large-field decay in our estimates for $K$.
The results of Sections~\ref{sec:mass_derivative_exp}--\ref{sec:regulator}
are of a general nature and are potentially
applicable to any RG analysis of hierarchical spin models.
In Section~\ref{sec:prelim}, we prove Theorem~\ref{thm:Phi^U_estimate}.
Finally, in Section~\ref{sec:prelim2} we obtain estimates on the building blocks of the
RG map $\Phi_+^K$.
These estimates will be assembled in Section~\ref{sec:pf2} where the proof
of Theorem~\ref{thm:Phi^K_estimate} is concluded.

{As usual, we often omit labels for the scale $j$ and write $+$ for scale $j+1$,  when only two scales are at play.

\subsection{The extended norm}
\label{sec:extended-norm}

To facilitate the control of $\Phi_+^K$ and $\Phi_+^U$ together with its
derivatives with respect to $\varphi,V,K,\ka$,
as in \cite[Chapter~10]{BBS-brief}
we use an extended version of the semi-norm employed in Section~\ref{sec:pf1}.
This extended norm is again defined by \eqref{eq:T_z_seminorm_definition}, but
with the space $\cZ$ now given in terms of a new space $\cY (\vec{\lambda})$ by $\cZ= \R^n \times \cY (\vec{\lambda})$.
The space $\cY (\vec{\lambda})$ is the subset of
$\cV \times \cF \times \R$ consisting of $y = (V, K,\ka)$ such that $\norm{y}_{\cY (\vec{\lambda}) } < \infty$,
where
\begin{align}
	\label{eq:extended_norm}
\norm{(V, K ,\ka)}_{\cY (\vec{\lambda})}
= \max \Big\{  \frac{\norm{V (b)}_{\cV (\ell)}}{\lambda_V} , \frac{\norm{K}_{\cW^\kappa}}{\lambda_K} , \frac{|\ka|}{\lambda_{\ka}}  \Big\}.
\end{align}
Here $\vec{\lambda} = (\lambda_V, \lambda_K,\lambda_{\ka} )$ is a triple of nonnegative real numbers,
and the choice $\lambda_V =0$ or $\lambda_K = 0$ or $\lambda_{\ka} = 0$ simply means the derivative in the corresponding coordinate is not considered.
For $F$ a smooth function of $\varphi,V,K,\ka$,
we use the extended seminorm $\norm{F}_{T_{\varphi, y} (\mathfrak{h}, \vec{\lambda})}$ given by \eqref{eq:T_z_seminorm_definition} with $\cZ = \R^n \times \cY (\vec{\lambda})$ and $|(\varphi, y)|_{\cZ} = \max\{ \frac{\norm{\varphi}_2 }{\mathfrak{h}},  \norm{y}_{\cY} \}$.
The norm
contains derivatives  of all orders with respect to $\varphi,V,K$, but we restrict
to an arbitrary but fixed finite number
$p_{\ka}$ of $\ka$-derivatives in \eqref{eq:T_z_seminorm_definition}.
For $y = (V, K, \ka) \in \cD \times \cF \times \II$, we
extend the definition of the norm in \eqref{eq:Wkappa-norm-def} by setting
\begin{align}
\label{eq:Gextended}
 \norm{F}_{T^G_{y} (h, \vec{\lambda})} & =  \sup_{\varphi \in \R^{n}}  \Gj (b, \varphi)^{-1} \norm{F(b)}_{T_{\varphi, y} (h, \vec{\lambda})},
\\
\label{eq:Wextended}
 \norm{ F }_{\cW_{y}^\kappa (\vec{\lambda})}
 &= \norm{F}_{T_{0, y}(\ell, \vec{\lambda})}
 + \tilde{g}^{9/4} \scale^{\kb} \norm{F}_{T^G_{y} (h, \vec{\lambda})}
 ,
\end{align}
with the exponent $\kb$ defined in \eqref{eq:kbd-bis}.

\subsection{Mass derivative of the expectation}
\label{sec:mass_derivative_exp}

In this section, we discuss how to control the mass derivative of Gaussian expectations whose covariance depends on $\ka$.
We have already observed around \eqref{eq:massnorm} that derivatives with respect
to $\ka$ lead to factors of order $L^{2j}$, and we have compensated for this
by equipping the mass interval $\II_j$ with norm $|x| L^d L^{2j}$.
This is equivalent to taking
\begin{align}
	\lambda_{\ka, j} = k_{\ka, j}^{-1}, \qquad k_{\ka, j} = L^{d} L^{2j}
\label{eq:k_j_definition}
\end{align}
in \eqref{eq:extended_norm},  and we often drop the scale $j$ from the notation as usual.
Throughout this section, when we are concerned with the expectation
$\E_+=\E_{j+1}$ with covariance $C_+=C_{j+1}=\gamma_+P_+$
(recall \eqref{eq:Cjdef}), we always assume that
\begin{equation}
\label{eq:mass-8}
    \ka \ge - \frac 12 L^{-2j}
\end{equation}
so that $\gamma_+ \le \frac 12 L^{2j}$.
We assume \eqref{eq:mass-8} explicitly in the statement of
Lemma~\ref{lemma:mass_derivative_of_expectation} for emphasis,
but afterwards we assume it tacitly.

In the following, as in Definition~\ref{def:thetaF}
we write $\theta F (\varphi) = F(\varphi + \zeta)$, where $\varphi \in \R^n$ and
 $\zeta$ is a $(\R^n)^{\cB}$-valued
 Gaussian random variable with covariance $C_+$.
The next lemma indicates how the extended norm, which in particular
includes derivatives with respect to the mass in the covariance in $\E_+$, can
be moved under the expectation in an upper bound.

\begin{lemma}
\label{lemma:mass_derivative_of_expectation}
Let $B\in \blocksplus$ and let
$F^{B} = \prod_{b\in \blocks (B)} F(b)$ be a product of scale-$j$ polymer activities
which may depend on  $(V,K)\in \cV \times \cF$ but not on
$\ka$.
There is a constant $C$ (depending on $p_a$ and $n$ but not on $L$ or $\ka$) such that,
with $\lambda_{\ka} = k_{\ka}^{-1}$, and
for all $\mathfrak{h}>0$ and all $\ka \geq -\frac{1}{2} L^{-2 j}$,
	\begin{align}
		\norm{\E_{+} \theta F^{B}} _{T_{\varphi,y} (\mathfrak{h}, \vec{\lambda})}
		\leq C \,
        \Big[
        \E_{+}
        \textstyle{\prod_{b\in \blocks (B)} }\norm{F(b)}_{T_{\varphi+\zeta_b,y} (\mathfrak{h}, \vec{\lambda})}^2  \Big]^{1/2}
        .
        \label{eq:mass_derivative_of_expectation_new}
	\end{align}

\end{lemma}

\begin{proof}
Let $\vec \lambda_0 = (\lambda_V,\lambda_K,0)$.
To simplify the notation, we write $\tilde F = \prod_{b\in \blocks (B) }\norm{F(b)}_{T_{\varphi+\zeta_b,y} (\mathfrak{h}, \vec{\lambda_0})}$.
	By definition of the $T_{\varphi,y}(h, \vec \lambda)$-seminorm, and by pushing the
norm inside the expectation (justified by \cite[(7.3.3)]{BBS-brief}), we have
	\begin{equation}
		\label{eq:mass_deriv_bound_starting_point_proof_new}
		\norm{ \E_+  \theta F^{B}  }_{T_{\varphi,  y} (\mathfrak{h}, \vec \lambda) }
		= \sum_{r=0}^{p_{\ka}} \frac{1}{r !}
		\big\| D_{\ka}^r \E_+  \theta F^{B} \big\|_{T_{\varphi, y} (\mathfrak{h}, \vec \lambda_0)}
		\leq \sum_{r=0}^{p_{\ka}} \frac{1}{r!}
		\frac{1}{k_a^r} \frac{d^r}{da^r}
		\E_+ \tilde F.
	\end{equation}
To proceed, we require good control of mass derivatives of
$\E_+ \tilde F$.  The proof does not use the specific form of $\tilde F$ until
the final step in \eqref{eq:mass_deriv_theta_F_bound_intermed-v2_new}, and until
then $\tilde F$ can be an arbitrary function of $\zeta$ (with dependence on $\varphi,V,K$ permitted).

	The derivative with respect to $\ka$ of the Gaussian expectation $\E_+$
acts on $\exp(-\frac12(\zeta,C_+^{-1}\zeta))$.
    Recall the covariance decomposition \eqref{eq:PBCdecomp}.
    By \eqref{eq:gamma_j_per},  $C_+^{-1}=\gamma_+^{-1}P_{+}$ depends on $\ka$ only through
	the prefactor $\gamma_+^{-1} = L^{-2j}+\ka \geq \frac{1}{2} L^{-2j}$.  It follows that
	\begin{align}
		\frac{d}{da}\E_+ \tilde F
		&=- \E_+[X_+ \tilde F ],
	\end{align}
	where
	\begin{equation}
		X_+(\ka)
		= Y_+ - \E_+ Y_+
    \qquad \text{with} \qquad Y_+
    = \frac12(\zeta,
    P_+
    \zeta).
	\end{equation}
    The derivative of $X_+$ is
	\begin{align}
			\frac{d}{da}X_+(\ka)
			&=-\E_+X_+Y_+.
	\end{align}
    It can then be seen from induction that there are constants $c_p$ such that
		\begin{align}
    \label{eq:EYprod}
			\frac{d^r}{d\ka^r} \E_+ \tilde F
			&=   \sum_{p} c_{p} \E_+ [Y_+^{p_1} \tilde F ] \prod_{i=2}^{m_p} \E_+ [ Y_+^{p_i} ],
		\end{align}
    where the sum ranges over $p = (p_i)_{i=1}^{m_p}$ with nonnegative $p_i$,
    for some $m_p \geq 1$, such that $\sum_{i=1}^{m_p} p_i = r$.

    By definition of $Y_+$, and with the integrals over $(\R^n)^{L^d-1}$,
    the expectation in the product over $i$ in \eqref{eq:EYprod} is given by
    \begin{equation}
        \gamma_+^{-p_i} \E_+\big[ Y_+^{p_i} \big]
        =
        (-1)^{p_i} \frac{d^{p_i}}{ds^{p_i}}\Big|_{s=1}
        \frac{  \int e^{-\frac{1}{2} s (\zeta, C_{+}^{-1} \zeta)  }d\zeta }
        {\int e^{-\frac{1}{2} (\zeta, C_{+}^{-1} \zeta)  }d\zeta}
        = (-1)^{p_i} \frac{d^{p_i}}{d s^{p_i}}
        \Big|_{s =1} s^{-\frac{1}{2} n (L^d - 1)}.
    \end{equation}
    The right-hand side is bounded above by $(\frac12 n L^d+ p_i )^{p_i}$, so
    the product over $i$ in \eqref{eq:EYprod} is bounded above by $O_{n,r}(L^d\gamma_+)^{r-p_1}$.
    Therefore,
		\begin{equation}
			\label{eq:mass_deriv_theta_F_bound-v2_new}
			\Big|\frac{d^r}{da^r}\E_+ \tilde F \Big| \leq
			O_{r,n}(1) (L^{d} \gamma_+)^r
    \max_{0 \le p_1 \le r} (L^{d}\gamma_+)^{-p_1} \E_+ \big[ |Y_+|^{p_1} \tilde F \big]   \qquad (r \geq 0).
		\end{equation}
    Since $k_{\ka}^{-r}(L^d \gamma_+)^r \le (2 k_{\ka}^{-1}L^d L^{2j})^r= 2^r$, we can now bound the
    right-hand side of
		\eqref{eq:mass_deriv_bound_starting_point_proof_new} to obtain
		\begin{equation}
			\label{eq:mass_deriv_theta_F_bound_intermed-v2_new}
			\norm{ \E_+ \theta F }_{T_{\varphi,  y} (h, \vec \lambda) }
			\leq
			O_{r,n} (1)\max_{0 \leq p_1 \leq p_{\ka}} ( L^{d}\gamma_+)^{- p_1}
    \E_+ [ |Y_+|^{p_1} \tilde F ]  	.
		\end{equation}
    By the Cauchy--Schwarz inequality and the bound obtained previously for the expectation
    of powers of $Y_+$, the quantity under the maximum is bounded above by
    an $L$-independent multiple of
    $\big( \E_{+} \tilde{F}^2 \big)^{1/2}$.
    Since $F$ has been assumed to be independent of $\ka$, the
    norm with $\vec{\lambda}_0$ in $\tilde F$ is equal to the norm with
     $\vec{\lambda}$, so the proof is complete.
\end{proof}

We need good estimates on polynomials.
The following lemma shows that the $T_{\varphi,y}$-norm of an $\ka$-independent polynomial
can be bounded in terms of the $T_{0,y}$-norm.
The lemma combines \cite[Exercise 7.5.2]{BBS-brief} and \cite[Lemma 7.2.2]{BBS-brief}
into one.
For its statement, given $\mathfrak{h}>0$ we define
\begin{equation}
	\label{eq:P_h_phi_def}
	P_{\mathfrak{h}}(\varphi) = 1 + |\varphi| / \mathfrak{h}.
\end{equation}

\begin{lemma}
	\label{lemma:T_z_deriv_and_phi_to_0}
With $z=(\varphi,y)$, suppose that $F(z)$ is an $\ka$-independent polynomial in $\varphi$ of degree $k$,
with coefficients that are functions of $y$. Then
	\begin{equation}
		\label{eq:T_phi_to_0}
		\|F\|_{T_{\varphi,y} (\mathfrak{h}, \vec{\lambda})}
        \leq \|F\|_{T_{0,y}(\mathfrak{h}, \vec{\lambda}) }
        P_{\mathfrak{h}}^k(\varphi).
	\end{equation}
Also, for $r\leq k$, the norm of the derivative of $F$ obeys
	\begin{equation}
		\label{eq:T_deriv}
		\|D_\varphi^r F\|_{\R^n \to T_{\varphi,y} (\mathfrak{h}, \vec{\lambda}) }
    \leq 2^k \frac{r!}{\mathfrak{h}^r}\|F\|_{T_{\varphi,y}(\mathfrak{h}, \vec{\lambda}) }.
	\end{equation}
\end{lemma}

The next lemma is analogous to what appears in
\cite[Section 7.6]{BBS-brief}, but it needs adaptation since our norm includes
the mass derivative.
We exclude $p=0$ in its statement because
\eqref{eq:E_on_theta_V1-V1_new}--\eqref{eq:cov_on_theta_V1-V2_new}
hold trivially in this case.  We write
\begin{equation}
\label{eq:cplusdef}
    \mathfrak{c}_+^2 = C_{+;00}
    = \frac{L^{-(d-2)j}}{1+ \ka L^{2j}} (1-L^{-d})
\end{equation}
for the diagonal element of the covariance $C_+$.

\begin{lemma}
	\label{lemma:polynomial_expectation_mass_derivative}
	Let $V_1 ,V_2$ be even $\ka$-independent polynomials on $B \in \cB_+$,
of degrees at most $2p$ for some $p \ge 1$.
Let $\mathfrak{h} \geq \mathfrak{c}_+$.
There is an $L$-independent constant $C$ such that
	\begin{align}
\label{eq:E_on_theta_V1-V1_new}
		\norm{\E_+ (\theta V_1 - V_1 ) (B)}_{T_{\varphi,y} (\mathfrak{h}, \vec{\lambda})}
		& \leq C \Big( \frac{\mathfrak{c}_+}{\mathfrak{h}} \Big)^2 \norm{V_1 (B)}_{T_{0,y} (\mathfrak{h}, \vec{\lambda})}P_{\mathfrak{h}}^{2p-2}(\varphi)
		, \\
\label{eq:cov_on_theta_V1-V2_new}
		\norm{\cov_+ [ \theta V_1 (B) , \theta V_2 (B) ]}_{T_{\varphi,y} (\mathfrak{h}, \vec{\lambda})}
		& \leq C \Big( \frac{\mathfrak{c}_+}{\mathfrak{h}} \Big)^4 \norm{V_1 (B)}_{T_{0,y} (\mathfrak{h}, \vec{\lambda})} \norm{V_2 (B)}_{T_{0,y} (\mathfrak{h}, \vec{\lambda})}
		P_{\mathfrak{h}}^{2p}(\varphi)
		.
	\end{align}
\end{lemma}

\begin{proof}
For the norm which does not measure $\ka$-derivatives,
\eqref{eq:E_on_theta_V1-V1_new} is proved in \cite[Exercise~7.6.2]{BBS-brief},
and we adapt that proof to our norm which does measure $\ka$-derivatives.
We use the notation $\alpha = (\alpha_1, \cdots, \alpha_n)$ for a multi-index,
and we say that $\alpha$ is even if each $\alpha_i$ is even.
By Taylor's theorem,
\begin{equation}
	\E_+(\theta V_1 - V_1)(B) = \sum_{x \in B}\sum_{2\leq |\alpha| \leq 2p} \frac{1}{\alpha!} V_{1,x}^{(\alpha)}(\varphi) \E_{+}\zeta_x^{\alpha},
\end{equation}
with the sum restricted to even multi-indices $\alpha$ since otherwise
$\E_{+}\zeta_x^{\alpha}=0$.
We use Lemma~\ref{lemma:mass_derivative_of_expectation}
(with $F = \zeta_x^{\alpha}$
and $\varphi=0$, so $F$ is independent of $\varphi,K,V,a$) to move the norm inside
the expectation, apply the triangle inequality and the product property of the $T_{\varphi,y}$-seminorm, and
finally use Lemma~\ref{lemma:T_z_deriv_and_phi_to_0} to conclude that there is an
$L$-independent
constant $C$ such that
\begin{align}
		\|\E_+(\theta V_1 - V_1)(B)\|_{T_{\varphi,y}(\mathfrak{h},\vec \lambda)}
		&\leq C
		 \sum_{x \in B}\sum_{2\leq |\alpha| \leq 2p} \mathfrak{h}^{-|\alpha|}
    \|V_{1,x}\|_{T_{0,y}(\mathfrak{h},\vec \lambda)} P_{\mathfrak{h}}^{2p-|\alpha|}(\varphi)
    (\E_{+}\|\zeta_x^{\alpha}\|^2_{T_{\varphi,y}(\mathfrak{h},\vec \lambda)})^{1/2}
    .
\end{align}
The norm $\|\zeta_x^{\alpha}\|^2_{T_{\varphi,y}(\mathfrak{h},\vec \lambda)}$ is simply
$\zeta_x^{2\alpha}$ for even $\alpha$, and
$\E_{+}\zeta_x^{2\alpha}$
is bounded by a constant times $\mathfrak{c}_+^{2|\alpha|}$ by
\cite[Exercise 2.1.7]{BBS-brief}.
With the worst power of $P_{\mathfrak{h}}$, this gives
\begin{align}
	\|\E_+(\theta V_1 - V_1)(B)\|_{T_{\varphi,y}(\mathfrak{h},\vec \lambda)}
	&\leq C\,P_{\mathfrak{h}}^{2p-2}(\varphi)\|V_{1}(B)\|_{T_{0,y}(\mathfrak{h},\vec \lambda)}
	\sum_{2\leq |\alpha| \leq 2p} \Big(\frac{\mathfrak{c}_+}{\mathfrak{h}}\Big)^{|\alpha|}
    ,
\end{align}
and then by taking the worst power of $\mathfrak{c}_+/\mathfrak{h}$
(which is at most $1$) gives
\eqref{eq:E_on_theta_V1-V1_new}.

The proof of \eqref{eq:cov_on_theta_V1-V2_new} is
a similar adaptation of the proof of \cite[Lemma~7.6.1]{BBS-brief}, and we omit
the details.
\end{proof}

\subsection{Estimates for the regulator}
\label{sec:regulator}

Decay of the nonperturbative coordinate $K_j$ for large fields is controlled
by the regulator $G_j(\varphi)=e^{-\kappa|\varphi/h_j|^4}$ defined in
\eqref{eq:G_j_definition}, where
$h_j  = k_0 \tilde{g}_{j}^{-1/4} L^{ - dj/4}$ as in \eqref{eq:h_j_definition}.
As has been emphasised by Brydges,
a good regulator should have a ``supermartingale'' property
(see, e.g., \cite[Section~6.5]{Bryd09}) which permits the propagation of
estimates as the scale is advanced.  In our context, the supermartingale property
is the inequality \eqref{eq:supermartingale} in the next lemma.
We choose to implement quartic exponential decay in the regulator since this
is easy to do for the hierarchical model.
Quadratic decay would also be sufficient, and may be best possible in
a Euclidean setting.

\begin{lemma}
\label{lemma:E_G_j}
Fix any $\bar{\kappa} \geq 0$.
For any $\varphi \in \R^{n}$ and $B\in \blocksplus$,
\begin{align}
	\label{eq:regul_a_s}
	\prod_{b\in \blocks (B)} e^{- \bar{\kappa}  |\varphi + \zeta_b|^4/h^4}
	\leq e^{- \bar{\kappa}  |\varphi/h_+|^4}
	\quad {\rm a.s.}\; [\E_+].
\end{align}
In particular, for any $t \ge 0$ and with $\bar\kappa = t\kappa$, this gives
\begin{align}
\label{eq:supermartingale}
	\Eplus \prod_{b\in \blocks (B)} \Gj^{t} (b, \varphi+\zeta_b)
    \leq \Gplus^{t} (B, \varphi)
	.
\end{align}
\end{lemma}

\begin{proof}
It follows by expansion of
$|\varphi+\zeta_b|^4 = [|\varphi|^2 +(2\varphi \cdot \zeta_b + |\zeta_b|^2) ]^2$
that
\begin{equation}
    |\varphi + \zeta_b|^4 \ge |\varphi|^4 + 4 |\varphi|^2 \varphi \cdot \zeta_b.
\end{equation}
Since
$\sum_{b\in \blocks (B)} \zeta_b = 0$ with probability 1 by \eqref{eq:zeta-zero-sum},
this implies that
\begin{align}
	\prod_{b\in \blocks (B)} e^{- \bar{\kappa}  |\varphi + \zeta_b|^4/h^4}
	\leq e^{- L^d \bar{\kappa}  |\varphi/h|^4}
	\quad \text{a.s.}
\end{align}
Now we use the definition $h_j= k_0 \tilde{g}_{j}^{-1/4} L^{ - dj/4}$,
together with the fact that $\tilde{g}_+\le \tilde{g}$, to see that
\begin{equation}
\label{eq:hhplus}
    L^{d} h^{-4} = h_+^{-4} \tilde{g}/\tilde{g}_+ \ge h_+^{-4},
\end{equation}
which proves \eqref{eq:regul_a_s}.
The supermartingale property \eqref{eq:supermartingale} then follows by taking
the expectation.
\end{proof}

For $\mathfrak{h}>0$, we recall the definition
\begin{equation}
    P_{\mathfrak{h}}(\varphi) = 1 + |\varphi| / \mathfrak{h}
\end{equation}
from \eqref{eq:P_h_phi_def}.
One usage of the regulator is to control polynomial factors of the form
$P_\mathfrak{h}^k(\varphi)$.  This is done via the observation
that for any $\mathfrak{h},k,\bar\kappa >0$
and uniformly in $\varphi \in \R^n$,
\begin{equation}
\label{eq:PG}
    P_{\mathfrak{h}}^{k} ( \varphi )
    e^{- \bar{\kappa}  |\varphi/\mathfrak{h}|^4}
	\leq O_{\bar\kappa,k} (1)
	,
\end{equation}
since $1+t$ is bounded above by a multiple of $e^{k^{-1}\bar\kappa t^4}$ for $t \ge 0$.
A related domination of polynomial factors is provided by the next lemma.

\begin{lemma}
\label{lem:G2G}
 Let $B\in \cB_+$, $b\in \cB( B)$, $p \ge 1$, and $t >0$.
 If $\tilde{g}$ is sufficiently small (depending on $L$) then there exists $c_{p,t, \kappa} > 0$
 such that
 \begin{align}
    \Eplus
    \Big(
    G^t (b,\varphi+\zeta_b) P^{p}_{h_+} (\varphi+\zeta_b)
    \prod_{b' \in \cB(B), \,  b' \neq b} G^{2t} (b',\varphi+\zeta_{b'}) \Big)
    \le  c_{p,t,\kappa} \,  G_+^{t}(B,\varphi)
    .
    \label{eq:G2G}
\end{align}
\end{lemma}

\begin{proof}
We write $c$ for a $(p,t, \kappa)$-dependent constant whose value may change from line to line.  Let $k$ be the smallest integer such that $4k \geq p$.
We first observe that $P_{h_+}^p (\varphi + \zeta_b) \leq c ( 1 + |\varphi / h_+|^p + |\zeta_b / h_+|^p )$.
We use the fact that $h_+^{-4} \leq 2 L^{d} h^{-4}$
together with the fact that $\varphi$ is constant in $B$
 (in the second inequality of \eqref{eq:G2G-pf})
 to obtain
\begin{align}
	1+ \Big| \frac{\varphi }{h_+} \Big|^{p}
    & \leq
    c \Big( 1 + \Big| \frac{\varphi}{h_+} \Big|^4 \Big)^k
	\leq c\Big( 1+  \sum_{b' \neq b} \Big| \frac{\varphi}{h} \Big|^4 \Big)^k
	\nnb
	& \leq
	c\Big( 1 +\sum_{b' \neq b} \Big| \frac{\varphi + \zeta_{b'}}{h}  \Big|^4
    + \sum_{b' \neq b} \Big| \frac{ \zeta_{b'}}{h}  \Big|^4  \Big)^k
	\nnb
	& \leq
	c \Big( e^{\kappa t \sum_{b' \neq b} | (\varphi + \zeta_{b'} ) /h |^4}
    +  \sum_{b' \neq b} \Big| \frac{ \zeta_{b'}}{h}  \Big|^{4k} \Big).
\label{eq:G2G-pf}
\end{align}
From this, we see that
\begin{align}
	P_{h_{+}}^p (\varphi + \zeta_b)
	\leq c \Big( \prod_{b' \neq b} G (b', \varphi + \zeta_{b'})^{-t}
    + \sum_{b' \neq b} \Big| \frac{\zeta_{b'}}{h} \Big|^{4k}
    + \Big| \frac{\zeta_b}{h_+} \Big|^p
     \Big).
\label{eq:P_h_new_bound_in_v2}
\end{align}

When multiplied by the regulators on the left-hand side of \eqref{eq:G2G},
the $G(b')^{-t}$ cancels with $G(b')^{2t}$ to produce $G(b')^t$, and we can then apply
Lemma~\ref{lemma:E_G_j} to get the desired upper bound for this term.
In preparation for the other two terms on the right-hand side of
\eqref{eq:P_h_new_bound_in_v2},
let $b'' \in \cB (B)$, let $q$ be a nonnegative integer, and let $\mathfrak{h} > 0$.
By the Cauchy--Schwarz inequality,
Lemma~\ref{lemma:E_G_j}, and \cite[Lemma~10.3.1]{BBS-brief}
(to bound the last factor in the second line),
\begin{align}
	& \Eplus \Big[ \Big(  \prod_{b' \neq b} G(b',\varphi+\zeta_{b'})^{2t} \Big)
    G(b,\varphi+\zeta_b)^t \Big| \frac{\zeta_{b''}}{\mathfrak{h}} \Big|^{q}  \Big]
    \nnb
    & \qquad \leq
	\Big( \frac{\ell_+}{\mathfrak{h}} \Big)^{q}	
	\Big[ \Eplus  \Big(  \prod_{b' \neq b} G(b',\varphi+\zeta_{b'})^{4t} \Big)
    G(b,\varphi+\zeta_b)^{2t}  \Big]^{1/2}
    \Big[ \Eplus \Big( \frac{\zeta_{b''}}{\ell_+} \Big)^{2q}	  \Big]^{1/2}
    \nnb
    & \qquad \leq
    c \Big( \frac{\ell_+}{\mathfrak{h}} \Big)^{q} G_+^t (B, \varphi) .
\end{align}
For the term $|\zeta_b/h_+ |^p$ (so $q=p$ and $\mathfrak{h} = h_+$), this gives the desired bound immediately since $\ell_+ \leq h_+$ for sufficiently small $\tilde{g}$
(since $d \ge 4$).
Finally,  for the term $\sum_{b' \neq b} | \frac{\zeta_{b'}}{h} |^{4k}$
(so $q=4k$ and $\mathfrak{h} = h$), there are $L^d-1$
terms but we have $h$ not $h_+$, so  for sufficiently small $\tilde{g}$
 we gain $(\ell_+ / h)^{4k} \leq  L^{-2 (d-2) k}
L^{-(d-4)jk}\leq  L^{-2 (d-2) }$.
The product of this beneficial factor with the dangerous factor $L^{d} -1$
is bounded for $d \ge 4$, and the proof is complete.
\end{proof}

The next lemma gathers and extends estimates
from \cite{BBS-brief} on how norms interact with the expectation.
It uses a general inequality
from \cite[(10.4.5)]{BBS-brief},
namely
\begin{align}
\label{eq:KKK}
	\norm{F(b)}_{T_{\varphi,y} (\ell_+, \vec{\lambda})}
    \leq P_{\ell_+}^{10} (\varphi) \norm{F(b)}_{\cW_{y,+}^0 (\vec{\lambda})}
	,
\end{align}
which permits the $T_{\varphi,y} (\ell_+, \vec{\lambda})$ norm to be bounded in terms of
the $T_{0,y} (\ell_+, \vec{\lambda})$ and $T_{\varphi,y} (h_+, \vec{\lambda})$ norms.

As usual, the fluctuation field $\zeta \sim \cN(0, C_+)$
is constant on blocks $b\in \cB (B)$ and $\varphi$ is constant on $B$,
and we assume $\ka \ge -\frac{1}{2} L^{-2j}$ as in \eqref{eq:mass-8}.
Note that on the right-hand side of \eqref{eq:EFB4} we have the scale-$j$ regulator $G$
which is defined with $h$, but that the $T_\varphi$ norm used to define
the $T^G_{y} (h_+, \vec{\lambda})$ norm instead uses
$h_+$.

\begin{lemma} \label{lemma:E_F^B}
Let $B\in \blocksplus$, let $F$ be a polymer activity at scale $j$, and let $F^{B} = \prod_{b\in \blocks (B)} F(b)$.
There is an $L$-independent constant $C$ and an $L$-dependent constant
$C_L$ such that
for any $\lambda_V, \lambda_K \ge 0$,
\begin{align}
\label{eq:EFB3}
	& \norm{\Eplus \theta F^B}_{T_{0, y} (\ellplus , \vec{\lambda})} \leq
C_L \textstyle{\prod_{b\in \cB (B)}}
\norm{F(b)}_{\cW^{0}_{y , +}
(\vec{\lambda})} ,
\\
\label{eq:EFB4}
	& \norm{ \Eplus \theta F^B }_{T_{\varphi,y} (h_{+}, \vec{\lambda})} \leq
	C \Gplus (B, \varphi) \textstyle{\prod_{b\in \cB (B)}} \norm{F(b)}_{T^G_{y} (h_+, \vec{\lambda})}
	.
\end{align}
\end{lemma}

\begin{proof}
For \eqref{eq:EFB3},
we combine
\eqref{eq:KKK}
with \eqref{eq:mass_derivative_of_expectation_new}. With $\mathfrak{h}=\ell_+$ and $\varphi=0$,
this gives
 \begin{align}
 		\norm{ \Eplus \theta F^B }_{T_{0,y} (\ell_+, \vec{\lambda})}
 		&\leq
 		C  \prod_{b\in \cB(B)} \norm{F(b)}_{\cW_{y,+}^0 (\vec{\lambda})}
        \Big( \Eplus\Big[\prod_{b\in \cB(B) }P_{\ell_+}^{20}(\zeta_b) \Big] \Big)^{1/2} .
 \end{align}
 The expectation is bounded by an $L$-dependent constant, as in \cite[(10.4.10)]{BBS-brief},
 and the proof of \eqref{eq:EFB3} is complete.
Finally, for \eqref{eq:EFB4}, we use \eqref{eq:mass_derivative_of_expectation_new} and
\eqref{eq:supermartingale}
to obtain
\begin{align}
	\norm{\Eplus \theta F^B}_{T_{\varphi,y} (h_+, \vec{\lambda})} & \leq C \Big( \Eplus \Big[ \prod_{b\in \cB (B)} \norm{F(b)}_{T_{y}^G (h_+, \vec{\lambda}) }^2 G^2 (b, \varphi + \zeta_b) \Big] \Big)^{1/2} \nnb
	& \leq C G_+ (B, \varphi) \prod_{b\in \cB(B) } \norm{F(b)}_{T_y^G (h_+, \vec{\lambda})}.
\end{align}
This completes the proof.
\end{proof}

\subsection{The perturbative RG map: proof of Theorem~\ref{thm:Phi^U_estimate}}
\label{sec:prelim}

We now prove Theorem~\ref{thm:Phi^U_estimate}, which contains the main
estimates for the perturbative RG map.
According to its definition in \eqref{eq:remaining_RG_coords_definition},
the remainder for the flow of coupling constants is given by
$R^U_{+} (V, K) = \Phi_{+}^{U} (V, K) - \Phi_{\pt} (V)$.
After some algebra (see \cite[(9.1.5)]{BBS-brief}), this can be written as
\begin{align}
	R_+^U (B) = - \E_+ \theta Q(B) + \cov_+ \Big[ \theta ( V(B) - \frac{1}{2} Q(B) ) , \theta Q (B)  \Big]
	\label{eq:R_+^U_expansion}
\end{align}
where
\begin{align}
\label{eq:QVK}
	Q(b) = \operatorname{Loc} ( e^{V} K )(b), \qquad  Q(B) = \sum_{b\in \cB (B)} Q(b).
\end{align}

In Lemma~\ref{lemma:Q_bound}, we bound $Q$, which does not depend on $\ka$.
The $\ka$-dependence of $R_+^U$ arises after application of $\E_+$ or $\cov_+$,
and is handled by our extended norm in Lemma~\ref{lemma:R_+_mass_derivative_bound}.

\begin{lemma}
\label{lemma:Q_bound}
For $(V,K) \in \domRG$,
\begin{align}
\label{eq:Q0}
	\norm{Q(b)}_{T_{0,y} (\ell,\vec{\lambda})}
    \leq e^{\norm{V(b)}_{T_0 (\ell)} + \lambda_V} \big( \norm{K}_{T_0 (\ell)} + \lambda_K \big).
\end{align}
If $\mathfrak{h} \geq \ell$,  $\lambda_V \leq 1$, and $\lambda_K \leq \tilde{g} \scale$, then for an $L$-independent constant $C$,
\begin{align}
\label{eq:Qh}
	\norm{Q(b)}_{T_{0,y} (\mathfrak{h},\vec{\lambda})}
    \leq C \Big( \frac{\mathfrak{h}}{h} \Big)^4
    .
\end{align}
\end{lemma}

\begin{proof}
The first bound is \cite[(9.3.20)]{BBS-brief}.
For \eqref{eq:Qh}, we first apply \cite[(9.3.24)]{BBS-brief} and then use
\eqref{eq:Q0} to obtain
\begin{align}
	\norm{Q(b)}_{T_{0,y} (\mathfrak{h},\vec{\lambda})}
	\leq  \Big( \frac{\mathfrak{h}}{\ell} \Big)^4 \norm{Q(b)}_{T_{0,y} (\ell, \vec{\lambda})}
	\leq  \Big( \frac{\mathfrak{h}}{\ell} \Big)^4 e^{\norm{V(b)}_{T_0 (\ell)} + \lambda_V} \big( \norm{K}_{T_0 (\ell)} + \lambda_K \big).
\end{align}
The assumptions on $V$ and $\lambda_V$ imply that
$e^{\norm{V(b)}_{T_0 (\ell)} + \lambda_V} \leq O(1)$.
Also, by the assumptions on $K$ and $\lambda_K$, and since $\kaa > 2$,
\begin{align}
	\norm{K}_{T_0 (\ell)} + \lambda_K
	&\leq
	C_{\rm RG} \tilde g^3\scale^{\kaa} + \tilde g \scale
	\leq  2 \tilde g \scale
\end{align}
for $\tilde g$ small enough compared to $C_{\rg}$ (which for now is an arbitrary
but fixed $L$-dependent constant, to be specified in \eqref{eq:C_RG_def}).
Since $\tilde g \scale \ell^{-4} \le O(1) h^{-4}$,
this completes the proof.
\end{proof}

\begin{lemma}
\label{lemma:R_+_mass_derivative_bound}
Let $(V, K) \in \domRG$,  $r_1 = \norm{V (b)}_{T_0 (\ell)} + \lambda_V$,
and $r_2 = \norm{K (b)}_{T_0 (\ell)} + \lambda_K$ with $\lambda_V \leq 1$.
There is an $L$-dependent $O_L (1)$ such that,
for $\ka \geq -\frac 12 L^{-2j}$ and $\lambda_{\ka} = k_{\ka}^{-1}$,
\begin{align}
	\norm{R_+^U (B)}_{T_{0,y} (\ell_+, \vec{\lambda})} = O_L (1) (1 + r_1 + r_2) r_2 .
\end{align}
\end{lemma}

\begin{proof}
We bound the two terms on the right-hand side of \eqref{eq:R_+^U_expansion} separately.
For the first term, we use
Lemma~\ref{lemma:Q_bound}, the assumptions on $V$ and $\lambda_V$,
and $\ell_+ \leq \ell$ (with monotonicity of the norm in $\ell$), to obtain
\begin{align}
	\norm{Q(B)}_{T_{0,y} (\ell_+, \vec{\lambda})} \leq O_L (1) r_2
    .
\end{align}
Therefore, by Lemma~\ref{lemma:polynomial_expectation_mass_derivative},
\begin{align}
	\norm{\E_+ \theta Q (B)}_{T_{0,y} (\ell_+, \vec{\lambda})}
    & \le \norm{Q (B) }_{T_{0,y} (\ell_+ , \vec{\lambda})}
    +
    \norm{\E_+ (\theta Q -Q) (B)}_{T_{0,y} (\ell_+, \vec{\lambda})}
    \nnb&
    \leq O (1) \norm{Q (B) }_{T_{0,y} (\ell_+ , \vec{\lambda})}
    \le O_L(1)r_2.
	\label{eq:E_Q_bound}
\end{align}
Also, again by Lemma~\ref{lemma:polynomial_expectation_mass_derivative},
\begin{align}
	\Big\| \cov_+ \big[ \theta ( V(B) - \frac{1}{2} Q(B) ) , \theta Q (B)  \big] \Big\|_{T_{0,y} (\ell_+, \vec{\lambda})} & \leq O_L (1) \norm{V (b) - \frac{1}{2} Q(b)}_{T_{0,y} (\ell_+, \vec{\lambda})}  \norm{Q (b)}_{T_{0,y} (\ell_+, \vec{\lambda})} \nnb
	& \leq O_L (1) (r_1 + r_2) r_2 .
	\label{eq:Cov_V_Q_bound}
\end{align}
Together, these bounds give the desired estimate and complete the proof.
\end{proof}

\begin{proof}[Proof of Theorem~\ref{thm:Phi^U_estimate}]
Let $d \ge 4$.
We wish to prove that under the hypotheses of Theorem~\ref{thm:Phi^U_estimate},
the RG remainder obeys the bounds
\begin{align}
	\norm{ D^p_V D^q_K D_{\ka}^r  R^U_{+} }_{\domRG  \rightarrow \cU_{+}(\ell_+)}
	\leq
	\begin{array}{ll}
	\begin{cases}
	M_{p, 0}  \tilde{\vartheta}_{+}^3  \tilde{g}_+^3 \scale_+ ^{\kaa} & ( p\geq 0,\, q=0) \\
	M_{p,q}  & ( p\geq 0,\, q \geq 1) \\
	0 & ( p\geq 0,\, q\geq 3).
	\end{cases}
	\end{array}
	\label{eq:R^U_derivative_estimate-pf}
\end{align}
For $q\ge 3$, the derivatives vanish because
by definition $R_+^U$ is only quadratic in $K$.
For the other cases, we fix $r \le p_a$ and
control derivatives using the extended norm as in \cite[(9.2.7)]{BBS-brief}, which
gives
\begin{align}
\label{eq:DR}
	\norm{ D^p_V D^q_K D_{\ka}^r  R^U_{+} }_{\domRG  \rightarrow \cU_{+}(\ell_+)}
	\leq
    \frac{p!q! r!}{\lambda_V^p\lambda_K^q}
    \norm{R_+^U}_{T_{0,y} (\ell_+, \vec{\lambda})}
    ,
\end{align}
(there is no $\lambda_{\ka}^{-r}$ for the $\ka$-derivative because this factor is included inside $D_{\ka}^{-r}$).
We apply Lemma~\ref{lemma:R_+_mass_derivative_bound} to bound the right-hand side by $(1+r_1+r_2)r_2$, with
$r_1 = \norm{V(b)}_{T_0 (\ell)} + \lambda_V$ and $r_2 = \norm{K(b)}_{T_0 (\ell)} + \lambda_K$.

For the case $q=0$, we apply
Lemma~\ref{lemma:R_+_mass_derivative_bound} with $(\lambda_V, \lambda_K) =(1,0)$.
The fact that $(V,K)$ is in the RG domain implies that
$r_2=\|K(b)\|_{T_0(\ell)} \le C_{\rm RG}\tilde\vartheta^3 \tilde g^3 \scale^{\kaa}$,
and also that $r_1= \norm{V}_{T_{0} (\ell)} + \lambda_K $ is bounded.
The conversion of $\tilde\vartheta^3 \tilde g^3 \scale^{\kaa}$ to
$\tilde\vartheta_+^3 \tilde g_+^3 \scale_+^{\kaa}$ costs an unimportant
multiple of $\scale/\scale_+=L^{d-4}$, so this gives the desired estimate for $q=0$,
with an $L$-dependent constant $M_{p,0}$.

For $q \ge 1$ and arbitrary $p \geq 0$, we instead use $(\lambda_V, \lambda_K) = (1,1)$.
Then $r_1 = \norm{V}_{T_{0} (\ell)} + \lambda_K$ is again bounded, but
now $r_2 = \norm{K}_{T_{0} (\ell)} + \lambda_K$ is  $O(1)$, so we obtain an $O(1)$ bound
on the norm on the right-hand side of \eqref{eq:DR}.  This completes the proof.
\end{proof}

\subsection{Preliminary bounds for the non-perturbative RG map}
\label{sec:prelim2}

Now we turn our attention to the building blocks of $\Phi^K_+$.
The point of departure is to
rewrite $\Phi_{+}^K$ as a sum of two terms.  To do so, we first
recall the definition of $Q$ in \eqref{eq:QVK} and define
\begin{align}
\label{eq:hatV_hatK_definition}
	\hat{V} = V - Q
    ,
	\qquad
	\hat{K} = K - (e^{-\hat{V}} - e^{-V}),
\end{align}
and, for $X \subset \cB (B)$ with $B \in \cB_+$,
\begin{align}
\hat{V} (B \backslash X) = \sum_{b\in \cB (B\backslash X)} \hat{V} (b), \qquad \hat{K}^X = \prod_{b\in X} \hat{K} (b).
\end{align}
Then, exactly as in \cite[(10.1.6)]{BBS-brief}, basic algebra leads to
\begin{align}
& K_+ = \Phi_+^K (V,K,\ka) = S_0 + S_1
\label{eq:Phi_+_is_S_0+S_1}
\end{align}
where
\begin{align}
    S_0 &= e^{u_{+} |B|}  \Eplus \Big( e^{- \theta \hat{V}(B)} - e^{-U_{+} (B)} \Big) 	, 	\\
    S_1 &= e^{u_{+} |B|} \sum_{X\subset \cB (B), |X| \geq 1} \Eplus \theta \Big( e^{-\hat{V} (B \backslash X)} \hat{K}^X \Big).
\end{align}
Here $|X|$ denotes the number of blocks in $X$.
In this section, we provide bounds on the
constituents of $S_0$ and $S_1$.

For situations in which $V$ and $K$ are themselves measured
using the norm $\norm{\cdot}_{T_{\varphi, y} (h, \vec{\lambda})}$, to avoid confusion
we use the notation $V^*$ and $K^*$ for the functions
\begin{align}
	V^* : (V, K) \mapsto V, \qquad K^* : (V, K) \mapsto K.
\end{align}
Bounds on the norms of these functions are given in
\cite[(9.3.9) ff.]{BBS-brief}.
We write $C_L$ for a generic $L$-dependent constant whose value is unimportant and
may change from line to line.
The real numbers $u_{\pt}$ and $u_+$ are respectively the vacuum energy components
(constant terms) of
the perturbative map $\Phi_{\pt}$ and the RG map $\Phi_+$ defined
in Definitions~\ref{def:Phipt}--\ref{def:RGmap}.

\begin{lemma} \label{lemma:V_bound}
Let $V\in \cD$,  $\lambda_V \leq  \tilde{g} \scale$,  $b\in \cB$,  and
let $L$ be sufficiently large.  Then
\begin{align}
	\norm{V^* (b)}_{T_{0,y} (h, \vec{\lambda})} \leq \frac{11}{8} k_0^3.
\end{align}
\end{lemma}

\begin{proof}
The proof is a small modification of the proof of \cite[Lemma~9.3.2]{BBS-brief}
to include dimensions $d>4$.
Let $\tau = \frac{1}{2} |\varphi|^2$, so that $V(b)=g\tau^2(b) + \nu\tau(b)$.
It suffices to prove that
\begin{align}
	\norm{ (g \tau^2)^* (b) }_{T_{0,y} (h, \vec{\lambda})} \leq \frac{3}{8} \frac{g}{\tilde{g}} k_0^4, \qquad \norm{(\nu \tau)^* (b)}_{T_{0,y} (h, \vec{\lambda})} \leq k_0^4.
\end{align}	
As in the bound on \cite[(9.3.16)]{BBS-brief},
\begin{align}
	\sup_{\dot{V} \in \cV, \dot{\nu}
    =0} \frac{\norm{\dot{V} (b)}_{T_0 (h)}}{\norm{\dot{V}(b)}_{T_0 (\ell)}}
    =
    h^4 \ell^{-4}
    =
    \ell_0^{-4} k_0^4 (\tilde{g}\scale)^{-1},
\end{align}
so that \cite[(9.3.17)]{BBS-brief} gets replaced by
\begin{align}
	\norm{ (g \tau^2)^* (b) }_{T_{0,y} (h, \vec{\lambda})} \leq \Big( \frac{g}{4 \tilde{g}} + \frac{\scale^{-1} \lambda_V}{\tilde{g} \ell_0^4} \Big) k_0^4 \leq \frac{3 g}{8 \tilde{g}} k_0^4,
\end{align}
using $\lambda_V \le \tilde g \scale$ and large $L$.
The bound on $\norm{(\nu \tau)^* (b)}_{T_{0,y} (h, \vec{\lambda})}$ follows similarly.	
\end{proof}

In the following, when we make reference to results in \cite{BBS-brief} proved for $d=4$,
we in fact use the result generalised to $d\geq 4$ using Lemma~\ref{lemma:V_bound}
instead of \cite[Lemma~9.3.2]{BBS-brief}.
This generalisation is possible because of the stronger assumptions
$\lambda_V,\lambda_K \leq \tilde{g} \scale$ (instead of $\lambda_V,\lambda_K \leq \tilde{g}$
in \cite{BBS-brief}).

\begin{remark}
Henceforth we
assume that the constant $k_0$ of \eqref{eq:h_j_definition} is sufficiently small
depending only on $n$ (as in \cite[Proposition 10.2.1]{BBS-brief}).
In particular, $k_0 \le [24(n+2)]^{-1}$.
In the important constant $\cst$ of \eqref{eq:V-Q_bound1-h}, the label ``st'' stands
for stability; the $\cst$ in the exponential decay is what enables us to include
the regulator in our norm.  This $\cst$ depends only on $k_0$.
We choose
	$\kappa$ small compared to $\cst$,
consistent with our assumption on $\kappa$ in
\ref{quote:assumPhi} and the discussion below \eqref{eq:G_j_definition}.
\end{remark}

\begin{lemma}
\label{lemma:hat_K_bound}
Assume \ref{quote:assumPhi} and $\lambda_V, \lambda_K \leq \tilde{g} \scale$.
There is a constant $C$ and an $L$-dependent constant $C_L$ such
that the following hold for all $\varphi\in \R^n$, $b\in \cB$, and $B \in \cB_+$:
\begin{enumerate}
\item For $\mathfrak{h} \in \{ \ell, h \}$,
\begin{align}
	\norm{V (b)}_{T_{\varphi} (\mathfrak{h})} & \leq \begin{array}{ll}
	\begin{cases}	
	C_L  \tilde{g} \scale P^4_{\ell} (\varphi) &   (\mathfrak{h} = \ell) \\
	C P^4_{h} (\varphi) &   (\mathfrak{h} = h),
	\end{cases}
	\end{array}
	\label{eq:V_bound0}
\\
	\norm{\hat V (B)}_{T_{0} (\mathfrak{h_+},\vec{\lambda})}
    & \leq \begin{array}{ll}
	\begin{cases}	
	C_L  \tilde{g}_+ \scale   &   (\mathfrak{h} = \ell) \\
	1 & (\mathfrak{h} = h).
	\end{cases}
	\end{array}
	\label{eq:Vhat_bound0}
\end{align}
\item
Let $k_0$ be sufficiently small.
Let
$t\geq 0$ and
$s \in [0,1]$,
There exists a constant $\cst >0$ (depending on $k_0$ only)
such that
\begin{alignat}{2}
\label{eq:V-Q_bound1-h}
    \norm{ e^{-t(V^* - s Q) (b)} }_{T_{\varphi, y} (\mathfrak{h}, \vec{\lambda})}
    &\leq
	2^{t/4}
	e^{-4 t\cst |\varphi / h |^4}
    &&
    (\hf \le h)
    ,
    \\
\label{eq:V-Q_bound1-hplus}
	\norm{ e^{-t(V^* - s Q) (b)} }_{T_{\varphi, y} (\mathfrak{h}_+, \vec{\lambda})}
    &\leq
	(2^{t/4}
	e^{-4 t\cst |\varphi / h_+ |^4})^{L^{-d}}
    \quad &&
    (\hf_+ \le h_+).
\end{alignat}

\item  For all $\kappa \leq 2 \cst$,
\begin{align}
& \norm{\hat{K} (b) }_{T_{0, y} (\ell ,\vec{\lambda})} \leq
C
\big( \norm{K(b)}_{\Wkappa} + \lambda_K \big) ,
\label{eq:hat_K_bound2}
\\
& \norm{\hat{K} (b)}_{T_{\varphi, y} (h, \vec{\lambda})} \leq
C
\big( \norm{K(b)}_{\Wkappa} + \lambda_K \big) (\tilde{g}^{9/4}\scale^{\kb} )^{-1} \Gj (b, \varphi).
\label{eq:hat_K_bound1}
\end{align}
\end{enumerate}
\end{lemma}

\begin{proof}
(i)
The inequality \eqref{eq:V_bound0}
follows from combining \cite[Lemma~9.3.1]{BBS-brief} and \eqref{eq:T_phi_to_0},
and \eqref{eq:Vhat_bound0} is proved as in \cite[Lemma~9.3.4]{BBS-brief}.

\smallskip\noindent (ii)
We first prove \eqref{eq:V-Q_bound1-h}.
For $s\in [0,1]$, it follows from the product property of the norm that
\begin{equation}
	\norm{ e^{-t(V^* - s Q) (b)} }_{T_{\varphi, y} (\mathfrak{h}, \vec{\lambda})} \leq
	\norm{e^{-tV^*(b)}}_{T_{\varphi, y}}e^{t\norm{Q(b)}_{T_{\varphi, y}}}.
\end{equation}
The first factor is bounded by the generalisation to $d\geq 4$ of
\cite[Proposition~10.2.1]{BBS-brief} which does hold for $\mathfrak{h} \leq h$.
For the second factor, we follow \cite[Lemma 10.2.2]{BBS-brief}
which is stated for $\mathfrak{h} \leq h_+$,
so we verify that the proof actually permits $\mathfrak{h} \leq h$. Indeed, the proof relies on the estimate \cite[(9.3.21)]{BBS-brief} which is our Lemma~\ref{lemma:Q_bound}, and requires $\mathfrak{h} \geq \ell$ (both here and in \cite{BBS-brief}). We
therefore obtain the desired bound for $\mathfrak{h} \in [\ell,h]$, but since $\norm{\cdot}_{T_{\varphi}(\mathfrak{h})}$ is monotone in $\mathfrak{h}$,
it holds also for $\mathfrak{h} \in [0,h]$.

To prove \eqref{eq:V-Q_bound1-hplus}, we recall from
\cite[Lemma~10.2.3]{BBS-brief} that, for $\hf_+\le h_+$, $t' \ge 0$, and $s \in [0,1]$,
\begin{align}
    \|e^{-t'(V^*-sQ)(B)}\|_{T_{\varphi,y}(\hf_+,\lambda)}
    \le2^{t'/4}e^{-4t'\cst |\varphi/h_+|^4}.
    \label{eq:V-Q_bound1plusplus}
\end{align}
We rewrite $t(V^*-sQ)(b)$ as $L^{-d}t(V^*-sQ)(B)$, and then apply
\eqref{eq:V-Q_bound1plusplus}
with $t'=tL^{-d}$.
This gives \eqref{eq:V-Q_bound1-hplus}.

\smallskip\noindent (iii)
The first inequality is contained in \cite[(10.4.18), (10.4.19)]{BBS-brief}.

For the second inequality, which incorporates our regulator, we first recall from
\cite[(10.4.16), (10.4.17)]{BBS-brief} that
\begin{align}
\norm{\hat{K} }_{T_{\varphi, y} (h, \vec{\lambda})} \leq  \norm{K^* }_{T_{\varphi, y} (h, \vec{\lambda})}
+ 2
P_{h}^4 (\varphi) \norm{K^*}_{T_{0, y} (h, \vec{\lambda})}
\sup_{s\in [0,1]} \norm{e^{-(V^* - sQ)(b)}}_{T_{\varphi, y} (h, \vec{\lambda})}
.
\end{align}
We insert \eqref{eq:V-Q_bound1-h} into this bound, choose $\kappa \le 2 \cst $, and
recall from \eqref{eq:PG} that
$P_{h}^4 (\varphi)  \Gj (b, \varphi) \leq O(1)$.
This gives
\begin{align}
\norm{\hat{K} }_{T_{\varphi, y} (h, \vec{\lambda})}
&\leq \norm{K^* }_{T_{\varphi, y} (h, \vec{\lambda})} +
4
P_{h}^4 (\varphi)
\norm{K^*}_{T_{0, y} (h, \vec{\lambda})} \Gj (b, \varphi)^2
\nnb
& \leq \norm{K^*}_{T_{\varphi, y} (h, \vec{\lambda})} +
C
\Gj (b, \varphi) \norm{K^*}_{T_{0,y} (h, \vec{\lambda})}
.
\label{eq:hat_K_bound3}
\end{align}
It follows as in \cite[(9.3.10), (9.3.13)]{BBS-brief},
together with the definition of the extended norm in \eqref{eq:extended_norm}, that
\begin{equation}
    \norm{K^*}_{T_{\varphi, y} (h, \vec{\lambda})}
    \le \norm{K}_{T_{\varphi} (h)} +
    \lambda_K ( \tilde{g}^{9/4} \scale^{\kb})^{-1} G(b,\varphi) ,
\end{equation}
which gives
\begin{align}
\label{eq:Khat4}
    \norm{\hat{K} }_{T_{\varphi, y} (h, \vec{\lambda})}
    &\leq
    \norm{K}_{T_{\varphi} (h)}
    +
    C
    \norm{K}_{T_{0} (h)} G(b,\varphi)
    +
    C
    \lambda_K ( \tilde{g}^{9/4} \scale^{\kb})^{-1} G(b,\varphi) 	
    .
\end{align}
By definition of the norm in \eqref{eq:Wkappa-norm-def},
$\norm{K}_{T_{\varphi} (h)} \le (\tilde{g}^{9/4} \scale^{\kb})^{-1}G(b,\varphi)
\norm{K}_{\Wkappa}$.  Since $G(b,0)=1$, this gives
the desired estimate \eqref{eq:hat_K_bound1} and completes the proof.
\end{proof}

The following lemma is an extension of \cite[Lemma~10.2.4]{BBS-brief}
which includes mass derivatives in the norm.

\begin{lemma}
\label{lemma:stability_estimate-mass_derivative}
Let $k_0$ be sufficiently small.
Assume \ref{quote:assumPhi},
$\lambda_V, \lambda_K \leq \tilde{g} \scale$,
and $\lambda_{\ka} = k_{\ka}^{-1}$.
There exists
an $L$-dependent constant $C_L$,
such that for all
$B \in \cB_+$,  $\mathfrak{h}_+ \leq h_+$, $t \ge 0$,
and $r \le p_{\ka}$,
\begin{align}
	\big\| e^{-t U_+ (B)} \big\|_{T_{\varphi, y} (\mathfrak{h}_+, \vec{\lambda})} & \leq 2^{t/2} e^{ -2 t \cst |\varphi/ h_+|^4 },
	\label{eq:stability_estimate-mass_derivative}\\
	\exp\big( \norm{ u_{+}|B|}_{T_{\varphi, y} (\mathfrak{h}, \vec{\lambda})} \big) & \leq 2^{1/2},
	\label{eq:vacuum_energy_estimate-mass_derivative}
\\
	 \big| D^{r}_{\ka} (u_{+} - u_{\pt} ) \big|  |B|
    & \leq C_L
    \norm{K}_{\Wkappa}
	.
	\label{eq:u_+_minus_u_pt}
\end{align}
\end{lemma}

\begin{proof}
For \eqref{eq:stability_estimate-mass_derivative},
by monotonicity of the norm in $\mathfrak{h}_+$, it is sufficient to set $\mathfrak{h}_+ =h_+$.
We make the decomposition $e^{-t U_+} = e^{-t (U_+ - V)} e^{-t V}$, and use \cite[Proposition~10.2.1]{BBS-brief} (which assumes that $k_0$ is sufficiently small)
to bound the norm of $e^{-tV^*}$.  This gives
\begin{align}
	\norm{e^{-tU_+}}_{T_{\varphi, y} (h_+, \vec{\lambda})}
    &\leq
    \norm{ e^{-t (U_+  - V^*) } }_{T_{\varphi, y} (h_+, \vec{\lambda})}
    \norm{ e^{ - t V^* }}_{T_{\varphi} (h_+, \vec \lambda)}
    \nnb
	&\leq
    \exp \big [ t \norm{ U_+ - V^* }_{T_{\varphi, y} (h_+, \vec{\lambda})} \big]
    \,
    2^{t/ 8} e^{-8 t \cst   |\varphi/ h_+|^4}.
\label{eq:etU}
\end{align}
According to the definition of $U_+$ in \eqref{eq:U+def},
$U_+=\Phi_{\pt}(\hat V)$ with $\Phi_{\pt}$ given by \eqref{eq:Phi_pt_definition}.
We write $U_+ - V = \Phi_{\pt} (\hat{V}) - \hat{V} + Q$ and obtain
\begin{align}
	(U_+
    - \hat{V} )(B) = \Eplus [\theta \hat{V} (B) - \hat{V}(B)]
    - \frac{1}{2} \Loc \var_+ (\theta \hat{V} (B)).
\end{align}
By Lemma~\ref{lemma:polynomial_expectation_mass_derivative},  by
\eqref{eq:Vhat_bound0},
and by the fact that $\Loc$ is a bounded operator
$\norm{\cdot}_{T_{\varphi} (\mathfrak{h})} \rightarrow \norm{\cdot}_{T_{\varphi} (\mathfrak{h})}$ (see \cite[(7.5.3)]{BBS-brief} with $k=4$), we find that
\begin{align}
	\norm{(U_{+}   - \hat{V} )(B)}_{T_{0,y} (h_+, \vec{\lambda})} \leq C \Big( \frac{\mathfrak{c}_+}{h_+} \Big)^2 	.
\end{align}
With the bound on $Q$ from \eqref{eq:Qh} (for which we assume that
$\tilde g_+$ is small depending on $L$ to accommodate the hypothesis $h_+ \geq \ell$),
and by \eqref{eq:T_phi_to_0}, this leads to
\begin{align}
	\norm{U_+ - V^*}_{T_{\varphi,y} (h_+, \vec{\lambda})} \leq \norm{U_+ - V^*}_{T_{0,y} (h_+, \vec{\lambda})} P_{h_+}^4 (\varphi) \leq O(1)  \big(L^{-d} + \tilde{g}^{1/2} L^{-(d-4)/2} \big)  P_{h_+}^4 (\varphi).
\end{align}
This produces a small multiple of $P_{h_+}^4 (\varphi)$ in the exponent,
and serves to reduce the $8$ in \eqref{eq:etU} to (say) $2$.
This completes the proof of
\eqref{eq:stability_estimate-mass_derivative}.

To obtain \eqref{eq:vacuum_energy_estimate-mass_derivative}, we simply use
\begin{align}
	\norm{u_+}_{T_{\varphi, y} (h_+,\vec{\lambda})} \leq \norm{U_+ - V^*}_{T_{0,y} (h_+, \vec{\lambda})}.
\end{align}
Finally,  for \eqref{eq:u_+_minus_u_pt},
since $R_+^U (V,K) = \Phi_+^U (V,K)-\Phi_+^U (V,0)$,
it follows from the Mean Value Theorem  and the case $(p,q)=(0,1)$ of Theorem~\ref{thm:Phi^U_estimate} (with its constant $M_{0,1}$) that
\begin{align}
    | D^{r}_{\ka} (u_{+} - u_{\pt} ) | |B|
    &\leq
	\norm{D^{r}_{\ka} R_+^U (V, K)}_{\cU_+} \leq M_{0, 1}
    \norm{K}_{\Wkappa}.
\end{align}
This completes the proof.
\end{proof}

\section{The non-perturbative RG map:  proof of Theorem~\ref{thm:Phi^K_estimate}}
\label{sec:pf2}

In this section, we prove Theorem~\ref{thm:Phi^K_estimate}, which
we restate here as Theorem~\ref{thm:Phi^K_estimate-bis} for easy reference, along
with its assumption \ref{quote:assumPhi}:
\begin{quote}
\begin{itemize}
\item[\customlabel{quote:assumPhi-bis}{$\assumPhi$}]
Let $\tilde{\ka} \ge 0$, let $L$ be sufficiently large, and let $\kappa$
be sufficiently small
(the latter depending only on $n$).
At scale $j$, let $\tilde{g} = \tilde{g}_j$ be sufficiently small depending on $L$,
and let $(V, K) \in \domRG = \domRG_j$.
\end{itemize}
\end{quote}

\begin{theorem} \label{thm:Phi^K_estimate-bis}
Let $d \ge 4$.
At scale $j \in \{0,\cdots, N - 1\}$, assume \ref{quote:assumPhi-bis} and let
$r \in \{0, \ldots, p_{\ka} \}$.
There exists a positive constant $C_{\rg}$ such that
$\Phi_{+}^K$ is well-defined as a map $\domRG \rightarrow \Wkappa$ (i.e., the integral \eqref{eq:K+B} converges), and there exist positive $M_{p,q}$ (for
integers $p,q \ge 0$) and $L$-independent $M_{0,1}'$ such that
\begin{align}
	\norm{ D^p_V D^q_K D_{\ka}^r \Phi_{+}^K}_{\domRG  \rightarrow \Wkappa_{+}} \leq
	\begin{array}{ll}
	\begin{cases}
	C_{\rg} \tilde{\vartheta}_{+}^3  \tilde{g}_{+}^3 \scale_+^{\kaa} & (p=0,\, q=0) \\
	M_{p, 0} \tilde{\vartheta}_{+}^{3} \tilde{g}_{+}^{3-p} \scale_+^{\kaa-p} & (p > 0,\, q=0) \\
	M_{0,1}' L^{-\max\{2,  (d-4) \kaa\}}    & (p = 0,\, q=1) \\
	M_{p,q} \tilde{g}_{+}^{-p- \frac{9}{4}(q-1)}
    \scale_+^{-p-\kb (q-1)} & (p\geq 0,\, q\geq 1, \; (p,q) \neq (0,1)).
	\end{cases}
	\end{array}
	\label{eq:Phi^K_derivatives_estimates-mass_derivative-bis}
\end{align}
\end{theorem}

We also repeat from \eqref{eq:kbd-bis} the definitions:
	\begin{align}
\label{eq:kab}
		\kaa (d) =&
		\begin{cases}
			3 & (4 \le d < 6) \\
			\frac{2d- 6}{d-4} (1- \epsilon(d))  & (d \ge 6),
		\end{cases}
	\qquad
		\kb (d) = \begin{cases}
                \kaa(d) - \frac 34 & (4\le d <12) \\
			\frac{3}{2} (1-\epsilon (d))   & (d \ge 12),
		\end{cases}
	\end{align}
with $\epsilon(d) \in (0,\frac{1}{d-3})$.
These obey
\begin{align}
	\kaa (d) \in (2, 3]  \qquad
	\kaa (d) - \kb(d) \in ( 1/2,  3/4 ] ,
\end{align}
and for $d \geq 5$,
\begin{align}
\label{eq:abp-bds}
	\kaa (d) < \frac{2d-6}{d-4} , \qquad
    \kaa (d) - \kb (d) <  \frac{d}{2(d-4)} .
\end{align}

\subsection{Structure of proof}
\label{sec:Phi_K_estimate}

Recall the decomposition \eqref{eq:Phi_+_is_S_0+S_1}, which states that
\begin{align}
	& K_+ = \Phi_+^K (V,K,\ka) = S_0 + S_1,
	\label{eq:Phi_+_is_S_0+S_1-bis}
\end{align}
where
\begin{align}
    S_0 &= e^{u_{+} |B|}  \Eplus \Big( e^{- \theta \hat{V}(B)} - e^{-U_{+} (B)} \Big) 	, 	\\
    S_1 &= e^{u_{+} |B|} \sum_{X\subset \cB (B), |X| \geq 1} \Eplus \theta \Big( e^{-\hat{V} (B \backslash X)} \hat{K}^X \Big),
\end{align}
with $|X|$ the number of blocks in $X$.

We follow the strategy of \cite[Section~10.1]{BBS-brief},
and prove Theorem~\ref{thm:Phi^K_estimate-bis}
using the following three propositions.
These propositions improve
\cite[Lemma~10.1.4]{BBS-brief}, \cite[Lemma~10.1.5]{BBS-brief},
and the crucial contraction
(the $(p,q)=(0,1)$ case of \cite[(10.1.2)]{BBS-brief}),
by replacing the $\cW$-norm by the $\cW^\kappa$-norm and including dimensions $d>4$.
We repeatedly use the fact that \ref{quote:assumPhi-bis} implies
\begin{equation}
\label{eq:K_rg_induct_bound}
		\|K\|_{\Wkappa} \le C_{\rg} \tilde{\vartheta}^3\tilde{g}^{3} \scale^{\kaa}
\end{equation}
where the important constant $C_{\rm RG}$ is specified below in \eqref{eq:C_RG_def}.
In the extended norm \eqref{eq:extended_norm}, we always take
$\lambda_{\ka}= k_{\ka}^{-1} = (L^dL^{2j})^{-1}$ as in \eqref{eq:k_j_definition}
and fix a finite number $p_{\ka}\ge 2$ of mass derivatives,
but we will vary our choices of $\lambda_V$ and $\lambda_K$.

\begin{proposition} \label{prop:S0}
Assume \ref{quote:assumPhi-bis} and suppose that
$\lambda_V, \lambda_K \leq  \tilde{g} \scale$.
Then
\begin{align}
	& \norm{S_0}_{T_{0,y} (\ell_+, \vec{\lambda})} \leq C_L \tilde{\vartheta}_{+}^3 ( \tilde{g}_{+} \scale_+)^{3} 	,
	\label{eq:S0_bound1}
	\\
	& \norm{S_0}_{T_{y}^{G_{ +}} (h_+, \vec{\lambda})} \leq C_L \tilde{\vartheta}_{+}^3 ( \tilde{g}_{+} \scale_+)^{3/4}	.
	\label{eq:S0_bound2}	
\end{align}
\end{proposition}

\begin{remark} \label{remark:C_RG_def}
The constant $C_{\rg}$ occurs in the definition of the domain $\domRG$
that is part of \ref{quote:assumPhi-bis}, so $C_L$
in Proposition~\ref{prop:S0} potentially depends on $C_{\rg}$.
We therefore cannot use $C_L$ to define $C_{\rg}$.
We avoid any such circularity as follows.  When we specialise
Proposition~\ref{prop:S0} by setting $K=0$ on the left-hand sides
of its two bounds, the constant $C_L$ that occurs for this restricted case
cannot depend on $C_{\rg}$
since there is no $K$-dependence.
We write this $L$-dependent constant as $C_{\pt}$, which satisfies
\begin{align}
	& \norm{S_0 (V, K=0)}_{T_{0,y} (\ell_+, \vec{\lambda})}
	\leq C_{\pt} \tilde{\vartheta}_{+}^3 ( \tilde{g}_{+} \scale_+)^{3} ,	 \\
	& \norm{S_0 (V, K=0)}_{T_{y}^{G_{+}} (h_+, \vec{\lambda})} \leq C_{\pt} \tilde{\vartheta}_{+}^3 ( \tilde{g}_{+} \scale_+)^{3/4}.
\end{align}
Then we define
\begin{equation}
\label{eq:C_RG_def}
	C_{\rm RG} = 4 p_a! C_{\pt}	,
\end{equation}
where $p_{\ka}$ is the number of mass derivatives included in our norm.
\end{remark}

\begin{proposition} \label{prop:S1}
Assume \ref{quote:assumPhi-bis} and suppose that
$\lambda_V \leq \tilde{g} \scale $ and
$\lambda_K \leq \tilde{g}^{9/4} \scale^{\kb}$.  Then
\begin{align}
\label{eq:S1bd}
	& \norm{S_1}_{\Wkappa_{y,+} (\vec{\lambda})}
    \leq C_L
    \tilde{\vartheta}_+^3 ( \tilde{g}_+^3 \scale^{\kaa} + \lambda_K).
\end{align}
\end{proposition}

As we show below,
Propositions~\ref{prop:S0} and \ref{prop:S1} imply Theorem~\ref{thm:Phi^K_estimate-bis} except for
$(p,q)$ equal to $(0,0)$ or $(0,1)$.  The case $(p,q) = (0,1)$ provides the crucial contraction of the RG map, and the case $(p,q) = (0,0)$ is proved using this contraction.
The crucial contraction requires special attention and we isolate it now
as the following proposition, which, as we will see
in the proof of Theorem~\ref{thm:Phi^K_estimate-bis}, is slightly stronger
than its counterpart in Theorem~\ref{thm:Phi^K_estimate-bis}.  It does not use the extended norm.

\begin{proposition} \label{prop:crucial}
Assume \ref{quote:assumPhi-bis}.  There is an $L$-independent $M_{0,1}'$ such that,
uniformly in $(V,K,\ka) \in \domRG$ and for  $0 \le r \le  p_{\ka}$,
\begin{align}
	\norm{D_{\ka}^r D_K \Phi_{+}^K  }_{\domRG \rightarrow \Wkappa_+} \leq M_{0,1}'
	\max\{ L^{-(2d -6)} ,  L^{-\frac{d}{2} - (d-4) \kb }  \}
    .
\end{align}
\end{proposition}

Our restrictions on the powers $\kaa$ and $\kb$ are used in the following proof.
As usual, throughout this section
$C_L$ denotes a generic $L$-dependent constant whose value may change from one line to another.

\begin{proof}[Proof of Theorem~\ref{thm:Phi^K_estimate-bis}]
Consider first the cases $(p,q) \notin  \{ (0,0), (0,1) \}$.
By  \cite[9.2.8)]{BBS-brief},
\begin{align}
	\norm{D_V^{p} D_K^q D_{\ka}^r \Phi_+^K}_{\domRG
\rightarrow \Wkappa_+} \leq \frac{p! q! r !}{\lambda_V^p \lambda_K^q} \norm{\Phi_+^K}_{\Wkappa_{y,+} (\vec{\lambda})}		
\label{eq:DVDK_bound}
\end{align}
(the absence of a factor $\lambda_{\ka}^r$ is as in \eqref{eq:DR}).
We bound $\Phi_+^K = S_0 + S_1$ by bounding $S_0$ and $S_1$ separately using Propositions~\ref{prop:S0} and \ref{prop:S1}.

For $q=0$ and $p\ge 1$, we apply \eqref{eq:DVDK_bound} with
$(\lambda_V, \lambda_K) = (\tilde{g} \scale,  0)$,
together with
\begin{align}
	\norm{\Phi_+^K}_{\Wkappa_{y,+} (\vec{\lambda})}
	&\leq \|S_0\|_{\Wkappa_{y,+} (\vec{\lambda})}
		+\|S_1\|_{\Wkappa_{y,+} (\vec{\lambda})} \nnb
	&\leq C_L\,\tilde{\vartheta}_+^3
\Big[
		( \tilde{g}_{+} \scale_+)^{3}
		+ \tilde g_+^{9/4}\scale_+^{\kb} ( \tilde{g}_{+} \scale_+)^{3/4}
		+ \tilde{g}_+^3 \scale_+^{\kaa}  \Big] \nnb
	& \leq C_L \tilde{\vartheta}_+^3 \tilde{g}_+^3 \scale_+^{\kaa}	.
\end{align}
The above holds also for $p=0$, but it is not sufficient for
this case of \eqref{eq:Phi^K_derivatives_estimates-mass_derivative-bis} because
it does not produce the constant $C_{\rm RG}$.
For $q>0$, we apply \eqref{eq:DVDK_bound} with
$(\lambda_V, \lambda_K)
= (\tilde{g} \scale,  \tilde{g}^{9/4} \scale^{\kb})$,
together with
\begin{align}
	\norm{\Phi_+^K}_{\Wkappa_{y,+} (\vec{\lambda})}
	&\leq \|S_0\|_{\Wkappa_{y,+} (\vec{\lambda})}
		+\|S_1\|_{\Wkappa_{y,+} (\vec{\lambda})} \nnb
	&\leq C_L\,\tilde{\vartheta}_+^3 \Big[
		( \tilde{g}_{+} \scale_+)^{3}
		+ \tilde g_+^{9/4}  \scale_+^{\kb}  ( \tilde{g}_{+} \scale_+)^{3/4}
		+ \tilde{g}_+^3 \scale_+^{\kaa} + \lambda_K \Big] \nnb
	&\leq C_L\,  \tilde{\vartheta}_+^3 \tilde g_+^{9/4}   \scale_+^{\kb}
    .
\end{align}
For $(p,q) = (0,1)$, by Proposition~\ref{prop:crucial}
it suffices to observe that
\begin{equation}
	\max\{L^{-(2d-6)},L^{-\frac{d}{2}-(d-4)\kb}\} \leq L^{-\max\{2,(d-4)\kaa\}}.
\end{equation}
Indeed, this is
an equality for $d=4$, while for $d>4$ our assumptions \eqref{eq:abp-bds}
on $\kaa,\kb$ imply that $\min\{ 2d-6 , \frac{d}{2}+(d-4)\kb\}  > (d-4)\kaa$.
Thus Proposition~\ref{prop:crucial} is slightly stronger than what is required for $d>4$.

Finally, we consider the remaining case $(p,q) = (0,0)$, whose proof uses the
crucial contraction.  We make the decomposition
\begin{equation}
\label{eq:DKPhih}
    \norm{D_{\ka}^r \Phi_+^K (V,K)}_{\Wkappa_+}
    \le
    \norm{D_{\ka}^r \Phi_+^K (V,0)}_{\Wkappa_+}
    +
    \norm{D_{\ka}^r ( \Phi_+^K (V,K) - \Phi_+^K (V,0) )}_{\Wkappa_+}
    ,
\end{equation}
and bound each term separately.
For the first term in \eqref{eq:DKPhih}, by definition $S_1 (V,0) = 0$, so
it follows from Proposition~\ref{prop:S0}, Remark~\ref{remark:C_RG_def} and
the definition of the norm in \eqref{eq:Wextended} that
\begin{align}
	\label{eq:Phi_K_bd_at_zero_pf}
	\norm{D_{\ka}^r \Phi_+^K (V,0)}_{\Wkappa_+}
	&\leq
	r!C_{\pt} [\tilde{\vartheta}_{+}^3 ( \tilde{g}_{+} \scale_+)^{3}
	+ \tilde{g}_+^{9/4} \scale_+^{\kb} \tilde{\vartheta}_{+}^3 ( \tilde{g}_{+} \scale_+)^{3/4}]\nnb
	&\leq  \frac12 C_{\rg}[ \tilde{\vartheta}_+^3 \tilde{g}_+^3 \scale_+^{3}
	+ \tilde{\vartheta}_{+}^3 \tilde{g}_{+}^3 \scale_+^{\kb+3/4}]\nnb
	&\leq  \frac12 C_{\rg} \tilde{\vartheta}_+^3 \tilde{g}_+^3 \scale_+^{\kaa} ,
\end{align}
where in the last line we used $\kaa \leq \min \{ 3, \kb + 3/4 \}$.
For the second term in \eqref{eq:DKPhih}, by the
Fundamental Theorem of Calculus,
the bound on $D_K \Phi_+^K$
in Proposition~\ref{prop:crucial}, and the bound on $K$ from \eqref{eq:K_rg_induct_bound},
\begin{align}
\label{eq:DKPhi-5}
	\norm{D_{\ka}^r ( \Phi_+^K (V,K) - \Phi_+^K (V,0) )}_{\Wkappa_+} & \leq \norm{K}_{\Wkappa} \sup_{t\in [0,1]} \norm{D_{\ka}^r D_K \Phi_+^K (V, tK)}_{\domRG \rightarrow  \Wkappa_+}
\nnb
	 &\leq C_{\rg}\tilde\vartheta^3 \tilde g^3 \scale^\kaa
     M_{0,1}'  \max\{ L^{-(2d-6)} ,  L^{-\frac{d}{2} - (d-4) \kb} \}	.
\end{align}
It suffices now to verify that the right-hand side of \eqref{eq:DKPhi-5}
is bounded
above by $\frac12 C_{\rg} \tilde{\vartheta}_+^3 \tilde{g}_+^3 \scale_+^{\kaa}$.
This requires some bookkeeping in order to advance the scale.  For $\tilde{\vartheta}$
and $\tilde{g}$, the advancement to $\tilde{\vartheta}_+$
and $\tilde{g}_+$ is at the cost of an $L$-independent constant.  For $d=4$, in which
case $\scale=1$, this cost is overcome by the $L^{-2}$ on the right-hand side
of \eqref{eq:DKPhi-5} with $L$ taken sufficiently large (recall that $M_{0,1}'$ is
independent of $L$) to reduce the right-hand side of \eqref{eq:DKPhi-5}
to $\frac12 C_{\rg} \tilde{\vartheta}_+^3 \tilde{g}_+^3 \scale_+^{\kaa}$.
For $d>4$, we need the powers of $L$ on the right-hand side of \eqref{eq:DKPhi-5}
also to overcome the ratio $(\scale/\scale_+)^{\kaa} = L^{(d-4)\kaa}$.
For this, it is sufficient if we have both
\begin{equation}
    2d-6 > (d-4)\kaa  \qquad \text{and} \qquad \frac d2 + (d-4)\kb > (d-4)\kaa.
\end{equation}
Both of these are guaranteed by \eqref{eq:abp-bds},
and the proof is complete.
\end{proof}

In the remainder of Section~\ref{sec:pf2}, we present the proofs of Propositions~\ref{prop:S0}, \ref{prop:S1}, \ref{prop:crucial}.
We start in Section~\ref{sec:S0} with the short proof of the bound on $S_0$, Proposition~\ref{prop:S0}. The bound on $S_1$,
Proposition~\ref{prop:S1}, is obtained in Section~\ref{sec:S1}.
In Section~\ref{sec:crucial}, we obtain the crucial contraction estimate, Proposition~\ref{prop:crucial}, which relies on a careful decomposition of $S_1$ and the computation of its $K$-derivative.
Throughout, we rely on the estimates of Section~\ref{sec:prelim2}.

\subsection{Bound on \texorpdfstring{$S_0$}{S0}: proof of Proposition~\ref{prop:S0}}
\label{sec:S0}

\begin{proof}[Proof of Proposition~\ref{prop:S0}]
Assume \ref{quote:assumPhi-bis} and suppose that
$\lambda_V, \lambda_K \leq \tilde{g} \scale$.
We proceed as in \cite[Section~10.3.2]{BBS-brief} with the important
improvement that our norm includes the regulator, which is
absent in \cite{BBS-brief}.

We start with the formula \cite[(10.3.10)]{BBS-brief}, which states that
\begin{align}
\label{eq:S0EA3-mass_derivative}
	S_0
	= e^{-U_{+} (B) + u_{+} |B|}
	\Big( \frac{1}{8} \big( \operatorname{Var}_+ \theta \hat{V} (B) \big)^2
	+ \E_+ A_3 (B) \Big),
\end{align}
where $A_3 (B) = -\frac{1}{2} (\delta \hat{V} (B))^3 \int_0^1 e^{-t \delta \hat{V} (B)} (1-t)^2 dt$
with $\delta \hat{V} = \theta \hat{V} (B) - U_+ (B)$.
It suffices to prove that
\begin{align}
\label{eq:S0pf1}
	\norm{e^{-U_{+}(B) + u_{+}|B|} }_{T_{\varphi, y} ( \mathfrak{h}_{+} , \vec{\lambda})}
    & \leq 2 e^{ -2 \cst |\varphi/ h_{+}|^4},
    \\
\label{eq:S0pf2}
    \norm{\operatorname{Var}_+ \theta \hat{V} (B)}_{T_{\varphi, y} ( \mathfrak{h}_{+} , \vec{\lambda})}
    & \le
	\begin{cases}
	C_L \tilde{\vartheta}_+^2 \tilde{g}_+^2 \scale_+^2
	P_{\ell_+}^4(\varphi)  & (\mathfrak{h}_{+} = \ell_{+}) \\
	C_L \tilde{\vartheta}_+^2
    \tilde{g}_+  \scale_+  P_{h_+}^4(\varphi) & (\mathfrak{h}_{+} = h_{+}),
	\end{cases}
\\
\label{eq:S0pf3}
	\norm{e^{-U_{+}(B) + u_{+}|B|} \E_{+} A_3}_{T_{\varphi, y} (\mathfrak{h}_{+}, \vec{\lambda})}
    & \le
	\begin{cases}
	C_L \tilde{\vartheta}_+^3 \tilde{g}_+^3 \scale_+^3
	P^{12}_{\ell_+}(\varphi) & (\mathfrak{h}_{+} = \ell_{+}) \\
	C_L \tilde{\vartheta}_+^3
    \tilde{g}_+^{3/4} \scale_+^{3/4} G_{+}(B,\varphi) & (\mathfrak{h}_{+} = h_{+}).
	\end{cases}
\end{align}
Indeed, for $\hf_+ =h_+$
the product of the exponential factor in \eqref{eq:S0pf1} and the polynomial
factor $P_{h_+}$ in \eqref{eq:S0pf2} combine to produce
a regulator,  since, as in \eqref{eq:PG},
\begin{equation}
	P_{h_+}^{4} ( \varphi )e^{ - 2 \cst  |\varphi/h_+|^4} \leq CG_+(B,\varphi).
\end{equation}
For $\hf_+ = \ell_+$, to prove \eqref{eq:S0_bound1}
we only need to consider $\varphi = 0$
and no role is played by $P_{\ell_+}(0)=G_+(B,0)=1$.

The bound \eqref{eq:S0pf1} follows from  Lemma~\ref{lemma:stability_estimate-mass_derivative}.
Also,  by \eqref{eq:Vhat_bound0}
 and Lemma~\ref{lemma:polynomial_expectation_mass_derivative}, the bound
 \eqref{eq:S0pf2} holds for $\varphi=0$, and hence it also holds in general
 by Lemma~\ref{lemma:T_z_deriv_and_phi_to_0} because
 $\operatorname{Var}_+ \theta \hat{V} (B)$ is a polynomial of degree four.

For \eqref{eq:S0pf3}, we observe that
\begin{align}
	& \norm{e^{-U_{+}(B) + u_{+}|B|} \E_{+} A_3}_{T_{\varphi, y} (\mathfrak{h}_{+}, \vec{\lambda})} \nnb
	& \qquad \leq \frac{1}{6} \sup_{t\in [0,1]} \Big\| \E_+ \Big( \delta \hat{V} (B)^3 e^{-(U_+ - u_+) (B) - t \delta \hat{V} (B)} \Big) \Big\|_{T_{\varphi, y} (\mathfrak{h}_+, \vec{\lambda})} \nnb
	& \qquad \leq C_L \sup_{t\in [0,1]} \sup_{b_1, b_2,b_3 \in B} \Big\| \E_+ \delta V (b_1) \delta V (b_2) \delta V (b_3) \prod_{b \in B} e^{- (U_+ - u_+) (b) - t \delta \hat{V} (b)} \Big\|_{T_{\varphi, y} (\mathfrak{h}_+, \vec{\lambda})}
	\nnb
	& \qquad \leq C_L \sup_{t\in [0,1]}\sup_{b' \in B}
\Big( \E_+ \norm{\delta \hat{V} (b')}^6_{T_{\varphi, y}  (\mathfrak{h}_+, \vec{\lambda})}
	\prod_{b \in B}  \norm{ e^{- (U_+ - u_+) (b) - t \delta \hat{V} (b)}  }^2_{T_{\varphi, y} (\mathfrak{h}_+, \vec{\lambda})} \Big)^{1/2}
,
\label{eq:A3bd}
\end{align}
where we used
$\delta \hat{V} (B)= \sum_{b \in B} \delta \hat{V} (b)$
for the second inequality and
Lemma~\ref{lemma:mass_derivative_of_expectation}
for the third inequality.
The exponent in the product over $b$ on the right-hand side of \eqref{eq:A3bd} can
be rewritten as
\begin{equation}
    -(U_+-u_+)(b) - t\delta\hat V(b) = u_+(b)+(1-t)U_+(b)-t\theta\hat V(b).
\end{equation}
The term $u_+(b)$ is insignificant, even after taking the product over blocks $b$,
due to \eqref{eq:vacuum_energy_estimate-mass_derivative}.  For the
other terms, we use \eqref{eq:stability_estimate-mass_derivative} with $t'= (1-t)L^{-d}$,
and also \eqref{eq:V-Q_bound1-hplus}, to see that
\begin{align}
    &\norm{ e^{- (1-t)U_+ (b) } }_{T_{\varphi, y} (\mathfrak{h}_+, \vec{\lambda})}
    \norm{ e^{ - t \theta\hat V(b)} }_{T_{\varphi, y} (\mathfrak{h}_+, \vec{\lambda})}
    \nnb & \qquad
    \le
    \big(2^{(1-t)L^{-d}/2}
    e^{-2(1-t)L^{-d}\cst|\varphi/h_+|^{4}}\big)
    \big(
    2^{tL^{-d}/4}e^{-4t\cst L^{-d}|(\varphi+\zeta_b)/h_+|^4} \big).
\end{align}
The product over $b$ of the above expression is bounded above by a multiple of
\begin{align}
    e^{-2(1-t)\cst|\varphi/h_+|^{4}}
    e^{-4t\cst L^{-d}\sum_b |(\varphi+\zeta_b)/h_+|^4} .
\end{align}
Since $L^{-d}h_+^{-4} \geq \frac12  h^{-4}$, it follows from the almost sure
bound of Lemma~\ref{lemma:E_G_j} that the above two factors combine in an upper
bound to produce the regulator $G_+^4(B,\varphi)$ (for $\kappa$ small
compared to $\cst$).  This can be factored out of the
expectation and after the square root it gives $G_+^2(B,\varphi)$.
What remains is the square root of
\begin{align}
	\label{eq:A3pf3}
	\E_+ \norm{\delta \hat{V} (b')}^6_{T_{\varphi, y}  (\mathfrak{h}_+, \vec{\lambda})}
	& \le
	C \big( \frac{\mathfrak{c}_+}{\hf_+} \big)^6 \|V(b')\|^6_{T_{0,y}(\hf_+)} P_{\hf_+}^{24}(\varphi)
	\; \E_+  P_{\ell_+}^{24}(\zeta_b),
\end{align}
where we used an adaptation of \cite[Proposition~7.6.4]{BBS-brief} for the inequality.
The factor $\E_+  P_{\ell_+}^{24}(\zeta_b)$ is bounded by a constant by \cite[Lemma 10.3.1]{BBS-brief}.
By \eqref{eq:V_bound0} and the definitions of $\mathfrak{c}_+$ (recall \eqref{eq:cplusdef})
and $h_+$,
the combination $( \frac{\mathfrak{c}_+}{\hf_+} )^6 \|V(b)\|^6_{T_{0,y}(\hf_+)}$
gives $(\tilde\vartheta_+ \tilde g_+  \scale_+)^6$ for $\hf_+=\ell_+$ and gives
$(\tilde\vartheta_+\tilde g_+^{1/4} \scale_+^{1/4})^6$ for $\hf_+=h_+$ which, after taking the square root, is what appears  in \eqref{eq:S0pf3}.
Finally, if $\hf_+ = h_+$ then we use one of our two regulators to control $P^{12}_{h_+}(\varphi)G_+(B,\varphi) \leq C$, whereas if $\hf_+ = \ell_+$ then we neglect the regulators and simply conserve the polynomial $P^{12}_{\ell_+}(\varphi)$.
This completes the proof.
\end{proof}

\subsection{Bound on \texorpdfstring{$S_1$}{S1}: proof of Proposition~\ref{prop:S1}}
\label{sec:S1}

\begin{proof}[Proof of Proposition~\ref{prop:S1}]

For $|X| \geq 1$, we define
\begin{align}
	F (B, X)  = e^{u_{+} |B|} \Eplus \theta \big( e^{-\hat{V} (B\backslash X)} \hat{K}^X \big)
\end{align}
so that $S_1 = \sum_{X\subset B, |X| \geq 1} F(B, X)$.
We first prove the estimates
\begin{align}
\label{eq:bound_on_S1_key_inequality-l}
	\norm{F(B, X)}_{T_{0, y} (\ell_{+}, \vec{\lambda})}
	&\leq C_L \big( \tilde{\vartheta}_+^{3} \tilde{g}_+^3 \scale^{\kaa}  + \lambda_K \big)^{|X|} , \\
	\norm{F(B, X)}_{T_{\varphi, y} (h_{+}, \vec{\lambda})}
	& \leq C_L \Gplus (B, \varphi) \big( \tilde{\vartheta}^{3} \tilde{g}_+^{3/4} \scale^{\kaa - \kb} + \lambda_K (\tilde{g}^{9/4} \scale^{\kb})^{-1}  \big)^{|X|}.
\label{eq:bound_on_S1_key_inequality-h}
\end{align}
By \eqref{eq:vacuum_energy_estimate-mass_derivative} the norm of $\exp(u_+|B|)$ is bounded by a constant so we only have to bound the norm of
$\Eplus \theta (e^{-\hat{V} (B \backslash X)} \hat{K}^X)$.

We start with \eqref{eq:bound_on_S1_key_inequality-h}.
Since $\hat{V} = V^* - Q$, and since
$L^{-d}h_+^{-4} \ge \frac12 h^{-4}$, \eqref{eq:V-Q_bound1-h} gives
\begin{align}
	\norm{e^{-\hat{V} (b)} }_{T_{\varphi, y} (h_+ , \vec{\lambda})} \leq C \Gj (b, \varphi)   \qquad (b\in \blocks).
 	\label{eq:exp_hat_V_is_bounded_by_G-mass_derivative}
\end{align}
This bound, together with \eqref{eq:EFB4},
\eqref{eq:hat_K_bound1}, and
our hypothesis that $\norm{K}_{\Wkappa} \leq C_{\rg} \tilde{\vartheta}^3 \tilde{g}^3 \scale^{\kaa}$,  imply
\begin{align}
	\norm{\Eplus \theta e^{-\hat{V} (B\backslash X)} \hat{K}^X}_{T_{\varphi, y} (h_{+}, \vec{\lambda})}
		& \leq C_L
    \Gplus (B, \varphi) \prod_{b \in B \backslash X}  \norm{e^{- \hat{V} (b)}}_{T^G_{y} (h_+, \vec{\lambda})} \prod_{b \in X}   \norm{\hat{K} (b) }_{T^G_{y} (h_+, \vec{\lambda})}  \nnb
		& \leq C_L
    \Gplus (B, \varphi) (\tilde{\vartheta}^{3} \tilde{g}^{3/4} \scale^{\kaa-\kb}
    + \lambda_K (\tilde{g}^{9/4} \scale^{\kb} )^{- 1})^{|X|} .
\end{align}
Thus we have \eqref{eq:bound_on_S1_key_inequality-h}.
Similarly, use of \eqref{eq:EFB3} instead of \eqref{eq:EFB4} gives
\begin{align}
	\norm{ \Eplus \theta e^{-\hat{V} (B\backslash X)} \hat{K}^X}_{T_{0, y} (\ell_{+}, \vec{\lambda})} &\leq C_L \prod_{b \in B\backslash X} \norm{e^{-V(b)}}_{\cW^{0}_{y,+}} \prod_{b\in X} \norm{\hat{K} (b)}_{\cW^{0}_{y,+}} \nnb
	& \leq  C_L (\tilde{\vartheta}^3 \tilde{g}^{3} \scale^{\kaa} + \lambda_K )^{|X|},
\end{align}
where we used \eqref{eq:V-Q_bound1-hplus} and \eqref{eq:hat_K_bound2} in the second inequality.
This gives \eqref{eq:bound_on_S1_key_inequality-l}.

Next, we bound the sum $S_1 (B) = \sum_{X\in \blocks (B), |X| \geq 1} F(B, X)$.
For this, we give a name to factors occurring on the right-hand sides of
\eqref{eq:bound_on_S1_key_inequality-l}--\eqref{eq:bound_on_S1_key_inequality-h}:
\begin{align}
	f(\mathfrak{h}_{+}) = \begin{array}{ll}
	\begin{cases}
	\tilde \vartheta^3 \tilde{g}^3 \scale^{\kaa} + \lambda_K
	& (\mathfrak{h}_{+} = \ell_{+}) \\
	\tilde{\vartheta}^{3} \tilde{g}^{3/4} \scale^{\kaa-\kb} + \lambda_K (\tilde{g}^{9/4} \scale^{\kb} )^{- 1}
	& (\mathfrak{h}_{+} = h_{+}).
	\end{cases}
	\end{array}
\end{align}
Then
\begin{align}
\label{eq:S1ell}
\norm{S_1 (B)}_{T_{0, y} (\ell_{+}, \vec{\lambda})} \leq C_L \sum_{X\subset \blocks (B),\, |X| \geq 1} f(\ell_{+})^{|X|} = C_L \big( (1+ f( \ell_{+}) )^{|\blocks (B)|} - 1 \big).
\end{align}
Since $\lambda_K \leq \tilde{g}^{9/4} \scale^{\kb} \leq 1$, we have $f(\ell_{+}) \leq 2$, so
there is a constant $C_L$ such that
\begin{align}
 (1+ f(\ell_{+}) )^{|\blocks (B)|} - 1 \leq e^{L^d f( \ell_{+}) } -1 \leq C_L f(\ell_{+}),
\end{align}
and hence
\begin{align}
    \norm{S_1 (B)}_{T_{0, y} (\ell_{+}, \vec{\lambda})} \leq C_L f(\ell_{+}).
\end{align}
A similar computation applies with $\norm{\cdot}_{T_{0,y}(\ell_{+}, \vec{\lambda})}$ replaced by $\norm{\cdot}_{T_{\varphi, y}(h_{+}, \vec{\lambda})}$, but
with a factor $\Gplus (B, \varphi)$, so that
\begin{align}
\label{eq:S1h}
	\norm{S_1 (B)}_{T_{\varphi, y} (h_{+}, \vec{\lambda})}
	\leq
	C_L G_+(B,\varphi) f(h_{+}).
\end{align}

Finally, we bundle the bounds \eqref{eq:S1ell} and \eqref{eq:S1h} into
a bound on the $\cW^\kappa_{y,+}(\vec\lambda)$ norm
(defined by \eqref{eq:Wextended}).
The result is the desired conclusion \eqref{eq:S1bd}.
\end{proof}

\subsection{Contraction of \texorpdfstring{$\Phi^K_+$}{PhiK+}: proof of Proposition~\ref{prop:crucial}}
\label{sec:crucial}

We now prove Proposition~\ref{prop:crucial}, which is the
contraction estimate that is crucial for the repeated iteration of the
RG map when started at or near
the critical point.
Our goal is to prove that
the $K$-derivative of $\Phi_+^K (V,K)$ is bounded above by an $L$-independent
multiple of the maximum of $L^{-(2d-6)}$ and $L^{-\frac d2 - (d-4)\kb}$ (each
is $L^{-2}$ when $d=4$).
We again use the decomposition
\begin{align}
	& K_+ = \Phi_+^K (V,K,\ka) = S_0 + S_1  \label{eq:PhiS}
\end{align}
of \eqref{eq:Phi_+_is_S_0+S_1}.
In the following lemma we see that the $K$-derivative of $S_0$ is negligible in the sense that it can be made smaller than any $L$-dependent constant (upon choosing $\tilde g$ small enough).

\begin{lemma}
\label{lemma:S_0_contraction}
Let $\eps_L$ be any small $L$-dependent constant and assume \ref{quote:assumPhi-bis}.  Then for
$0 \le r \le p_{\ka}$,
\begin{equation}
	\norm{D_{\ka}^r D_K S_0}_{\domRG   \rightarrow \Wkappa_{+}}
	\leq \eps_L.
\end{equation}
\end{lemma}

\begin{proof}
By Proposition~\ref{prop:S0} with the choice $\lambda_K =  \tilde{g}\scale$,
and by using the norm to bound the $K$-derivative as in \eqref{eq:DVDK_bound},
\begin{align}
\norm{D_{\ka}^r D_K S_0}_{\Wkappa_+} \leq C_L \lambda_K^{-1} \tilde{\vartheta}^3_{+} \tilde{g}_{+}^3 \scale_+^{3}
= C_L \tilde{\vartheta}^3_{+} \tilde{g}_+^2 \scale_+^{2} .
\end{align}
Since we assume that $\tilde g$ is small (depending on $L$),
and since $\tilde g_+ \le \tilde g$
by the recursion and Theorem~\ref{thm:Phi^U_estimate}, we have the desired conclusion.
\end{proof}

\subsubsection{Extraction of linear term}

It remains to bound the $K$-derivative of $S_1$.
For this, we extract from $S_1$ its linear
term in $K$, as in \cite[(10.5.9)--(10.5.10)]{BBS-brief}, namely
\begin{align}
	S_1 =   e^{u_{\pt} |B|} \Eplus \theta \cL K + \cE ,
	\label{eq:K_+_linear_nonlinear_decompoisition}
\end{align}
with
\begin{align}
\label{eq:LK}
	\cL K (B) &= \sum_{b\in \blocks (B)} \Big( e^{-V (B)} (1- \Loc) (e^{V(b)} K(b))  \Big)
\end{align}
and with $\cE$ then implicitly defined.
By definition, $S_1$ is zero when $K$ is zero, as is $\cE$.

While the optimal bound on $D_{K} \Eplus \theta \cL K$ with correct
$L$ dependence requires careful treatment,
the contraction of $\cE$ is relatively easy because $\cE$ comprises terms that are at least second order in $K$.
The bounds we need are summarised in the following lemmas, whose proofs are given respectively in Sections~\ref{sec:1-Loc_contraction} and \ref{sec:cEbd}.
As we show below the statement of the lemmas, they quickly imply Proposition~\ref{prop:crucial}.

\begin{lemma} \label{lemma:S_1_contraction-v4}
Assume \ref{quote:assumPhi-bis} and $\norm{\dot{K}}_{\Wkappa_j } < \infty$.
There is an $L$-independent constant $C>0$ such that, for $0 \le r \le p_{\ka}$,
\begin{align}
	\norm{D_{\ka}^r \Eplus \theta \cL  \dot{K} (B)}_{\Wkappa_+} \leq C
	\max\{ L^{-(2d -6)} ,  L^{-\frac{d}{2} - (d-4) \kb }  \}
	\norm{\dot{K}}_{\Wkappa }.
\end{align}
\end{lemma}

\begin{lemma} \label{lemma:crucial_contraction-v2}
Assume \ref{quote:assumPhi-bis}.
For $0 \le r \le p_{\ka}$,
the Fr{\'e}chet derivative of $D^r_{\ka} \cE (V, \cdot):\Wkappa\to\Wkappa_+$ vanishes at $K =0$.
\end{lemma}

\begin{proof}[Proof of Proposition~\ref{prop:crucial}]
	We seek a bound on the norm of $D_K \Phi_+^K$.
	By the Fundamental Theorem of Calculus,
\begin{align}
	\label{eq:MVT_pf_contrac}
	&\norm{D_a^rD_K \Phi_+^K (V, K) }_{\domRG \rightarrow \Wkappa_+}
	\nnb & \qquad \leq
	\norm{D_a^rD_K \Phi^K_+ (V,0)}_{\domRG \rightarrow \Wkappa_+}
	+
	\sup_{t\in [0,1]}
	\norm{D_a^r D^2_K \Phi^K_+ (V,  t K) }_{\domRG^2 \rightarrow \Wkappa_+} \norm{K}_{\Wkappa}.
\end{align}
For the second term we
apply Theorem~\ref{thm:Phi^K_estimate-bis} with $(p,q) = (0,2)$
(recall that this follows from Propositions~\ref{prop:S0} and \ref{prop:S1}, so our
argument is not circular here), and obtain
\begin{align}
	\label{eq:D_K_2Phi_contrac_pf}
	\norm{D_a^r D^2_K \Phi_+^K (V,  t K) }_{\domRG^2 \rightarrow \Wkappa_+}\norm{K}_{\Wkappa}
		\leq M_{0,2}\tilde g_+^{-9/4}\scale_+^{-\kb}  C_{\rg}\tilde g^3\scale^{\kaa}
	\leq O_L\big( \tilde g^{3/4} \scale^{\kaa-\kb}  \big).
\end{align}
The right-hand side is smaller than any small fixed $L$-dependent constant,
for $\tilde g$
sufficiently small.
For the first term in \eqref{eq:MVT_pf_contrac} we use
$\Phi_+ = S_0 +S_1$,
Lemma~\ref{lemma:S_0_contraction},
and Lemma~\ref{lemma:crucial_contraction-v2}, to see that
\begin{align}
	\norm{D_a^rD_K \Phi_+^K (V, 0) }_{\domRG \rightarrow \Wkappa_+}
	&\leq
	\norm{D_a^rD_K S_0(V,0) }_{\domRG \rightarrow \Wkappa_+}  + 	
	\norm{D_a^r D_K S_1(V,0) }_{\domRG \rightarrow \Wkappa_+} \nnb
	&\leq \eps_L + \norm{D_{\ka}^r D_Ke^{u_{\pt} |B|} \Eplus \theta \cL  \dot{K} (B)}_{\domRG \rightarrow \Wkappa_+},
\end{align}
where $\epsilon_L$ is as small as desired.
Since  $\cL K$ is linear in $K$, the Fréchet $K$-derivative of $\E_+ \theta\cL K$ in direction $\dot K$
is simply $\E_+ \theta\cL \dot K$.
Then, since $u_{\pt}$ does not depend on $K$, and using the estimates \eqref{eq:vacuum_energy_estimate-mass_derivative}--\eqref{eq:u_+_minus_u_pt} to bound $e^{u_{\pt} |B| }$, together with
Lemma~\ref{lemma:S_1_contraction-v4}, we obtain finally that
\begin{equation}
	\norm{D_a^rD_K \Phi_+^K (V, K) }_{\domRG \rightarrow \Wkappa_+}
	\leq C
	\max\{ L^{-(2d -6)} ,  L^{-\frac{d}{2} - (d-4) \kb }  \},
\end{equation}
once we take
$\eps_L \le \max\{ L^{-(2d -6)} ,  L^{-\frac{d}{2} - (d-4) \kb }  \}$.
This completes the proof.
\end{proof}

\subsubsection{Contraction of \texorpdfstring{$\cL$}{L}: proof of Lemma~\ref{lemma:S_1_contraction-v4}}
\label{sec:1-Loc_contraction}

The following lemma from \cite{BBS-brief}
is at the heart of the proof of contraction.  It is used to implement the
idea  that
$(1-\Loc)F$ contains only irrelevant contributions which contract under
change of scale.  Note that this good operator $1-\Loc$ appears explicitly in \eqref{eq:LK}.

\begin{lemma}\cite[(10.5.11)]{BBS-brief}
\label{lemma:Loc_contraction}
Let $F: \R^n \rightarrow \R$ be $O(n)$-invariant and $\mathfrak{h} \geq \mathfrak{h}' > 0$.
Then
\begin{align}
	\norm{(1-\Loc) F}_{T_{\varphi} (\mathfrak{h}')}
	\leq 2 \Big( \frac{\mathfrak{h}'}{\mathfrak{h}} \Big)^6  P_{\mathfrak{h}'} (\varphi)^6  \sup_{0\leq t \leq 1} \norm{F}_{T_{t\varphi}(\mathfrak{h})}
	.
	\label{eq:Loc_contraction-v1}
\end{align}
\end{lemma}

We decompose the operator $\cL$ from \eqref{eq:LK} as
\begin{align}
	\label{eq:L_decomp}
	\cL \dot{K}  = \sum_{q=0}^{2} \frac{1}{q !} \cL_1 (V^q \dot{K} ) + \cL_2 \dot{K}
	,	
\end{align}
where
\begin{align}
	\cL_1 (V^q \dot{K}) (B)
    &=
    \sum_{b\in \blocks (B)} \Big( e^{-V  (B)} (1- \Loc) (V^q \dot{K}) (b) \Big),
	\\
	\cL_2  \dot{K} (B) &=
	\sum_{b\in \blocks (B)} \Big( e^{-V  (B)} (1- \Loc)
    \Big[\big( e^{V   }  -1 - V  - \frac 12 V^2   \big) \dot{K}  \Big](b) \Big).
\end{align}
(This decomposition is different from that in \cite[(10.5.16), (10.5.17)]{BBS-brief}.)
We prove
contraction estimates for $\cL_1 (V^q \dot{K})$ and $\cL_2 \dot{K}$
in Lemmas~\ref{lemma:S_1_contraction-v1} and \ref{lemma:S_1_contraction-v2}, respectively.
These estimates then imply Lemma~\ref{lemma:S_1_contraction-v4}, as we show at the
end of this section.
The next two lemmas,
Lemmas~\ref{lemma:S_1_contraction-v1}--\ref{lemma:S_1_contraction-v2}, involve
extensions of and small corrections to \cite[Section~10.5]{BBS-brief}.
The extensions are improvements due to the inclusion of our regulator in the norm.

\begin{lemma} \label{lemma:S_1_contraction-v1}
Assume \ref{quote:assumPhi-bis} and $\norm{\dot{K}}_{\Wkappa } < \infty$.
There is an $L$-independent $C$ such that for
$0 \le r \le p_{\ka}$
and for bounded $q \geq 0$,
\begin{align}
	& \norm{D_{\ka}^r \Eplus \theta \cL_1 (V^q \dot{K} ) (B)}_{T_0 (\ell_+)}
	\leq C L^{-(2d-6)} \norm{\dot{K}}_{\Wkappa }
    ,
	\label{eq:S_1_contraction-a}	
	\\
	& \norm{D_{\ka}^r \Eplus \theta \cL_1 (V^q \dot{K} ) (B)}_{T^{G_+} (h_+) }
	\leq C L^{-d/2} \norm{\dot{K}}_{T^G (h) }  .
	\label{eq:S_1_contraction-b}	
\end{align}
\end{lemma}

\begin{proof}
By the triangle inequality and the definition of $\cL_1$,
\begin{align}
    \norm{D_{\ka}^r \Eplus \theta \cL_1 (V^q\dot{K}) (B)}_{T_{\varphi}  (\hf_{+})}
    \le
    L^d \sup_b
    \norm{D_{\ka}^r \Eplus \theta   e^{-V  (B)} (1- \Loc)  (V^q\dot{K}) (b)}_{T_{\varphi}  (\hf_{+})}.
\end{align}
We write the integrand on the right-hand side
as a product over blocks in $B$, and apply
Lemma~\ref{lemma:mass_derivative_of_expectation}.  This eliminates
the mass derivative and yields
\begin{align}
    &\norm{D_{\ka}^r \Eplus \theta \cL_1 (V^q\dot{K}) (B)}_{T_{\varphi}  (\hf_{+})}
    \nnb & \qquad
    \le
    O( L^d ) \sup_b
    \Big[\Eplus \Big(   \prod_{b'} \norm{e^{-V  (b')}}^2_{T_{\varphi+\zeta_{b'}}(\hf_{+})}
    \norm{(1- \Loc)  (V^q\dot{K}) (b)}^2_{T_{\varphi+\zeta_b}  (\hf_{+})}
    \Big)\Big]^{1/2}.
\end{align}
The potentially dangerous entropic factor $L^d$ is more than cancelled
by good powers of $L$ arising from $1-\Loc$.
To see this, we first note that it follows from
\eqref{eq:Loc_contraction-v1} that
\begin{align}
\label{eq:1-Loc}
	\norm{ (1- \operatorname{Loc}) (V^q\dot{K})(b) }_{T_{\varphi+\zeta_b} (\mathfrak{h}_{+})}
	&\leq O \Big( \frac{\mathfrak{h}_+}{\mathfrak{h}} \Big)^6 P_{\mathfrak{h}_+}^6 (\varphi+\zeta_b)
    \sup_{t \in  [0,1]}
    \norm{\dot{K}(b)}_{T_{t(\varphi+\zeta_b)} (\mathfrak{h}  )},
\end{align}
where we used Lemma~\ref{lemma:hat_K_bound}(i) to bound the norm of $V^q$
by an $L$-independent constant.
Also, by \eqref{eq:V-Q_bound1-hplus} (with $\kappa \leq 2 \cst$)
\begin{equation}
\label{eq:eVG}
    \norm{e^{-V  (b')}}_{T_{\varphi+\zeta_{b'}}(\hf_{+})}
    \le
    2^{L^{-d}} G^2(b',\varphi+\zeta_{b'})
    .
\end{equation}
Since $(\frac{\ell_+}{\ell})^6 = O(L^{-3(d-2)})$ and $(\frac{h_+}{h})^6 = O(L^{-3 d/2})$, this leads to
\begin{align}
\label{eq:cL1a}
    & \norm{D_{\ka}^r \Eplus \theta \cL_1 \dot{K} (B)}_{T_{\varphi}  (\hf_{+})}  \nnb
	& \qquad \le
    O( L^{-p(\hf, \hf_+)})
    \, \sup_{b} \;
    \Big[
    \Eplus \Big(   \prod_{b'} G(b',\varphi+\zeta_{b'})^4 P_{\mathfrak{h}_+}^{12} (\varphi + \zeta_b)
    \sup_{t \in  [0,1]}
    \norm{\dot{K}(b)}^2_{T_{t(\varphi+\zeta_b)} (\mathfrak{h}  )}
    \Big)\Big]^{1/2},
\end{align}
where $p(\ell, \ell_+) = 2d-6$ and $p(h,h_+) = d/2$.

For the case $\hf=h$, since
$\norm{\dot{K}}_{T_{t(\varphi+\zeta_b)} (h)}  \le \norm{\dot{K}}_{T^G(h)}$
by definition of the norm, it follows that
\begin{align}
	\norm{D_{\ka}^r  \Eplus \theta \cL_1 \dot{K} (B)}_{T_{\varphi}  (h_{+})}
    &
    \le
    O(L^{-d/2}) \norm{\dot{K}}_{T^G(h)}
   	\,\sup_{b}\;\Big[\Eplus \Big(   \prod_{b'} G(b',\varphi+\zeta_{b'})^4
   	P_{h_+}^{12} (\varphi + \zeta_b )
    \Big)\Big]^{1/2}
    \nnb
    &
    \leq
    O(L^{-d/2}) \norm{\dot{K}}_{T^G(h)} G_+ (B, \varphi)
    \label{eq:S_1_contraction-1-proof-4},	
\end{align}
where the second inequality follows from Lemma~\ref{lem:G2G}.
This proves \eqref{eq:S_1_contraction-b}.

Finally, for the case $\hf=\ell$ with $\varphi=0$,  we bound all regulators by 1.
Also, by \eqref{eq:KKK} (at scale $j$),
\begin{align}
\label{eq:KKK-1}
    \sup_{t \in  [0,1]} \| \dot{K}\|_{T_{t\zeta_b}(\ell)}
    &\le
    P^{10}_\ell( \zeta_b)
    \norm{\dot{K}}_{\cW^0} \le P^{10}_{\ell_+}( \zeta_b) \norm{\dot{K}}_{\Wkappa}
    .
\end{align}
Then \eqref{eq:cL1a} becomes
\begin{align}
    \norm{D_{\ka}^r \Eplus \theta \cL_1 \dot{K} (B)}_{T_{0}  (\ell_{+})}
    \le
    O(L^{-(2d-6)})\norm{\dot{K}}_{\Wkappa}
    \Big[\Eplus  P^{32}_{\ell_+} (\zeta_b)
    \Big]^{1/2}.
\end{align}
Since
the expectation of $P_{\ell_+}^{32}(\zeta_b)$
is bounded by \cite[Lemma~10.3.1]{BBS-brief},
we obtain
\eqref{eq:S_1_contraction-a} and the proof is complete.
\end{proof}

\begin{lemma}
\label{lemma:S_1_contraction-v2}
Assume \ref{quote:assumPhi-bis} and $\norm{\dot{K}}_{\cW^\kappa  } < \infty$.
There is an $L$-independent $C$ such that for
$0 \le r \le p_{\ka}$,
\begin{align}
	& \norm{D_{\ka}^r \Eplus \theta \cL_2 \dot{K} (B)}_{T_0 (\ell_+)}
    \leq C L^{-(2d-6)} \norm{\dot{K}}_{\cW^\kappa } \label{eq:S_1_contraction-z}
    ,
    \\
    & \norm{D_{\ka}^r \Eplus \theta \cL_2 \dot{K} (B)}_{T^{G_+} (h_+)}
    \leq CL^{-d/2} \norm{\dot{K}}_{T^G (h)} \label{eq:S_1_contraction-y}
	.
\end{align}
\end{lemma}

\begin{proof}
Let $F_V  = (e^{V } - 1- V  - V^2 / 2)$.
Since the Taylor expansion of $F_V$ in $\varphi$ vanishes up to degree five,
$1 - \Loc$ acts as the identity
on $F_V \dot{K}$ and $e^{-V } F_V $, and thus
\begin{align}
	\cL_2 \dot{K} (B)
	&= \sum_{b\in\blocks(B)} e^{-V  (B )} F_V  (b)  \dot{K}(b)
	\nnb
	&= \sum_{b\in\blocks(B)} e^{-V  (B \backslash b)} \dot{K}(b) (1- \Loc) \big( e^{-V  (b)} F_V  (b)  \big)
	.
\end{align}
By the triangle inequality and Lemma~\ref{lemma:mass_derivative_of_expectation},
\begin{align}
    &\norm{D_{\ka}^r  \Eplus \theta \cL_2 \dot{K} (B)}_{T_{\varphi}  (\hf_+)}    \nnb
	& \le
    O(L^d)
    \sup_b
    \Big[ \Eplus
    \Big(   \norm{\dot{K}(b)}^2_{T_{\varphi + \zeta_b} (\hf_+)}
    \norm{(1- \Loc)  e^{-V  (b)} F_V  (b)}^2_{T_{\varphi+\zeta_b}  (\hf_+)}
    \prod_{b' \neq b} \norm{e^{-V  (b')}}^2_{T_{\varphi+\zeta_{b'}}(\hf_{+})}
    \Big)\Big]^{1/2}
    .
\end{align}
By Lemma~\ref{lemma:Loc_contraction} and Lemma~\ref{lemma:hat_K_bound}(i,ii),
\begin{align}
	\| (1- \Loc) e^{-V} F_V   \|_{T_{\varphi+\zeta_b} (\mathfrak{h}_+)}
	& \leq 2 \Big( \frac{\mathfrak{h}_+}{\mathfrak{h}} \Big)^6
    P^6_{\mathfrak{h}_+} (\varphi+\zeta_b)
    \sup_{t\in [0,1] } \|  e^{-V  } F_V   \|_{T_{t (\varphi+\zeta_b)} (\mathfrak{h})}
	\nnb
	& \leq O(L^{-d-p(\mathfrak{h}, \mathfrak{h}_+)}) P_{\mathfrak{h}_+}^6 (\varphi + \zeta_b),
	\label{eq:1-Loc_of_v2}
\end{align}
for either $(\mathfrak{h}, \mathfrak{h}') = (\ell, \ell_+)$ or $(h, h_+)$,
where again $p(\ell, \ell_+) = 2d-6$ and $p(h,h_+) = d/2$.
With \eqref{eq:eVG}, this gives
\begin{align}
\label{eq:cL2a}
    & \norm{D_{\ka}^r \Eplus \theta \cL_2 \dot{K} (B)}_{T_{\varphi}  (\hf_+)} \nnb
    & \qquad \leq
    O(L^{-p(\mathfrak{h}, \mathfrak{h}_+)} )  \sup_b
    \Big[ \Eplus \Big(
    \norm{\dot{K}}_{T_{\varphi+\zeta_b} (\hf_+)}^2
    P_{\hf_+}^{12} (\varphi+\zeta_b)
    \prod_{b' \neq b} G^4(b',\varphi+\zeta_{b'})
    \Big)\Big]^{1/2}
    .
\end{align}

For $\hf=h$,
it follows from  $h_+\le  h$ and
the monotonicity of the $T_\varphi(\hf)$-norm in $\hf$ that
$\norm{\dot{K}}_{T_{\varphi+\zeta_b} (h_+)}
\le \norm{\dot{K}}_{T_{\varphi+\zeta_b} (h)}
\le \norm{\dot{K}}_{T^G(h)}G(b,\varphi+\zeta_b)$, so
\begin{align}
\label{eq:cL2b}
    &\norm{D_{\ka}^r \Eplus \theta \cL_2 \dot{K} (B)}_{T_{\varphi} (h_+)}
    \nnb & \quad \le
    O(L^{-p(h,h_+)})  \norm{\dot{K}}_{T^G(h)} \sup_b
    \Big[ \Eplus \Big(
    G(b,\varphi+\zeta_b)^2
    P_{h_+}^{12} (\varphi+\zeta_b)
    \prod_{b' \neq b} G(b',\varphi+\zeta_{b'})^4
    \Big)\Big]^{1/2}
    .
\end{align}
The supremum on the right-hand side is at most $G_+(B,\varphi)$
by Lemma~\ref{lem:G2G}, so this proves
\eqref{eq:S_1_contraction-y}.

For $\hf=\ell$ and $\varphi=0$,
we apply \eqref{eq:KKK} to see that
\begin{align}
    \| \dot{K}\|_{T_{\zeta_b} (\ell_+)}
    & \le
    \| \dot{K}\|_{T_{\zeta_b} (\ell)}
    \le
    P^{10}_\ell( \zeta_b)  \norm{\dot{K}}_{\cW^0}
    \le P^{10}_\ell( \zeta_b)
    \norm{\dot{K}}_{\Wkappa}.
\end{align}
After substitution of this into in \eqref{eq:cL2a},
we bound all regulators by $1$, and then
the expectation is bounded.
This  proves
\eqref{eq:S_1_contraction-z}.
\end{proof}

\begin{proof}[Proof of Lemma~\ref{lemma:S_1_contraction-v4}]
By definition of the $\cW^\kappa$ norm in \eqref{eq:Wextended},
\begin{align}
    \norm{D_{\ka}^r \E_+ \theta \cL \dot K(B)}_{\cW_+^\kappa}
    & =
    \norm{D_{\ka}^r \E_+ \theta \cL \dot K(B)}_{T_0(\ell_+)}
    +
    \tilde g_+^{9/4}\scale_+^{\kb} \norm{D_{\ka}^r \E_+ \theta \cL \dot K(B)}_{T_+^G(h_+)}.
\end{align}
With the decomposition \eqref{eq:L_decomp},
Lemmas~\ref{lemma:S_1_contraction-v1} and Lemma~\ref{lemma:S_1_contraction-v2}
imply that there is an $L$-independent $C$ such that
	\begin{align}
		\norm{D_{\ka}^r \E_+ \theta \cL \dot K(B)}_{\cW_+^\kappa}
		\leq
        C\big(
		L^{-(2d-6)} + L^{-d/2} (\tilde g_+/\tilde g)^{9/4}(\scale_+/\scale)^\kb
    \big) \|\dot K\|_{\cW^\kappa}  .
	\end{align}
Since $\tilde g_+/\tilde g\le 1$ and $\scale_+/\scale = L^{-(d-4)}$,
this gives the desired result with its maximum of $L^{-(2d-6)}$ and $L^{-\frac d2
-(d-4)\kb}$.
\end{proof}

\subsubsection{Vanishing derivative of \texorpdfstring{$\cE$}{E}:
proof of Lemma~\ref{lemma:crucial_contraction-v2}}
\label{sec:cEbd}

Let
\begin{align}
	A(b) = e^{-V (b)} \big( 1+ Q (b) - e^{Q  (b)} \big),
    \qquad
    Q(b) = \Loc(e^{V(b)}K(b)).
\end{align}
It follows from the definition of $\cE$ in \eqref{eq:K_+_linear_nonlinear_decompoisition}
and basic algebra, exactly as in \cite[(10.5.10)]{BBS-brief}, that
\begin{align}
	\cE
    =   \Delta_1 + \Delta_2 + \Delta_3,
	\label{eq:nonlinear_terms_definition}
\end{align}
with
\begin{align}
	\Delta_1 & = (e^{u_{+} |B|} - e^{u_{\pt} |B|} ) \Eplus \theta \cL K ,
	\\
	\Delta_2 & = e^{u_{+} |B|} \sum_{b\in \blocks (B)} \Eplus \theta \big( e^{-\hat{V} (B\backslash b)} A (b)  \big),
	\\
	\Delta_3 & = e^{u_{+} |B|} \sum_{X\subset \blocks (B),\, |X| \geq 2} \Eplus \theta \big( e^{-\hat{V} (B\backslash X)} \hat{K}^X \big).
\end{align}
Although the formula for $\cE$ appears complicated, only crude estimates are needed.

\begin{proof}[Proof of Lemma~\ref{lemma:crucial_contraction-v2}]
Since $\cE=0$ when $K=0$, it suffices to prove that
\begin{equation}
\label{eq:ccpf}
    \norm{D_{\ka}^r  \cE(V,K)}_{\Wkappa_+}
    \le C
    \norm{K}_{\Wkappa}^2
\end{equation}
for $\norm{K}_{\Wkappa}$ sufficiently small.  Since we are letting $K$ tend to
zero with $L$ and $\tilde g$ held fixed, we can and do permit the above constant $C$
to depend (badly) on $L$ and $\tilde g$, and we use $C$ throughout the proof as a generic
such constant.
For \eqref{eq:ccpf},
it  suffices to prove that, for
$i \in \{1,2,3\}$,
\begin{align}
	\norm{D_{\ka}^{r} \Delta_{i} (V,K) }_{T_{0} (\ell_{+})}
    & \leq C \norm{K}_{\Wkappa}^2 ,
\label{eq:Deltas_differentiability_inequality-0}	
    \\
    \norm{D_{\ka}^r  \Delta_{i} (V,K) }_{T^{G_+} (h_{+})}
    & \leq C \norm{K}_{\Wkappa} \norm{K}_{T^G (h)}.
\label{eq:Deltas_differentiability_inequality}
\end{align}
We proceed as in the proof of  \cite[(10.5.29)--(10.5.30)]{BBS-brief},
with adaptations in particular to accommodate our regulator.

For $\Delta_1$, we see from \eqref{eq:S_1_contraction-a}
that $\norm{D_{\ka}^r \Eplus \theta \cL K }_{T_0 (\ell_{+})} \le
C
\norm{K}_{\Wkappa}$
and $\norm{D_{\ka}^r \Eplus \theta \cL K }_{T^{G_+} (h_{+})} \le
C
\norm{K}_{T^G(h)}$.
Also, \eqref{eq:vacuum_energy_estimate-mass_derivative}--\eqref{eq:u_+_minus_u_pt}
give $|e^{u_{+}|B|} - e^{u_{\pt} |B|}|\le C\|K\|_{\Wkappa}$.
This proves the $\Delta_1$ case of
\eqref{eq:Deltas_differentiability_inequality-0}--\eqref{eq:Deltas_differentiability_inequality}.

For $\Delta_2$, we first apply Lemma~\ref{lemma:E_F^B}
and then use \eqref{eq:V-Q_bound1-hplus}
to see that
\begin{align}
\label{eq:Del2pf0}
	& \norm{D^r_{\ka} \Delta_{2} (V, K) }_{T_0 (\ell_{+})}
    	\le C \sup_b \norm{A(b)}_{\cW^0_+}
    \prod_{b' \neq b} \norm{e^{-\hat V(b')}}_{\cW^0_+}
		\le C \sup_b \norm{A(b)}_{\cW^0_+}
    ,
    	\\
\label{eq:Del2pf}
    & \norm{D^r_{\ka}  \Delta_{2} (V, K) }_{T^{G_+}  (h_{+})}
    	\le C \sup_b \norm{A(b)}_{T^{G}(h_+)} \prod_{b' \neq b} \norm{e^{-\hat V(b')}}_{T^G (h_+)}
    	\le C \sup_b \norm{A(b)}_{T^{G}(h_+)},
\end{align}
for sufficiently small $\tilde g$.
Note that the regulator is $G$ in \eqref{eq:Del2pf}, not $G_+$.
By Taylor's Theorem,
\begin{align}
    A(b) = - e^{-V(b)}Q(b)^2 \int_0^1 (1-s) e^{sQ(b)} ds,
\end{align}
so
\begin{align}
    \|A(b)\|_{T_\varphi(\hf_+)}
    & \le
    \|A(b)\|_{T_\varphi(\hf)}
    \le
    \|Q(b)\|_{T_\varphi(\hf)}^2
    \sup_{s\in[0,1]}\| e^{-(V-sQ)(b)}\|_{T_\varphi(\hf)}.
\end{align}
By \cite[(9.3.19)]{BBS-brief}, $\|Q(b)\|_{T_\varphi(\hf)}
\le C\|K(b)\|_{T_0(\hf)}P_{\hf}^4(\varphi)$.
With \eqref{eq:V-Q_bound1-h} for the exponential factor, this gives
\begin{align}
	\norm{{A} (b)}_{T_{\varphi  (\mathfrak{h}_+)} }
	& \leq C \norm{K (b)}^2_{T_0 (\mathfrak{h})}
    P^8_{\mathfrak{h}} (\varphi) e^{-4\cst |\varphi / h|^4}
	\leq
    C\norm{K(b)}^2_{T_0 (\mathfrak{h})}e^{-2\cst |\varphi / h|^4}
    .
\end{align}
Thus for $\kappa \leq  2\cst$,
\begin{align}
	\norm{{A} (b)}_{T_{0 }(\ell_+)}
	& \leq C \norm{K}_{T_{0 }(\ell)}^2 ,
\qquad
    \norm{{A} (b)}_{T^{G}  (h_+)}
	\leq
    C
    \norm{K}_{T_{0 }(h)}^2 .
\end{align}
This proves the required bounds on $\Delta_2$,
since we are not concerned with the factor $(\tilde g \scale^\kb)^{-1}$ required
to bound $\norm{K}_{T_{0 }(h)}$ by $\norm{K}_{\Wkappa}$.

Finally, $\Delta_3$ is similar to $\Delta_2$, but with
$A(b)$ on the right-hand sides of \eqref{eq:Del2pf0}--\eqref{eq:Del2pf} now
replaced by $\hat K^X$ with $|X|\ge 2$.
Since $\|K\|_{\cW^0_+}  \le \|K\|_{\cW^0} \le \|K\|_{\Wkappa}$, the analogue
of \eqref{eq:Del2pf0} gives the $i=3$ version of \eqref{eq:Deltas_differentiability_inequality-0}.
Since $\|K\|_{T^G(h_+)}  \le \|K\|_{T^G(h)}$, the analogue of
\eqref{eq:Del2pf} gives the $i=3$ version of
\eqref{eq:Deltas_differentiability_inequality}.
The sum over $X$ has an $L$-dependent number of terms, but this does not matter,
and the sum is dominated by $|X|=2$ as $K\to 0$.  Also, the factor
$(\tilde g \scale^{\kb})^{-1}$ needed to convert the $T^G(h)$ norm
to the $\Wkappa$ norm does not matter.
This proves the $\Delta_3$ case of \eqref{eq:Deltas_differentiability_inequality}
and completes the proof.
\end{proof}

\appendix
\setcounter{section}{0}
\renewcommand{\thesection}{\Alph{section}}

\section{Monotonicity of the universal profile \texorpdfstring{$\uf_n$}{fn}}
\label{sec:ufmono}

The universal profile
\begin{equation}
    \uf_n(s)
    =
    \frac{\int_{\R^n} |x|^2 e^{-\frac 14 |x|^4 - \frac s2 |x|^2} dx}
    {n\int_{\R^n} e^{-\frac 14 |x|^4 - \frac s2 |x|^2} dx}
    =
    \frac{I_{n+1}(s)}{nI_{n-1}(s)}
    \qquad
    (n>0,\, s\in\R)
\end{equation}
is introduced in Section~\ref{sec:profile}.
It extends to $n \in [-2,\infty)$ using the recursion \eqref{eq:ufn-recursion}.
Plots of $\uf_n$ are given in
Figure~\ref{fig:ufplots}.

\begin{figure}[h]
    \centering
    \begin{subfigure}[h]{0.5\textwidth}
        \includegraphics[scale=0.6]{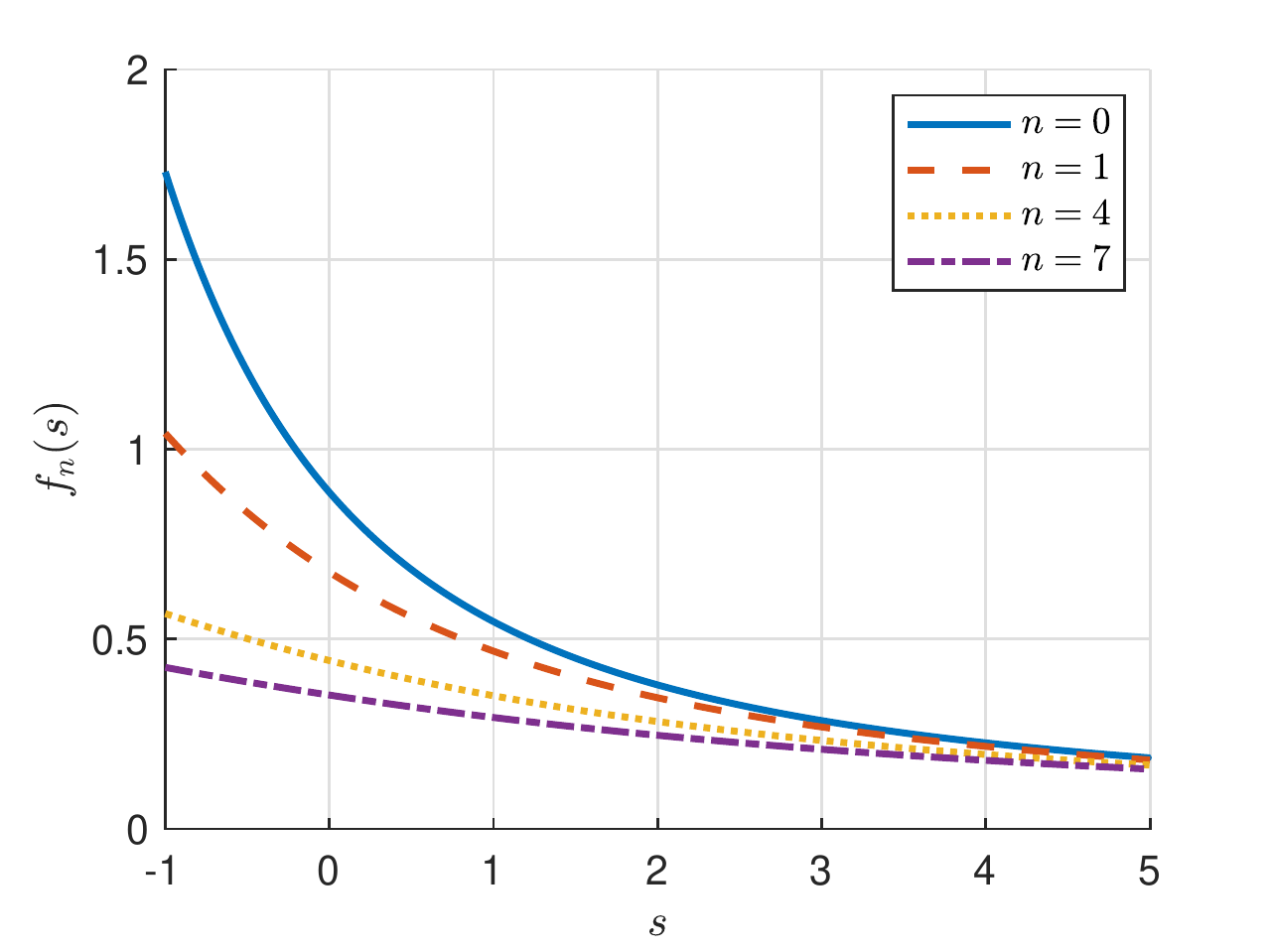}
    \end{subfigure}
    ~
    \begin{subfigure}[h]{0.5\textwidth}
        \includegraphics[scale=0.6]{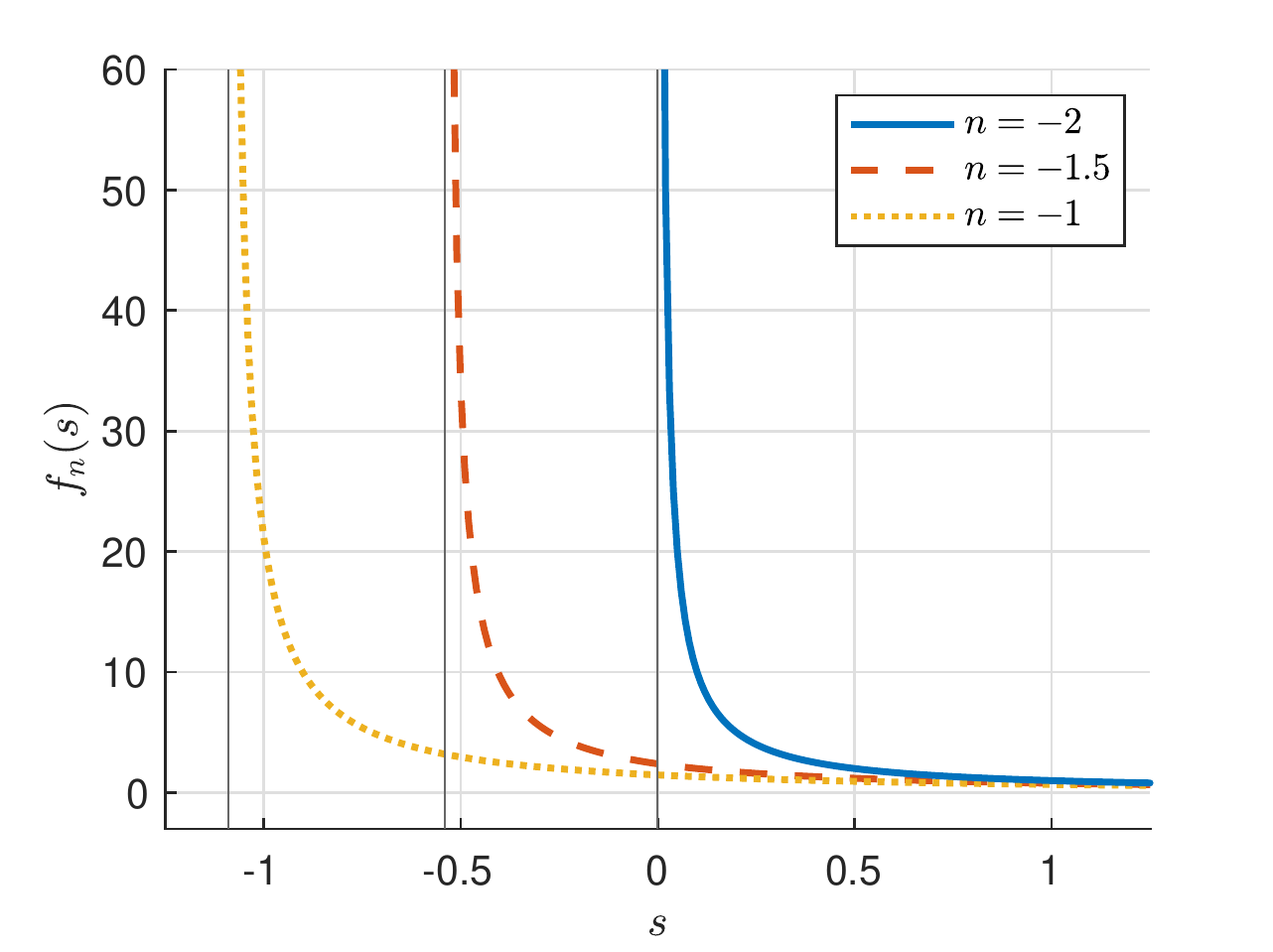}
    \end{subfigure}
    \caption{Plots of the universal profile $\uf_n$ \label{fig:ufplots} }
\end{figure}

Lemma~\ref{lemma:fn} proves elementary monotonicity properties of the universal profile $\uf_n$.
For its statement we define $s_n^*=-\infty$ if $n \ge 0$,
and $s_n^* = \sup\{ s : (n+2) f_{n+2} (s) + s =0 \}$ for $n\in [-2,0)$.
The divergence of $f_n$ at $s_n^*$ for $n \in [-2,0)$, visible in Figure~\ref{fig:ufplots}, follows from \eqref{eq:ufn-recursion}.

\begin{lemma}
\label{lemma:fn}
For $n \in [-2,\infty)$ and
$s > s_n^*$,
the universal profile
$\uf_n(s)$ is strictly decreasing both as a function of $s$ and as a function of $n$.
\end{lemma}

\begin{proof}
We first prove monotonicity in $s$.
For $n=0,-2$, the functions $f_0(s)=I_1(s)$ and $f_{-2} (s) = s^{-1}$ are monotone in $s$ by definition.
For $k>0$,
we define the probability measure
$d\mu_{k,s}(x) \propto x^{k-1} e^{-\frac{1}{4} x^4 - \frac12 s x^2 } dx$ on $(0,\infty)$.
Then for $n>0$ we have $n\uf_n (s) =  \int_0^\infty x^2 d\mu_{n,s}$, so
\begin{align}
2n\frac{d}{ds} \uf_n (s)  =
-\int x^4 d\mu_{n,s} + \left(\int x^2 d\mu_{n,s}\right)^2
.
\label{eq:nf_n_derivative}
\end{align}
Notice that, for any strictly increasing functions $f$ and $g$,
\begin{align}
\mu_{k,s} (fg) - \mu_{k,s}  (f) \mu_{k,s}  (g) = \frac{1}{2} \iint d\mu_{k,s}(x) d\mu_{k,s}(y) \big(f (x) - f(y) \big) \big( g(x)- g(y) \big) > 0
,
\label{eq:FKG_strict}
\end{align}
(this is the FKG inequality with strict inequality)
so we obtain $\frac{d}{ds} f_n (s) < 0$.
For $n \in (-2,0)$, by \eqref{eq:ufn-recursion} it is sufficient to prove that
\begin{align}
\frac{d}{ds} (n+2) \uf_{n+2} (s) > - 1
\quad \text{if} \;\; s > s_n^*
.
\label{eq:nf_n_derivative_lower_bound}
\end{align}
By \eqref{eq:nf_n_derivative} and \eqref{eq:nI_n-1_integration_by_parts}, we have
\begin{align}
2(n+2) \frac{d}{ds} \uf_{n+2} (s)
&=
-\frac{I_{n+5}(s)}{I_{n+1} (s)} + \Big( \frac{I_{n+3} (s)}{I_{n+1} (s)} \Big)^2 \nnb
& = - (n+2) + s (n+2) \uf_{n+2} (s) + \big((n+2) \uf_{n+2}(s) \big)^2 .
\end{align}
Since $s + (n+2) f_{n+2} (s) > 0$ for $s> s_n^*$, we see that indeed
$(n+2) \frac{d}{ds} \uf_{n+2} (s) > - \frac{n+2}{2} > -1$.

Finally, for the monotonicity in $n$, for
$k>-1$ and $q>0$ we have
\begin{align}
\frac{d}{dk} \frac{I_{q+k} (s)}{I_{k} (s)}
&= \int x^{q} \log x \, d\mu_{k,s}
- \left(\int x^q d\mu_{k,s} \right)\left(\int \log x \, d \mu_{k,s}  \right)
> 0
\label{eq:I_k_ratio_increasing}
\end{align}
where the $1$-dimensional integral $I_k(s)$ is defined in \eqref{eq:Ik_definition}, and
the final inequality follows from \eqref{eq:FKG_strict}.
(Note that this proves that the moments
$\Sigma_{n,k}(s)$ in \eqref{eq:Signks} are
monotone
\emph{increasing} in $n>0$.)
Therefore, by \eqref{eq:newuf} and by \eqref{eq:I_k_ratio_increasing},  for all $n >-2$ we have
\begin{align}
\frac{d}{dn} \frac{1}{\uf_n (s)}
= \frac{d}{dn} \Big( \frac{I_{n+3} (s)}{I_{n+1} (s)} + s \Big) > 0.
\end{align}
This proves the desired monotonicity in $n$,
and the proof is complete.
\end{proof}

A cursory examination of
Figure~\ref{fig:ufplots} may suggest
that $\uf_n$ is convex but this is misleading: Figure~\ref{fig:ufnotconvex}
shows that $s\mapsto \uf_1(s)+s$ is not convex and hence neither is $\uf_1$.

\begin{figure}[h]
	\begin{center}
	\includegraphics[scale=0.6]{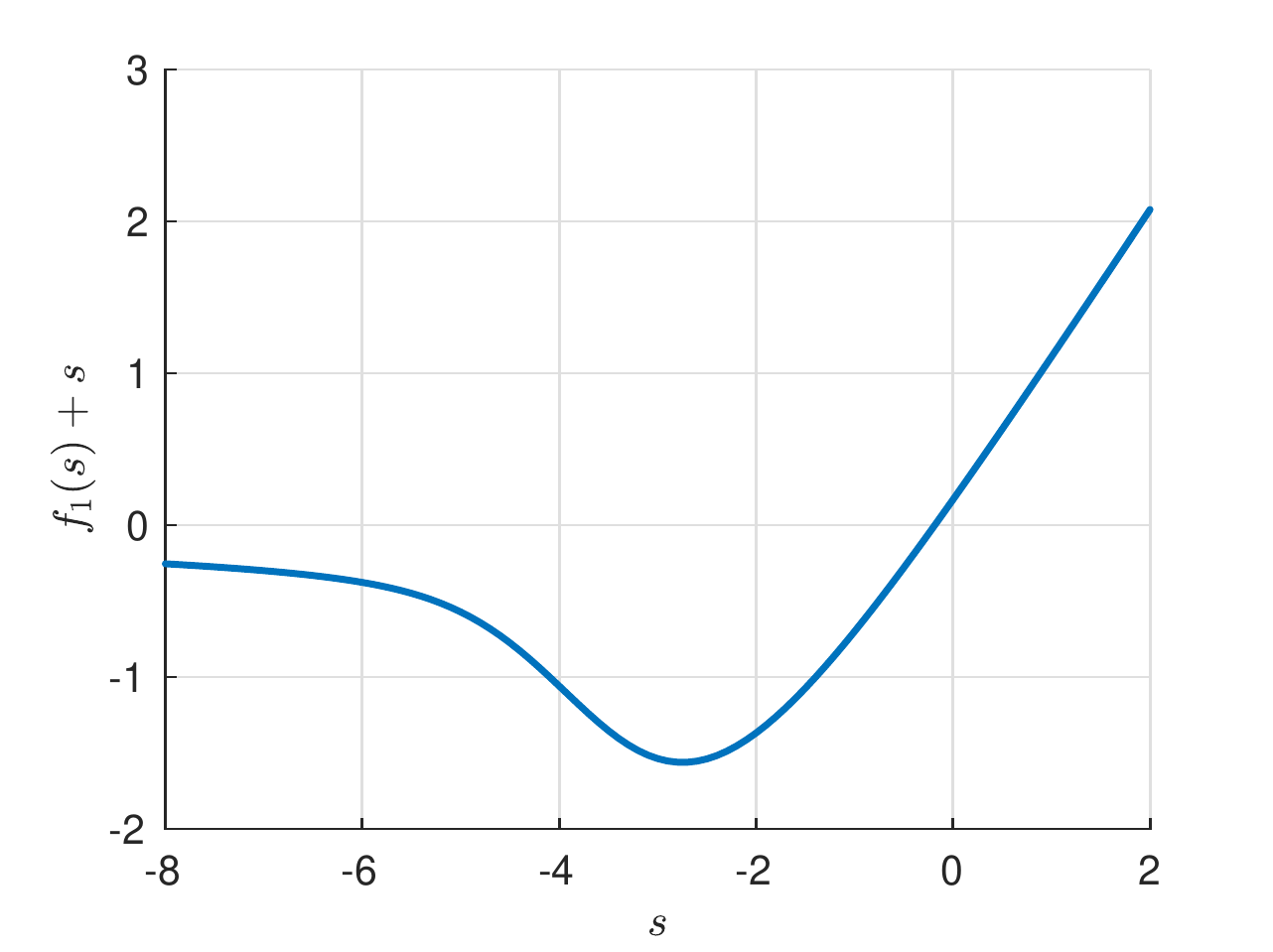}
	\caption{
\label{fig:ufnotconvex}
Plot of $f_1 (s )+s$ illustrating lack of convexity of $f_1(s)+s$ and hence of $f_1(s)$.
 }
	\end{center}
\end{figure}

\section{Self-avoiding walk: proof of Propositions~\ref{prop:SAW-KN}--\ref{prop:WSAW}}
\label{sec:SAW}

In this section we prove Propositions~\ref{prop:SAW-KN}--\ref{prop:WSAW},
which we restate here as Propositions~\ref{prop:SAW-KN-bis}--\ref{prop:WSAW-bis}.

\begin{proposition}
\label{prop:SAW-KN-bis}
For self-avoiding walk on the complete graph, as $N \to \infty$,
\begin{align}
\label{eq:chiKn-bis}
     \chi_N^{\mathbb{K}}(N^{-1}(1- s(2N)^{-1/2}))
    &\sim (2N)^{1/2} f_0( s)
    \qquad
    (s \in \R).
\end{align}
\end{proposition}

\begin{proof}
It is proved in \cite[Theorem~1.1]{Slad20}
on the basis of an exact rewriting of the susceptibility in terms
of the incomplete Gamma function, that, as $N \to \infty$,
\begin{align}
\label{eq:chicrit}
    \chi_N^{\mathbb{K}}(N^{-1}(1- s(2N)^{-1/2}))
    &\sim  N^{1/2}  \sqrt{\frac{\pi}{2}} e^{s^2/4}   {\rm erfc}(s/2),
\end{align}
where the complementary error function is defined for $x \in \R$ by
${\rm erfc}(x)= \frac{2}{\sqrt{\pi}} \int_x^\infty e^{-t^2}dt$.
With the change of variable $y = x^2 /2$ in the definition of $I_1$ in
\eqref{eq:Ik_definition}, we see that
\begin{align}
\label{eq:f0}
f_0 (s) &= I_1(s)
= e^{\frac{1}{4} s^2 } \int_0^{\infty} e^{-(y + s / 2)^2} dy
= \frac{\sqrt{\pi}}{2} e^{s^2/4} \operatorname{erfc} (s / 2) .
\end{align}
This proves \eqref{eq:chiKn-bis}.
\end{proof}

A detailed analysis of self-interacting walk on the complete graph is
given in \cite{BS20}.  Although the emphasis in \cite{BS20} is on elucidating
the phase diagram for a model with tricritical behaviour, much of
\cite{BS20} applies more generally and in particular to the
continuous-time weakly self-avoiding walk.
The critical point $\nu_c=\nu_c(g)$ is characterised in \cite[Definition~1.2]{BS20} as follows.
First, the effective potential $V: [0,\infty) \to \R$ is defined by
\begin{align}
    V(t) &= t - \log(1+v(t)),
\qquad
    v(t)  =
    \int_0^\infty e^{-gs^2-\nu s}e^{-s} \sqrt{\frac{t}{s}} \tilde{I}_1(2\sqrt{st}) \, ds.
\label{eq:Veff}
\end{align}
The critical point is then the choice of $\nu_c$ for which $V_c'(0)=0$, $V_c''(0)>0$
(the subscript $c$ denotes evaluation at $\nu_c$).
The function $\tilde{I}_1$ is the modified Bessel function of the first kind
(the tilde is used to avoid confusion with $I_1$ of \eqref{eq:Ik_definition}), and $V(0)=0$
since $\tilde{I}_1(0)=0$.  See Figure~\ref{fig:Veg} for an illustration.

In the following theorem, we
apply the results of \cite{BS20} to prove that the profile $f_0(s)$
arises for the susceptibility in this model in its critical window.
Note that the profile is independent of the value of $g>0$, as we have also seen for
$n \ge 1$ in Corollary~\ref{cor:suscept_crit_window}.

\begin{figure}[ht]
\centering{
\includegraphics[scale=0.79]{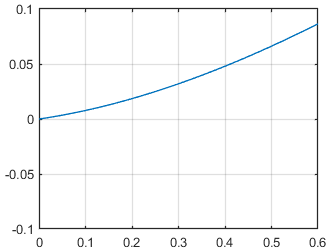}
\includegraphics[scale=0.79]{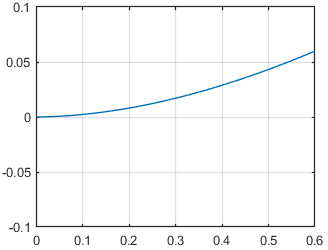}
\includegraphics[scale=0.79]{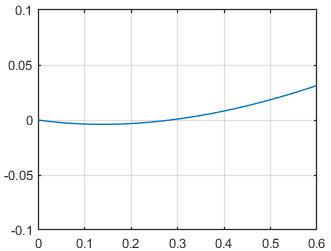}

\caption{Effective potential $V$ vs $t$ for $g=1$ and
$\nu=-1.1, -1.2, -1.3$.
The critical value is close to $-1.2$.}
\label{fig:Veg}}
\end{figure}

\begin{proposition}
\label{prop:WSAW-bis}
There are positive constants $\lambda_1,\lambda_2$ such that
\begin{align}
    \lim_{N \to \infty} N^{-1/2} \chi_N^{\rm WSAW}(\nu_c+sN^{-1/2}) & =
    \lambda_1 f_0(\lambda_2 s)
    \qquad
    (s \in \R).
\end{align}
\end{proposition}

\begin{proof}
At the critical point $\nu_c$, it is proved in \cite[Theorem~1.3]{BS20} that
$G_{00}(\nu_c) \to 1$ as $N \to \infty$, whereas $G_{01}(\nu_c)$ is of order $N^{-1/2}$.
Thus, at the critical point, $G_{00}$ does not contribute to the leading order of
$\chi$, which is $\chi(\nu_c) \sim N G_{01}$.  We now extend this to the scaling window
$\nu(s)=\nu_c+sN^{-1/2}$.

The combination of \cite[(3-5))]{BS20} and \cite[(3-17)]{BS20}
gives the formula
\begin{equation}
    G_{01}(\nu) = \int_0^\infty e^{-NV(t)}(NV'(t)F(t,t) - \partial_1 F(t,t)) \, dt,
\end{equation}
with $F(t_1,t_2) = (1-V'(t_1))^2 t_2$.  Integration by
parts then gives
\begin{align}
    G_{01}(\nu)
    &  =
    \int_0^\infty e^{-NV(t)}\Big(\frac{d}{dt}F(t,t) - \partial_1 F(t,t) \Big) \, dt
    \nnb & =
    \int_0^\infty e^{-NV(t)} \partial_2 F(t,t)  \, dt
    \nnb & =
    \int_0^\infty e^{-NV(t)} (1-V'(t))^2    dt
    .
\label{eq:G01}
\end{align}
(This integration by parts was not done in \cite{BS20}; its use would simplify some
computations in \cite{BS20}.)
We use prime and dot to denote derivatives
with respect to $t$ and $\nu$, respectively.
By Taylor expansion in the variable $s$,
as $N \to \infty$ we have
\begin{align}
    V_{\nu(s)}(t)
    & = V_c(t) + \dot{V}_c(t) sN^{-1/2} +O(s^2N^{-1}).
\label{eq:Vwindow}
\end{align}
The integral on the right-hand side  of \eqref{eq:G01}
is dominated by its behaviour near $t=0$
because $V$ grows at least linearly as $t \to\infty$ by \cite[Proposition~1.1]{BS20},
so we may restrict the integral to a finite interval
and use \eqref{eq:Vwindow}
and \cite[Ex.~4.1, p.333]{Olve97} to evaluate the asymptotic form of $G_{01}$
in the window.  In the notation of \cite[Ex.~4.1, p.333]{Olve97}, and with \eqref{eq:Vwindow}, we have
\begin{align}
    &p(t)  \sim Pt^2, \quad s(t) = St, \quad r(x,t)=O(1),
    \quad
    q(N,t)=(1-V_c'(t))^2 \sim 1,
    \nnb &
     P=\frac 12 V_c''(0), \quad S = -\dot{V}_c'(0) s,
     \quad \mu=2, \quad \sigma = 1,  \quad \lambda=1.
\end{align}
With $\Fi$ the Fax\'en integral \eqref{eq:Faxen}, we conclude that
\begin{align}
    \chi_N^{\rm WSAW}(\nu_c+sN^{-1/2}) & \sim NG_{01}(\nu_c+sN^{-1/2})
    \sim N \Fi ({\textstyle \frac{1}{2}, \frac{1}{2}; SP^{-1/2} }) \frac{1}{(PN)^{1/2}}.
\end{align}
According to \eqref{eq:IFi},
$\Fi ({\textstyle \frac{1}{2}, \frac{1}{2} ; u }) =2I_1 (-u)=2f_0(-u)$,
and therefore we arrive at
\begin{align}
    \chi_N^{\rm WSAW}(\nu_c+sN^{-1/2}) & \sim
    N^{1/2} \frac{2}{(\frac 12 V_c''(0))^{1/2}}
    f_0\Big( s\frac{\dot{V}_c'(0)}{(\frac 12 V_c''(0))^{1/2}} \Big),
\end{align}
which identifies the constants $\lambda_1,\lambda_2$ and completes the proof.
\end{proof}

\section*{Acknowledgements}
We thank Tom Hutchcroft for a preliminary version of
\cite{Hutc25-PLMS} and for discussions about boundary conditions for the hierarchical lattice.
We are grateful to Roland Bauerschmidt for discussions, to
Yucheng Liu for comments concerning Appendix~\ref{sec:SAW}, and to Tim Garoni
for comments on a preliminary draft.  We
are grateful to the late
Ralph Kenna for
comments on a preliminary draft and for
advice concerning the physics literature.
The work of EM and GS was supported in part by NSERC of Canada.
JP was partially supported by the European Research Council under the
European Union's Horizon 2020 research and innovation programme
(grant agreement No.~851682 SPINRG),
and gratefully acknowledges the support and hospitality of the
University of British Columbia and of the Pacific Institute for the Mathematical Sciences in Vancouver during part of this work.


\end{document}